%% file: art.tex
\documentclass[onecolumn]{aa}

\usepackage[tbtags,fleqn]{amsmath}
\usepackage{amsfonts}
\usepackage{graphics}
\usepackage{natbib}
\newcommand{\e}{\mathrm e}
\newcommand{\diff}{\mathrm d}

\newcommand{\Lp}{\mathcal L}
\newcommand{\R}{\mathbb R}
\newcommand{\mincir}{\raise
  -2.truept\hbox{\rlap{\hbox{$\sim$}}\raise5.truept \hbox{$<$}\ }}
\newcommand{\magcir}{\raise
  -2.truept\hbox{\rlap{\hbox{$\sim$}}\raise5.truept \hbox{$>$}\ }}

\DeclareMathOperator{\Cov}{Cov}

\begin{document}

\title{Smooth maps from clumpy data: Covariance analysis}
\author{Marco Lombardi and Peter Schneider}
\offprints{M. Lombardi}
\mail{lombardi@astro.uni-bonn.de}
\institute{%
  Instit\"ut f\"ur Astrophysik und Extraterrestrische Forschung,
  Universit\"at Bonn, Auf dem H\"ugel 71, D-53121 Bonn, Germany}
\date{Received ***date***; Accepted ***date***}
\abstract{%
  Interpolation techniques play a central role in Astronomy, where one
  often needs to smooth irregularly sampled data into a smooth map.
  In a previous article (\citealp{2001A&A...373..359L}, hereafter
  Paper~I), we have considered a widely used smoothing technique and
  we have evaluated the expectation value of the smoothed map under a
  number of natural hypotheses.  Here we proceed further on this
  analysis and consider the variance of the smoothed map, represented
  by a two-point correlation function.  We show that two main sources
  of noise contribute to the total error budget and we show several
  interesting properties of these two noise terms.  The expressions
  obtained are also specialized to the limiting cases of low and high
  densities of measurements.  A number of examples are used to show in
  practice some of the results obtained.  }

\maketitle

%
%_____________________________________________________________________

\section{Introduction}
\label{sec:introduction}

Raw astronomical data are very often discrete, in the sense that
measurements are performed along a finite number of directions on the
sky.  In many cases, the discrete data are believed to be single
measurements of a smooth underlying field.  In such cases, it is
desirable to reconstruct the original field using interpolation
techniques.  A typical example of the general situation just described
is given by weak lensing mass reconstructions in clusters of galaxies.
In this case, thousands of noisy estimates of the tidal field of the
cluster (shear) can be obtained from the observed shapes of background
galaxies whose images are distorted by the gravitational field of the
cluster.  All these measurements can then be combined to produce a
smooth map of the cluster shear, which in turn is subsequently
converted into a projected density map of the cluster mass
distribution.

One of the most widely used interpolation techniques in Astronomy is
based on a weighted average.  More precisely, a positive weight
function, describing the relative weight of a datum at the position
$\vec \theta + \vec\phi$ on the point $\vec\theta$, is introduced.
The weight function is often chosen to be of the form $w \bigl( |
\vec\phi | \bigr)$, i.e.\ depends only on the separation $| \vec\phi
|$ of the two points considered.  Normally, $w$ is also a decreasing
function of $| \vec\phi |$ in order to ensure that the largest
contributions to the interpolated value at $\vec\theta$ comes from
nearby measurements.  Then, the data are averaged using a weighted
mean with the weights given by the function $w$.  More precisely,
calling $\hat f_n$ the $n$-th datum obtained at the position
$\vec\theta_n$, the smooth map is defined as
\begin{equation}
  \label{eq:1}
  \tilde f(\vec\theta) \equiv \dfrac{\sum_{n=1}^N \hat f_n
  w(\vec\theta - \vec\theta_n)}{\sum_{n=1}^N w(\vec\theta -
  \vec\theta_n)} \; , 
\end{equation}
where $N$ is the total number of objects.  In a previous paper
(\citealp{2001A&A...373..359L}, hereafter Paper~I) we have evaluated
the expectation value for this expression under the following
hypothesis:
\begin{itemize}
\item The measured values $\{ \hat f_n \}$ are independent random
  variables with expectation value
  \begin{equation}
    \label{eq:2}
    \bigl\langle \hat f_n \bigr\rangle = f(\vec\theta_n) \; .
  \end{equation}
  In other words, the $\{ \hat f_n \}$ are \textit{unbiased\/}
  measurements of a field $f(\vec\theta)$.
\item The positions $\{ \vec\theta_n \}$ are independent random
  variables with uniform distribution and density $\rho$.  In Paper~I
  we initially considered a fixed number $N$ of positions inside a
  field $\Omega$ of finite area $A$; then, we took the
  \textit{continuous limit\/} letting $N$ go to infinity with $\rho =
  N/A$ constant.  Equivalently, we considered $N$ to be a Poisson
  random variable with average $\rho A$:
  \begin{equation}
    \label{eq:3}
    p_N(N) = \e^{-\rho A} \frac{(\rho A)^N}{N!} \; ,
  \end{equation}
  and each location $\vec\theta_n$ to be uniformly distributed inside
  $A$:
  \begin{equation}
    \label{eq:4}
    p_\theta(\vec\theta_n) = \frac{1}{A} \; .
  \end{equation}
\end{itemize}
In Paper~I we have shown that
\begin{equation}
  \label{eq:5}
  \bigl\langle \tilde f(\vec\theta) \bigr\rangle = \int f(\vec\theta')
  w_\mathrm{eff}(\vec\theta - \vec\theta') \, \diff^2 \theta' \; .
\end{equation}
Thus, $\bigl\langle \tilde f \bigr\rangle$ is the convolution of the
unknown field $f$ with an \textit{effective weight\/} $w_\mathrm{eff}$
which, in general, differs from the weight function $w$.  We also have
shown that $w_\mathrm{eff}$ has a ``similar'' shape as $w$ and
converges to $w$ when the object density $\rho$ is large; however for
finite $\rho$, $w_\mathrm{eff}$ is broader than $w$.

Here we proceed further with the statistical analysis by obtaining an
expression for the two-point correlation function (covariance) of this
estimator.  More precisely, given two points $\vec\theta_A$ and
$\vec\theta_B$, we consider the two-point correlation function of
$\tilde f$, defined as
\begin{equation}
  \label{eq:6}
  \Cov(\tilde f; \vec\theta_A, \vec\theta_B) \equiv \bigl\langle
  \tilde f(\vec\theta_A) \tilde f(\vec\theta_B) \bigr\rangle -
  \bigl\langle \tilde f(\vec\theta_A) \bigr\rangle \bigl\langle
  \tilde f(\vec\theta_B) \bigr\rangle
\end{equation}
In our calculations, similarly to Paper~I, we assume that $\hat f_n$
are \textit{unbiased and mutually independent\/} estimates of
$f(\vec\theta_n)$ [cf.\ Eq.~\eqref{eq:2}].  We also assume that the
$\{ \hat f_n \}$ have fixed variance $\sigma^2$, so that
\begin{equation}
  \label{eq:7}
  \bigl\langle \bigl[ \hat f_n - f(\vec\theta_n) \bigr]
  \bigl[ \hat f_m - f(\vec\theta_m) \bigr] \bigr\rangle = \sigma^2
  \delta_{nm} \; .
\end{equation}

The paper is organized as follows.  In Sect.~\ref{sec:summary} we
summarize the results obtained in this paper.  In
Sect.~\ref{sec:eval-covar} we derive the general expression for the
covariance of the interpolating techniques and we show that two main
noise terms contribute to the total error.  These results are then
generalized in Sect.~\ref{sec:vanishing-weights} to include the case
of weight functions that are not strictly positive.  A useful
expansion at high densities $\rho$ of the covariance is obtained in
Sect.~\ref{sec:moments-expansion}.  Section~\ref{sec:properties} is
devoted to the study of several interesting properties of the
expressions obtained in the paper.  Finally, in
Sect.~\ref{sec:examples} we consider three simple weight functions and
derive (analytically or numerically) the covariance for these cases.
Four appendixes on more technical topics complete the paper.

\section{Summary}
\label{sec:summary}

As mentioned in the introduction, the primary aim of this paper is the
evaluation of the covariance (two-point correlation function) of the
smoothing estimator \eqref{eq:1} under the hypotheses that
measurements $\hat f_n$ are unbiased estimates of a field
$f(\vec\theta)$ [Eq.~\eqref{eq:2}] and that location measurements $\{
\vec\theta_n \}$ are independent, uniformly distributed variables with
density $\rho$.  Hence, we do not allow for angular clustering on the
positions $\{ \vec\theta_n \}$, and we also do not include the effects
of a finite field in our calculations (these effects are expected to
play a role on points close to the boundary of the region where data
are available).  Moreover, we suppose that the noise on the
measurements $\{ \hat f_n \}$ is uncorrelated with the signal (i.e.,
that variance $\sigma^2$ is constant on the field), and that
measurements are uncorrelated to each other.  Finally, we stress that
in the whole paper we assume a non-negative (i.e., positive or
vanishing) weight function $w(\vec\theta) \ge 0$.  Surprisingly,
weight functions with arbitrary sign cannot be studied in our
framework (see discussion at the end of
Sect.~\ref{sec:vanishing-weights}).

The results obtained in this paper can be summarized in the following
points.
\begin{enumerate}
\item We evaluate analytically the two-point correlation function of
  $\tilde f(\vec\theta)$, showing that it is composed of two main
  terms:
  \begin{equation}
    \label{eq:8}
    \Cov(\tilde f; \vec\theta_A, \vec\theta_B) \equiv \bigl\langle
    \tilde f(\vec\theta_A) \tilde f(\vec\theta_B) \bigr\rangle -
    \bigl\langle \tilde f(\vec\theta_A) \bigr\rangle \bigl\langle
    \tilde f(\vec\theta_B) \bigr\rangle = T_\sigma + T_\mathrm{P} \; .
  \end{equation}
  The term $T_\sigma$ is proportional to $\sigma^2$ and can thus be
  interpreted as the contribution to the covariance from measurement
  errors; the term $T_\mathrm{P}$ depends on the signal
  $f(\vec\theta)$ and can be interpreted as Poisson noise.  These
  terms can be evaluated using the following set of equations:
  \begin{align}
    \label{eq:9}
    Q(s_A, s_B) = {} & \int_\Omega \bigl[ \e^{-s_A w_A(\vec\theta) -
      s_B w_B(\vec\theta)} - 1 \bigr] \, \diff^2
    \theta \; , \\
    \label{eq:10}
    Y(s_A, s_B) = {} & \exp \bigl[ \rho Q(s_A, s_B) \bigr] \; . \\
    \label{eq:11}
    C(w_A, w_B) = {} & \rho^2 \int_0^\infty \diff s_A \int_0^\infty
    \diff s_B \, \e^{-s_A w_A - s_B w_B} Y(s_A, s_B) = \rho^2
    \Lp[Y](w_A, w_B) \; , \\
    \label{eq:12}
    T_\sigma = {} & \frac{\sigma^2}{\rho} \int_\Omega \diff^2 \theta 
    w_A(\vec\theta) w_B(\vec\theta) C \bigl( w_A(\vec\theta),
    w_B(\vec\theta) \bigr) \; , \\
    \label{eq:13}
    T_\mathrm{P} = {} & \frac{1}{\rho} \int_\Omega \diff^2 \theta
    \bigl[ f(\vec\theta) \bigr]^2 w_A(\vec\theta) w_B(\vec\theta) C
    \bigl( w_A(\vec\theta), w_B(\vec\theta) \bigr) \notag\\
    & {} + \int_\Omega \diff^2 \theta_1 \int_\Omega \diff^2 \theta_2
    \, f(\vec\theta_1) f(\vec\theta_2) w_A(\vec\theta_1)
    w_B(\vec\theta_2) \Bigl[ C\bigl( w_A(\vec\theta_1) +
    w_A(\vec\theta_2), w_B(\vec\theta_1) +
    w_B(\vec\theta_2) \bigr) \notag\\
    & \phantom {{} + \int_\Omega \diff^2 \theta_1 \int_\Omega \diff^2
      \theta_2 \, f(\vec\theta_1) f(\vec\theta_2) w_A(\vec\theta_1)
      w_B(\vec\theta_2) \Bigl[ } {} - C_A\bigl( w_A(\vec\theta_1)
    \bigr) C_B\bigl( w_B(\vec\theta_2) \bigr) \Bigr] \; .
  \end{align}
  In the last two equations we used the notation $w_A(\vec\theta) =
  w(\vec\theta_A - \vec\theta)$, and similarly for $w_B(\vec\theta)$;
  moreover the two functions $C_A$ and $C_B$ can be obtained from the
  following limits:
  \begin{align}
    \label{eq:14}
    C_A(w_A) & {} = \frac{1}{\rho} \lim_{w_B \rightarrow \infty}
    w_B C \bigl( w_A, w_B \bigr) \; , &
    C_B(w_B) & {} = \frac{1}{\rho} \lim_{w_A \rightarrow \infty} 
    w_A C \bigl( w_A, w_B  \bigr) \; .    
  \end{align}
\item We show that the quantity $C(w_A, w_B)$ of Eq.~\eqref{eq:11}, in
  the limit of high density $\rho$, converges to 
  \begin{equation}
    \label{eq:15}
    C(w_A, w_B) \simeq \frac{\rho^2}{(\rho + w_A) (\rho + w_B)} +
    \frac{\rho^3 S_{20}}{(\rho + w_A)^3 (\rho + w_B)} +
    \frac{\rho^3 S_{11}}{(\rho + w_A)^2 (\rho + w_B)^2} +
    \frac{\rho^3 S_{02}}{(\rho + w_A) (\rho + w_B)^3} \; .
  \end{equation}
  where $S_{ij}$ are the moments of the functions $(w_A, w_B)$:
  \begin{equation}
    \label{eq:16}
    S_{ij} \equiv \int_\Omega \diff^2 \theta \bigl[ w_A(\vec\theta)
    \bigr]^i \bigl[ w_B(\vec\theta) \bigr]^j \; .
  \end{equation}
\item We derive a number of properties for the noise terms and the
  function $C(w_A, w_B)$.  In particular, we show (1) that
  $w_A(\vec\theta) w_B(\vec\theta) C\bigl( w_A(\vec\theta),
  w_B(\vec\theta) \bigr) \le \rho^2$ in every point $\vec\theta$; (2)
  that the measurement error has as upper bound $T_\sigma \le
  \sigma^2$; (3) that the same error has as lower bound the
  convolution $\sigma^{-1} \int w_{\mathrm{eff} A}(\vec\theta)
  w_{\mathrm{eff} B}(\vec\theta) \, \diff^2 \theta$ of the two
  \textit{effective weights\/} $w_{\mathrm{eff}A}(\vec\theta) =
  w_A(\vec\theta) C\bigl( w_A(\vec\theta), \infty \bigr)$ and
  $w_{\mathrm{eff}B}(\vec\theta) = w_B(\vec\theta) C\bigl( \infty,
  w_B(\vec\theta) \bigr)$ \citep[cf.][]{2001A&A...373..359L}; (4) that
  the measurement noise converges to $T_\sigma \simeq \sigma^2$ at low
  densities ($\rho \rightarrow 0$) and to $T_\sigma \simeq \sigma^2
  S_{11} / \rho$ at high densities ($\rho \rightarrow \infty$).
\end{enumerate}

\section{Evaluation of the covariance}
\label{sec:eval-covar}

\subsection{Preliminaries}
\label{sec:preliminaries}

Before starting the analysis, let us introduce a simpler notation.  In
the following we will often drop the arguments $\vec\theta_A$ and
$\vec\theta_B$ in $\Cov(\tilde f; \vec\theta_A, \vec\theta_B)$ and
other related quantities.  Note, in fact, that the problem is
completely defined with the introduction of the two ``shifted'' weight
functions $w_A(\vec\theta) \equiv w(\vec\theta_A - \vec\theta)$ and
$w_B(\vec\theta) \equiv w(\vec\theta_B - \vec\theta)$.  We also call
$\tilde f_A \equiv \tilde f(\vec\theta_A)$ and $\tilde f_B \equiv
\tilde f(\vec\theta_B)$ the values of $\tilde f$ at the two points of
interest $\vec\theta_A$ and $\vec\theta_B$, so that
\begin{equation}
  \label{eq:17}
  \tilde f_X = \dfrac{\sum_{n=1}^N \hat f_n
  w_X(\vec\theta_n)}{\sum_{n=1}^N w_X(\vec\theta_n)} \; .
\end{equation}
Hence, Eq.~\eqref{eq:6} can be rewritten in this notation as
\begin{equation}
  \label{eq:18}
  \Cov(\tilde f) = \bigl\langle \tilde f_A \tilde f_B \bigr\rangle -
  \bigl\langle \tilde f_A \bigr\rangle \bigl\langle \tilde f_B
  \bigr\rangle \; .
\end{equation}
Note that, using this notation, we are not taking advantage of the
invariance upon translation of $w(\vec\theta)$ in Eq.~\eqref{eq:1}; in
other words, we are not using the fact that $w_A$ and $w_B$ are
basically the same function shifted by $\vec\theta_A - \vec\theta_B$.
Actually, all calculations can be carried out without using this
property; however, we will explicitly point out simplifications that
can be made using the invariance upon translation.

We would also like to spend a few words about averages.  Note that, as
anticipated in Sect.~\ref{sec:introduction}, we need to carry out two
averages, one with respect to $\{ \hat f_n \}$ [Eqs.~\eqref{eq:2} and
\eqref{eq:7}], and one with respect to $\{ \vec\theta_n \}$
[Eqs.~\eqref{eq:3} and \eqref{eq:4}].  Taking $\{ \vec\theta_n \}$ to
be random variables is often reasonable because in Astronomy one does
not have a direct control over the positions where observations are
made (this happens because measurements are normally performed in the
direction of astronomical objects such as stars and galaxies, and thus
at ``almost random'' directions); it also has the advantage of letting
us obtain general results, independent of any particular configuration
of positions.  Note, however, that taking $\{ \vec\theta_n \}$ to be
\textit{independent\/} variables is a strong simplification which
might produce inaccurate results in some context (e.g., in case of a
direction dependent density, or in case of clustering; see
\citealp{LPM}).  Finally, since the number of observations $N$ is
itself a random variable, we need to perform first the average on $\{
\hat f_n \}$ and then the one on $\{ \vec\theta_m \}$.

In closing this section, we observe that in this paper, similarly to
Paper~I, we will almost always consider the smoothing problem on the
plane, i.e.\ we will assume that the positions $\{ \vec\theta_n \}$
are vectors of $\R^2$.  We proceed this way because in Astronomy the
smoothing process often takes places on small regions of the celestial
sphere, and thus on sets that can be well approximated with subsets of
the plane.  However, we stress that all the results stated here can be
easily applied to smoothing processes that takes places on different
sets, such as the real axis $\R$ or the space $\R^3$.

\subsection{Analytical solution}
\label{sec:analytical-solution}

Let us now focus on the first term on the r.h.s.\ of
Eq.~\eqref{eq:18}.  We have
\begin{equation}
  \label{eq:19}
  \bigl\langle \tilde f_A \tilde f_B \bigr\rangle = \frac{1}{A^N}
  \int_\Omega \diff^2 \theta_1 \int_\Omega \diff^2 \theta_2
  \dotsi \int_\Omega \diff^2 \theta_N \dfrac{ \bigl\langle \bigl[
    \sum_n \hat f_n w_A(\vec\theta_n) \bigr] \bigl[ \sum_m \hat f_m
    w_B(\vec\theta_m) \bigr] \bigr\rangle}{\bigl[ \sum_n
    w_A(\vec\theta_n) \bigr] \bigl[ \sum_m w_B(\vec\theta_m) \bigr]}
  \; .
\end{equation}
Note that the average in the r.h.s.\ of this equation is only with
respect to $\{ \hat f_n \}$.  Expanding the numerator in the integrand
of this equation, we obtain $N^2$ terms, $N$ of which have $n = m$ and
$N (N - 1)$ have $n \neq m$.  We can then rewrite Eq.~\eqref{eq:19}
above as
\begin{equation}
  \label{eq:20}
  \bigl\langle \tilde f_A \tilde f_B \bigr\rangle = T_1 + T_2 \; ,
\end{equation}
where
\begin{align}
  \label{eq:21}
  T_1 \equiv {} & \frac{1}{A^N} \int_\Omega \diff^2 \theta_1
  \int_\Omega \diff^2 \theta_2 \dotsi \int_\Omega \diff^2 \theta_N
  \dfrac{\sum_n \bigl\langle \hat f_n^2 \bigr\rangle w_A(\vec\theta_n)
    w_B(\vec\theta_n)}{\bigl[ \sum_n w_A(\vec\theta_n) \bigr] \bigl[
    \sum_m w_B(\vec\theta_m) \bigr]} \; , \\
  \label{eq:22}
  T_2 \equiv {} & \frac{1}{A^N} \int_\Omega \diff^2 \theta_1
  \int_\Omega \diff^2 \theta_2 \dotsi \int_\Omega \diff^2 \theta_N
  \dfrac{\sum_{n \neq m} \bigl\langle \hat f_n \hat f_m \bigr\rangle
    w_A(\vec\theta_n) w_B(\vec\theta_m)}{\bigl[ \sum_n
    w_A(\vec\theta_n) \bigr] \bigl[ \sum_m w_B(\vec\theta_m) \bigr]}
  \; .
\end{align}
Despite the apparent differences, these two terms can be simplified in
a similar manner.  Let us consider first $T_1$.  Using
Eq.~\eqref{eq:7}, we can evaluate the average $\langle \hat f^2_n
\rangle = \sigma^2 + \bigl[ f(\vec\theta_n) \bigr]^2$.  Since the
positions $\{ \vec\theta_n \}$ appear as ``dummy variables'' in
Eq.~\eqref{eq:21}, we can relabel them as follows
\begin{equation}
  \label{eq:23}
  T_1 = \frac{N}{A^N} \int_\Omega \diff^2 \theta_1 \int_\Omega
  \diff^2 \theta_2 \dotsi \int_\Omega \diff^2 \theta_N \dfrac{\bigl[
    f^2(\vec\theta_1) + \sigma^2 \bigr] w_A(\vec\theta_1)
    w_B(\vec\theta_1)}{\bigl[ \sum_n w_A(\vec\theta_n) \bigr] \bigl[
    \sum_m w_B(\vec\theta_m) \bigr]} \; .
\end{equation}
In order to simplify this equation, we use a technique similar to the
one adopted in Paper~I.  More precisely, we split the two sums in the
denominator of the integrand of Eq.~\eqref{eq:23}, taking away the
terms $w_A(\vec\theta_1)$ and $w_B(\vec\theta_1)$.  Hence, we write
\begin{equation}
  \label{eq:24}
  T_1 = \frac{1}{\rho} \int_\Omega \diff^2 \theta_1 \bigl[
  f^2(\vec\theta_1) + \sigma^2 \bigr] w_A(\vec\theta_1)
  w_B(\vec\theta_1) C \bigl( w_A(\vec\theta_1), w_B(\vec\theta_1)
  \bigr) \; ,
\end{equation}
where $C(w_A, w_B)$ is a \textit{corrective factor\/} given by
\begin{equation}
  \label{eq:25}
  C(w_A, w_B) \equiv \frac{N^2}{A^{N+1}} \int_\Omega \diff^2
  \theta_2 \dotsi \int_\Omega \diff^2 \theta_N \dfrac{1}{\bigl[ w_A +
  \sum_{n=2}^N w_A(\vec\theta_n) \bigr] \bigl[w_B + \sum_{m=2}^N
  w_B(\vec\theta_n) \bigr]} \; . 
\end{equation}
The additional factor $\rho = N/A$ has been introduced to simplify
some of the following equations.  Note that in the definition of $C$
$w_A$ and $w_B$ are formally taken to be two real variables (instead
of two real functions of argument $\vec\theta_1$).

The definition of $C$ above suggests to define two new random
variables $y_A$ and $y_B$:
\begin{equation}
  \label{eq:26}
  y_X \equiv \sum_{n=2}^N w_X(\vec\theta_n) \; , \qquad \text{with $X
  = \{ A, B \}$} \; .
\end{equation}
Note that the sum runs from $n = 2$ to $n = N$.  If we could evaluate
the \textit{combined\/} probability distribution function $p_y(y_A,
y_B)$ for $y_A$ and $y_B$, we would have solved our problem: In fact
we could use this probability to write $C(w_A, w_B)$ as follows
\begin{equation}
  \label{eq:27}
  C(w_A, w_B) = \rho^2 \int_0^\infty \diff y_A \int_0^\infty \diff y_B
  \frac{p_y(y_A, y_B)}{(w_A + y_A) (w_B + y_B)} \; .
\end{equation}
To obtain the probability distribution $p_y(y_A, y_B)$, we need to use
the combined probability distribution $p_w(w_A, w_B)$ for $w_A$ and
$w_B$.  This distribution is implicitly defined by saying that the
probability that $w_A(\vec\theta)$ be in the range $[ w_A, w_A + \diff
w_A ]$ and $w_B(\vec\theta)$ be in the range $[ w_B, w_B + \diff w_B
]$ is $p_w(w_A, w_B) \, \diff w_A \, \diff w_B$.  We can evaluate
$p_w(w_A, w_B)$ using
\begin{equation}
  \label{eq:28}
  p_w(w_A, w_B) = \frac{1}{A} \int_\Omega \delta\bigl( w_A -
  w_A(\vec\theta) \bigr) \delta\bigl( w_B - w_B(\vec\theta) \bigr) \,
  \diff^2 \theta \; .
\end{equation}
Turning back to $(y_A, y_B)$, we can write a similar expression for
$p_y$:
\begin{equation}
  \label{eq:29}
  p_y(y_A, y_B) = \frac{1}{A^{N-1}} \int_\Omega \diff^2
  \theta_2 \dotsi \int_\Omega \diff^2 \theta_N \delta( y_A - w_{A2}
  - \dots - w_{AN} ) \delta( y_B - w_{B2} - \dots - w_{BN} ) \; ,
\end{equation}
where for simplicity we have called $w_{Xn} = w_X(\vec\theta_n)$.
Note that inserting this equation into Eq.~\eqref{eq:27} we recover
Eq.~\eqref{eq:25}, as expected.  Actually, for our purposes it is more
useful to consider $y_X$ to be the sum of $N$ random variables $\{
w_{Xn} \}$.  In other words, we consider the set of couples $\bigl\{
(w_{An}, w_{Bn}) \bigr\}$, made of the two weight functions at the
various positions, as a set of $N$ \textit{independent\/}
two-dimensional random variables $(w_A, w_B)$ with probability
distribution $p_w(w_A, w_B)$.  [Hence, similarly to Eq.~\eqref{eq:25},
we consider the weight functions $w_X$ to be real variables instead of
real functions; the independence of the positions $\vec\theta_n$ then
implies the independence of the \textit{couples\/} $(w_{An},
w_{Bn})$.]\@ Taking this point of view, we can rewrite
Eq.~\eqref{eq:29} as
\begin{align}
  \label{eq:30}
  p_y(y_A, y_B) = \int_0^\infty & {} \diff w_{A2} \int_0^\infty \diff
  w_{B2} \, p_w( w_{A2}, w_{B2} ) \dotsi \int_0^\infty \diff w_{AN}
  \int_0^\infty \diff w_{BN} \, p_w( w_{AN}, w_{BN} ) \notag\\
  & {} \times \delta( y_A - w_{A2} - \dots - w_{AN} ) \delta( y_B -
  \dots - w_{BN} ) \; .
\end{align}
It is well known in Statistics that the sum of independent random
variables with the same probability distribution can be better studied
using Markov's method (see, e.g., \citealp{1989QB461.C47......}; see
also \citealp{1987PhRvL..59.2814D} for an application to microlensing
studies).  This method is based on the use of Fourier transforms for
the probability distributions $p_w$ and $p_y$.  However, since we are
dealing with non negative quantities (we recall that we assumed
$w(\vec\theta) \ge 0$), we can replace the Fourier transform with
Laplace transform which turns out to be more appropriate in for our
problem (see Appendix~\ref{sec:prop-lapl-transf} for a summary of the
properties of Laplace transforms).  Hence, we define $W(s_A, s_B)$ and
$Y(s_A, s_B)$ to be the Laplace transforms of $p_w(w_A, w_B)$ and
$p_y(w_A, w_B)$, respectively.  Note that, since both functions $p_w$
and $p_y$ have two arguments, we need two arguments for the Laplace
transforms as well:
\begin{align}
  \label{eq:31}
  W(s_A, s_B) & {} \equiv \Lp[p_w](s_A, s_B) = \int_0^\infty \diff w_A
  \int_0^\infty \diff w_B \e^{-s_A w_A - s_B w_B} p_w(w_A, w_B) \; , \\  
  \label{eq:32}
  Y(s_A, s_B) & {} \equiv \Lp[p_y](s_A, s_B) = \int_0^\infty \diff y_A
  \int_0^\infty \diff y_B \e^{-s_A y_A - s_B y_B} p_y(y_A, y_B) \; .
\end{align}
We use now in these expressions the Eq.~\eqref{eq:28} for $p_w$ and
Eq.~\eqref{eq:30} for $p_y$, thus obtaining
\begin{align}
  \label{eq:33}
  W(s_A, s_B) = {} & \frac{1}{A} \int_\Omega \e^{-s_A w_A(\vec\theta)
    - s_B w_B(\vec\theta)} \, \diff^2 \theta \; , \\
  \label{eq:34}
  Y(s_A, s_B) = {} & \frac{1}{A^{N-1}} \int_\Omega \! \diff^2
  \theta_2 \dotsi \int_\Omega \! \diff^2 \theta_N \exp \biggl[ - s_A
  \sum_{n=2}^N w_{An} - s_B \sum_{m=2}^N w_{Bm} \biggr] = \bigl[
  W(s_A, s_B) \bigr]^{N-1} \; .
\end{align}
Hence, $p_y$ can in principle be obtained from the following scheme.
First, we evaluate $W(s_A, s_B)$ using Eq.~\eqref{eq:33}, then we
calculate $Y(s_A, s_B)$ from Eq.~\eqref{eq:34}, and finally we
back-transform this function to obtain $p_y(y_A, y_B)$.

Actually, another, more convenient, technique is viable.  Following
the path of Paper~I, we now take the ``continuous limit'' and treat
$N$ as a random variable.  As explained in
Sect.~\ref{sec:introduction}, we can take this limit using two
equivalent approaches:
\begin{itemize}
\item We keep the area $A$ fixed and consider $N$ to be a random
  variable with Poisson distribution given by Eq.~\eqref{eq:3}.  We
  then average over all possible configurations obtained.
\item We take the limit $N \rightarrow \infty$ taking the density
  $\rho = N/A$ fixed.
\end{itemize}
The equivalence of the two methods can be shown as follows.  Let us
consider a large area $A' \supset A$, and let us suppose that the
number $N' = \rho A'$ of objects inside $A'$ is fixed.  Since objects
are randomly distributed inside $A'$, the probability for each object
to fall inside $A$ is just $A / A'$.  Hence $N$, the number of objects
inside $A$, follows a binomial distribution:
\begin{equation}
  \label{eq:35}
  p_N(N) = \binom{N'}{N} \left( \frac{A}{A'} \right)^N \left( \frac{A'
  - A}{A'} \right)^{N' - N} \; . 
\end{equation}
If we now let $N'$ go to infinity with $N' / A' = \rho$ fixed, the
probability distribution for $N$ converges \citep[see, e.g.][]{Eadie}
to the Poisson distribution in Eq.~\eqref{eq:3}.

We will follow here the second strategy, i.e.\ we will take the limit
$N, A \rightarrow \infty$ keeping $\rho = N/A$ constant.  In the limit
$A \rightarrow \infty$ the quantity $W(s_A, s_B)$ goes to unity and
thus is not useful for our purposes.  Instead, it is convenient to
define
\begin{equation}
  \label{eq:36}
  Q(s_A, s_B) \equiv \int_\Omega \bigl[ \e^{-s_A w_A(\vec\theta)
  - s_B w_B(\vec\theta)} - 1 \bigr] \, \diff^2 \theta = A \bigl[
  W(s_A, s_B) - 1 \bigr] \; .
\end{equation}
This definition is sensible because, this way, $Q$ remains finite for
$A \rightarrow \infty$.  In the continuous limit, Eq.~\eqref{eq:34}
becomes
\begin{equation}
  \label{eq:37}
  Y(s_A, s_B) = \lim_{N \rightarrow \infty} \left[ 1 +  \frac{\rho
      Q(s_A, s_B)}{N} \right]^{N-1} = \e^{Q(s_A, s_B)} \; .
\end{equation}
In order to evaluate $C(w_A, w_B)$, we rewrite its definition
\eqref{eq:27} as
\begin{equation}
  \label{eq:38}
  C(w_A, w_B) = \rho^2 \int_0^\infty \diff x_A \int_0^\infty \diff x_B
  \frac{\zeta_w(x_A, x_B)}{x_A x_B} \; ,
\end{equation}
where $x_X \equiv y_X + w_X$ and
\begin{equation}
  \label{eq:39}
  \zeta_w(x_A, x_B) \equiv \mathrm{H}(x_A - w_A) \mathrm{H}(x_B - w_B)
  p_y(x_A - w_A, x_B - w_B) \; .
\end{equation}
Here $\mathrm{H}(x_X - w_X)$ are Heaviside functions at the positions
$w_X$, i.e.
\begin{equation}
  \label{eq:40}
  \mathrm{H}(x) =
  \begin{cases}
    0 & \text{if $x < 0 \; ,$} \\
    1 & \text{otherwise.}
  \end{cases}
\end{equation}
Note that $\zeta_w$ is basically a ``shifted'' version of $p_y$.
Looking back at Eq.~\eqref{eq:38}, we can interpret the integration
present in this equation as a \textit{very\/} particular case of
Laplace transform with vanishing argument.  In other words, we can
write
\begin{equation}
  \label{eq:41}
  C(w_A, w_B) = \rho^2 \Lp[\zeta_w/x_A x_B](0, 0) \; .
\end{equation}
Thus our problem is solved if we can obtain the Laplace transform of
$\zeta_w / x_A x_B$ evaluated at $s_A = s_B = 0$.  From the properties
of Laplace transform [cf.\ Eq.~\eqref{eq:160}] we find
\begin{equation}
  \label{eq:42}
  \Lp[\zeta_w(x_A, x_B) / x_A x_B](s_A, s_B) = \int_{s_A}^\infty \diff
  s'_A \int_{s_B}^\infty \diff s'_B \, Z_w(s'_A, s'_B) \; ,
\end{equation}
where $Z_w$ is the Laplace transform of $\zeta_w$:
\begin{equation}
  \label{eq:43}
  Z_w(s_A, s_B) \equiv \Lp[\zeta_w](s_A, s_B) = \e^{-s_A w_A - s_B w_B}
  Y(s_A, s_B) \; .
\end{equation}
Combining together Eqs.~\eqref{eq:41}, \eqref{eq:42}, and
\eqref{eq:43} we finally obtain
\begin{equation}
  \label{eq:44}
  C(w_A, w_B) = \rho^2 \int_0^\infty \diff s_A \int_0^\infty
  \diff s_B \, \e^{-s_A w_A - s_B w_B} Y(s_A, s_B) = \rho^2
  \Lp[Y](w_A, w_B) \; .
\end{equation}
In summary, the set of equations that can be used to evaluate $T_1$
are
\begin{align}
  \label{eq:45}
  Q(s_A, s_B) & {} = \int_\Omega \bigl[ \e^{-s_A w_A(\vec\theta)
    - s_B w_B(\vec\theta)} - 1 \bigr] \, \diff^2 \theta \; , \\
  \label{eq:46}
  Y(s_A, s_B) & {} = \exp \bigl[ \rho Q(s_A, s_B) \bigr] \; , \\
  \label{eq:47}
  C(w_A, w_B) & {} = \rho^2 \int_0^\infty \diff s_A \int_0^\infty
  \diff s_B \, \e^{-s_A w_A - s_B w_B} Y(s_A, s_B) = \rho^2
  \Lp[Y](w_A, w_B) \; , \\
  \label{eq:48}
  T_1 & {} = \frac{1}{\rho} \int_\Omega \diff^2 \theta \bigl[
  f^2(\vec\theta) + \sigma^2 \bigr] w_A(\vec\theta) w_B(\vec\theta) C
  \bigl( w_A(\vec\theta), w_B(\vec\theta) \bigr) \; .
\end{align}
These equations solve completely the first part of our problem,
the determination of $T_1$.

Let us now consider the second term of Eq.~\eqref{eq:20}, namely $T_2$
[see Eq.~\eqref{eq:22}].  We first evaluate the average in $\{ \hat
f_n \}$ that appears in the numerator of the integrand of
Eq.~\eqref{eq:22}, obtaining $\langle \hat f_n \hat f_m \rangle =
f(\vec\theta_n) f(\vec\theta_m)$ [cf.\ Eq.~\eqref{eq:7} with $n \neq
m$].  Then we relabel the ``dummy'' variables $\{ \vec\theta_n \}$
similarly to what has been done for $T_1$, thus obtaining
\begin{equation}
  \label{eq:49}
  T_2 = \frac{N (N - 1)}{A^N} \int_\Omega \diff^2 \theta_1 \,
  \int_\Omega \diff^2 \theta_2 \dotsi \int_\Omega \diff^2 \theta_N
  \dfrac{f(\vec\theta_1) w_A(\vec\theta_1) f(\vec\theta_2)
  w_B(\vec\theta_2)}{\bigl[ \sum_n w_A(\vec\theta_n) \bigr] \bigl[
  \sum_m w_B(\vec\theta_m) \bigr]} \; .
\end{equation}
We now split, in the two sums in the denominator, the terms
$w_A(\vec\theta_1) + w_A(\vec\theta_2)$ and $w_B(\vec\theta_1) +
w_B(\vec\theta_2)$ and define the new random variables
\begin{equation}
  \label{eq:50}
  z_X \equiv \sum_{n=3}^N w_X(\vec\theta_n) \; , \qquad \text{with $X
  = \{ A, B \}$}\; .
\end{equation}
Again, if we know the \textit{combined\/} probability distribution
$p_z(z_A, z_B)$ of $z_A$ and $z_B$ our problem is solved, since we can
write [cf.\ Eqs.~\eqref{eq:24} and \eqref{eq:27}]
\begin{align}
  \label{eq:51}
  T_2 = {} & \frac{N (N - 1)}{A^2} \int_\Omega \diff^2 \theta_1
  \int_\Omega \diff^2 \theta_2 \, f(\vec\theta_1) f(\vec\theta_2)
  w_A(\vec\theta_1) w_B(\vec\theta_2) \int_0^\infty \diff z_A
  \int_0^\infty \diff z_B \, p_z(z_A, z_B) \frac{1}{w_A(\vec\theta_1)
  + w_A(\vec\theta_2) + z_A} \notag\\
  & \phantom{{} \times \int_0^\infty \diff z_A \int_0^\infty \diff z_B
  \, } \times \frac{1}{w_B(\vec\theta_1) + w_B(\vec\theta_2) + z_B} \;
  .
\end{align}
Actually, in the continuous limit, $z_X$ is indistinguishable from
$y_X$ ($z_X$ differs from $y_X$ only on the fact that it is the sum of
$N-2$ ``weights'' instead of $N-1$; however, $N$ goes to infinity in
the continuous limit and thus $y_X$ and $z_X$ converge to the same
quantity).  Thus we can rewrite Eq.~\eqref{eq:51} as
\begin{equation}
  \label{eq:52}
  T_2 = \int_\Omega \diff^2 \theta_1 \int_\Omega \diff^2 \theta_2
  \, f(\vec\theta_1) f(\vec\theta_2) w_A(\vec\theta_1) w_B(\vec\theta_2)
  C\bigl( w_A(\vec\theta_1) + w_A(\vec\theta_2),
  w_B(\vec\theta_1) + w_B(\vec\theta_2) \bigr) \; ,
\end{equation}
where $C$ is still given by Eq.~\eqref{eq:47}.

Finally, in order to evaluate $\Cov(\tilde f)$, we still need the
simple averages $\bigl\langle \tilde f_A \bigr\rangle$ and
$\bigl\langle \tilde f_B \bigr\rangle$.  These can be obtained
directly using the technique described in Paper~I, where we have shown
that the set of equations to be used is
\begin{align}
  \label{eq:53}
  Q_X(s_X) & {} \equiv \int_\Omega \bigl[ \e^{-s_X
  w_X(\vec\theta)} - 1 \bigr] \; \, \diff^2 \theta \; , \\
  \label{eq:54}
  Y_X(s_X) & {} \equiv \exp \bigl[ \rho Q_X(s_X) \bigr] \; , \\
  \label{eq:55}
  C_X(w_X) & {} \equiv \rho \int_0^\infty \diff s_X \e^{-s_X w_X}
  Y_X(s_X) \; , \\
  \label{eq:56}
  \bigl\langle \tilde f_X \bigr\rangle & {} = \int_\Omega \diff^2
  \theta \, f(\vec\theta) w_X(\vec\theta) C_X \bigl( w_X(\vec\theta)
  \bigr) \; .
\end{align}
We recall that in Paper~I we called the combination
$w_{\mathrm{eff}X}(\vec\theta) = w_X(\vec\theta) C_X \bigl(
w_X(\vec\theta) \bigr)$ \textit{effective weight\/} [cf.
Eq.~\eqref{eq:5} in the introduction].  Alternatively, we can use the
quantities $Y(s_A, s_B)$ and $C(w_A, w_B)$ to calculate the correcting
factors $C_A$ and $C_B$.  From Eqs.~\eqref{eq:53} and \eqref{eq:54} we
immediately find
\begin{align}
  \label{eq:57}
  Q_A(s_A) = {} & Q(s_A, 0) \; , & 
  Y_A(s_A) = {} & Y(s_A, 0) \; , \\
  \label{eq:58}
  Q_B(s_B) = {} & Q(0, s_B) \; , & 
  Y_B(s_B) = {} & Y(0, s_B) \; .
\end{align}
Then, using the properties of Laplace transforms [cf.\ 
Eq.~\eqref{eq:163}], and comparing the definition of $C(w_A, w_B)$
[Eq.~\eqref{eq:44}] with the one of $C_X(w_X)$ [Eq.~\eqref{eq:55}] we
find
\begin{align}
  \label{eq:59}
  C_A(w_A) & {} = \frac{1}{\rho} \lim_{w_B \rightarrow \infty}
  w_B C \bigl( w_A, w_B \bigr) \; , &
  C_B(w_B) & {} = \frac{1}{\rho} \lim_{w_A \rightarrow \infty} 
  w_A C \bigl( w_A, w_B  \bigr) \; .
\end{align}
We now have at our disposal the complete set of equations that can be
used to determine the covariance of $\tilde f$.

In closing this subsection we makes a few comments on the translation
invariance for $w_X$ (see Sect.~\ref{sec:preliminaries}).  Since
$w_A(\vec\theta)$ and $w_B(\vec\theta)$ differ by an angular shift
only, the two functions $Q_A$ and $Q_B$ are the same, so that $C_A$
coincides with $C_B$.  Not surprisingly, the two effective weights
$w_{\mathrm{eff}A}$ and $w_{\mathrm{eff}B}$ differ also only by a
shift.

\subsection{Noise contributions}
\label{sec:noise-contributions}

A simple preliminary analysis of the Eqs.~\eqref{eq:48} and
\eqref{eq:52} allows us to recognize two main sources of noise.  In
fact, a term in Eq.~\eqref{eq:48} is proportional to $\sigma^2$, and
is clearly related to measurement errors of $f$, namely
\begin{equation}
  \label{eq:60}
  T_\sigma \equiv \frac{\sigma^2}{\rho} \int_\Omega \diff^2 \theta
  w_A(\vec\theta) w_B(\vec\theta) C \bigl( w_A(\vec\theta),
  w_B(\vec\theta) \bigr) \; .
\end{equation}
Other factors entering $\Cov(\tilde f)$ can be interpreted as Poisson
noise.  Hence, we call $T_\mathrm{P1} \equiv T_1 - T_\sigma$,
$T_\mathrm{P2} \equiv T_2$, and $T_\mathrm{P3} \equiv \bigl\langle
\tilde f_A \bigr\rangle \bigl\langle \tilde f_B \bigr\rangle$, so that
the total Poisson noise is $T_\mathrm{P} \equiv T_\mathrm{P1} +
T_\mathrm{P2} - T_\mathrm{P3}$.  Note that the Poisson noise
$T_\mathrm{P}$, in contrast with the measurement noise $T_\sigma$,
strongly depends on the signal $f(\vec\theta)$.

The noise term $T_\sigma$ is quite intuitive and does not require a
long explanation.  We note here only that this term is independent of
the field $f(\vec\theta)$ because we assumed measurements $\hat f_n$
with fixed variance $\sigma^2$ [see Eq.~\eqref{eq:7}].

The Poisson noise $T_\mathrm{P}$ can be better understood with a
simple example.  Suppose that $f(\vec\theta)$ is \textit{not\/}
constant and let us focus on a point where this function has a strong
gradient.  Then, when measuring $\tilde f$ in this point, we could
obtain an excess of signal because of an overdensity of objects in the
region where $f(\vec\theta)$ is large; the opposite happens if we have
an overdensity of objects in the region where $f(\vec\theta)$ is
small.  This noise source, called Poisson noise, vanishes if the
function $f(\vec\theta)$ is flat.  

In the rest of this paper we will study the properties of the
two-point correlation function.  Before proceeding, however, we need
to consider an important generalization of the results obtained here
to the case of vanishing weights.

\section{Vanishing weights}
\label{sec:vanishing-weights}

So far we have implicitly assumed that both $w_A$ and $w_B$ are always
positive.  In some cases, however, it might be interesting to consider
vanishing weight functions (for example, functions with finite
support).  We need then to modify accordingly our equations.

When using vanishing weights, we might encounter situations where the
denominator of Eq.~\eqref{eq:1} vanishes because all weight functions
$w(\vec\theta - \vec\theta_n)$ vanish as well.  In this case, the
estimator $\tilde f(\vec\theta)$ cannot be even defined (we encounter
the ratio $0 / 0$), and any further statistical analysis is
meaningless.  In practice, when smoothing data using a vanishing
weight function, one could just ignore the points $\vec\theta$ where
the smoothed function $\tilde f(\vec\theta)$ is not defined, i.e.\ the
points $\vec\theta$ for which $w(\vec\theta - \vec\theta_n) = 0$ for
every $n$.  This simple approach leads to smoothed maps with
``holes'', i.e.\ defined only on subsets of the plane.  Hence, if we
choose this approach we need to modify accordingly the statistical
analysis that we carry out in this paper.

This problem was already encountered in Paper~I, where we used the
following prescription.  When using a finite-field weight function, we
discard, for every configuration of measurement points $\{
\vec\theta_n \}$, the points $\vec\theta$ on the plane for which the
smoothing $\tilde f(\vec\theta)$ is not defined.  Then, when taking
the average with respect to all possible configurations $\{
\vec\theta_n \}$ of $\tilde f(\vec\theta)$, we just exclude these
configurations.  We stress that, this way, the averages $\bigl\langle
\tilde f(\vec\theta) \bigr\rangle$ and $\bigl\langle \tilde
f(\vec\theta') \bigr\rangle$ of the smoothing \eqref{eq:1} at two
different points $\vec\theta$ and $\vec\theta'$ are effectively
carried out using different ensembles: In one case we exclude the
``bad configurations'' for $\vec\theta$, in the other case the ``bad
configurations'' for $\vec\theta'$.

The same prescription is also adopted here to evaluate the covariance
of our estimator.  Hence, when performing the ensemble average to
estimate the covariance $\Cov(\tilde f; \vec\theta_A, \vec\theta_B)$,
we explicitly exclude configurations where either $\tilde f_A$ or
$\tilde f_B$ cannot be evaluated.  This is implemented with a slight
change in the definition of $p_y$, which in turn implies a change in
Eq.~\eqref{eq:46} for $Y$.  A rigorous generalization of the relevant
equations can now be carried out without significant difficulties.
However, the equations obtained are quite cumbersome and present some
technical peculiarities.  Hence, we prefer to postpone a complete
discussion of vanishing weights until
Appendix~\ref{sec:vanishing-weights-1}; we report here only the main
results.

As mentioned above, the basic problem of having vanishing weights is
that in some cases the estimator is not defined.  Hence, it is
convenient to define three probabilities, namely $P_A$ and $P_B$, the
probabilities, respectively, that $\tilde f_A$ and $\tilde f_B$ are
not defined, and $P_{AB}$, the probability that both quantities are
not defined.  Note that, because of the invariance upon translation
for $w$, we have $P_A = P_B$.  These probabilities can be estimated
without difficulties.  In fact, the quantity $\tilde f_X$ is not
defined if and only if there is no object inside the support of $w_X$.
Since the number of points inside the support of $w_X$ follows a
Poisson probability, we have $P_X = \exp(-\rho \pi_X)$, where $\pi_X$
is the area of the support of $w_X$.  Similarly, calling $\pi_{A \cup
  B}$ the area of the union of the supports of $w_A$ and $w_B$, we
find $P_{AB} = \exp(-\rho \pi_{A \cup B})$.  Using Eqs.~\eqref{eq:45}
and \eqref{eq:46} we can also verify the following relations
\begin{align}
  \label{eq:61}
  P_{AB} = {} & \lim_{\substack{s_A \rightarrow \infty \\
      s_B \rightarrow \infty}} Y(s_A, s_B) \; , &
  1 = {} & \lim_{\substack{s_B \rightarrow 0^+ \\
      s_B \rightarrow 0^+}} Y(s_A, s_B) \; . \\
  \label{eq:62}
  P_A = {} & \lim_{\substack{s_A \rightarrow 0^+ \\
      s_B \rightarrow \infty}} Y(s_A, s_B) \; , &  
  P_B = {} & \lim_{\substack{s_B \rightarrow \infty \\
      s_B \rightarrow 0^+}} Y(s_A, s_B) \; ,
\end{align}
Appendix~\ref{sec:vanishing-weights-1} better clarifies the
relationship between the limiting values of $Y$ and the probabilities
defined above.  In the following we will use a simplified notation for
limits, and we will write something like $P_A = Y(0^+, \infty)$ for
the left equation in \eqref{eq:62}.

The only significant modification to the equations obtained above for
vanishing weights is an overall factor in Eq.~\eqref{eq:47}, which now
becomes
\begin{equation}
  \label{eq:63}
  C(w_A, w_B) = \frac{\rho^2}{1 - P_A - P_B + P_{AB}} \Lp[Y](w_A,
  w_B) \; .
\end{equation}
The factor $1/(1 - P_A - P_B + P_{AB})$ is basically a
renormalization; more precisely, it is introduced to take into account
the fact that we are discarding cases where either $\tilde f_A$ or
$\tilde f_B$ are not defined.  Note, in fact, that in agreement with
the inclusion-exclusion principle, $(1 - P_A - P_B + P_{AB})$ is the
probability that the both $\tilde f_A$ and $\tilde f_B$ are defined.
Since the combination $(1 - P_A - P_B + P_{AB})$ enters several
equations, we define
\begin{equation}
  \label{eq:64}
  \nu \equiv \frac{1}{1 - P_A - P_B + P_{AB}} \; .
\end{equation}

Equation~\eqref{eq:63} is the most important correction to take into
account for vanishing weights.  Actually, there are also a number of
peculiarities to consider when dealing with the probability $p_y$ and
its Laplace transform $Y$.  Fortunately, however, these peculiarities
have no significant consequence for our purpose and thus we can still
safely use Eqs.~\eqref{eq:45} and \eqref{eq:46}.  Again, we refer to
Appendix~\ref{sec:vanishing-weights-1} for a complete explanation.

In closing this section, we spend a few words on weight functions with
arbitrary sign [i.e., functions $w(\vec\theta)$ that can be positive,
vanishing, or positive depending on $\vec\theta$].  As mentioned in
Sect.~\ref{sec:summary}, in this case a statistical study of the
smoothing \eqref{eq:1} cannot be carried out using our framework.  In
order to understand why this happens, let us consider the weight
function
\begin{equation}
  \label{eq:65}
  w(\vec\theta) = \bigl( 1 - \lvert \vec\theta \rvert^2 \bigr) \exp
  \bigl( - \lvert \vec\theta \rvert^2 \bigr) \; .
\end{equation}
This function is continuous, positive for $\lvert \vec\theta \rvert <
1$, and quickly vanishes for large $\lvert \vec\theta \rvert$.  Let us
then consider separately the numerator and denominator of
Eq.~\eqref{eq:1}.  The denominator can clearly be positive or
negative; more precisely, the denominator is positive for points
$\vec\theta$ close to at least one of the locations $\vec\theta_n$,
and negative for points $\vec\theta$ which are in ``voids'' (i.e., far
away from the locations $\{ \vec\theta_n \}$).  Hence, the lines where
the denominator vanishes separate the regions of high density of
locations from the regions of low density.  Note that, even for very
large average densities $\rho$, we still expect to find ``voids'' of
arbitrary size (in other words, for every finite density $\rho$, there
is a non-vanishing probability of having no point inside an
arbitrarily large region).  As a result, there will be always regions
where the denominator vanishes.  The discussion for the numerator is
similar but, in this case, we also need to take into account the field
$f(\vec\theta)$.  Hence, we still expect to have regions where the
numerator is positive and regions where it is negative but, clearly,
these regions will in general be different from the analogous regions
for the denominator.  As a result, when evaluating the ratio between
the numerator and the denominator, we will obtain arbitrarily large
values close to the lines where the denominator vanishes.  Note also
that these lines will change for different configurations of locations
$\{ \vec\theta_n \}$.  In summary, if the weight function is allowed
to be negative, the denominator of Eq.~\eqref{eq:1} is no longer
guaranteed to be positive, and infinities are expected when performing
the ensemble average.

\section{Moments expansion}
\label{sec:moments-expansion}

In most applications, the density of objects is rather large.  Hence,
it is interesting to obtain an expansion for $C(w_A, w_B)$ valid at
high densities.

In Paper~I we already obtained an expansion for $C_A(w_A)$ (or,
equivalently, $C_B(w_B)$) for $\rho \rightarrow \infty$:
\begin{equation}
  \label{eq:66}
  C_A(w_A) \simeq \frac{\rho}{\rho + w_A} + \frac{\rho^2
    S_{20}}{(\rho + w_A)^3} - \frac{\rho^2 S_{30}}{(\rho + w_A)^4} +
  \frac{\rho^2 S_{40} + 3 \rho^3 S_{20}^2}{(\rho + w_A)^5} \; .
\end{equation}
In this equation, $S_{ij}$ are the moments of the functions $(w_A,
w_B)$, defined as
\begin{equation}
  \label{eq:67}
  S_{ij} \equiv \int_\Omega \diff^2 \theta \bigl[ w_A(\vec\theta)
  \bigr]^i \bigl[ w_B(\vec\theta) \bigr]^j \; .
\end{equation}
Clearly, in Eq.~\eqref{eq:66} enter only the moments $S_{i0}$, since the
form of $w_B$ is not relevant for $C_A(w_A)$.  Similarly, the
expression for $C_B(w_B)$ contains only the moments $S_{0j}$.  Note
that for weight functions invariant upon translation we have $S_{ij} =
S_{ji}$.

\begin{figure}[!t]
  \parbox[t]{0.49\hsize}{%
    \resizebox{\hsize}{!}{\input fig1.tex}
    \caption{The moment expansion of $C(w_A, w_B)$ for
      1-dimensional Gaussian weight functions $w_A(x) = w_B(x)$
      centered on $0$ and with unit variance.  The plot shows the
      various order approximations obtained using the method
      described in Sect.~\ref{sec:moments-expansion} (equations for
      the orders $n=3$ and $n=4$ are not explicitly reported in the
      text; see however Table~\ref{tab:1} in
      Appendix~\ref{sec:moments-expansion-1}).  The density used is
      $\rho = 1$.}%
    \label{fig:1}}
  \hfill
  \parbox[t]{0.49\hsize}{%
    \resizebox{\hsize}{!}{\input fig5.tex}
    \caption{The function $C(w_A, w_B)$ is
      monotonically decreasing with $w_A$ and $w_B$, while $w_A w_B
      C(w_A, w_B)$ (scaled in this plot) is monotonically
      increasing.  The parameters used for this figure are the same
      as Fig.~\ref{fig:1}.  Note that, since $P_A = P_B = 0$, we
      have $\lim_{w_A \rightarrow 0^+} w_A w_B C(w_A, w_B) =
      \lim_{w_B \rightarrow 0^+} w_A w_B C(w_A, w_B) = 0$ in
      agreement with Eqs.~\eqref{eq:82} and \eqref{eq:83}; moreover
      $w_A w_B C(w_A, w_B) < \rho^2 = 1$ as expected from
      Eq.~\eqref{eq:84}.}%
    \label{fig:2}}
\end{figure}

A similar expansion can be obtained for $C(w_A, w_B)$.  Calculations
are basically a generalization of what was done in Paper~I for $C(w)$
and can be found in Appendix~\ref{sec:moments-expansion-1}.  Here we
report only the final result obtained:
\begin{align}
  \label{eq:68}
  C(w_A, w_B) \simeq {} & \frac{\rho^2}{(\rho + w_A) (\rho + w_B)} +
  \frac{\rho^3 S_{20}}{(\rho + w_A)^3 (\rho + w_B)} +
  \frac{\rho^3 S_{11}}{(\rho + w_A)^2 (\rho + w_B)^2} +
  \frac{\rho^3 S_{02}}{(\rho + w_A) (\rho + w_B)^3} \; .
\end{align}
We note that using this expansion and Eqs.~\eqref{eq:59} we can
recover the first terms of Eq.~\eqref{eq:66}, as expected.
Figure~\ref{fig:1} left shows the results of applying this expansion
to a Gaussian weight.  For clarity, we have considered in this figure
(and in others shown below) a 1-dimensional smoothing instead of the
2-dimensional case discussed in the text, and we have used $x$ as
spatial variable instead of $\vec\theta$.  The figure refers to two
identical Gaussian weight functions with vanishing average and unit
variance.  A comparison of this figure with Fig.~2 of Paper~I shows
that the convergence here is much slower.  Nevertheless,
Eq.~\eqref{eq:68} will be very useful to investigate some important
limiting cases in the next section.

\section{Properties}
\label{sec:properties}

In this section we will study in detail the two noise terms $T_\sigma$
and $T_\mathrm{P}$ introduced in Sect.~\ref{sec:noise-contributions},
showing their properties and considering several limiting cases.  The
results obtained are of clear interest of themselves; for example, we
will derive here upper and lower limits for the measurement error
$T_\sigma$ that can be used at low and high densities.  Moreover, this
section helps us understand the results obtained so far, and in
particular the peculiarities of vanishing weights.

\subsection{Normalization}
\label{sec:normalization}

A simple normalization property for $C(w_A, w_B)$ can be derived,
similarly to what we have already done for the average of $\tilde f$
in Paper~I.  Suppose that $f(\vec\theta) = 1$ and that no errors are
present on the measurements, so that $\sigma^2 = 0$.  In this case we
will always measure $\tilde f(\vec\theta) = 1$ [see Eq.~\eqref{eq:1}],
so that $\bigl\langle \tilde f_A \bigr\rangle = \bigl\langle \tilde
f_B \bigr\rangle = 1$, $\bigl\langle \tilde f_A \tilde f_B
\bigr\rangle = 1$, and no error is expected on $\tilde f$.  This
result can be also recovered using the analytical expressions obtained
so far.  Let us first consider the simpler case of non-vanishing
weights.

Using Eq.~\eqref{eq:47} and \eqref{eq:48}, we can write the term
$T_\mathrm{P1}$ in the case $f(\vec\theta) = 1$ as
\begin{equation}
  \label{eq:69}
  T_\mathrm{P1} = \rho \int_0^\infty \diff s_A \int_0^\infty
  \diff s_B \, \e^{\rho Q(s_A, s_B)} \int_\Omega \diff^2 \theta \,
  w_A(\vec\theta) w_B(\vec\theta) \e^{-s_A w_A(\vec\theta) - s_B
    w_B(\vec\theta)} \; .
\end{equation}
The last integrand in this equation can be rewritten as $\partial^2 Q
/ \partial s_A \, \partial s_B$ [cf.\ the definition of $Q$,
Eq.~\eqref{eq:45}]:
\begin{equation}
  \label{eq:70}
  T_\mathrm{P1} = \rho \int_0^\infty \diff s_A \int_0^\infty \diff s_B
    \, \e^{\rho Q(s_A, s_B)} \frac{\partial^2 Q(s_A, s_B)}{\partial s_A
    \partial s_B} \; .
\end{equation}
Analogously, for $T_\mathrm{P2}$ we obtain [cf.\ Eq.~\eqref{eq:52}]
\begin{align}
  \label{eq:71}
  T_\mathrm{P2} = {} & \rho^2 \int_0^\infty \diff s_A \int_0^\infty
  \diff s_B \, \e^{\rho Q(s_A, s_B)} \int_\Omega \diff^2 \theta_1 \,
  w_A(\vec\theta_1) \e^{-s_A w_A(\vec\theta_1) - s_B
    w_B(\vec\theta_1)} \int_\Omega \diff^2 \theta_2 \,
  w_B(\vec\theta_2) \e^{-s_A w_A(\vec\theta_2) - s_B
    w_B(\vec\theta_2)} \notag\\
  {} = {} & \rho^2 \int_0^\infty \diff s_A \int_0^\infty \diff s_B \,
  \e^{\rho Q(s_A, s_B)} \frac{\partial Q(s_A, s_B)}{\partial s_A}
  \frac{\partial Q(s_A, s_B)}{\partial s_B} \; .
\end{align}
We can integrate this expression by parts taking $\rho \e^{\rho Q} \,
(\partial Q / \partial s_B) = \bigl[ \partial \exp(\rho Q) / \partial
s_B \bigr]$ as differential term:
\begin{equation}
  \label{eq:72}
  T_\mathrm{P2} = \rho \int_0^\infty \diff s_A \biggl\{ \biggl[
  \e^{\rho Q(s_A, s_B)} \frac{Q(s_A, s_B)}{\partial s_A}
  \biggr]_{s_B = 0}^\infty - \int_0^\infty \diff s_B \, \e^{\rho
  Q(s_A, s_B)} \frac{\partial^2 Q(s_A, s_B)}{\partial s_A \partial
  s_B} \biggr\} \; .
\end{equation}
We now observe that the last term in Eq.~\eqref{eq:72} is identical to
what we founded in Eq.~\eqref{eq:70}.  Hence, the sum $T_\mathrm{P1} +
T_\mathrm{P2}$ is
\begin{align}
  \label{eq:73}
  T_\mathrm{P1} + T_\mathrm{P2} & {} = \biggr[ \rho \int_0^\infty
  \diff s_A \, \e^{\rho Q(s_A, s_B)} \frac{\partial Q(s_A,
    s_B)}{\partial s_A} \biggr]_{s_B = 0}^\infty = \biggl[ \biggl[
  \rho \e^{\rho Q(s_A, s_B)} \biggr]_{s_A = 0}^\infty \biggr]_{s_B =
  0}^\infty \notag\\
  & {} = Y(\infty, \infty) - Y(\infty, 0^+) - Y(0^+, \infty) + Y(0^+,
  0^+) = 1 \; .
\end{align}
The last equation holds because, for non-vanishing weights, $Y(0^+,
0^+) = 1$ and all other terms vanishes [cf.\ 
Eqs.~(\ref{eq:61}--\ref{eq:62})].  Hence, as expected, $\bigl\langle
\tilde f_A \tilde f_B \bigr\rangle = T_\mathrm{P1} + T_\mathrm{P2} = 1
= \bigl\langle \tilde f_A \bigr\rangle \bigl\langle \tilde f_B
\bigr\rangle$.

In case of vanishing weights, we can still use Eqs.~\eqref{eq:70}
and \eqref{eq:72} with an additional factor $\nu$ [due to the
extra factor in Eq.~\eqref{eq:63}].  The last step in
Eq.~\eqref{eq:73} thus now becomes
\begin{equation}
  \label{eq:74}
  T_\mathrm{P1} + T_\mathrm{P2} = \nu \bigl[ Y(\infty, \infty) -
  Y(\infty, 0^+) - Y(0^+, \infty) + Y(0^+, 0^+) \bigr] = 1 \; .
\end{equation}
The last equality holds since now $Y$ does not vanishes for large
$(s_A, s_B)$ [see again Eqs.~(\ref{eq:61}--\ref{eq:62})].

\subsection{Scaling}
\label{sec:scaling}

Similarly to what was already shown in Paper~I, for all expressions
encountered so far some scaling invariance properties hold.

First, we note that, although we have assumed that the weight
functions $w_A$ and $w_B$ are normalized to unity, all results are
clearly independent of their actual normalization.  Hence, a trivial
scaling property holds: All results (and in particular the final
expression for $\Cov(\tilde f)$) are left unchanged by the
transformation $w(\vec\theta) \mapsto k w(\vec\theta)$ or,
equivalently,
\begin{align}
  \label{eq:75}
  w_A(\vec\theta) & {} \mapsto k w_A(\vec\theta) \; , &
  w_B(\vec\theta) & {} \mapsto k w_B(\vec\theta) \; .
\end{align}

A more interesting scaling property is the following.  Consider the
transformation
\begin{align}
  \label{eq:76}
  w(\vec\theta) & {} \mapsto k^2 w(k \vec\theta) \; , &
  \rho & {} \mapsto k^2 \rho \; ,
\end{align}
where both factors $k^2$ must be changed according to the dimension of
the $\vec\theta$ vector space.  If we apply this transformation, then
the expression for $\Cov(\tilde f)$ is transformed according to
\begin{equation}
  \label{eq:77}
  \Cov(\tilde f; \vec\theta_A, \vec\theta_B) \mapsto \Cov(\tilde f;
  k \vec\theta_A, k \vec\theta_B) \; .
\end{equation}
This invariance suggests that the shape of $\Cov(\tilde f)$ is
controlled by the expected number of objects for which the two weight
functions are significantly different from zero.  Hence, similarly to
what done in Paper~I, we define the two weight areas $\mathcal{A}_A$
and $\mathcal{A}_B$ as
\begin{equation}
  \label{eq:78}
  \mathcal{A}_X \equiv \biggl[ \int_\Omega \bigl[
  w_X(\vec\theta) \bigr]^2 \, \diff^2 \theta \biggr]^{-1} =
  \begin{cases}
    S_{20}^{-1} & \text{if $X = A \; ,$} \\
    S_{02}^{-1} & \text{if $X = B \; .$}
  \end{cases}
\end{equation}
For weight functions invariant upon translation we have $\mathcal{A}_A
= \mathcal{A}_B$.  We call $\mathcal{N}_X \equiv \rho \mathcal{A}_X$
the \textit{weight number of objects\/} (again, $\mathcal{N}_A =
\mathcal{N}_B$ because of the invariance upon translation).  Note that
this quantity is left unchanged by the scaling \eqref{eq:76}.  Similar
definitions hold for the \textit{effective weight\/}
$w_{\mathrm{eff}X}(\vec\theta) \equiv w_X(\vec\theta) C_X\bigl(
w_X(\vec\theta) \bigr)$ and the \textit{effective number of objects\/}
$\mathcal{N}_{\mathrm{eff}X} \equiv \rho \mathcal{A}_{\mathrm{eff}X}$.

\subsection{Behavior of $C$}
\label{sec:behavior-c}

\begin{figure}[!t]
  \begin{center}
  \end{center}
\end{figure}

In order to better understand the properties of $C$, it is useful to
briefly consider its behavior as a function of the weights $w_A$ and
$w_B$.

We observe that, since $Y(s_A, s_B) > 0$ for every $(s_A, s_B)$ [see
Eq.~\eqref{eq:46}], $C(w_A, w_B)$ decreases if either $w_A$ or $w_B$
increase.  In order to study the behavior of the quantity $w_A w_B
C(w_A, w_B)$ that enters the noise term $T_1$, we consider the
quantity $w_A C(w_A, w_B)$:
\begin{equation}
  \label{eq:79}
  w_A C(w_A, w_B) = \nu \rho^2 \int_{0^-}^\infty \diff s_B \,
  \biggl[ Y(0^-, s_B) + \int_{0^-}^\infty \diff s_A \, \biggl(
  \frac{\partial Y(s_a, s_B)}{\partial s_A} \biggr) \e^{- s_A w_A}
  \biggr] \e^{- s_B w_B} \; .
\end{equation}
This equation can be shown by integrating by parts the integral over
$s_A$.  The partial derivative required in Eq.~\eqref{eq:79} can be
evaluated from Eq.~\eqref{eq:46}:
\begin{equation}
  \label{eq:80}
  \frac{\partial Y(s_A, s_B)}{\partial s_A} = \rho \frac{\partial
  Q(s_A, s_B)}{\partial s_A} \e^{\rho Q(s_A, s_B)} \le 0 \; .
\end{equation}
Since this derivative is negative, we can deduce that the integral
over $s_A$ in Eq.~\eqref{eq:79} increases with $w_A$, and thus $w_A
C(w_A, w_B)$ also increases as $w_A$ increases.  Similarly, it can be
shown that $w_B C(w_A, w_B)$ increases as $w_B$ increases.  In
summary, the quantity $w_A w_B C(w_A, w_B)$ behaves as $w_A w_B$, in
the sense that its partial derivatives have the same sign as the
partial derivatives of $w_A w_B$ (see Fig.~\ref{fig:2}).  Also,
since $C(w_A, w_B)$ decreases if either $w_A$ or $w_B$ increase, we
can deduce that $w_A w_B C(w_A, w_B)$ is ``broader'' than $w_A w_B$.

Since $C\bigl( w_A(\vec\theta), w_B(\vec\theta) \bigr)$ is positive,
the function $w_A(\vec\theta) w_B(\vec\theta) C\bigl( w_A(\vec\theta),
w_B(\vec\theta) \bigr)$ shares the same support as $w_A(\vec\theta)
w_B(\vec\theta)$.  It is also interesting to study the limits of $w_A
w_B C(w_A, w_B)$ at high and low values for $w_A$ and $w_B$.  From the
properties of Laplace transform [see Eqs.~\eqref{eq:163}], we have
\begin{equation}
  \label{eq:81}
  \lim_{\substack{w_A \rightarrow 0^+ \\ w_B \rightarrow 0^+}} w_A w_B
  C(w_A, w_B) = \nu \rho^2 \lim_{\substack{s_A \rightarrow \infty \\
  s_B \rightarrow \infty}} Y(s_A, s_B) = \nu \rho^2 P_{AB} \; ,
\end{equation}
where Eq.~\eqref{eq:61} has been used in the second equality.  Hence,
the quantity $w_A w_B C(w_A, w_B)$ goes to zero only if $P_{AB} = 0$.
In other cases, we expect a discontinuity at $w_A = w_B = 0$.
Similarly, using Eqs.~(\ref{eq:61}--\ref{eq:62}) we find
\begin{align}
  \label{eq:82}
  \lim_{\substack{w_A \rightarrow \infty \\ w_B \rightarrow 0^+}} w_A w_B
  C(w_A, w_B) & {} = \nu \rho^2 \lim_{\substack{s_A \rightarrow 0^+ \\
      s_B \rightarrow \infty}} Y(s_A, s_B) = \nu \rho^2 P_A \; , \\
  \label{eq:83}
  \lim_{\substack{w_A \rightarrow 0^+ \\ w_B \rightarrow \infty}} w_A w_B
  C(w_A, w_B) & {} = \nu \rho^2 \lim_{\substack{s_A \rightarrow \infty \\
      s_B \rightarrow 0^+}} Y(s_A, s_B) = \nu \rho^2 P_B \; , \\
  \label{eq:84}
  \lim_{\substack{w_A \rightarrow \infty \\ w_B \rightarrow \infty}} w_A w_B
  C(w_A, w_B) & {} = \nu \rho^2 \lim_{\substack{s_A \rightarrow 0^+ \\
      s_B \rightarrow 0^+}} Y(s_A, s_B) = \nu \rho^2 \; .
\end{align}
Since $w_A w_B C(w_A, w_B)$ increases with both $w_A$ and $w_B$, the
last equation above puts a superior limit for this quantity:
\begin{equation}
  \label{eq:85}
  w_A w_B C(w_A, w_B) \le \nu \rho^2 \; .
\end{equation}

\subsection{Large separations}

Suppose that the two points $\vec\theta_A$ and $\vec\theta_B$ are far
away from each other, so that $w_A(\vec\theta) w_B(\vec\theta)$ is
very close to zero everywhere.  In this situation we can greatly
simplify our equations.

If $\vec\theta_A$ is far away from $\vec\theta_B$, then
$w_A(\vec\theta)$ and $w_B(\vec\theta)$ are never significantly
different from zero at the same position $\vec\theta$.  In this case,
the integral in the definition of $Q(s_A, s_B)$ [see
Eq.~\eqref{eq:45}] can be split into two integrals that corresponds to
$Q_A$ and $Q_B$ [Eq.~\eqref{eq:53}]:
\begin{align}
  \label{eq:86}
  Q(s_A, s_B) & {} \simeq Q_A(s_A) + Q_B(s_B) \; , &
  Y(s_A, s_B) & {} \simeq Y_A(s_A) Y_B(s_B) \; , &
  C(w_A, w_B) & {} \simeq C_A(w_A) C_B(w_B) \; .
\end{align}
Hence, if the two weight functions $w_A$ and $w_B$ do not have
significant overlap, the function $C(w_A, w_B)$ reduces to the product
of the two correcting functions $C_A$ and $C_B$.

In general, it can be shown that $C(w_A, w_B) \ge C_A(w_A) C_B(w_B)$.
In fact, we have
\begin{equation}
  \label{eq:87}
  C(w_A, w_B) - C_A(w_A) C_B(w_B) = \rho^2 \int_0^\infty \diff s_A
  \int_0^\infty \diff s_B \, \e^{-s_A w_A - s_B w_B} \bigl[ \e^{\rho
  Q(s_A, s_B)} - \e^{\rho Q_A(s_A) + \rho Q_B(s_B)} \bigr] \; .
\end{equation}
We now observe that
\begin{equation}
  \label{eq:88}
  Q(s_A, s_B) - Q_A(s_A) - Q_B(s_B) = \int_\Omega \bigl[ \e^{-s_A
  w_A(\vec\theta)} - 1 \bigr] \bigl[ \e^{-s_B w_B(\vec\theta)} - 1
  \bigr] \ge 0 \; .
\end{equation}
Hence, $Q(s_A, s_B) \ge Q_A(s_A) + Q_B(s_B)$ and the difference
between the two terms of this inequality is an indication of overlap
between the two weight functions $w_A$ and $w_B$.  Since the
exponential function is monotonic, we find $Y(s_A, s_B) \ge Y_A(s_A)
Y_B(s_B)$ and thus
\begin{equation}
  \label{eq:89}
  C(w_A, w_B) \ge C_A(w_A) C_B(w_B) \; .
\end{equation}

\subsection{Upper and lower limits for $T_{\protect\sigma}$}
\label{sec:upper-limit-t_sigma}

The normalization property shown in Sect.~\ref{sec:normalization} can
also be used to obtain an upper limit for $T_\sigma$.  We observe, in
fact, that $T_\sigma$ is indistinguishable from $\sigma^2
T_\mathrm{P1}$ for a constant function $f(\vec\theta) = 1$.  This
case, however, has already been considered above in
Sect.~\ref{sec:normalization}: There we have shown that $T_\mathrm{P1}
+ T_\mathrm{P2} = 1$.  Since $T_\mathrm{P2} \ge 0$, we find the
relation $T_\sigma \le \sigma^2$.

The property just obtained has a simple interpretation.  As shown by
Eq.~\eqref{eq:60}, $T_\sigma$ is proportional to $1/\rho$ and thus we
would expect that this quantity is unbounded superiorly.  In reality,
even when we are dealing with a very small density of objects, the
estimator \eqref{eq:1} ``forces'' us to use at least one object.  This
point has already been discussed in Paper~I, where we showed that the
number of effective objects, $\mathcal{N}_\mathrm{eff}$, is always
larger than unity.  The upper limit found for $T_\sigma$ can be
interpreted using the same argument.  Note that this result also holds
for weight functions with finite support.

A lower limit for $T_\sigma$, instead, can be obtained from the
inequality \eqref{eq:89}:
\begin{equation}
  \label{eq:90}
  T_\sigma \ge \frac{\sigma^2}{\rho} \int_\Omega w_A(\vec\theta)
  w_B(\vec\theta) C_A\bigl( w_A(\vec\theta) \bigr) C_B\bigl(
  w_B(\vec\theta) \bigr) \; \, \diff^2 \theta = \frac{\sigma^2}{\rho}
  \int_\Omega w_{\mathrm{eff}A}(\vec\theta)
  w_{\mathrm{eff}B}(\vec\theta) \, \diff^2 \theta \; .
\end{equation}
Hence, the error $T_\sigma$ is larger than a convolution of the two
effective weight functions.  In case of finite-field weight functions,
the limit just obtained must be corrected with a factor $\nu$.  The
argument to derive Eq.~\eqref{eq:90} is then slightly more complicated
because of the presence of the $P_X$ probabilities.  However, using
the relation $P_A P_B \le P_{AB}$, we can recover Eq.~\eqref{eq:90}
with the aforementioned corrective factor.

\subsection{Limit of low and high densities}
\label{sec:limit-high-low}

In the limit $\rho \rightarrow 0$ we can obtain simple expressions for
the noise terms.  If $\rho$ vanishes, we have $Y(s_A, s_B) = 1$ [cf.\ 
Eq.~\eqref{eq:46}] and thus
\begin{equation}
  \label{eq:91}
  C(w_A, w_B) \simeq \frac{\nu \rho^2}{w_A w_B} \; .
\end{equation}
In this equation we have assumed $w_A w_B > 0$.  Note that we have
reached here the superior limit indicated by Eq.~\eqref{eq:85}.  In
the same limit, $\rho \rightarrow 0$, $P_X \simeq 1 - \pi_X \rho$ and
$\nu \simeq 1 / \rho \pi_{A \cap B}$, where $\pi_{A \cap B} = \pi_{A}
+ \pi_B - \pi_{A \cup B}$ is the area of the intersection of the
supports of $w_A$ and $w_B$.  Hence we find
\begin{equation}
  \label{eq:92}
  C(w_A, w_B) \simeq \frac{\rho}{\pi_{A \cap B} w_A w_B} \; .
\end{equation}
Analogously, in the same limit, we have found in Paper~I
\begin{equation}
  \label{eq:93}
  C_X(w_X) \simeq \frac{1}{\pi_X w_X} \; ,
\end{equation}
where $w_X > 0$ has been assumed.  We can then proceed to evaluate the
various terms.  For $T_\sigma$ we obtain the expression
\begin{equation}
  \label{eq:94}
  T_\sigma \simeq \frac{\sigma^2}{\rho} \int_{\pi_{A \cap B}}
  \frac{\rho}{\pi_{A \cap B}} \, \diff^2 \theta = \sigma^2 \; .
\end{equation}
Note that the integral has been evaluated only on the subset of the
plane where $w_A w_B > 0$; the case where this product vanishes, in
fact, need not to be considered because the quantity $w_A w_B C(w_A,
w_B)$ vanishes as well.  Exactly the same result holds for weight
functions with infinite support.  Hence, when $\rho \rightarrow 0$ we
reach the superior limit discussed in
Sect.~\ref{sec:upper-limit-t_sigma} for $T_\sigma$.

Equation~\eqref{eq:94} can be better appreciated with the following
argument.  As the density $\rho$ approaches zero, the probability of
having two objects on $\pi_{A \cup B}$ vanishes.  Because of the
prescription regarding vanishing weights (cf.\ beginning of
Sect.~\ref{sec:vanishing-weights}), the ensemble average in our limit
is performed with one and only one object in $\pi_{A \cap B}$.  Since
we have only one object, this is basically used with unit weight in
the average \eqref{eq:17}, and thus the measurement noise is just given
by $T_\sigma = \sigma^2$.

Let us now consider the limit at low densities of the Poisson noise,
which, we recall, has been split into three terms, $T_\mathrm{P1}$,
$T_\mathrm{P2}$, and $T_\mathrm{P3}$ (see
Sect.~\ref{sec:noise-contributions}).  Inserting Eq.~\eqref{eq:92}
into Eq.~\eqref{eq:24}, we find for $T_\mathrm{P1}$
\begin{equation}
  \label{eq:95}
  T_\mathrm{P1} \simeq \frac{1}{\rho} \int_{\pi_{A \cap B}}
  f^2(\vec\theta) \frac{\rho}{\pi_{A \cap B}} \, \diff^2 \theta =
  \bigl\langle f^2 \bigr\rangle_{\pi_{A \cap B}} \; ,
\end{equation}
where $\bigl\langle f^2 \bigr\rangle_{\pi_{A \cap B}}$ denotes the
simple average of $f^2$ on the set $\pi_{A \cap B}$.  Hence,
$T_\mathrm{P1}$ converges to the average of $f^2$ on the intersection
of the supports of $w_A$ and $w_B$.  Again, we can explain this result
using an argument similar to the one used for Eq.~\eqref{eq:94}.
Regarding $T_\mathrm{P2} \equiv T_2$, we observe that this term is of
first order in $\rho$ because $C(w_A, w_B)$ is of first order [cf.\ 
Eqs.~\eqref{eq:92} and \eqref{eq:52}].  We can then safely ignore this
term in our limit $\rho \rightarrow 0$.  Finally, as shown in Paper~I,
at low densities the expectation value for $\tilde f_X$ is a simple
average of $f$ on the support of $w_X$, i.e.\ $\langle \tilde f_X
\rangle \simeq \langle f \rangle_{\pi_X}$.  Hence, $T_\mathrm{P3} =
\langle f \rangle_{\pi_A} \langle f \rangle_{\pi_B}$ and the Poisson
noise in the limit of small densities is given by
\begin{equation}
  \label{eq:96}
  T_\mathrm{P} = \bigl\langle f^2 \bigr\rangle_{\pi_{A \cap B}} - \langle
  f \rangle_{\pi_A} \langle f \rangle_{\pi_B} \; .
\end{equation}
In case of a constant function $f(\vec\theta)$, this expression
vanishes as expected.  Surprisingly, in general, we cannot say that
$T_\mathrm{P} \ge 0$.  Rather, if $w_A \neq w_B$, and if in particular
the two weight functions have different supports, we might have a
negative $T_\mathrm{P}$.  Suppose, for example, that $f$ vanishes on
the intersection of the two supports $\pi_{A \cap B}$, but is
otherwise positive.  In this case, the first term in the r.h.s.\ of
Eq.~\eqref{eq:96} vanishes, while the second term contributes with a
negative sign, and thus $T_\mathrm{P} < 0$.  On the other hand, if
$w_A = w_B$ then $T_\mathrm{P}$ has to be non-negative.

We now consider the opposite limiting case, namely high density.  In
this case, it is useful to use the moment expansion \eqref{eq:68}.
Since $T_\sigma$ and $T_\mathrm{P1}$ have an overall factor $1/\rho$
in its definition [cf.\ Eq.~\eqref{eq:60}], we can simply take the
$0$-th order for $C(w_A, w_B)$, thus obtaining
\begin{align}
  \label{eq:97}
  T_\sigma \simeq {} & \frac{\sigma^2}{\rho} \int w_A(\vec\theta)
  w_B(\vec\theta) \, \diff^2 \theta = \frac{\sigma^2 S_{11}}{\rho} \;
  . &
  T_\mathrm{P1} \simeq {} & \frac{1}{\rho} \int w_A(\vec\theta)
  w_B(\vec\theta) f^2(\vec\theta) \, \diff^2 \theta \; .
\end{align}
For $T_\mathrm{P2}$ and $T_\mathrm{P3}$, instead, we need to use a
first order expansion in $1/\rho$ for $C(w_A, w_B)$.  This can be done
by using the first terms in series \eqref{eq:66}, and by expanding all
fractions in terms of powers of $1/\rho$.  Inserting the result into
the definitions of $T_\mathrm{P2}$ and $T_\mathrm{P3}$ we obtain
\begin{align}
  \label{eq:98}
  T_\mathrm{P2} & {} \simeq \int_\Omega \diff^2 \theta_1 \int_\Omega \diff^2
  \theta_2 \, f(\vec\theta_1) f(\vec\theta_2) w_A(\vec\theta_1)
  w_B(\vec\theta_2) \biggl[ 1 - \frac{w_A(\vec\theta_1)}{\rho} -
  \frac{w_A(\vec\theta_2)}{\rho} - \frac{w_B(\vec\theta_1)}{\rho} -
  \frac{w_B(\vec\theta_2)}{\rho} + \frac{S_{20} + S_{11} +
    S_{02}}{\rho} \biggr] \; , \\
  \label{eq:99}
  T_\mathrm{P3} & {} \simeq \int_\Omega \diff^2 \theta_1 \int_\Omega
  \diff^2 \theta_2 \, f(\vec\theta_1) f(\vec\theta_2) w_A(\vec\theta_1)
  w_B(\vec\theta_2) \biggl[ 1 - \frac{w_A(\vec\theta_1)}{\rho} -
  \frac{w_B(\vec\theta_2)}{\rho} + \frac{S_{20} + S_{02}}{\rho}
  \biggr] \; .
\end{align}
Note that we have dropped, in these equations, terms of order higher
than $1/\rho$.  The difference $T_\mathrm{P2} - T_\mathrm{P3}$ is
\begin{equation}
  \label{eq:100}
  T_\mathrm{P2} - T_\mathrm{P3} \simeq \frac{1}{\rho}
  \int_\Omega \diff^2 \theta_1 \int_\Omega \diff^2 \theta_2 \,
  f(\vec\theta_1) f(\vec\theta_2) w_A(\vec\theta_1) w_B(\vec\theta_2)
  \bigl[ S_{11} - w_A(\vec\theta_2) - w_B(\vec\theta_1) \bigr] \; .
\end{equation}
Using Eqs.~\eqref{eq:100} and \eqref{eq:97}, we can verify that
$T_\mathrm{P}$ vanishes if $f$ is constant, as expected:
\begin{align}
  \label{eq:101}
  T_\mathrm{P1} + T_\mathrm{P2} - T_\mathrm{P3} & {} \simeq
  \frac{S_{11}}{\rho} + \frac{1}{\rho} \int_\Omega \diff \theta_1
  \int_\Omega \diff \theta_2 \, \bigl[ S_{11} w_A(\theta_1)
  w_B(\theta_2) - w_A(\theta_1) w_A(\theta_2) w_B(\theta_2) -
  w_A(\theta_1) w_B(\theta_2) w_B(\theta_1) \bigr]
  \notag\\
  & {} = \frac{S_{11}}{\rho} + \frac{1}{\rho} \bigl[ S_{11} - S_{11} -
  S_{11} \bigr] = 0 \; ,
\end{align}
where the normalization of $w$ has been used.  Also, it is apparent
that all noise sources, including Poisson noise, are proportional to
$1/\rho$ at high densities.

In order to further investigate the properties of Poisson noise at
high densities, we write it in a more compact form.  Let us define the
average of a function $g(\vec\theta)$ weighted with $q(\vec\theta)$ as
\begin{equation}
  \label{eq:102}
  \langle g \rangle_q \equiv \biggl[ \int_\Omega g(\vec\theta)
  q(\vec\theta) \, \diff^2 \theta \biggr] \biggm/ \biggl[ \int_\Omega
  q(\vec\theta) \, \diff^2 \theta \biggr] \; .
\end{equation}
Using this definition we can rearrange Eqs.~\eqref{eq:97} and
\eqref{eq:100} in the form
\begin{equation}
  \label{eq:103}
  T_\mathrm{P} = \frac{S_{11}}{\rho} \Bigl[ \bigl\langle f^2
  \bigr\rangle_{w_A w_B} - \langle f \rangle_{w_A
    w_B} \langle f \rangle_{w_A w_B} + \bigl( \langle f \rangle_{w_A
  w_B} - \langle f \rangle_{w_A} \bigr) \bigl( \langle f \rangle_{w_A
  w_B} - \langle f \rangle_{w_B} \bigr) \Bigr] \; .
\end{equation}
This expression suggests that the Poisson noise is actually made of
two different terms, $\bigl\langle f^2 \bigr\rangle_{w_A w_B} -
\langle f \rangle_{w_A w_B} \langle f \rangle_{w_A w_B}$ and $\bigl(
\langle f \rangle_{w_A w_B} - \langle f \rangle_{w_A} \bigr) \bigl(
\langle f \rangle_{w_A w_B} - \langle f \rangle_{w_B} \bigr)$.  The
first term is proportional to the difference between two averages of
$f^2$ and $f$; both averages are performed using $w_A w_B$ as weight.
Hence, this term is controlled by the ``internal scatter'' of $f$ on
points where both weight functions are significantly different from
zero; it is always positive.  The second term is made of averages $f$
using different weight functions.  It can be either positive or
negative if $w_A \neq w_B$.  Actually, as already seen in the limiting
case $\rho \rightarrow 0$, the overall Poisson noise does not need to
be positive, and anti-correlation can be present in some cases.

\subsection{Limit of high and low frequencies}
\label{sec:limit-high-low-1}

The strong dependence of the Poisson noise on the function
$f(\vec\theta)$ makes an analytical estimate of this noise
contribution extremely difficult in the general case.  However, it is
still possible to study the behavior of $T_P$ in two important
limiting cases, that we now describe.

Suppose that the function $f(\vec\theta)$ does not change
significantly on the scale length of the weight functions
$w_A(\vec\theta)$ and $w_B(\vec\theta)$ (or, in other words, that the
power spectrum of $f$ has a peak at significantly lower frequencies
than the power spectra of $w_A$ and $w_B$).  In this case, we can take
the function $f$ as a constant in the integrals of Eq.~\eqref{eq:13},
and apply the results of Sect.~\ref{sec:normalization}.  Hence, in the
limit of low frequencies, the Poisson noise vanishes.

Suppose now, instead, that the function $f(\vec\theta)$ does not have
any \textit{general\/} trend on the scale length of the weight
functions, but that instead changes at significantly smaller scales
(again, this behavior is better described in terms of power spectra:
We require here that the power spectrum of $f$ has a peak at high
frequencies, while it vanishes for the frequencies where the power
spectra of $w_A$ and $w_B$ are significantly different from zero).
In this case, we can assume that integrals such as
\begin{align}
  \label{eq:104}
  \int_\Omega f(\vec\theta) w_X(\vec\theta) \, \diff^2 \theta & {}
  \simeq 0 &
  X = \{ A, B \}
\end{align}
vanish approximately, because the average of $f$ on large scales
vanishes (remember that we are assuming that $f$ has no general trend
on large scales).  Similarly, the integrals that appear in $T_{P2}$
and $T_{P3}$ vanish as well.  In this case, then, the only
contribution to the Poisson noise arises from $T_{P1}$.  This can be
easily evaluated
\begin{equation}
  \label{eq:105}
  T_P \simeq T_{P1} \simeq \frac{\bigl\langle | f |^2
  \bigr\rangle}{\rho} \int_\Omega \diff^2 \theta \, w_A(\vec\theta)
  w_B(\vec\theta) C\bigl( w_A(\vec\theta), w_B(\vec\theta) \bigr) \; ,
\end{equation}
where we have denoted with $\bigl\langle | f |^2 \bigr\rangle$ the
average of $|f|^2$ on large scales.  Hence we finally obtain
\begin{equation}
  \label{eq:106}
  T_P \simeq \frac{\bigl\langle | f |^2 \bigr\rangle}{\sigma^2}
  T_\sigma \; .
\end{equation}
The results discussed in this section can also be numerically verified
in simple cases.  Figure~\ref{fig:8}, for example, shows the Poisson
noise expected in the measurement of a periodic field when using
two Gaussian weight functions (see Sect.~\ref{sec:gaussian} for details).
From this figure, we see that the Poisson noise increases with the
frequency of the field $f$, and quickly attains a maximum value at
high frequencies.  Moreover, the same figure shows that, in agreement
with Eq.~\eqref{eq:106}, the Poisson noise at the maximum is simply
related to the measurement noise $T_\sigma$ (cf.\ Fig.~\ref{fig:7} for
$\rho = 2$).

\section{Examples}
\label{sec:examples}

Similarly to what has been done in Paper~I, in this section we
consider three typical weight functions, namely a top-hat, a Gaussian,
and a parabolic weight.  For simplicity, we will consider
1-dimensional cases only; this will have also some advantages when
representing the results obtained with figures.  Hence, we will use
$x$ instead of $\vec\theta$ as spatial variable.

\subsection{Top-hat}
\label{sec:top-hat}

\begin{figure}[!t]
  \parbox[t]{0.49\hsize}{%
    \resizebox{\hsize}{!}{\input fig6.tex}
    \caption{The value of $C(1,1)/\rho$ for top-hat weights as a
      function of the density $\rho$.  Both weight functions $w_A$ and
      $w_B$ are top-hats [see Eq.~\eqref{eq:107}] centered on zero.
      Using Eq.~\eqref{eq:108}, we can use this graph to obtain
      $T_\sigma$ as a function of the density and the point separation
      $x_A - x_B$.}
    \label{fig:3}}%
  \hfill
  \parbox[t]{0.49\hsize}{%
    \resizebox{\hsize}{!}{\input fig9.tex}
    \caption{The noise term $T_\sigma$ for two top-hat weights as a
      function of the point separation $\delta = | x_A - x_B |$ for
      two densities, $\rho = 2$ and $\rho = 5$.  The plot also shows
      the quantity $S_{11} / \rho$, which at high densities
      approximates $T_\sigma$ (since then $C \simeq 1$).  Note that
      $S_{11}$ for a top-hat function is just given by $S_{11} =
      ( 2 - \delta ) / 4$.}
    \label{fig:4}}
\end{figure}

The simplest weight that we can consider is a top-hat function,
defined as
\begin{equation}
  \label{eq:107}
  w(x) =
  \begin{cases}
    1 & \text{if $|x| < 1/2 \; ,$} \\
    0 & \text{otherwise$\; .$}
  \end{cases}
\end{equation}
Since $w$ is either $1$ or $0$, we just need to consider $C(1,1)$ to
evaluate $T_\sigma$.  Regarding the Poisson noise, from
Eq.~\eqref{eq:52} we deduce that $C(1,2)$, $C(2,1)$, and $C(2,2)$ are
also required.

Figure~\ref{fig:3} shows $C(1,1)$ and $C(1,1)/\rho$ as functions of
the density $\rho$ for two identical top-hat weight functions centered
on the origin.  From this plot we can recognize some of the limiting
cases studied above.  In particular, the fact that $C(1,1)/\rho$ goes
to unity at low densities is related to Eq.~\eqref{eq:92}; similarly,
the limit of $C(1,1)$ at high densities is consistent with
Eq.~\eqref{eq:68}.  The same figure shows also the moments expansion
of $C(1,1)$ up to forth order.  As expected, the expansion completely
fails at low densities, while is quite accurate for $\rho > 5$.

Curves in Fig.~\ref{fig:3} have been calculated using the standard
approach described by Eqs.~\eqref{eq:45}, \eqref{eq:46} and
\eqref{eq:63}.  Actually, in the simple case of top-hat weight
functions, we can evaluate $C(1,1)$ using a more direct statistical
argument.  We start by observing that in our case, for $x_A = x_B$, we have 
\begin{equation}
  \label{eq:108}
  T_\sigma = \sigma^2 C(1,1) / \rho \; .
\end{equation}
On the other hand, a top-hat weight function is basically acting by
taking simple averages for all objects that fall inside its support.
This suggests that, for $x_A = x_B$, we can evaluate its measurement
noise as
\begin{equation}
  \label{eq:109}
  T_\sigma = \sigma^2 \sum_{N=1}^\infty \frac{p(N)}{N} \; ,
\end{equation}
where $p(N)$ is the probability of having $N$ objects inside the
support.  This probability is basically a Poisson probability
distribution with average $\rho$.  However, since we are adopting the
prescription of ``avoiding'' weight functions without objects in their
support, we must explicitly discard the case $N = 0$ and consequently
renormalize the probability.  In summary, we have
\begin{equation}
  \label{eq:110}
  p(N) = \frac{\e^{-\rho} \rho^N}{N!} \biggm/ \bigl[ 1 - \e^{-\rho}
  \bigr] \; .
\end{equation}
This expression combined with Eq.~\eqref{eq:109} allows us to evaluate
$C(1,1) = \rho T_\sigma / \sigma^2$:
\begin{equation}
  \label{eq:111}
  C(1,1) = \frac{\e^{-\rho}}{1 - \e^{-\rho}} \sum_{N=1}^\infty
  \frac{\rho^{N + 1}}{N! \, N} \; .
\end{equation}
We can directly verify this result using Eqs.~\eqref{eq:45},
\eqref{eq:46} and \eqref{eq:63}.  In fact, for the top-hat function we
find
\begin{align}
  \label{eq:112}
  Q(s_A, s_B) = {} & \bigl[ \e^{-s_A -s_B} - 1 \bigr] \; , \\
  \label{eq:113}
  Y(s_A, s_B) = {} & \e^{\rho Q(s_A, s_B)} = \e^{-\rho}
  \sum_{k=0}^\infty \frac{\e^{-k (s_A + s_B)} \rho^k}{k!} \; , \\
  \label{eq:114}
  C(1, 1) = {} & \frac{\rho^2}{1 - \e^{-\rho}} \int_0^\infty \!\diff
  s_A \int_0^\infty \!\diff s_B Y(s_A, s_B) \e^{-s_A - s_B} =
  \frac{\rho^2 \e^{-\rho}}{1 - \e^{-\rho}} \sum_{k=0}^\infty
  \int_0^\infty \!\diff s_A \int_0^\infty \!\diff s_B
  \frac{\e^{-(k+1)(s_A + s_B)} \rho^k}{k!} \notag\\
  {} = {} & \frac{\e^{-\rho}}{1 - \e^{-\rho}} \sum_{k=0}^\infty
  \frac{\rho^{k + 2}}{k! \, (k+1)^2} \; . 
\end{align}
Finally, with a change of the dummy variable $k \mapsto n - 1$ we
recover Eq.~\eqref{eq:111}.

The other terms needed for the Poisson noise can be evaluated using a
calculation similar to the one performed in Eq.~\eqref{eq:114}.
Actually, it can be shown that for any positive integers $w_A$ and
$w_B$ we have
\begin{equation}
  \label{eq:115}
  C(w_A, w_B) = \frac{\e^{-\rho}}{1 - \e^{-\rho}} \sum_{k=0}^\infty
  \frac{\rho^{k + 2}}{k! \, (k + w_A) (k + w_B)} \; .
\end{equation}

Figure~\ref{fig:4} shows the expected measurement noise $T_\sigma$ as
a function of the point separation $\delta = |x_A - x_B|$.  Note that,
for densities of order $\rho = 5$ or larger, a good approximation is
obtained by just taking $C(w_A, w_B) = 1$ [cf.\ the moments expansion
\eqref{eq:68}], so that $T_\sigma \simeq S_{11} / \rho$; we also
observe that, for a top-hat weight function, $S_{11}$ is a linear
function.

\subsection{Gaussian}
\label{sec:gaussian}

\begin{figure}[!t]
  \parbox[t]{0.49\hsize}{%
    \resizebox{\hsize}{!}{\input fig2.tex}
    \caption{Numerical calculations  for 1-dimensional
      Gaussian weight functions $w_A = w_B$ centered on $0$ and with
      unit variance.  The various curves shows the function $w_A w_B
      C(w_A, w_B)$ for different densities $\rho$.  Note that, as
      expected, $C(w_A, w_B)$ approaches unity for large densities.}
    \label{fig:5}}
  \hfill
  \parbox[t]{0.49\hsize}{%
    \resizebox{\hsize}{!}{\input fig3.tex}
    \caption{Same as Fig.~\ref{fig:5}, but for two Gaussian weight
      functions centered on $0$ and $1$ and with unit variance.}
    \label{fig:6}}%
\end{figure}

\begin{figure}[!t]
  \parbox[t]{0.49\hsize}{%
    \resizebox{\hsize}{!}{\input fig8.tex}
    \caption{The noise term $T_\sigma$ for two Gaussian weights (of
      unit variance) as function of their separation.  Similarly to
      Fig.~\ref{fig:4}, the plot also shows the high-density
      approximations $S_{11} / \rho$.  Note that in this case $S_{11}$
      is also a Gaussian (with double variance).}
    \label{fig:7}}
  \hfill
  \parbox[t]{0.49\hsize}{%
    \resizebox{\hsize}{!}{\input fig11.tex}
    \caption{The Poisson noise $T_P$ for two Gaussian weights (of unit
      variance) for a periodic function of the form $f(x) = \sin k x$
      as a function of the weight separation $\delta = | x_A - x_B |$,
      for a density $\rho = 2$.  Note that, as expected, the Poisson
      noise increases with $k$ and approaches the limit discussed in
      Sect.~\ref{sec:limit-high-low-1} for high frequencies.  More
      precisely, since for a sine function we have $\langle \sin^2 x
      \rangle = 1/2$, Eq.~\eqref{eq:106} gives $T_P \simeq T_\sigma /
      (2 \sigma^2)$ (this can indeed be verified by a comparison with
      Fig.~\ref{fig:7}).  Note also that, while $T_P$ is strictly
      positive for $\delta = 0$, it can became negative (see curve for
      $k = 0.5$) at larger separations.}
    \label{fig:8}}%
\end{figure}

Frequently, a Gaussian weight function of the form
\begin{equation}
  \label{eq:116}
  w(x) = \frac{1}{\sqrt{2 \pi}} \e^{-x^2/2}
\end{equation}
is used.  Although it is not possible to carry out analytical
calculations and obtain $C(w_A, w_B)$, numerical integrations do not
pose any problem.  Figure~\ref{fig:5} shows, for different densities,
the function $w_A w_B C(w_A, w_B)$ for two identical weights $w_A =
w_B$ centered in zero; Fig.~\ref{fig:6} shows the same quantity when
one of the weight function is centered at unity.  Note that, in this
last figure, the largest covariance is at $x = 0.5$, as expected.

Figure~\ref{fig:7} shows the expected measurement noise $T_\sigma$ as
a function of the weight separation.  Similarly to the top-hat weight,
an approximation valid for high density is $T_\sigma = S_{11}/\rho$.
Figure~\ref{fig:8} shows, instead, the Poisson noise $T_P$ expected
for a field $f$ of the form $f(x) = \sin k x$, for different values of
$k$.  Note that the noise, as expected, increases with $k$, and
quickly reaches the ``saturation'' value discussed in
Sect.~\ref{sec:limit-high-low-1}.  Note also that the noise is, at lowest
lowest density, negative for $\delta \simeq 2.5$.

\subsection{Parabolic weight}
\label{sec:parabolic}

\begin{figure}[!t]
  \parbox[t]{0.49\hsize}{%
    \resizebox{\hsize}{!}{\input fig4.tex}
    \caption{Numerical calculations  for 1-dimensional
      parabolic weight functions $w_A = w_B$ centered on $0$ and with
      unit variance.  The various curves shows the function $w_A w_B
      C(w_A, w_B)$ for different densities $\rho$.}
    \label{fig:9}}
  \hfill
  \parbox[t]{0.49\hsize}{%
    \resizebox{\hsize}{!}{\input fig10.tex}
    \caption{The noise term $T_\sigma$ for two parabolic weights as a
      function of their separation (see also Figs.~\ref{fig:4} and
      \ref{fig:7}).}
    \label{fig:10}}%
\end{figure}

Finally, we study of a parabolic weight function of the form
\begin{equation}
  \label{eq:117}
  w(x) =
  \begin{cases}
    3 x^2 / 4 & \text{if $|x| < 1 \; ,$} \\
    0 & \text{otherwise$\; .$}
  \end{cases}
\end{equation}
This function illustrates well some of the peculiarities of finite
support weight functions.  Figure~\ref{fig:9} shows the results of
numerical integrations for $w_A w_B C(w_A, w_B)$ at different
densities $\rho$.  A first interesting point to note is the
discontinuity observed at $x = 1$, which is in agreement with
Eq.~\eqref{eq:81}.  Moreover, as expected from Eq.~\eqref{eq:92}, the
function plotted clearly approaches a constant at low densities
$\rho$.  Finally, the measurement noise $T_\sigma$ is plotted in
Fig.~\ref{fig:10}.

\section{Conclusions}
\label{sec:conclusions}

In this article we have studied in detail the covariance of a widely
used smoothing technique.  The main results obtained are summarized in
the following items.
\begin{enumerate}
\item The covariance is composed of two main terms, $T_\sigma$ and
  $T_\mathrm{P}$, representing measurement errors and Poisson noise,
  respectively; the latter one depends on the field $f$ on which the
  smoothing is performed.
\item Expressions to compute $T_\sigma$ and $T_\mathrm{P}$ have been
  provided.  In particular, it has been shown that both terms can be
  obtained in term of a kernel $C(w_A, w_B)$, which in turn can be
  evaluated from the weight function $w(\vec\theta)$.
\item We have obtained an expansion of the kernel $C(w_A, w_B)$ valid
  at high densities $\rho$.
\item We have shown that $T_\sigma$ has an upper limit, given by
  $\sigma^2$, and a lower limit, provided by Eq.~\eqref{eq:90}.
\item We have evaluated the form of the noise contributions in the
  limiting cases of high and low densities.
\item We have considered three typical cases of weight functions and
  we have evaluated $C(w_A, w_B)$ for them.
\end{enumerate}

Finally, we note that although the smoothing technique considered in
this paper is by far the most widely used in Astronomy, alternative
methods are available.  A statistical characterization of these
methods, using a completely different approach, will be presented in a
future paper (Lombardi \& Schneider, in preparation).

\acknowledgements{This work was partially supported by a grant from
  the Deutsche Forschungsgemeinschaft, and the TMR Network
  ``Gravitational Lensing: New constraints on Cosmology and the
  Distribution of Dark Matter.''}

\appendix

\section{Vanishing weights}
\label{sec:vanishing-weights-1}

In Sect.~\ref{sec:analytical-solution} we have obtained the solution
of the covariance problem under the hypothesis that the weight
function $w(\vec\theta)$ is strictly positive.  In this appendix we
will generalize the results obtained there to non-negative weight
functions (see also Sect.~\ref{sec:vanishing-weights}).

If $w_A$ is allowed to vanish, then we might have a finite probability
that $y_A$ vanishes, i.e.\ a finite probability that no point
$\vec\theta_n$ is inside the support of $w_A$.  A finite probability
in a probability distribution function appears as a Dirac's delta
distribution.  Since this point is quite important for our discussion,
let us make a simple example.  Suppose that $\xi$ is a real random
variable with the following characteristics:
\begin{itemize}
\item $\xi$ has probability $1/3$ to vanish.
\item $\xi$ has probability $2/3$ to be in the range $(0,\infty)$; in
  this range $\xi$ has an exponential distribution.
\end{itemize}
Then we can write the probability distribution function for $\xi$
as
\begin{equation}
  \label{888}
  p_\xi(\xi) = \frac{1}{3} \delta(\xi) + \frac{2}{3} \mathrm{H}(\xi)
  \exp(-\xi) \; ,
\end{equation}
where $\mathrm{H}$ is the Heaviside function [see Eq.~\eqref{eq:40}].
In other words, the probability distribution for $\xi$ includes the
contribution from a Dirac's delta distribution centered on $\xi = 0$.
If $p_\xi$ is known, the probability that $\xi$ is exactly zero ($1/3$
in this example) can be obtained using
\begin{equation}
  \label{889}
  P(\xi = 0) = \int_{0^-}^{0^+} p_\xi(\xi') \, \diff \xi' = \lim_{\xi
  \rightarrow 0^+} \int_0^\xi p_\xi(\xi') \, \diff \xi' \; .
\end{equation}

Let us now turn to our problem.  As mentioned above, for vanishing
weights we expect that $y_A$ might vanish, i.e.\ its probability might
include the contribution from a delta distribution centered on $y_A =
0$; similarly, if $w_B$ is allowed to vanish, the probability
distribution for $y_B$ might include a delta centered in $y_B = 0$.
For a given $y_B$, the probability $P_A(y_B)$ that $y_A$ vanishes is
given by
\begin{equation}
  \label{eq:118}
  P_A(y_B) \equiv \lim_{y_A \rightarrow 0^+} \int_{0^-}^{y_A} p_y(y'_A, y_B) \,
  \diff y'_A = \lim_{s_A \rightarrow \infty} \Lp_A\bigl[ p_y( \cdot, y_B)
  \bigr](s_A) \; ,
\end{equation}
where the properties of Laplace transform have been used in the last
equality (see Appendix~\ref{sec:prop-lapl-transf}).  A similar
equation holds for the probability that $y_B$ vanishes, $P_B(y_A)$.
Note that the Laplace transform in Eq.~\eqref{eq:118} is performed only
with respect to the first variable.  The joint probability $P_{AB}$
that both $y_A$ and $y_B$ vanish is [cf.\ Eq.~\eqref{eq:61}]
\begin{equation}
  \label{eq:119}
  P_{AB} \equiv \lim_{\substack{y_A \rightarrow 0^+ \\ y_B
      \rightarrow 0^+}} \int_{0^-}^{y_A} \diff y'_A \int_{0^-}^{y_B}
  \diff y'_B \, p_y(y'_A, y'_B) = \lim_{\substack{s_A \rightarrow
      \infty \\ s_B \rightarrow \infty}} \Lp[p_y](s_A, s_B) =
  Y(\infty, \infty) \; .
\end{equation}
We then also define [cf.\ Eqs.~\eqref{eq:62}]
\begin{align}
  \label{eq:120}
  P_A & {} \equiv \int_0^\infty P_A(y_B) \, \diff y_B =
  \Lp[p_y](\infty, 0^+) = Y(\infty, 0^+) \; , \\
  \label{eq:121}
  P_B & {} \equiv \int_0^\infty P_B(y_A) \, \diff y_A = \Lp[p_y](0^+,
  \infty) = Y(0^+, \infty) \; .
\end{align}
Using Eq.~\eqref{eq:45}, we find $P_A = \exp (-\rho \pi_A)$, $P_B =
\exp (-\rho \pi_B)$, and $P_{AB} = \exp (-\rho \pi_{A \cup B})$, where
$\pi_A$ is the area of the support of $w_A$, $\pi_B$ is the area of
the support of $w_B$, and $\pi_{A \cup B}$ is the area of the union of
the two supports.  This result is of course not surprising and has
been already derived in the paragraph before Eq.~\eqref{eq:61} using a
different approach.

For vanishing weights, we decided to use the following prescription:
We discard, in the ensemble average for $\Cov(\tilde f; \vec\theta_A,
\vec\theta_B)$ , the configurations $\{ \vec\theta_n \}$ for which the
function $\tilde f$ is not defined either at $\vec\theta_A$ or at
$\vec\theta_B$.  In order to implement this prescription, we can
explicitly modify the probability distribution $p_y$ and exclude ``by
hand'' cases where the denominator of Eq.~\eqref{eq:19} vanishes; for
the purpose, we consider separately cases where $w_A$ or $w_B$ vanish.
We define a new probability distribution for $(y_A, y_B)$ which
accounts for vanishing weights:
\begin{equation}
  \label{eq:122}
  \bar p_y(y_A, y_B) \equiv
  \begin{cases}
    p_y(y_A, y_B)
    & \text{if $w_A \neq 0$, $w_B \neq 0$} \; , \\
    \bigl[ p_y(y_A, y_B) - P_A(y_B) \delta(y_A) \bigr] / (1 - P_A)
    & \text{if $w_A = 0$, $w_B \neq 0$} \; , \\
    \bigl[ p_y(y_A, y_B) - P_B(y_A) \delta(y_B) \bigr] / (1 - P_B)
    & \text{if $w_A \neq 0$, $w_B = 0$} \; , \\
    \bigl[ p_y(y_A, y_B) - P_A(y_B) \delta(y_A) - P_B(y_A) \delta(y_B) 
    + P_{AB} \delta(y_A) \delta(y_B) \bigr] \nu
    & \text{if $w_A = 0$, $w_B = 0$} \; .
  \end{cases}
\end{equation}
We recall that $\nu = 1/(1 - P_A - P_B + P_{AB})$.  In constructing
this probability, first we have explicitly removed the degenerate
situations, then we have renormalized the resulting probability.  Note
that the normalization factor in the last case, namely $1 - P_A - P_B
+ P_{AB}$, comes from the so-called ``inclusion-exclusion principle''
($1 - P_A - P_B + P_{AB}$ is the probability that \textit{both\/}
$f_A$ \textit{and\/} $f_B$ are defined).  Using this new probability
distribution in the definition \eqref{eq:32} for $Y$ we obtain
\begin{equation}
  \label{eq:123}
  Y(s_A, s_B) =
  \begin{cases}
    \e^{\rho Q(s_A, s_B)}
    & \text{if $w_A \neq 0$, $w_B \neq 0$} \; , \\
    \bigl[ \e^{\rho Q(s_A, s_B)} - \e^{\rho Q(\infty, s_B)} \bigr]
    / (1 - P_A)
    & \text{if $w_A = 0$, $w_B \neq 0$} \; , \\
    \bigl[ \e^{\rho Q(s_A, s_B)} - \e^{\rho Q(s_A, \infty)} \bigr]
    / (1 - P_B)
    & \text{if $w_A \neq 0$, $w_B = 0$} \; , \\
    \bigl[ \e^{\rho Q(s_A, s_B)} - \e^{\rho Q(s_A, \infty)} - \e^{\rho
    Q(\infty, s_B)} + \e^{\rho Q(\infty, \infty)} \bigr] \nu 
    & \text{if $w_A = 0$, $w_B = 0$} \; .
  \end{cases}
\end{equation}
Finally, we need to change the normalization factor in
Eq.~\eqref{eq:47} in order to account for cases where $y_A$ or $y_B$
are vanishing.  Indeed, the correcting factor $C(w_A, w_B)$ has been
obtained by assuming that all objects can populate all the plane with
uniform probability distribution [cf.\ Eq.~\eqref{eq:25}]; now,
however, a fraction $(P_A + P_B - P_{AB})$ of configurations have been
excluded.  Hence we have
\begin{equation}
  \label{eq:124}
  C(w_A, w_B) = \frac{\rho^2}{1 - P_A - P_B + P_{AB}} \Lp[Y](w_A, w_B)
  \; .
\end{equation}
This complete the discussion of vanishing weights.

\section{Moments expansion}
\label{sec:moments-expansion-1}

In Sect.~\ref{sec:moments-expansion} we have written the moments
expansion for $C(w_A, w_B)$.  Here we complete the discussion by
providing a proof for that result.

\begin{table*}[!t]
  \centering
  \begin{tabular}{cccccc}
    Order & \multicolumn{5}{c}{$M_{ij}$} \\
    ($i+j$) & $j=0$ & $j=1$ & $j=2$ & $j=3$ & $j=4$ \\
    \hline
    0 & 1 & -- & -- & -- & -- \\
    1 & 0 & 0 & -- & -- & --\\
    2 & $\rho S_{20}$ & $\rho S_{11}$ & $\rho S_{02}$ & -- & --\\
    3 & $\rho S_{30}$ & $\rho S_{21}$ & $\rho S_{12}$ & $\rho S_{03}$
    & -- \\
    4 & $\rho S_{40} + 3 \rho^2 S_{20}$ & $\rho S_{30} + 3 \rho^2
    S_{20} S_{11}$ & $\rho S_{22} + \rho^2 S_{02} S_{20} + 2 \rho^2
    S_{11} S_{11}$ & $\rho S_{03} + 3 \rho^2
    S_{02} S_{11}$ & $\rho S_{04} + 3 \rho^2 S_{02}$
  \end{tabular}
  \caption{Moments $M_{ij}$ up to the fourth order.  The table shows,
    for each row, the values of $M_{ij}$ with $(i + j)$, the order,
    fixed.  Hence, for example, the row for order 2 shows $M_{20}$,
    $M_{11}$, and $M_{02}$ in sequence.}
  \label{tab:1}
\end{table*}

At high densities, $y_A$ and $y_B$ are basically Gaussian random
variables with average values $\bar y_A$ and $\bar y_B$ (we anticipate
here that these averages are given by the density $\rho$).  Hence, we
can expand them in the definition of $C(w_A, w_B)$:
\begin{align}
  \label{eq:125}
  C(w_A, w_B) & {} = \rho^2 \int_0^\infty \diff y_A \int_0^\infty
  \diff y_B \frac{p_y(y_A, y_B)}{(w_A + y_A) (w_B + y_B)} \notag\\
  & {} = \rho^2 \int_0^\infty \diff y_A \int_0^\infty \diff y_B
  \frac{p_y(y_A, y_B)}{(w_A + \bar y_A) (w_B + \bar y_B)}
  \biggl[ \sum_{i=0}^\infty \left( \frac{\bar y_A - y_A}{w_A + \bar
        y_A} \right)^i \biggr] \biggl[ \sum_{j=0}^\infty \left( \frac{\bar
        y_B - y_B}{w_B + \bar y_B} \right)^j \biggr] \notag\\
  & {} = \rho^2 \sum_{i, j=0}^\infty (-1)^{i+j} \frac{M_{ij}}{( \bar
    y_A + w_A)^{i+1} (\bar y_B + w_B)^{j+1}} \; ,
\end{align}
where $M_{ij}$ are the ``centered'' moments of $p_y$:
\begin{equation}
  \label{eq:126}
  M_{ij} \equiv \int_0^\infty \diff y_A \int_0^\infty \diff y_B
  p_y(y_A, y_B) \, (y_A - \bar y_A)^i (y_B - \bar y_B)^j \; .
\end{equation}
The centered moments can be expressed in terms of the ``un-centered''
ones, defined as
\begin{equation}
  \label{eq:127}
  \mathcal{M}_{ij} \equiv \int_0^\infty \diff y_A \int_0^\infty \diff
  y_B p_y(y_A, y_B) \, y_A^i y_B^j = (-1)^{i+j} Y^{(i,j)}(0, 0) \; .
\end{equation}
Here $Y^{(i,j)}(0, 0)$ is the $i$-th partial derivative on $s_A$ and
$j$-th partial derivative on $s_B$ of $Y(s_A, s_B)$, evaluated at $(0,
0)$.  These, in turn, can be expressed as derivatives of $Q$.  For the
first terms we have
\begin{align}
  \label{eq:128}
  Y^{(0,0)}(0,0) =& {} Y(0,0) = 1 \; , &
  Y^{(1,0)}(0,0) =& {} \rho Q^{(1,0)}(0,0) \; , &
  Y^{(0,1)}(0,0) =& {} \rho Q^{(0,1)}(0,0) \; , \\
  \label{eq:129}
  Y^{(2,0)}(0,0) =& {} \rho Q^{(2,0)}(0,0) + \bigl[ \rho Q^{(1,0)}
  (0,0) \bigr]^2 \; , \\
  \label{eq:130}
  Y^{(1,1)}(0,0) =& {} \rho Q^{(1,1)}(0,0) + \bigl[ \rho Q^{(1,0)}(0,0)
  \bigr] \bigl[ \rho Q^{(0,1)}(0,0) \bigr] \; , \\
  \label{eq:131}
  Y^{(0,2)}(0,0) =& {} \rho Q^{(0,2)}(0,0) + \bigl[ \rho
  Q^{(0,1)}(0,0) \bigr]^2 \; .
\end{align}
Finally, the derivatives of $Q$ can be evaluated as
\begin{equation}
  \label{eq:132}
  Q^{(i,j)}(0, 0) = (-1)^{i+j} S_{ij} \; ,
\end{equation}
where $S_{ij}$, we recall, is given by Eq.~\eqref{eq:67}.  Note that
$S_{01} = S_{10} = 1$ because of the normalization of $w_A$ and $w_B$,
and thus, as already anticipated, $\bar y_A = \bar y_B = \rho$.  In
summary, we find
\begin{align}
  \label{eq:133}
  M_{00} & {} = 1 \; , &
  M_{10} & {} = M_{01} = 0 \; , &
  M_{20} & {} = \mathcal{M}_{20} - (\mathcal{M}_{10})^2 = \rho S_{20}
  \; , \\
  \label{eq:134}
  M_{11} & {} = \mathcal{M}_{11} - \mathcal{M}_{10} \mathcal{M}_{01} =
  \rho S_{11} \; , &
  M_{02} & {} = \mathcal{M}_{20} - (\mathcal{M}_{01})^2 = \rho S_{20} \; .
\end{align}
We stress that, in general, it is not true that $M_{ij} = \rho S_{ij}$
(more complex expressions are encountered for higher order terms; cf.\ 
the last term in Eq.~\eqref{eq:66}).  Finally, we can write the
expansion of $C(w_A, w_B)$:
\begin{equation}
  \label{eq:135}
  C(w_A, w_B) \simeq \frac{\rho^2}{(\rho + w_A) (\rho + w_B)} +
  \frac{\rho^3 S_{20}}{(\rho + w_A)^3 (\rho + w_B)} + \frac{\rho^3
    S_{11}}{(\rho + w_A)^2 (\rho + w_B)^2} +
  \frac{\rho^3 S_{02}}{(\rho + w_A) (\rho + w_B)^3} \; .
\end{equation}
This is precisely Eq.~\eqref{eq:68}.  Using the same technique and a
little more perseverance, we can also obtain higher order terms.  In
particular, Table~\ref{tab:1} reports the moments $M_{ij}$ defined in
Eq.~\eqref{eq:126} up to the forth order.  This table, together with 
Eq.~\eqref{eq:125}, can be used to write an accurate moment expansion
of $C(w_a, w_B)$.

\section{Varying weights}
\label{sec:varying-weights}

In Paper~I we have considered a modified version of the estimator
\eqref{eq:1} which allows for the use of supplementary weights.
Suppose that we measure a given field $f(\vec\theta)$ at some
positions $\vec\theta_n$ of the sky.  Suppose also that we use a
weight $u_n$ for each object observed, so that we replace
Eq.~\eqref{eq:1} with
\begin{equation}
  \label{eq:136}
  \tilde f(\vec\theta) \equiv \dfrac{\sum_{n=1}^N \hat f_n
  u_n w(\vec\theta - \vec\theta_n)}{\sum_{n=1}^N u_n w(\vec\theta -
  \vec\theta_n)} \; .
\end{equation}
For example, if we have at our disposal some error estimate $\sigma_n$
for each object, we might use the weighting scheme $u_n =
1/\sigma_n^2$ in order to minimize the noise of the estimator
\eqref{eq:136}.

A statistical study of the expectation value of this estimator has
already been carried out in Paper~I.  Here we proceed further and
study its covariance under the same assumptions as the ones used for
the study of Eq.~\eqref{eq:1} in the main text.  However, since one of
the main reasons to use weights is some knowledge on the variance of
each object, we use a generalized form of Eq.~\eqref{eq:7}:
\begin{equation}
  \label{eq:137}
  \bigl\langle \bigl[ \hat f_n - f(\vec\theta_n) \bigr]
  \bigl[ \hat f_m - f(\vec\theta_m) \bigr] \bigr\rangle = \sigma^2(u_n)
  \delta_{nm} \; .
\end{equation}
Note, in particular, that the variance is assumed to depend on $u_n$
(or, equivalently, the weight is assumed to depend on the variance).
Similarly to Paper~I, we also assume that, for each object $n$, the
weight $u_n$ is independent of the position $\vec\theta_n$ and of the
measured signal $\hat f_n$, and that each $u_n$ follows a known
probability distribution $p_u$.

In Paper~I we have shown that the average value of $\tilde
f(\vec\theta)$ can be calculated using the equations
\begin{align}
  \label{eq:138}
  R_X(s_X) \equiv {} & \int_\Omega \diff^2 \theta \int_0^\infty \diff u
  \, p_u(u) \bigl[ \e^{-s_X u w_X(\vec\theta)} - 1 \bigr] = \int_0^\infty
  p_u(u) Q_X(u s_X) \, \diff u \; , \\
  \label{eq:139}
  Y_X(s_X) \equiv {} & \exp\bigl[ \rho R_X(s_X) \bigr] \; , \\
  \label{eq:140}
  B_X(v_X) \equiv {} & \rho \Lp[Y_X](v_X) \; , \\
  \label{eq:141}
  \langle \tilde f_X \rangle = {} & \int_\Omega \diff^2 \theta
  \int_0^\infty \diff u \, p_u(u) f(\vec\theta) B_X \bigl( u
  w_X(\vec\theta) \bigr) u w_X(\vec\theta) \; .
\end{align}
We now evaluate the covariance of the estimator \eqref{eq:136} using a
technique similar to the one used in Sect.~\ref{sec:eval-covar}.  We have
\begin{equation}
  \label{eq:142}
  \bigl\langle \tilde f_A \tilde f_B \bigr\rangle = \frac{1}{A^N}
  \int_\Omega \diff^2 \theta_1 \int_0^\infty \diff u_1 \, p_u(u_1)
  \dotsi \int_\Omega \diff^2 \theta_N \int_0^\infty \diff u_N \,
  p_u(u_N) \dfrac{ \bigl\langle \bigl[ \sum_n \hat f_n u_n
  w_A(\vec\theta_n) \bigr] \bigl[ \sum_m \hat f_m u_m
  w_B(\vec\theta_m) \bigr] \bigr\rangle}{\bigl[ \sum_n u_n
  w_A(\vec\theta_n) \bigr] \bigl[ \sum_m u_m w_B(\vec\theta_m) \bigr]}
  \; .
\end{equation}
As usual we consider separately the cases $n = m$ and $n \neq m$, thus
obtaining the two terms $T_1$ and $T_2$:
\begin{align}
  \label{eq:143}
  T_1 = {} & \frac{N}{A^N} \int_\Omega \diff^2 \theta_1 \int_0^\infty
  \diff u_1 \, p_u(u_1) \dotsi \int_\Omega \diff^2 \theta_N
  \int_0^\infty \diff u_N \, p_u(u_N) \dfrac{\bigl[
    f^2(\vec\theta_1) + \sigma^2(u_1) \bigr] u_1^2 w_A(\vec\theta_1)
    w_B(\vec\theta_1)}{\bigl[ \sum_n u_n w_A(\vec\theta_n) \bigr] \bigl[
    \sum_m u_m w_B(\vec\theta_m) \bigr]} \; , \\
  \label{eq:144}
  T_2 = {} & \frac{N (N - 1)}{A^N} \int_\Omega \diff^2 \theta_1
  \int_0^\infty \diff u_1 \, p_u(u_1) \dotsi \int_\Omega \diff^2 \theta_N
  \int_0^\infty \diff u_N \, p_u(u_N) \dfrac{f(\vec\theta_1) u_1
  w_A(\vec\theta_1) f(\vec\theta_2) u_2 w_B(\vec\theta_2)}{\bigl[
  \sum_n u_n w_A(\vec\theta_n) \bigr] \bigl[ \sum_m u_m
  w_B(\vec\theta_m) \bigr]} \; .
\end{align}
Let us introduce new variables $v_{Xn} = u_n w_X(\vec\theta_n)$ (with
$X \in \{ A, B \}$) for the combination of weights, and let us define
similarly to Eq.~\eqref{eq:26} $y_X \equiv \sum_{n=2}^N v_n$.  Then
the probability distributions for $v_{Xn}$ and $y_X$ can be evaluated
using the set of equations
\begin{align}
  \label{eq:145}
  p_v(v_A, v_B) = {} & \frac{1}{A} \int_\Omega \diff^2 \theta
  \int_0^\infty \diff u \, p_u(u) \delta \bigl( v_A - u
  w_A(\vec\theta) \bigr) \delta \bigl( v_B - u w_B(\vec\theta) \bigr)
  \; , \\
  \label{eq:146}
  p_y(y_A, y_B) = {} & \int_0^\infty \diff v_{A2} \int_0^\infty \diff
  v_{B2} \, p_v(v_{A2}, v_{B2}) \dotsi \int_0^\infty \diff v_{AN}
  \int_0^\infty \diff v_{BN} \, p_v(v_{AN}, v_{BN}) \delta(y_A -
  v_{A2} - \cdots - v_{AN}) \notag\\
  & \phantom{\int_0^\infty} {} \times \delta(y_B - v_{B2} - \cdots -
  v_{BN}) \; .
\end{align}
Again, it is convenient to consider the Laplace transforms of these
two probability distributions:
\begin{align}
  \label{eq:147}
  V(s_A, s_B) = {} & \Lp[p_v](s_A, s_B) = \frac{1}{A} \int_0^\infty
  \diff u \, p_u(u) \int_\Omega \diff^2 \theta \, \e^{-s_A u
  w_A(\vec\theta) - s_B u w_B(\vec\theta)} = \int_0^\infty p(u) W(u
  s_A, u s_B) \, \diff u \; , \\
  \label{eq:148}
  Y(s_A, s_B) = {} & \Lp[p_y](s_A, s_B) = \bigl[ V(s_A, s_B)
  \bigr]^{N-1} \; .
\end{align}
In the continuous limit we define instead
\begin{align}
  \label{eq:149}
  R(s_A, s_B) \equiv {} & \int_\Omega \diff^2 \theta \int_0^\infty
  \diff u \, p_u(u) \bigl[ \e^{- s_A u w_A(\vec\theta) - s_B u
    w_B(\vec\theta)} - 1 \bigr] = \int_0^\infty p_u(u) Q(u s_A, u s_B) \,
  \diff u \; , \\
  \label{eq:150}
  Y(s_A, s_B) = {} & \exp\bigl[ \rho R(s_A, s_B) \bigr] \; .
\end{align}
Finally, the equivalent of the correcting factor $C(w_A, w_B)$ [cf.\
Eq.~\eqref{eq:11}] is, in our case, the quantity
\begin{equation}
  \label{eq:151}
  B(v_A, v_B) \equiv \rho^2 \int_0^\infty \diff y_A \int_0^\infty
  \diff y_B \frac{p_y(y_A, y_B)}{(v_A + y_A) (v_B + y_B)} = \rho
  \Lp[Y](v_A, v_B) \; .
\end{equation}
The quantity $B$ can be used to evaluate $T_1$: In fact, we
have
\begin{align}
  \label{eq:152}
  T_1 = {} & \frac{N}{A} \int_\Omega \diff^2 \theta_1 \int_0^\infty
  \diff u_1 \, p_u(u_1) \int_0^\infty \diff y_A \int_0^\infty \diff
  y_B \, p_y(y_A, y_B) \frac{\bigl[ f^2(\vec\theta_1) + \sigma^2(u_1)
    \bigr] u_1^2 w_A(\vec\theta_1) w_B(\vec\theta_1)}{\bigl[ u_1
    w_A(\vec\theta_1) + y_A \bigr] \bigl[ u_1 w_B(\vec\theta_1) + y_B
  \bigr]} \notag\\
  {} = {} & \frac{1}{\rho} \int_\Omega \diff^2 \theta_1 \int_0^\infty
  \diff u_1 p_u(u_1) \bigl[ f^2(\vec\theta_1) + \sigma^2(u_1) \bigr]
  u_1^2 w_A(\vec\theta_1) w_B(\vec\theta_1) B \bigl(u_1
  w_A(\vec\theta_1), u_1 w_B(\vec\theta_1) \bigr) \; .
\end{align}
Similarly, for $T_2$ we obtain
\begin{align}
  \label{eq:153}
  T_2 = {} & \int_\Omega \diff^2 \theta_1 \int_0^\infty \diff u_1
  p_u(u_1) \int_\Omega \diff^2 \theta_2 \int_0^\infty \diff u_2
  p_u(u_2) f(\vec\theta_1) f(\vec\theta_2) u_1
  w_A(\vec\theta_1) u_2 w_B(\vec\theta_2) \notag\\
  & \phantom{\int_\Omega} {} \times B \bigl(u_1 w_A(\vec\theta_1) +
  u_2 w_A(\vec\theta_2), u_1 w_B(\vec\theta_1) + u_2 w_B(\vec\theta_2)
  \bigr) \; .
\end{align}
The final evaluation of $\Cov(\tilde f)$ then proceeds similarly to
what done in the main text for the estimator \eqref{eq:1}.

\section{Properties of the Laplace transform}
\label{sec:prop-lapl-transf}

For the convenience of the reader, we summarize in this appendix some
useful properties of the Laplace transform.  Proofs of the results
stated here can be found in any advanced analysis book
\citep[e.g.][]{Arfken}.  Although in this paper we have been dealing
mainly with Laplace transforms of two-argument functions, we write the
properties below for the case of a function of a single argument for
two main reasons: (i) The generalization to functions of several
arguments is in most cases trivial; (ii) Several properties can be
better understood in the simpler case considered here.

Suppose that a function $f(x)$ of a real argument $x$ is given.  Its
Laplace transform is defined as
\begin{equation}
  \label{eq:154}
  \Lp[f](s) \equiv {} \lim_{x \rightarrow 0^-} \int_x^\infty \diff
  x' \, f(x') \e^{-s x'} = \int_{0^-}^\infty \diff x \, f(x) \e^{-s x}
  \; .
\end{equation}
Note that we use $0^-$ as lower integration limit in this definition.

The Laplace transform is a \textit{linear operator\/}; hence, if
$\alpha$ and $\beta$ are two real numbers and $g(x)$ is a function of
real argument $x$, we have $\Lp[\alpha f + \beta g] = \alpha \Lp[f] +
\beta \Lp[g]$.

The Laplace transform of the derivative of $f$ can be expressed in
terms of the Laplace transform of $f$.  In particular, we have
\begin{equation}
  \label{eq:155}
  \Lp \bigl[ f' \bigr](s) = s \Lp[f](s) - f(0^-) \; .
\end{equation}
This equation can be generalized to higher order derivatives.  Calling
$f^{(n)}$ the $n$-th derivative of $f$, we have
\begin{equation}
  \label{eq:156}
  \Lp \bigl[ f^{(n)} \bigr](s) = s^n \Lp[f](s) - \sum_{i=0}^{n-1}
  s^{n-i-1} f^{(n)}(0^-) \; .
\end{equation}
Surprisingly, this equation holds if, for $n$ negative, we consider
$f^{(n)}$ to be the $-n$-th integral of $f$; note that in this case
the summation disappears.  Hence, for example, we have
\begin{equation}
  \label{eq:157}
  \Lp \biggl[ \int_{0^-}^x f(x') \, \diff x' \biggr](s) = \frac{1}{s}
  \Lp[f](s) \; .
\end{equation}

Often, properties of the Laplace transform come in pairs: For every
property there is a similar one where the role of $f$ and $\Lp[f]$ are
swapped.  Here is the ``dual'' of property \eqref{eq:155}:
\begin{equation}
  \label{eq:158}
  \Lp\bigl[ x f(x) \bigr](s) = - \frac{\diff \Lp[f](s)}{\diff s} \; ,
\end{equation}
or, more generally,
\begin{equation}
  \label{eq:159}
  \Lp\bigl[ x^n f(x) \bigr](s) = (-1)^n \frac{\diff^n \Lp[f](s)}{\diff
  s^n} \; .
\end{equation}
A similar equation holds for ``negative'' derivatives, i.e.\ integrals
of the Laplace transform.  In this case, however, it is convenient to
change the integration limits to $(s, \infty)$.  In summary, we can
write 
\begin{equation}
  \label{eq:160}
  \Lp\bigl[ f(x) / x \bigr](s) = \int_s^\infty \Lp[f](s') \, \diff s'
  \; .
\end{equation}

Given a positive number $a$, the Laplace transform of the function $f$
shifted by $a$ is given by
\begin{equation}
  \label{eq:161}
  \Lp \bigl[ f(x - a) \mathrm{H}(x) \bigr](s) = \Lp[f](s) \e^{- s a}
  \; ,
\end{equation}
where $\mathrm{H}$ is the Heaviside function defined in
Eq.~\eqref{eq:40}.  A dual of this property can also be written:
\begin{equation}
  \label{eq:162}
  \Lp \bigl[ f(x) \e^{b x} \bigr](s) = \Lp[f](s - b) \; .
\end{equation}

Finally, we consider two useful relationships between limiting values
of $f$ and $\Lp[f]$:
\begin{align}
  \label{eq:163}
  \lim_{x \rightarrow 0^+} f(x) & {} = \lim_{s \rightarrow \infty} s
  \Lp[f](s) \; , &
  \lim_{x \rightarrow \infty} f(x) & {} = \lim_{s \rightarrow 0^+} s
  \Lp[f](s) \; .
\end{align}

\bibliographystyle{aa}
\bibliography{../lens-refs}

\end{document}

%% file: fig1.tex
% GNUPLOT: LaTeX picture with Postscript
\begingroup%
  \makeatletter%
  \newcommand{\GNUPLOTspecial}{%
    \@sanitize\catcode`\%=14\relax\special}%
  \setlength{\unitlength}{0.1bp}%
{\GNUPLOTspecial{!
%!PS-Adobe-2.0
%%Title: fig1.tex
%%Creator: gnuplot 3.7 patchlevel 2
%%CreationDate: Sun Jun 16 23:24:35 2002
%%DocumentFonts: 
%%BoundingBox: 0 0 360 216
%%Orientation: Landscape
%%Pages: (atend)
%%EndComments
/gnudict 256 dict def
gnudict begin
/Color false def
/Solid false def
/gnulinewidth 5.000 def
/userlinewidth gnulinewidth def
/vshift -33 def
/dl {10 mul} def
/hpt_ 31.5 def
/vpt_ 31.5 def
/hpt hpt_ def
/vpt vpt_ def
/M {moveto} bind def
/L {lineto} bind def
/R {rmoveto} bind def
/V {rlineto} bind def
/vpt2 vpt 2 mul def
/hpt2 hpt 2 mul def
/Lshow { currentpoint stroke M
  0 vshift R show } def
/Rshow { currentpoint stroke M
  dup stringwidth pop neg vshift R show } def
/Cshow { currentpoint stroke M
  dup stringwidth pop -2 div vshift R show } def
/UP { dup vpt_ mul /vpt exch def hpt_ mul /hpt exch def
  /hpt2 hpt 2 mul def /vpt2 vpt 2 mul def } def
/DL { Color {setrgbcolor Solid {pop []} if 0 setdash }
 {pop pop pop Solid {pop []} if 0 setdash} ifelse } def
/BL { stroke userlinewidth 2 mul setlinewidth } def
/AL { stroke userlinewidth 2 div setlinewidth } def
/UL { dup gnulinewidth mul /userlinewidth exch def
      dup 1 lt {pop 1} if 10 mul /udl exch def } def
/PL { stroke userlinewidth setlinewidth } def
/LTb { BL [] 0 0 0 DL } def
/LTa { AL [1 udl mul 2 udl mul] 0 setdash 0 0 0 setrgbcolor } def
/LT0 { PL [] 1 0 0 DL } def
/LT1 { PL [4 dl 2 dl] 0 1 0 DL } def
/LT2 { PL [2 dl 3 dl] 0 0 1 DL } def
/LT3 { PL [1 dl 1.5 dl] 1 0 1 DL } def
/LT4 { PL [5 dl 2 dl 1 dl 2 dl] 0 1 1 DL } def
/LT5 { PL [4 dl 3 dl 1 dl 3 dl] 1 1 0 DL } def
/LT6 { PL [2 dl 2 dl 2 dl 4 dl] 0 0 0 DL } def
/LT7 { PL [2 dl 2 dl 2 dl 2 dl 2 dl 4 dl] 1 0.3 0 DL } def
/LT8 { PL [2 dl 2 dl 2 dl 2 dl 2 dl 2 dl 2 dl 4 dl] 0.5 0.5 0.5 DL } def
/Pnt { stroke [] 0 setdash
   gsave 1 setlinecap M 0 0 V stroke grestore } def
/Dia { stroke [] 0 setdash 2 copy vpt add M
  hpt neg vpt neg V hpt vpt neg V
  hpt vpt V hpt neg vpt V closepath stroke
  Pnt } def
/Pls { stroke [] 0 setdash vpt sub M 0 vpt2 V
  currentpoint stroke M
  hpt neg vpt neg R hpt2 0 V stroke
  } def
/Box { stroke [] 0 setdash 2 copy exch hpt sub exch vpt add M
  0 vpt2 neg V hpt2 0 V 0 vpt2 V
  hpt2 neg 0 V closepath stroke
  Pnt } def
/Crs { stroke [] 0 setdash exch hpt sub exch vpt add M
  hpt2 vpt2 neg V currentpoint stroke M
  hpt2 neg 0 R hpt2 vpt2 V stroke } def
/TriU { stroke [] 0 setdash 2 copy vpt 1.12 mul add M
  hpt neg vpt -1.62 mul V
  hpt 2 mul 0 V
  hpt neg vpt 1.62 mul V closepath stroke
  Pnt  } def
/Star { 2 copy Pls Crs } def
/BoxF { stroke [] 0 setdash exch hpt sub exch vpt add M
  0 vpt2 neg V  hpt2 0 V  0 vpt2 V
  hpt2 neg 0 V  closepath fill } def
/TriUF { stroke [] 0 setdash vpt 1.12 mul add M
  hpt neg vpt -1.62 mul V
  hpt 2 mul 0 V
  hpt neg vpt 1.62 mul V closepath fill } def
/TriD { stroke [] 0 setdash 2 copy vpt 1.12 mul sub M
  hpt neg vpt 1.62 mul V
  hpt 2 mul 0 V
  hpt neg vpt -1.62 mul V closepath stroke
  Pnt  } def
/TriDF { stroke [] 0 setdash vpt 1.12 mul sub M
  hpt neg vpt 1.62 mul V
  hpt 2 mul 0 V
  hpt neg vpt -1.62 mul V closepath fill} def
/DiaF { stroke [] 0 setdash vpt add M
  hpt neg vpt neg V hpt vpt neg V
  hpt vpt V hpt neg vpt V closepath fill } def
/Pent { stroke [] 0 setdash 2 copy gsave
  translate 0 hpt M 4 {72 rotate 0 hpt L} repeat
  closepath stroke grestore Pnt } def
/PentF { stroke [] 0 setdash gsave
  translate 0 hpt M 4 {72 rotate 0 hpt L} repeat
  closepath fill grestore } def
/Circle { stroke [] 0 setdash 2 copy
  hpt 0 360 arc stroke Pnt } def
/CircleF { stroke [] 0 setdash hpt 0 360 arc fill } def
/C0 { BL [] 0 setdash 2 copy moveto vpt 90 450  arc } bind def
/C1 { BL [] 0 setdash 2 copy        moveto
       2 copy  vpt 0 90 arc closepath fill
               vpt 0 360 arc closepath } bind def
/C2 { BL [] 0 setdash 2 copy moveto
       2 copy  vpt 90 180 arc closepath fill
               vpt 0 360 arc closepath } bind def
/C3 { BL [] 0 setdash 2 copy moveto
       2 copy  vpt 0 180 arc closepath fill
               vpt 0 360 arc closepath } bind def
/C4 { BL [] 0 setdash 2 copy moveto
       2 copy  vpt 180 270 arc closepath fill
               vpt 0 360 arc closepath } bind def
/C5 { BL [] 0 setdash 2 copy moveto
       2 copy  vpt 0 90 arc
       2 copy moveto
       2 copy  vpt 180 270 arc closepath fill
               vpt 0 360 arc } bind def
/C6 { BL [] 0 setdash 2 copy moveto
      2 copy  vpt 90 270 arc closepath fill
              vpt 0 360 arc closepath } bind def
/C7 { BL [] 0 setdash 2 copy moveto
      2 copy  vpt 0 270 arc closepath fill
              vpt 0 360 arc closepath } bind def
/C8 { BL [] 0 setdash 2 copy moveto
      2 copy vpt 270 360 arc closepath fill
              vpt 0 360 arc closepath } bind def
/C9 { BL [] 0 setdash 2 copy moveto
      2 copy  vpt 270 450 arc closepath fill
              vpt 0 360 arc closepath } bind def
/C10 { BL [] 0 setdash 2 copy 2 copy moveto vpt 270 360 arc closepath fill
       2 copy moveto
       2 copy vpt 90 180 arc closepath fill
               vpt 0 360 arc closepath } bind def
/C11 { BL [] 0 setdash 2 copy moveto
       2 copy  vpt 0 180 arc closepath fill
       2 copy moveto
       2 copy  vpt 270 360 arc closepath fill
               vpt 0 360 arc closepath } bind def
/C12 { BL [] 0 setdash 2 copy moveto
       2 copy  vpt 180 360 arc closepath fill
               vpt 0 360 arc closepath } bind def
/C13 { BL [] 0 setdash  2 copy moveto
       2 copy  vpt 0 90 arc closepath fill
       2 copy moveto
       2 copy  vpt 180 360 arc closepath fill
               vpt 0 360 arc closepath } bind def
/C14 { BL [] 0 setdash 2 copy moveto
       2 copy  vpt 90 360 arc closepath fill
               vpt 0 360 arc } bind def
/C15 { BL [] 0 setdash 2 copy vpt 0 360 arc closepath fill
               vpt 0 360 arc closepath } bind def
/Rec   { newpath 4 2 roll moveto 1 index 0 rlineto 0 exch rlineto
       neg 0 rlineto closepath } bind def
/Square { dup Rec } bind def
/Bsquare { vpt sub exch vpt sub exch vpt2 Square } bind def
/S0 { BL [] 0 setdash 2 copy moveto 0 vpt rlineto BL Bsquare } bind def
/S1 { BL [] 0 setdash 2 copy vpt Square fill Bsquare } bind def
/S2 { BL [] 0 setdash 2 copy exch vpt sub exch vpt Square fill Bsquare } bind def
/S3 { BL [] 0 setdash 2 copy exch vpt sub exch vpt2 vpt Rec fill Bsquare } bind def
/S4 { BL [] 0 setdash 2 copy exch vpt sub exch vpt sub vpt Square fill Bsquare } bind def
/S5 { BL [] 0 setdash 2 copy 2 copy vpt Square fill
       exch vpt sub exch vpt sub vpt Square fill Bsquare } bind def
/S6 { BL [] 0 setdash 2 copy exch vpt sub exch vpt sub vpt vpt2 Rec fill Bsquare } bind def
/S7 { BL [] 0 setdash 2 copy exch vpt sub exch vpt sub vpt vpt2 Rec fill
       2 copy vpt Square fill
       Bsquare } bind def
/S8 { BL [] 0 setdash 2 copy vpt sub vpt Square fill Bsquare } bind def
/S9 { BL [] 0 setdash 2 copy vpt sub vpt vpt2 Rec fill Bsquare } bind def
/S10 { BL [] 0 setdash 2 copy vpt sub vpt Square fill 2 copy exch vpt sub exch vpt Square fill
       Bsquare } bind def
/S11 { BL [] 0 setdash 2 copy vpt sub vpt Square fill 2 copy exch vpt sub exch vpt2 vpt Rec fill
       Bsquare } bind def
/S12 { BL [] 0 setdash 2 copy exch vpt sub exch vpt sub vpt2 vpt Rec fill Bsquare } bind def
/S13 { BL [] 0 setdash 2 copy exch vpt sub exch vpt sub vpt2 vpt Rec fill
       2 copy vpt Square fill Bsquare } bind def
/S14 { BL [] 0 setdash 2 copy exch vpt sub exch vpt sub vpt2 vpt Rec fill
       2 copy exch vpt sub exch vpt Square fill Bsquare } bind def
/S15 { BL [] 0 setdash 2 copy Bsquare fill Bsquare } bind def
/D0 { gsave translate 45 rotate 0 0 S0 stroke grestore } bind def
/D1 { gsave translate 45 rotate 0 0 S1 stroke grestore } bind def
/D2 { gsave translate 45 rotate 0 0 S2 stroke grestore } bind def
/D3 { gsave translate 45 rotate 0 0 S3 stroke grestore } bind def
/D4 { gsave translate 45 rotate 0 0 S4 stroke grestore } bind def
/D5 { gsave translate 45 rotate 0 0 S5 stroke grestore } bind def
/D6 { gsave translate 45 rotate 0 0 S6 stroke grestore } bind def
/D7 { gsave translate 45 rotate 0 0 S7 stroke grestore } bind def
/D8 { gsave translate 45 rotate 0 0 S8 stroke grestore } bind def
/D9 { gsave translate 45 rotate 0 0 S9 stroke grestore } bind def
/D10 { gsave translate 45 rotate 0 0 S10 stroke grestore } bind def
/D11 { gsave translate 45 rotate 0 0 S11 stroke grestore } bind def
/D12 { gsave translate 45 rotate 0 0 S12 stroke grestore } bind def
/D13 { gsave translate 45 rotate 0 0 S13 stroke grestore } bind def
/D14 { gsave translate 45 rotate 0 0 S14 stroke grestore } bind def
/D15 { gsave translate 45 rotate 0 0 S15 stroke grestore } bind def
/DiaE { stroke [] 0 setdash vpt add M
  hpt neg vpt neg V hpt vpt neg V
  hpt vpt V hpt neg vpt V closepath stroke } def
/BoxE { stroke [] 0 setdash exch hpt sub exch vpt add M
  0 vpt2 neg V hpt2 0 V 0 vpt2 V
  hpt2 neg 0 V closepath stroke } def
/TriUE { stroke [] 0 setdash vpt 1.12 mul add M
  hpt neg vpt -1.62 mul V
  hpt 2 mul 0 V
  hpt neg vpt 1.62 mul V closepath stroke } def
/TriDE { stroke [] 0 setdash vpt 1.12 mul sub M
  hpt neg vpt 1.62 mul V
  hpt 2 mul 0 V
  hpt neg vpt -1.62 mul V closepath stroke } def
/PentE { stroke [] 0 setdash gsave
  translate 0 hpt M 4 {72 rotate 0 hpt L} repeat
  closepath stroke grestore } def
/CircE { stroke [] 0 setdash 
  hpt 0 360 arc stroke } def
/Opaque { gsave closepath 1 setgray fill grestore 0 setgray closepath } def
/DiaW { stroke [] 0 setdash vpt add M
  hpt neg vpt neg V hpt vpt neg V
  hpt vpt V hpt neg vpt V Opaque stroke } def
/BoxW { stroke [] 0 setdash exch hpt sub exch vpt add M
  0 vpt2 neg V hpt2 0 V 0 vpt2 V
  hpt2 neg 0 V Opaque stroke } def
/TriUW { stroke [] 0 setdash vpt 1.12 mul add M
  hpt neg vpt -1.62 mul V
  hpt 2 mul 0 V
  hpt neg vpt 1.62 mul V Opaque stroke } def
/TriDW { stroke [] 0 setdash vpt 1.12 mul sub M
  hpt neg vpt 1.62 mul V
  hpt 2 mul 0 V
  hpt neg vpt -1.62 mul V Opaque stroke } def
/PentW { stroke [] 0 setdash gsave
  translate 0 hpt M 4 {72 rotate 0 hpt L} repeat
  Opaque stroke grestore } def
/CircW { stroke [] 0 setdash 
  hpt 0 360 arc Opaque stroke } def
/BoxFill { gsave Rec 1 setgray fill grestore } def
/Symbol-Oblique /Symbol findfont [1 0 .167 1 0 0] makefont
dup length dict begin {1 index /FID eq {pop pop} {def} ifelse} forall
currentdict end definefont
end
}}%
\begin{picture}(3600,2160)(0,0)%
{\GNUPLOTspecial{"
gnudict begin
gsave
0 0 translate
0.100 0.100 scale
0 setgray
newpath
1.000 UL
LTb
450 300 M
63 0 V
2937 0 R
-63 0 V
450 520 M
63 0 V
2937 0 R
-63 0 V
450 740 M
63 0 V
2937 0 R
-63 0 V
450 960 M
63 0 V
2937 0 R
-63 0 V
450 1180 M
63 0 V
2937 0 R
-63 0 V
450 1400 M
63 0 V
2937 0 R
-63 0 V
450 1620 M
63 0 V
2937 0 R
-63 0 V
450 1840 M
63 0 V
2937 0 R
-63 0 V
450 2060 M
63 0 V
2937 0 R
-63 0 V
450 300 M
0 63 V
0 1697 R
0 -63 V
1200 300 M
0 63 V
0 1697 R
0 -63 V
1950 300 M
0 63 V
0 1697 R
0 -63 V
2700 300 M
0 63 V
0 1697 R
0 -63 V
3450 300 M
0 63 V
0 1697 R
0 -63 V
1.000 UL
LTb
450 300 M
3000 0 V
0 1760 V
-3000 0 V
450 300 L
1.000 UL
LT0
3087 1947 M
263 0 V
450 1785 M
30 -2 V
31 -1 V
30 -2 V
30 -3 V
30 -2 V
31 -3 V
30 -4 V
30 -4 V
31 -4 V
30 -5 V
30 -6 V
30 -7 V
31 -7 V
30 -8 V
30 -9 V
31 -10 V
30 -9 V
30 -11 V
30 -11 V
31 -11 V
30 -12 V
30 -13 V
31 -12 V
30 -14 V
30 -13 V
30 -14 V
31 -15 V
30 -15 V
30 -15 V
31 -15 V
30 -17 V
30 -16 V
30 -17 V
31 -17 V
30 -17 V
30 -18 V
31 -17 V
30 -18 V
30 -19 V
31 -18 V
30 -19 V
30 -19 V
30 -19 V
31 -19 V
30 -19 V
30 -19 V
31 -20 V
30 -19 V
30 -20 V
30 -19 V
31 -19 V
30 -20 V
30 -19 V
31 -20 V
30 -19 V
30 -19 V
30 -19 V
31 -19 V
30 -19 V
30 -18 V
31 -19 V
30 -18 V
30 -18 V
30 -18 V
31 -17 V
30 -18 V
30 -17 V
31 -17 V
30 -16 V
30 -17 V
30 -16 V
31 -15 V
30 -16 V
30 -15 V
31 -15 V
30 -14 V
30 -14 V
30 -14 V
31 -14 V
30 -13 V
30 -13 V
31 -12 V
30 -12 V
30 -12 V
30 -12 V
31 -11 V
30 -11 V
30 -10 V
31 -10 V
30 -10 V
30 -9 V
30 -9 V
31 -8 V
30 -9 V
30 -8 V
31 -8 V
30 -8 V
30 -8 V
30 -8 V
1.000 UL
LT1
3087 1847 M
263 0 V
450 1195 M
30 -2 V
31 -2 V
30 -1 V
30 -2 V
30 -2 V
31 -3 V
30 -3 V
30 -3 V
31 -4 V
30 -4 V
30 -5 V
30 -6 V
31 -6 V
30 -7 V
30 -7 V
31 -8 V
30 -8 V
30 -9 V
30 -9 V
31 -9 V
30 -10 V
30 -10 V
31 -10 V
30 -11 V
30 -11 V
30 -11 V
31 -12 V
30 -12 V
30 -12 V
31 -12 V
30 -13 V
30 -13 V
30 -13 V
31 -13 V
30 -14 V
30 -13 V
31 -14 V
30 -14 V
30 -13 V
31 -14 V
30 -14 V
30 -14 V
30 -14 V
31 -14 V
30 -14 V
30 -14 V
31 -14 V
30 -14 V
30 -13 V
30 -14 V
31 -13 V
30 -14 V
30 -13 V
31 -13 V
30 -13 V
30 -12 V
30 -13 V
31 -12 V
30 -12 V
30 -12 V
31 -11 V
30 -12 V
30 -11 V
30 -11 V
31 -10 V
30 -10 V
30 -10 V
31 -10 V
30 -9 V
30 -9 V
30 -9 V
31 -9 V
30 -8 V
30 -8 V
31 -8 V
30 -7 V
30 -7 V
30 -7 V
31 -7 V
30 -6 V
30 -6 V
31 -6 V
30 -6 V
30 -5 V
30 -5 V
31 -5 V
30 -4 V
30 -5 V
31 -4 V
30 -3 V
30 -4 V
30 -3 V
31 -3 V
30 -3 V
30 -3 V
31 -3 V
30 -3 V
30 -2 V
30 -3 V
1.000 UL
LT2
3087 1747 M
263 0 V
450 1581 M
30 -2 V
31 -1 V
30 -3 V
30 -2 V
30 -3 V
31 -3 V
30 -4 V
30 -4 V
31 -5 V
30 -5 V
30 -6 V
30 -8 V
31 -7 V
30 -9 V
30 -9 V
31 -10 V
30 -11 V
30 -11 V
30 -11 V
31 -12 V
30 -13 V
30 -13 V
31 -13 V
30 -14 V
30 -15 V
30 -14 V
31 -15 V
30 -16 V
30 -16 V
31 -16 V
30 -17 V
30 -17 V
30 -17 V
31 -18 V
30 -18 V
30 -18 V
31 -19 V
30 -18 V
30 -19 V
31 -19 V
30 -19 V
30 -19 V
30 -19 V
31 -20 V
30 -19 V
30 -19 V
31 -20 V
30 -19 V
30 -19 V
30 -20 V
31 -19 V
30 -19 V
30 -19 V
31 -18 V
30 -19 V
30 -18 V
30 -18 V
31 -18 V
30 -18 V
30 -17 V
31 -18 V
30 -16 V
30 -17 V
30 -16 V
31 -16 V
30 -16 V
30 -15 V
31 -15 V
30 -14 V
30 -14 V
30 -14 V
31 -13 V
30 -13 V
30 -13 V
31 -12 V
30 -12 V
30 -11 V
30 -11 V
31 -11 V
30 -10 V
30 -10 V
31 -9 V
30 -9 V
30 -9 V
30 -8 V
31 -8 V
30 -8 V
30 -7 V
31 -6 V
30 -7 V
30 -6 V
30 -5 V
31 -6 V
30 -5 V
30 -5 V
31 -4 V
30 -5 V
30 -4 V
30 -5 V
1.000 UL
LT3
3087 1647 M
263 0 V
450 1521 M
30 -2 V
31 -1 V
30 -3 V
30 -2 V
30 -3 V
31 -3 V
30 -3 V
30 -4 V
31 -5 V
30 -5 V
30 -7 V
30 -7 V
31 -7 V
30 -9 V
30 -9 V
31 -9 V
30 -11 V
30 -10 V
30 -12 V
31 -11 V
30 -13 V
30 -12 V
31 -13 V
30 -14 V
30 -14 V
30 -14 V
31 -15 V
30 -15 V
30 -16 V
31 -16 V
30 -16 V
30 -16 V
30 -17 V
31 -17 V
30 -17 V
30 -18 V
31 -18 V
30 -18 V
30 -18 V
31 -18 V
30 -18 V
30 -19 V
30 -18 V
31 -19 V
30 -19 V
30 -18 V
31 -19 V
30 -18 V
30 -19 V
30 -18 V
31 -18 V
30 -19 V
30 -18 V
31 -17 V
30 -18 V
30 -18 V
30 -17 V
31 -17 V
30 -17 V
30 -16 V
31 -17 V
30 -16 V
30 -15 V
30 -15 V
31 -16 V
30 -14 V
30 -15 V
31 -13 V
30 -14 V
30 -13 V
30 -13 V
31 -13 V
30 -12 V
30 -11 V
31 -12 V
30 -11 V
30 -10 V
30 -11 V
31 -10 V
30 -9 V
30 -9 V
31 -9 V
30 -8 V
30 -8 V
30 -8 V
31 -7 V
30 -7 V
30 -7 V
31 -6 V
30 -6 V
30 -5 V
30 -6 V
31 -4 V
30 -5 V
30 -5 V
31 -4 V
30 -4 V
30 -4 V
30 -5 V
1.000 UL
LT4
3087 1547 M
263 0 V
450 1837 M
30 -2 V
31 -2 V
30 -2 V
30 -2 V
30 -3 V
31 -4 V
30 -3 V
30 -5 V
31 -5 V
30 -5 V
30 -7 V
30 -8 V
31 -8 V
30 -9 V
30 -9 V
31 -11 V
30 -11 V
30 -11 V
30 -12 V
31 -13 V
30 -13 V
30 -14 V
31 -14 V
30 -15 V
30 -15 V
30 -16 V
31 -16 V
30 -17 V
30 -17 V
31 -18 V
30 -18 V
30 -18 V
30 -19 V
31 -20 V
30 -19 V
30 -20 V
31 -20 V
30 -21 V
30 -21 V
31 -21 V
30 -21 V
30 -22 V
30 -21 V
31 -22 V
30 -22 V
30 -23 V
31 -22 V
30 -22 V
30 -22 V
30 -23 V
31 -22 V
30 -22 V
30 -23 V
31 -22 V
30 -22 V
30 -22 V
30 -22 V
31 -22 V
30 -21 V
30 -22 V
31 -21 V
30 -21 V
30 -20 V
30 -20 V
31 -20 V
30 -20 V
30 -19 V
31 -19 V
30 -19 V
30 -18 V
30 -18 V
31 -17 V
30 -17 V
30 -17 V
31 -16 V
30 -16 V
30 -15 V
30 -15 V
31 -14 V
30 -14 V
30 -13 V
31 -13 V
30 -13 V
30 -12 V
30 -12 V
31 -11 V
30 -10 V
30 -10 V
31 -10 V
30 -9 V
30 -8 V
30 -8 V
31 -8 V
30 -8 V
30 -7 V
31 -7 V
30 -6 V
30 -7 V
30 -7 V
1.000 UL
LT5
3087 1447 M
263 0 V
450 2051 M
30 -4 V
31 -5 V
30 -5 V
30 -5 V
30 -6 V
31 -6 V
30 -8 V
30 -9 V
31 -10 V
30 -12 V
30 -14 V
30 -15 V
31 -16 V
30 -18 V
30 -19 V
31 -20 V
30 -22 V
30 -22 V
30 -24 V
31 -24 V
30 -25 V
30 -25 V
31 -26 V
30 -27 V
30 -28 V
30 -27 V
31 -29 V
30 -28 V
30 -30 V
31 -29 V
30 -30 V
30 -30 V
30 -30 V
31 -30 V
30 -30 V
30 -31 V
31 -30 V
30 -30 V
30 -30 V
31 -30 V
30 -30 V
30 -29 V
30 -29 V
31 -29 V
30 -28 V
30 -28 V
31 -28 V
30 -27 V
30 -26 V
30 -26 V
31 -26 V
30 -25 V
30 -24 V
31 -24 V
30 -23 V
30 -22 V
30 -22 V
31 -22 V
30 -21 V
30 -20 V
31 -19 V
30 -19 V
30 -18 V
30 -17 V
31 -17 V
30 -17 V
30 -15 V
31 -15 V
30 -15 V
30 -14 V
30 -13 V
31 -12 V
30 -13 V
30 -11 V
31 -11 V
30 -11 V
30 -10 V
30 -9 V
31 -10 V
30 -8 V
30 -8 V
31 -8 V
30 -8 V
30 -7 V
30 -6 V
31 -6 V
30 -6 V
30 -6 V
31 -5 V
30 -4 V
30 -5 V
30 -4 V
31 -4 V
30 -3 V
30 -4 V
31 -3 V
30 -4 V
30 -3 V
30 -3 V
stroke
grestore
end
showpage
}}%
\put(3037,1447){\makebox(0,0)[r]{$w_A w_B$}}%
\put(3037,1547){\makebox(0,0)[r]{$n=4$}}%
\put(3037,1647){\makebox(0,0)[r]{$n=3$}}%
\put(3037,1747){\makebox(0,0)[r]{$n=2$}}%
\put(3037,1847){\makebox(0,0)[r]{$n=1$}}%
\put(3037,1947){\makebox(0,0)[r]{$w_A w_B C(w_A, w_B)$}}%
\put(1950,50){\makebox(0,0){$x$}}%
\put(100,1180){%
\special{ps: gsave currentpoint currentpoint translate
270 rotate neg exch neg exch translate}%
\makebox(0,0)[b]{\shortstack{$w_A w_B C(w_A, w_B)$}}%
\special{ps: currentpoint grestore moveto}%
}%
\put(3450,200){\makebox(0,0){$2.0$}}%
\put(2700,200){\makebox(0,0){$1.5$}}%
\put(1950,200){\makebox(0,0){$1.0$}}%
\put(1200,200){\makebox(0,0){$0.5$}}%
\put(450,200){\makebox(0,0){$0.0$}}%
\put(400,2060){\makebox(0,0)[r]{$0.16$}}%
\put(400,1840){\makebox(0,0)[r]{$0.14$}}%
\put(400,1620){\makebox(0,0)[r]{$0.12$}}%
\put(400,1400){\makebox(0,0)[r]{$0.10$}}%
\put(400,1180){\makebox(0,0)[r]{$0.08$}}%
\put(400,960){\makebox(0,0)[r]{$0.06$}}%
\put(400,740){\makebox(0,0)[r]{$0.04$}}%
\put(400,520){\makebox(0,0)[r]{$0.02$}}%
\put(400,300){\makebox(0,0)[r]{$0.00$}}%
\end{picture}%
\endgroup
 

%% file: fig5.tex
% GNUPLOT: LaTeX picture with Postscript
\begingroup%
  \makeatletter%
  \newcommand{\GNUPLOTspecial}{%
    \@sanitize\catcode`\%=14\relax\special}%
  \setlength{\unitlength}{0.1bp}%
{\GNUPLOTspecial{!
%!PS-Adobe-2.0
%%Title: fig5.tex
%%Creator: gnuplot 3.7 patchlevel 2
%%CreationDate: Sun Jun 16 23:24:38 2002
%%DocumentFonts: 
%%BoundingBox: 0 0 360 216
%%Orientation: Landscape
%%Pages: (atend)
%%EndComments
/gnudict 256 dict def
gnudict begin
/Color false def
/Solid false def
/gnulinewidth 5.000 def
/userlinewidth gnulinewidth def
/vshift -33 def
/dl {10 mul} def
/hpt_ 31.5 def
/vpt_ 31.5 def
/hpt hpt_ def
/vpt vpt_ def
/M {moveto} bind def
/L {lineto} bind def
/R {rmoveto} bind def
/V {rlineto} bind def
/vpt2 vpt 2 mul def
/hpt2 hpt 2 mul def
/Lshow { currentpoint stroke M
  0 vshift R show } def
/Rshow { currentpoint stroke M
  dup stringwidth pop neg vshift R show } def
/Cshow { currentpoint stroke M
  dup stringwidth pop -2 div vshift R show } def
/UP { dup vpt_ mul /vpt exch def hpt_ mul /hpt exch def
  /hpt2 hpt 2 mul def /vpt2 vpt 2 mul def } def
/DL { Color {setrgbcolor Solid {pop []} if 0 setdash }
 {pop pop pop Solid {pop []} if 0 setdash} ifelse } def
/BL { stroke userlinewidth 2 mul setlinewidth } def
/AL { stroke userlinewidth 2 div setlinewidth } def
/UL { dup gnulinewidth mul /userlinewidth exch def
      dup 1 lt {pop 1} if 10 mul /udl exch def } def
/PL { stroke userlinewidth setlinewidth } def
/LTb { BL [] 0 0 0 DL } def
/LTa { AL [1 udl mul 2 udl mul] 0 setdash 0 0 0 setrgbcolor } def
/LT0 { PL [] 1 0 0 DL } def
/LT1 { PL [4 dl 2 dl] 0 1 0 DL } def
/LT2 { PL [2 dl 3 dl] 0 0 1 DL } def
/LT3 { PL [1 dl 1.5 dl] 1 0 1 DL } def
/LT4 { PL [5 dl 2 dl 1 dl 2 dl] 0 1 1 DL } def
/LT5 { PL [4 dl 3 dl 1 dl 3 dl] 1 1 0 DL } def
/LT6 { PL [2 dl 2 dl 2 dl 4 dl] 0 0 0 DL } def
/LT7 { PL [2 dl 2 dl 2 dl 2 dl 2 dl 4 dl] 1 0.3 0 DL } def
/LT8 { PL [2 dl 2 dl 2 dl 2 dl 2 dl 2 dl 2 dl 4 dl] 0.5 0.5 0.5 DL } def
/Pnt { stroke [] 0 setdash
   gsave 1 setlinecap M 0 0 V stroke grestore } def
/Dia { stroke [] 0 setdash 2 copy vpt add M
  hpt neg vpt neg V hpt vpt neg V
  hpt vpt V hpt neg vpt V closepath stroke
  Pnt } def
/Pls { stroke [] 0 setdash vpt sub M 0 vpt2 V
  currentpoint stroke M
  hpt neg vpt neg R hpt2 0 V stroke
  } def
/Box { stroke [] 0 setdash 2 copy exch hpt sub exch vpt add M
  0 vpt2 neg V hpt2 0 V 0 vpt2 V
  hpt2 neg 0 V closepath stroke
  Pnt } def
/Crs { stroke [] 0 setdash exch hpt sub exch vpt add M
  hpt2 vpt2 neg V currentpoint stroke M
  hpt2 neg 0 R hpt2 vpt2 V stroke } def
/TriU { stroke [] 0 setdash 2 copy vpt 1.12 mul add M
  hpt neg vpt -1.62 mul V
  hpt 2 mul 0 V
  hpt neg vpt 1.62 mul V closepath stroke
  Pnt  } def
/Star { 2 copy Pls Crs } def
/BoxF { stroke [] 0 setdash exch hpt sub exch vpt add M
  0 vpt2 neg V  hpt2 0 V  0 vpt2 V
  hpt2 neg 0 V  closepath fill } def
/TriUF { stroke [] 0 setdash vpt 1.12 mul add M
  hpt neg vpt -1.62 mul V
  hpt 2 mul 0 V
  hpt neg vpt 1.62 mul V closepath fill } def
/TriD { stroke [] 0 setdash 2 copy vpt 1.12 mul sub M
  hpt neg vpt 1.62 mul V
  hpt 2 mul 0 V
  hpt neg vpt -1.62 mul V closepath stroke
  Pnt  } def
/TriDF { stroke [] 0 setdash vpt 1.12 mul sub M
  hpt neg vpt 1.62 mul V
  hpt 2 mul 0 V
  hpt neg vpt -1.62 mul V closepath fill} def
/DiaF { stroke [] 0 setdash vpt add M
  hpt neg vpt neg V hpt vpt neg V
  hpt vpt V hpt neg vpt V closepath fill } def
/Pent { stroke [] 0 setdash 2 copy gsave
  translate 0 hpt M 4 {72 rotate 0 hpt L} repeat
  closepath stroke grestore Pnt } def
/PentF { stroke [] 0 setdash gsave
  translate 0 hpt M 4 {72 rotate 0 hpt L} repeat
  closepath fill grestore } def
/Circle { stroke [] 0 setdash 2 copy
  hpt 0 360 arc stroke Pnt } def
/CircleF { stroke [] 0 setdash hpt 0 360 arc fill } def
/C0 { BL [] 0 setdash 2 copy moveto vpt 90 450  arc } bind def
/C1 { BL [] 0 setdash 2 copy        moveto
       2 copy  vpt 0 90 arc closepath fill
               vpt 0 360 arc closepath } bind def
/C2 { BL [] 0 setdash 2 copy moveto
       2 copy  vpt 90 180 arc closepath fill
               vpt 0 360 arc closepath } bind def
/C3 { BL [] 0 setdash 2 copy moveto
       2 copy  vpt 0 180 arc closepath fill
               vpt 0 360 arc closepath } bind def
/C4 { BL [] 0 setdash 2 copy moveto
       2 copy  vpt 180 270 arc closepath fill
               vpt 0 360 arc closepath } bind def
/C5 { BL [] 0 setdash 2 copy moveto
       2 copy  vpt 0 90 arc
       2 copy moveto
       2 copy  vpt 180 270 arc closepath fill
               vpt 0 360 arc } bind def
/C6 { BL [] 0 setdash 2 copy moveto
      2 copy  vpt 90 270 arc closepath fill
              vpt 0 360 arc closepath } bind def
/C7 { BL [] 0 setdash 2 copy moveto
      2 copy  vpt 0 270 arc closepath fill
              vpt 0 360 arc closepath } bind def
/C8 { BL [] 0 setdash 2 copy moveto
      2 copy vpt 270 360 arc closepath fill
              vpt 0 360 arc closepath } bind def
/C9 { BL [] 0 setdash 2 copy moveto
      2 copy  vpt 270 450 arc closepath fill
              vpt 0 360 arc closepath } bind def
/C10 { BL [] 0 setdash 2 copy 2 copy moveto vpt 270 360 arc closepath fill
       2 copy moveto
       2 copy vpt 90 180 arc closepath fill
               vpt 0 360 arc closepath } bind def
/C11 { BL [] 0 setdash 2 copy moveto
       2 copy  vpt 0 180 arc closepath fill
       2 copy moveto
       2 copy  vpt 270 360 arc closepath fill
               vpt 0 360 arc closepath } bind def
/C12 { BL [] 0 setdash 2 copy moveto
       2 copy  vpt 180 360 arc closepath fill
               vpt 0 360 arc closepath } bind def
/C13 { BL [] 0 setdash  2 copy moveto
       2 copy  vpt 0 90 arc closepath fill
       2 copy moveto
       2 copy  vpt 180 360 arc closepath fill
               vpt 0 360 arc closepath } bind def
/C14 { BL [] 0 setdash 2 copy moveto
       2 copy  vpt 90 360 arc closepath fill
               vpt 0 360 arc } bind def
/C15 { BL [] 0 setdash 2 copy vpt 0 360 arc closepath fill
               vpt 0 360 arc closepath } bind def
/Rec   { newpath 4 2 roll moveto 1 index 0 rlineto 0 exch rlineto
       neg 0 rlineto closepath } bind def
/Square { dup Rec } bind def
/Bsquare { vpt sub exch vpt sub exch vpt2 Square } bind def
/S0 { BL [] 0 setdash 2 copy moveto 0 vpt rlineto BL Bsquare } bind def
/S1 { BL [] 0 setdash 2 copy vpt Square fill Bsquare } bind def
/S2 { BL [] 0 setdash 2 copy exch vpt sub exch vpt Square fill Bsquare } bind def
/S3 { BL [] 0 setdash 2 copy exch vpt sub exch vpt2 vpt Rec fill Bsquare } bind def
/S4 { BL [] 0 setdash 2 copy exch vpt sub exch vpt sub vpt Square fill Bsquare } bind def
/S5 { BL [] 0 setdash 2 copy 2 copy vpt Square fill
       exch vpt sub exch vpt sub vpt Square fill Bsquare } bind def
/S6 { BL [] 0 setdash 2 copy exch vpt sub exch vpt sub vpt vpt2 Rec fill Bsquare } bind def
/S7 { BL [] 0 setdash 2 copy exch vpt sub exch vpt sub vpt vpt2 Rec fill
       2 copy vpt Square fill
       Bsquare } bind def
/S8 { BL [] 0 setdash 2 copy vpt sub vpt Square fill Bsquare } bind def
/S9 { BL [] 0 setdash 2 copy vpt sub vpt vpt2 Rec fill Bsquare } bind def
/S10 { BL [] 0 setdash 2 copy vpt sub vpt Square fill 2 copy exch vpt sub exch vpt Square fill
       Bsquare } bind def
/S11 { BL [] 0 setdash 2 copy vpt sub vpt Square fill 2 copy exch vpt sub exch vpt2 vpt Rec fill
       Bsquare } bind def
/S12 { BL [] 0 setdash 2 copy exch vpt sub exch vpt sub vpt2 vpt Rec fill Bsquare } bind def
/S13 { BL [] 0 setdash 2 copy exch vpt sub exch vpt sub vpt2 vpt Rec fill
       2 copy vpt Square fill Bsquare } bind def
/S14 { BL [] 0 setdash 2 copy exch vpt sub exch vpt sub vpt2 vpt Rec fill
       2 copy exch vpt sub exch vpt Square fill Bsquare } bind def
/S15 { BL [] 0 setdash 2 copy Bsquare fill Bsquare } bind def
/D0 { gsave translate 45 rotate 0 0 S0 stroke grestore } bind def
/D1 { gsave translate 45 rotate 0 0 S1 stroke grestore } bind def
/D2 { gsave translate 45 rotate 0 0 S2 stroke grestore } bind def
/D3 { gsave translate 45 rotate 0 0 S3 stroke grestore } bind def
/D4 { gsave translate 45 rotate 0 0 S4 stroke grestore } bind def
/D5 { gsave translate 45 rotate 0 0 S5 stroke grestore } bind def
/D6 { gsave translate 45 rotate 0 0 S6 stroke grestore } bind def
/D7 { gsave translate 45 rotate 0 0 S7 stroke grestore } bind def
/D8 { gsave translate 45 rotate 0 0 S8 stroke grestore } bind def
/D9 { gsave translate 45 rotate 0 0 S9 stroke grestore } bind def
/D10 { gsave translate 45 rotate 0 0 S10 stroke grestore } bind def
/D11 { gsave translate 45 rotate 0 0 S11 stroke grestore } bind def
/D12 { gsave translate 45 rotate 0 0 S12 stroke grestore } bind def
/D13 { gsave translate 45 rotate 0 0 S13 stroke grestore } bind def
/D14 { gsave translate 45 rotate 0 0 S14 stroke grestore } bind def
/D15 { gsave translate 45 rotate 0 0 S15 stroke grestore } bind def
/DiaE { stroke [] 0 setdash vpt add M
  hpt neg vpt neg V hpt vpt neg V
  hpt vpt V hpt neg vpt V closepath stroke } def
/BoxE { stroke [] 0 setdash exch hpt sub exch vpt add M
  0 vpt2 neg V hpt2 0 V 0 vpt2 V
  hpt2 neg 0 V closepath stroke } def
/TriUE { stroke [] 0 setdash vpt 1.12 mul add M
  hpt neg vpt -1.62 mul V
  hpt 2 mul 0 V
  hpt neg vpt 1.62 mul V closepath stroke } def
/TriDE { stroke [] 0 setdash vpt 1.12 mul sub M
  hpt neg vpt 1.62 mul V
  hpt 2 mul 0 V
  hpt neg vpt -1.62 mul V closepath stroke } def
/PentE { stroke [] 0 setdash gsave
  translate 0 hpt M 4 {72 rotate 0 hpt L} repeat
  closepath stroke grestore } def
/CircE { stroke [] 0 setdash 
  hpt 0 360 arc stroke } def
/Opaque { gsave closepath 1 setgray fill grestore 0 setgray closepath } def
/DiaW { stroke [] 0 setdash vpt add M
  hpt neg vpt neg V hpt vpt neg V
  hpt vpt V hpt neg vpt V Opaque stroke } def
/BoxW { stroke [] 0 setdash exch hpt sub exch vpt add M
  0 vpt2 neg V hpt2 0 V 0 vpt2 V
  hpt2 neg 0 V Opaque stroke } def
/TriUW { stroke [] 0 setdash vpt 1.12 mul add M
  hpt neg vpt -1.62 mul V
  hpt 2 mul 0 V
  hpt neg vpt 1.62 mul V Opaque stroke } def
/TriDW { stroke [] 0 setdash vpt 1.12 mul sub M
  hpt neg vpt 1.62 mul V
  hpt 2 mul 0 V
  hpt neg vpt -1.62 mul V Opaque stroke } def
/PentW { stroke [] 0 setdash gsave
  translate 0 hpt M 4 {72 rotate 0 hpt L} repeat
  Opaque stroke grestore } def
/CircW { stroke [] 0 setdash 
  hpt 0 360 arc Opaque stroke } def
/BoxFill { gsave Rec 1 setgray fill grestore } def
/Symbol-Oblique /Symbol findfont [1 0 .167 1 0 0] makefont
dup length dict begin {1 index /FID eq {pop pop} {def} ifelse} forall
currentdict end definefont
end
}}%
\begin{picture}(3600,2160)(0,0)%
{\GNUPLOTspecial{"
gnudict begin
gsave
0 0 translate
0.100 0.100 scale
0 setgray
newpath
1.000 UL
LTb
1.000 UL
LT0
1524 862 M
93 -55 V
162 -49 V
1524 862 M
12 -5 V
0 -1 R
3 0 R
0 -1 R
3 0 R
0 -1 R
3 0 R
0 -1 R
3 0 R
0 -1 R
3 0 R
0 -1 R
213 -86 R
0 -1 R
3 0 R
0 -1 V
0 -1 R
3 0 R
0 -1 R
3 0 R
6 -3 V
1.000 UL
LT0
1686 806 M
78 -41 V
3 -1 V
0 -1 R
3 -1 V
3 -1 V
1524 862 M
12 -5 R
3 -1 V
3 -1 V
3 -1 V
3 -1 V
3 -1 V
135 -46 V
-162 56 R
1.000 UL
LT0
1686 806 M
93 -48 R
159 -20 V
-252 68 R
3 0 R
0 -1 R
3 0 R
0 -1 R
3 0 R
0 -1 R
237 -63 R
0 -1 R
3 0 R
0 -1 R
3 0 R
1.000 UL
LT0
1431 920 M
93 -58 V
-93 58 R
9 -4 V
0 -1 R
3 0 R
0 -1 R
3 0 R
3 -2 V
0 -1 R
3 0 R
0 -1 R
3 0 R
0 -1 R
3 0 R
0 -1 R
6 -2 R
0 -1 R
3 0 R
0 -1 R
198 -88 R
0 -1 R
3 0 R
0 -1 R
3 -1 R
0 -1 R
3 0 R
0 -1 R
3 0 R
0 -1 R
1.000 UL
LT2
1363 455 M
93 -30 V
162 46 V
1363 455 M
3 0 V
3 0 R
246 16 R
3 0 V
1.000 UL
LT2
1524 500 M
90 -29 V
1363 455 M
3 0 R
159 45 V
1363 455 M
1.000 UL
LT2
1524 500 M
93 -29 R
162 36 V
-255 -7 R
252 7 R
3 0 V
1.000 UL
LT0
1846 783 M
87 -43 V
3 -1 V
3 -1 V
-253 68 R
6 -1 R
3 -1 V
150 -21 V
-159 23 R
1.000 UL
LT0
1846 783 M
94 -45 R
162 -8 V
-256 53 R
3 0 R
0 -1 R
246 -50 R
0 -1 R
3 0 R
0 -1 R
3 0 R
1.000 UL
LT0
1592 857 M
75 -41 V
3 -1 V
3 -2 V
3 -1 V
3 -1 V
0 -1 R
3 -1 V
1431 920 M
9 -4 R
3 -1 V
3 -1 V
3 -2 R
3 -1 V
3 -1 V
3 -1 V
6 -3 V
3 -1 V
123 -48 V
-159 63 R
1.000 UL
LT0
1338 985 M
93 -65 V
-93 65 R
6 -3 V
0 -1 R
3 0 R
3 -2 V
0 -1 R
3 0 R
0 -1 R
3 0 R
0 -1 V
0 -1 R
3 0 R
0 -1 R
3 0 R
0 -1 V
0 -1 R
3 0 R
0 -1 R
3 0 R
0 -1 R
3 -1 R
0 -1 R
1572 867 M
0 -1 R
3 -1 R
0 -1 R
3 0 R
0 -1 V
0 -1 R
3 0 R
0 -1 R
3 0 R
0 -1 R
1.000 UL
LT0
1592 857 M
94 -51 R
-94 51 R
3 0 R
0 -1 R
3 0 R
0 -1 R
3 0 R
0 -1 R
240 -69 R
0 -1 R
1.000 UL
LT2
1686 534 M
90 -27 V
-252 -7 R
3 0 R
159 34 V
1524 500 M
1.000 UL
LT2
1686 534 M
93 -27 R
159 32 V
-252 -5 R
249 5 R
3 0 V
1.000 UL
LT2
1269 485 M
93 -30 V
1 0 R
-94 30 R
3 0 V
3 0 R
1.000 UL
LT0
2008 773 M
87 -41 V
3 -1 V
3 -1 V
-255 53 R
3 0 R
159 -10 V
-162 10 R
1.000 UL
LT0
2008 773 M
93 -43 R
162 -2 V
-255 45 R
3 0 R
0 -1 R
249 -43 R
0 -1 R
3 0 R
1.000 UL
LT2
1431 528 M
90 -28 V
1269 485 M
3 0 R
159 43 V
1269 485 M
1.000 UL
LT2
1431 528 M
93 -28 R
-93 28 R
1.000 UL
LT2
1846 566 M
90 -27 V
-250 -5 R
3 0 R
156 32 V
1686 534 M
1.000 UL
LT2
1846 566 M
94 -27 R
162 30 V
-256 -3 R
252 3 R
3 0 V
1.000 UL
LT0
1754 831 M
87 -46 V
3 -1 V
3 -1 V
-255 74 R
6 -1 R
3 -1 V
153 -24 V
-162 26 R
1.000 UL
LT0
1754 831 M
92 -48 R
-92 48 R
3 0 R
0 -1 R
3 0 R
0 -1 R
243 -54 R
0 -1 R
3 0 R
0 -1 R
1.000 UL
LT0
1499 913 M
75 -46 V
3 -2 V
3 -1 V
0 -1 R
3 -1 V
3 -1 V
0 -1 R
3 -1 V
1338 985 M
6 -3 R
3 -1 V
3 -2 R
3 -1 V
3 -1 V
0 -1 R
3 -1 V
3 -1 V
0 -1 R
3 -1 V
3 -1 V
3 -2 V
126 -56 V
-159 72 R
1.000 UL
LT0
1244 1057 M
96 -72 V
-2 0 R
-94 72 R
3 -2 V
0 -1 R
3 0 R
3 -2 V
0 -1 R
3 0 R
0 -1 V
0 -1 R
3 0 R
3 -2 V
0 -1 R
3 0 R
0 -1 V
0 -1 R
3 0 R
0 -1 R
3 0 R
0 -1 V
0 -1 R
3 0 R
0 -1 R
3 -1 R
0 -1 R
3 0 R
0 -1 R
3 -1 R
0 -1 R
3 -1 R
0 -1 R
6 -2 R
0 -1 R
177 -99 R
0 -1 R
3 -1 R
0 -1 R
6 -2 R
0 -1 R
3 -1 R
0 -1 R
3 -1 R
0 -1 R
3 0 R
0 -1 V
0 -1 R
3 0 R
0 -1 R
3 -1 R
0 -1 R
3 0 R
0 -1 R
1.000 UL
LT0
1499 913 M
93 -56 R
-93 56 R
3 0 R
0 -1 R
3 0 R
0 -1 R
3 0 R
0 -1 R
3 0 R
0 -1 R
237 -76 R
0 -1 R
1.000 UL
LT0
2169 770 M
90 -41 V
3 -1 V
-254 45 R
3 0 R
159 -3 V
-162 3 R
1.000 UL
LT0
2169 770 M
94 -42 R
162 1 V
-256 41 R
3 0 R
0 -1 R
249 -39 R
0 -1 R
3 0 R
1.000 UL
LT2
1592 562 M
93 -28 V
-254 -6 R
3 0 R
156 34 V
1431 528 M
1.000 UL
LT2
1592 562 M
94 -28 R
-94 28 R
1.000 UL
LT2
2008 595 M
90 -26 V
-252 -3 R
3 0 R
159 29 V
1846 566 M
1.000 UL
LT2
2008 595 M
93 -26 R
162 28 V
-255 -2 R
252 2 R
3 0 V
1.000 UL
LT2
1176 514 M
93 -29 V
-93 29 R
3 0 V
3 0 R
1.000 UL
LT0
1915 818 M
87 -43 V
3 -1 V
3 -1 V
-254 58 R
3 0 R
3 -1 V
156 -12 V
-162 13 R
1.000 UL
LT0
1915 818 M
93 -45 R
-93 45 R
3 0 R
0 -1 R
249 -46 R
0 -1 R
1.000 UL
LT0
1406 973 M
63 -42 V
3 -2 V
6 -3 V
3 -2 V
3 -2 V
3 -1 V
0 -1 R
3 -1 V
0 -1 R
3 -1 V
-249 140 R
3 -2 R
3 -1 V
3 -2 R
3 -1 V
0 -1 R
3 -1 V
3 -2 R
3 -1 V
0 -1 R
3 -1 V
3 -1 V
0 -1 R
3 -1 V
3 -2 V
3 -1 V
3 -2 V
3 -2 V
6 -3 V
114 -58 V
-162 84 R
1.000 UL
LT0
1151 1140 M
93 -83 V
-93 83 R
9 -6 V
0 -1 R
3 0 R
0 -1 V
0 -1 R
3 0 R
0 -1 V
0 -1 R
3 0 R
0 -1 V
0 -1 R
3 0 R
0 -1 V
0 -1 R
3 0 R
0 -1 R
3 -1 R
0 -1 R
3 -1 R
0 -1 R
3 -1 R
0 -1 R
3 -1 R
0 -1 R
3 -1 R
0 -1 R
6 -2 R
0 -1 R
3 -1 R
0 -1 R
3 -1 R
0 -1 R
3 -1 R
0 -1 R
1373 996 M
0 -1 R
3 -1 R
0 -1 R
3 -1 R
0 -1 R
3 -1 R
0 -1 R
3 -1 R
0 -1 V
0 -1 R
3 0 R
0 -1 V
0 -1 R
3 0 R
0 -1 V
0 -1 R
3 0 R
0 -1 V
0 -1 R
3 0 R
0 -1 V
0 -1 R
3 0 R
0 -1 R
3 -1 R
0 -1 R
1.000 UL
LT0
1661 882 M
87 -49 V
3 -1 V
3 -1 V
-255 82 R
6 -1 R
3 -1 V
3 -1 V
150 -28 V
-162 31 R
1.000 UL
LT0
1661 882 M
93 -51 R
-93 51 R
3 0 R
0 -1 R
3 0 R
0 -1 R
243 -60 R
0 -1 R
3 0 R
0 -1 R
1.000 UL
LT2
1338 556 M
90 -28 V
1176 514 M
3 0 R
159 42 V
1176 514 M
1.000 UL
LT2
1338 556 M
93 -28 R
-93 28 R
1.000 UL
LT2
1754 592 M
90 -26 V
-252 -4 R
3 0 R
159 30 V
1592 562 M
1.000 UL
LT2
1754 592 M
92 -26 R
-92 26 R
1.000 UL
LT2
2169 622 M
90 -25 V
-251 -2 R
3 0 R
159 27 V
2008 595 M
1.000 UL
LT2
2169 622 M
94 -25 R
162 27 V
-256 -2 R
252 2 R
3 0 V
1.000 UL
LT0
1406 973 M
93 -60 R
-93 60 R
6 -2 R
0 -1 R
240 -84 R
0 -1 R
3 0 R
0 -1 R
3 0 R
0 -1 R
3 0 R
0 -1 R
1.000 UL
LT0
2331 770 M
90 -40 V
3 -1 V
-255 41 R
3 0 R
159 0 V
-162 0 R
1.000 UL
LT0
2331 770 M
93 -41 R
162 5 V
-255 36 R
3 0 R
0 -1 R
249 -34 R
0 -1 R
3 0 R
1.000 UL
LT0
2076 813 M
90 -42 V
3 -1 V
-254 48 R
3 0 R
159 -5 V
-162 5 R
1.000 UL
LT0
2076 813 M
93 -43 R
-93 43 R
3 0 R
0 -1 R
249 -41 R
0 -1 R
1.000 UL
LT2
1499 589 M
90 -27 V
-251 -6 R
3 0 R
156 33 V
1338 556 M
1.000 UL
LT2
1499 589 M
93 -27 R
-93 27 R
1.000 UL
LT2
1915 621 M
90 -26 V
-251 -3 R
3 0 R
159 29 V
1754 592 M
1.000 UL
LT2
1915 621 M
93 -26 R
-93 26 R
1.000 UL
LT2
1083 544 M
93 -30 V
-93 30 R
3 0 V
3 0 R
1.000 UL
LT2
2331 649 M
90 -25 V
-252 -2 R
3 0 R
159 27 V
2169 622 M
1.000 UL
LT2
2331 649 M
93 -25 R
162 26 V
-255 -1 R
252 1 R
3 0 V
1.000 UL
LT0
1313 1042 M
60 -46 V
3 -2 V
3 -2 V
3 -2 V
3 -2 V
0 -1 R
3 -1 V
0 -1 R
3 -1 V
0 -1 R
3 -1 V
0 -1 R
3 -1 V
0 -1 R
3 -1 V
-249 162 R
9 -6 R
3 -1 V
0 -1 R
3 -1 V
0 -1 R
3 -1 V
0 -1 R
3 -1 V
0 -1 R
3 -1 V
3 -2 V
3 -2 V
3 -2 V
3 -2 V
3 -2 V
6 -3 V
3 -2 V
3 -2 V
3 -2 V
108 -64 V
-162 98 R
1.000 UL
LT0
1058 1242 M
93 -102 V
-93 102 R
3 -3 V
0 -1 R
3 0 R
6 -6 V
0 -1 R
3 0 R
0 -1 V
0 -1 R
3 0 R
0 -1 V
0 -1 R
3 0 R
0 -2 V
0 -1 R
3 0 R
0 -1 V
0 -1 R
3 0 R
0 -1 V
0 -1 R
3 0 R
0 -1 V
0 -1 R
3 -1 R
0 -1 R
3 -1 R
0 -1 V
0 -1 R
3 0 R
0 -1 V
0 -1 R
3 -1 R
0 -1 R
3 -1 R
0 -1 R
3 -1 R
0 -1 R
3 -2 R
0 -1 R
3 -1 R
0 -1 R
168 -131 R
0 -1 R
3 -1 R
0 -1 R
3 -2 R
0 -1 R
3 -1 R
0 -1 R
3 -1 R
0 -1 V
0 -1 R
3 -1 R
0 -1 V
0 -1 R
3 0 R
0 -1 V
0 -1 R
3 0 R
0 -2 V
0 -1 R
3 0 R
0 -1 V
0 -1 R
3 0 R
0 -1 R
3 -1 R
0 -1 R
1.000 UL
LT0
1821 866 M
87 -46 V
3 -1 V
3 -1 V
-253 64 R
3 0 R
3 -1 V
156 -15 V
-162 16 R
1.000 UL
LT0
1821 866 M
94 -48 R
-94 48 R
3 0 R
0 -1 R
246 -50 R
0 -1 R
3 0 R
0 -1 R
1.000 UL
LT0
2492 774 M
93 -39 V
3 -1 V
-257 36 R
3 0 R
156 4 V
-159 -4 R
1.000 UL
LT0
2492 774 M
94 -40 R
159 6 V
-253 34 R
3 0 R
0 -1 R
249 -32 R
0 -1 R
3 0 R
1.000 UL
LT0
1567 937 M
84 -51 V
3 -1 V
3 -1 V
-251 89 R
3 -1 V
3 -1 V
156 -34 V
-162 36 R
1.000 UL
LT0
1567 937 M
94 -55 R
-94 55 R
3 0 R
0 -1 R
3 0 R
0 -1 R
246 -68 R
0 -1 R
1.000 UL
LT0
1313 1042 M
93 -69 R
-93 69 R
6 -2 R
0 -1 R
3 0 R
0 -1 R
3 0 R
0 -1 R
237 -97 R
0 -1 R
3 0 R
0 -1 R
3 0 R
0 -1 R
1.000 UL
LT2
1244 585 M
93 -29 V
1083 544 M
3 0 R
156 41 V
1083 544 M
1.000 UL
LT2
1244 585 M
94 -29 R
-94 29 R
1.000 UL
LT2
1661 618 M
90 -26 V
-252 -3 R
3 0 R
159 29 V
1499 589 M
1.000 UL
LT2
1661 618 M
93 -26 R
-93 26 R
1.000 UL
LT0
2238 812 M
90 -41 V
3 -1 V
-255 43 R
3 0 R
159 -1 V
-162 1 R
1.000 UL
LT0
2238 812 M
93 -42 R
-93 42 R
3 0 R
0 -1 R
246 -36 R
0 -1 R
1.000 UL
LT2
2076 647 M
90 -25 V
-251 -1 R
3 0 R
159 26 V
1915 621 M
1.000 UL
LT2
2076 647 M
93 -25 R
-93 25 R
1.000 UL
LT2
2492 674 M
93 -24 V
-254 -1 R
3 0 R
156 25 V
2331 649 M
1.000 UL
LT2
2492 674 M
94 -24 R
159 25 V
-253 -1 R
252 1 R
3 0 V
1.000 UL
LT0
1219 1123 M
60 -54 V
3 -2 V
3 -3 V
3 -2 V
3 -2 V
0 -1 R
3 -2 V
0 -1 R
3 -1 V
0 -1 R
3 -1 V
0 -2 R
3 -1 V
0 -1 R
3 -1 V
-248 194 R
3 -3 R
3 -1 V
6 -6 R
3 -1 V
0 -1 R
3 -1 V
0 -1 R
3 -1 V
0 -2 R
3 -1 V
0 -1 R
3 -1 V
0 -1 R
3 -1 V
0 -1 R
3 -2 V
3 -2 V
0 -1 R
3 -1 V
0 -1 R
3 -2 V
3 -2 V
3 -2 V
3 -3 V
3 -2 V
108 -78 V
-162 119 R
1.000 UL
LT0
965 1376 M
93 -134 V
-93 134 R
21 -21 V
0 -1 R
3 0 R
0 -2 V
0 -1 R
3 0 R
0 -2 V
0 -1 R
3 0 R
0 -2 V
0 -1 R
3 0 R
0 -2 V
0 -1 R
3 0 R
0 -2 V
0 -1 R
3 0 R
0 -2 V
0 -1 R
3 0 R
0 -1 V
0 -1 R
3 -1 R
0 -1 V
0 -1 R
3 -1 R
0 -1 V
0 -1 R
3 -1 R
0 -1 V
0 -1 R
3 -1 R
0 -1 V
0 -1 R
3 -1 R
0 -1 V
0 -1 R
3 -1 R
0 -1 R
3 -2 R
0 -1 R
3 -2 R
0 -1 R
3 -2 R
0 -1 R
3 -2 R
0 -1 R
3 -2 R
0 -1 R
3 -2 R
0 -1 R
129 -127 R
0 -1 R
3 -2 R
0 -1 R
3 -2 R
0 -1 R
3 -2 R
0 -1 R
3 -2 R
0 -1 R
3 -2 R
0 -1 R
3 -2 R
0 -1 V
0 -1 R
3 -1 R
0 -1 V
0 -1 R
3 -1 R
0 -1 V
0 -1 R
3 -1 R
0 -1 V
0 -1 R
3 -1 R
0 -1 V
0 -1 R
3 -1 R
0 -2 V
0 -1 R
3 0 R
0 -2 V
0 -1 R
3 0 R
0 -1 V
0 -1 R
3 -1 R
0 -1 R
3 -2 R
0 -1 R
1.000 UL
LT0
1983 859 M
87 -44 V
3 -1 V
3 -1 V
-255 53 R
3 0 R
159 -7 V
-162 7 R
1.000 UL
LT0
1983 859 M
93 -46 R
-93 46 R
3 0 R
0 -1 R
249 -45 R
0 -1 R
1.000 UL
LT0
2654 779 M
90 -38 V
3 -1 V
-255 34 R
3 0 R
159 5 V
-162 -5 R
1.000 UL
LT0
2654 779 M
93 -39 R
162 8 V
-255 31 R
3 0 R
0 -1 R
249 -29 R
0 -1 R
3 0 R
1.000 UL
LT2
1406 616 M
90 -27 V
-252 -4 R
3 0 R
159 31 V
1244 585 M
1.000 UL
LT2
1406 616 M
93 -27 R
-93 27 R
1.000 UL
LT2
990 574 M
93 -30 V
-93 30 R
3 0 V
3 0 R
1.000 UL
LT2
1821 646 M
90 -25 V
-250 -3 R
3 0 R
159 28 V
1661 618 M
1.000 UL
LT2
1821 646 M
94 -25 R
-94 25 R
1.000 UL
LT0
1729 918 M
90 -51 V
3 -1 V
-255 71 R
6 -1 R
156 -18 V
-162 19 R
1.000 UL
LT0
1729 918 M
92 -52 R
-92 52 R
3 0 R
0 -1 R
246 -56 R
0 -1 R
3 0 R
0 -1 R
1.000 UL
LT2
2238 673 M
90 -24 V
-252 -2 R
3 0 R
159 26 V
2076 647 M
1.000 UL
LT2
2238 673 M
93 -24 R
-93 24 R
1.000 UL
LT2
2654 699 M
90 -24 V
-252 -1 R
3 0 R
159 25 V
2492 674 M
1.000 UL
LT2
2654 699 M
93 -24 R
162 25 V
-255 -1 R
252 1 R
3 0 V
1.000 UL
LT0
2399 815 M
90 -40 V
3 -1 V
-254 38 R
3 0 R
156 3 V
-159 -3 R
1.000 UL
LT0
2399 815 M
93 -41 R
-93 41 R
3 0 R
0 -1 R
249 -34 R
0 -1 R
1.000 UL
LT0
1474 998 M
87 -58 V
3 -1 V
-251 103 R
3 -1 V
3 -1 V
3 -1 V
3 -1 V
150 -40 V
-162 44 R
1.000 UL
LT0
1474 998 M
93 -61 R
-93 61 R
3 0 R
0 -1 R
3 0 R
0 -1 R
1.000 UL
LT0
1219 1123 M
94 -81 R
-94 81 R
6 -2 R
0 -1 V
0 -1 R
3 0 R
0 -1 R
3 0 R
0 -1 R
237 -115 R
0 -1 V
0 -1 R
3 0 R
0 -1 R
3 0 R
0 -1 R
1.000 UL
LT0
1126 1222 M
45 -50 V
3 -3 V
3 -3 V
3 -3 V
3 -3 V
3 -3 V
3 -3 V
0 -1 R
3 -2 V
0 -1 R
3 -2 V
0 -1 R
3 -2 V
0 -1 R
3 -2 V
0 -1 R
3 -2 V
0 -2 R
3 -1 V
0 -2 R
3 -1 V
0 -2 R
3 -1 V
0 -2 R
3 -1 V
0 -2 R
3 -1 V
965 1376 M
21 -21 R
3 -1 V
0 -2 R
3 -1 V
0 -2 R
3 -1 V
0 -2 R
3 -1 V
0 -2 R
3 -1 V
0 -2 R
3 -1 V
0 -2 R
3 -1 V
0 -1 R
3 -2 V
0 -1 R
3 -2 V
0 -1 R
3 -2 V
0 -1 R
3 -2 V
0 -1 R
3 -2 V
0 -1 R
3 -2 V
3 -3 V
3 -3 V
3 -3 V
3 -3 V
3 -3 V
3 -3 V
84 -78 V
965 1376 M
1.000 UL
LT0
871 1577 M
93 -201 V
1 0 R
-94 201 R
9 -14 V
0 -1 R
3 0 R
0 -3 V
0 -1 R
3 0 R
0 -3 V
0 -1 R
3 0 R
0 -3 V
0 -1 R
3 0 R
0 -3 V
0 -1 R
3 0 R
0 -3 V
0 -1 R
3 -1 R
0 -1 V
0 -1 R
3 -2 R
0 -1 V
0 -1 R
3 -2 R
0 -1 V
0 -1 R
3 -2 R
0 -1 V
0 -1 R
3 -2 R
0 -1 V
0 -1 R
3 -2 R
0 -1 V
0 -1 R
3 -3 R
0 -1 R
3 -3 R
0 -1 R
3 -3 R
0 -1 R
3 -3 R
0 -1 R
3 -3 R
0 -1 R
3 -3 R
0 -1 R
162 -225 R
0 -1 R
3 -3 R
0 -1 R
3 -3 R
0 -1 V
0 -1 R
3 -3 R
0 -1 V
0 -1 R
3 -2 R
0 -1 V
0 -1 R
3 -2 R
0 -2 V
0 -1 R
3 -1 R
0 -2 V
0 -1 R
3 -1 R
0 -3 V
0 -1 R
3 0 R
0 -4 V
0 -1 R
3 0 R
0 -2 V
0 -1 R
3 -1 R
0 -1 V
0 -1 R
3 -2 R
0 -1 R
1.000 UL
LT0
2816 786 M
90 -37 V
3 -1 V
-255 31 R
3 0 R
159 7 V
-162 -7 R
1.000 UL
LT0
2816 786 M
93 -38 R
162 8 V
-255 30 R
3 0 R
0 -1 R
249 -28 R
0 -1 R
3 0 R
1.000 UL
LT0
2144 857 M
93 -44 V
3 -1 V
-257 47 R
3 0 R
156 -2 V
-159 2 R
1.000 UL
LT0
2144 857 M
94 -45 R
-94 45 R
3 0 R
0 -1 R
249 -40 R
0 -1 R
1.000 UL
LT2
1151 612 M
90 -27 V
990 574 M
3 0 R
156 38 V
990 574 M
1.000 UL
LT2
1151 612 M
93 -27 R
-93 27 R
1.000 UL
LT2
1567 644 M
90 -26 V
-251 -2 R
3 0 R
159 28 V
1406 616 M
1.000 UL
LT2
1567 644 M
94 -26 R
-94 26 R
1.000 UL
LT2
1983 672 M
90 -25 V
-252 -1 R
3 0 R
159 26 V
1821 646 M
1.000 UL
LT2
1983 672 M
93 -25 R
-93 25 R
1.000 UL
LT0
1033 1364 M
60 -94 V
3 -4 V
3 -4 V
0 -1 R
3 -4 V
0 -1 R
3 -3 V
0 -1 R
3 -3 V
0 -2 R
3 -2 V
0 -2 R
3 -2 V
0 -3 R
3 -1 V
0 -4 R
3 -1 V
0 -3 R
3 -1 V
871 1577 M
9 -14 R
3 -1 V
0 -3 R
3 -1 V
0 -3 R
3 -1 V
0 -3 R
3 -1 V
0 -3 R
3 -1 V
0 -3 R
3 -2 V
0 -1 R
3 -3 V
0 -1 R
3 -3 V
0 -1 R
3 -3 V
0 -1 R
3 -3 V
0 -1 R
3 -3 V
0 -1 R
3 -4 V
3 -4 V
3 -4 V
3 -4 V
3 -4 V
3 -4 V
102 -132 V
871 1577 M
1.000 UL
LT0
871 1577 M
69 47 R
90 -254 V
0 -2 R
3 -2 V
871 1577 M
69 47 V
1.000 UL
LT2
2399 698 M
90 -24 V
-251 -1 R
3 0 R
156 25 V
2238 673 M
1.000 UL
LT2
2399 698 M
93 -24 R
-93 24 R
1.000 UL
LT2
2816 723 M
90 -23 V
-252 -1 R
3 0 R
159 24 V
2654 699 M
1.000 UL
LT2
2816 723 M
93 -23 R
162 24 V
-255 -1 R
252 1 R
3 0 V
1.000 UL
LT0
1890 908 M
87 -47 V
3 -1 V
3 -1 V
-254 59 R
3 0 R
159 -10 V
-162 10 R
1.000 UL
LT0
1890 908 M
93 -49 R
-93 49 R
3 0 R
0 -1 R
243 -48 R
0 -1 R
3 0 R
0 -1 R
1.000 UL
LT0
2561 819 M
90 -39 V
3 -1 V
-255 36 R
3 0 R
159 4 V
-162 -4 R
1.000 UL
LT0
2561 819 M
93 -40 R
-93 40 R
3 0 R
0 -1 R
249 -31 R
0 -1 R
1.000 UL
LT0
1636 975 M
90 -56 V
3 -1 V
-255 80 R
6 -1 R
156 -22 V
-162 23 R
1.000 UL
LT0
1636 975 M
93 -57 R
-93 57 R
3 0 R
0 -1 R
3 0 R
0 -1 R
246 -64 R
0 -1 R
1.000 UL
LT2
1313 642 M
90 -26 V
-252 -4 R
3 0 R
159 30 V
1151 612 M
1.000 UL
LT2
1313 642 M
93 -26 R
-93 26 R
1.000 UL
LT0
1381 1068 M
87 -66 V
0 -1 R
3 -1 V
-252 123 R
6 -3 R
3 -1 V
3 -1 V
150 -50 V
-162 55 R
1.000 UL
LT0
1126 1222 M
93 -99 R
-93 99 R
3 -1 R
0 -1 R
3 -1 R
0 -1 R
6 -2 R
0 -1 R
237 -142 R
0 -1 V
0 -1 R
3 0 R
0 -1 R
1.000 UL
LT2
896 603 M
96 -29 V
-2 0 R
-94 29 R
3 0 V
3 0 R
1.000 UL
LT2
1729 671 M
90 -25 V
-252 -2 R
3 0 R
159 27 V
1567 644 M
1.000 UL
LT2
1729 671 M
92 -25 R
-92 25 R
1.000 UL
LT0
2977 794 M
90 -37 V
3 -1 V
-254 30 R
3 0 R
159 8 V
-162 -8 R
1.000 UL
LT0
1381 1068 M
93 -70 R
-93 70 R
249 -90 R
0 -1 R
3 0 R
0 -1 R
3 0 R
0 -1 R
1.000 UL
LT2
2144 697 M
93 -24 V
-254 -1 R
3 0 R
156 25 V
1983 672 M
1.000 UL
LT2
2144 697 M
94 -24 R
-94 24 R
1.000 UL
LT0
2306 858 M
90 -42 V
3 -1 V
-255 42 R
3 0 R
159 1 V
-162 -1 R
1.000 UL
LT0
2306 858 M
93 -43 R
-93 43 R
3 0 R
0 -1 R
249 -37 R
0 -1 R
1.000 UL
LT2
2561 722 M
90 -23 V
-252 -1 R
3 0 R
159 24 V
2399 698 M
1.000 UL
LT2
2561 722 M
93 -23 R
-93 23 R
1.000 UL
LT2
2977 747 M
90 -23 V
-251 -1 R
3 0 R
159 24 V
2816 723 M
1.000 UL
LT0
2051 904 M
87 -45 V
3 -1 V
3 -1 V
-254 51 R
3 0 R
156 -4 V
-159 4 R
1.000 UL
LT0
2051 904 M
93 -47 R
-93 47 R
3 0 R
0 -1 R
249 -44 R
0 -1 R
1.000 UL
LT0
2722 826 M
90 -39 V
3 -1 V
-254 33 R
3 0 R
159 7 V
-162 -7 R
1.000 UL
LT0
2722 826 M
94 -40 R
-94 40 R
3 0 R
0 -1 R
249 -30 R
0 -1 R
1.000 UL
LT2
1058 640 M
90 -28 V
896 603 M
3 0 R
159 37 V
896 603 M
1.000 UL
LT2
1058 640 M
93 -28 R
-93 28 R
1.000 UL
LT2
1474 669 M
90 -25 V
-251 -2 R
3 0 R
159 27 V
1313 642 M
1.000 UL
LT2
1474 669 M
93 -25 R
-93 25 R
1.000 UL
LT2
1890 696 M
90 -24 V
-251 -1 R
3 0 R
159 25 V
1729 671 M
1.000 UL
LT2
1890 696 M
93 -24 R
-93 24 R
1.000 UL
LT2
2306 721 M
90 -23 V
-252 -1 R
3 0 R
159 24 V
2144 697 M
1.000 UL
LT2
2306 721 M
93 -23 R
-93 23 R
1.000 UL
LT0
1797 962 M
90 -53 V
3 -1 V
-254 67 R
3 0 R
3 -1 V
156 -12 V
-162 13 R
1.000 UL
LT0
1797 962 M
93 -54 R
-93 54 R
3 0 R
0 -1 R
243 -55 R
0 -1 R
3 0 R
0 -1 R
1.000 UL
LT0
1288 1153 M
87 -80 V
0 -1 R
3 -1 V
-252 151 R
3 -2 R
3 -1 V
6 -3 V
150 -63 V
-162 69 R
1.000 UL
LT0
1033 1364 M
93 -142 R
-93 142 R
3 -1 R
0 -1 R
3 -2 R
0 -1 R
3 -1 R
0 -1 R
237 -195 R
0 -1 R
3 -2 R
0 -1 R
3 -1 R
0 -1 V
0 -1 R
1.000 UL
LT2
2722 746 M
27 -6 R
63 -17 V
-251 -1 R
3 0 R
87 14 V
3 0 R
6 1 V
-99 -15 R
1.000 UL
LT2
2722 746 M
94 -23 R
-94 23 R
1.000 UL
LT0
2467 862 M
90 -42 V
3 -1 V
-254 39 R
3 0 R
159 4 V
-162 -4 R
1.000 UL
LT0
2467 862 M
94 -43 R
-94 43 R
3 0 R
0 -1 R
249 -34 R
0 -1 R
1.000 UL
LT0
1542 1039 M
87 -61 V
3 -1 V
-251 91 R
3 -1 V
159 -28 V
-162 29 R
1.000 UL
LT0
1542 1039 M
94 -64 R
-94 64 R
3 0 R
0 -1 R
3 0 R
0 -1 R
246 -74 R
0 -1 R
1.000 UL
LT0
1288 1153 M
93 -85 R
-93 85 R
6 -2 R
0 -1 R
246 -109 R
0 -1 R
3 0 R
0 -1 R
1.000 UL
LT2
803 633 M
93 -30 V
-93 30 R
3 0 V
3 0 R
1.000 UL
LT2
1219 668 M
90 -26 V
-251 -2 R
3 0 R
159 28 V
1058 640 M
1.000 UL
LT2
1219 668 M
94 -26 R
-94 26 R
1.000 UL
LT0
2884 833 M
90 -38 V
3 -1 V
-255 32 R
3 0 R
159 7 V
-162 -7 R
1.000 UL
LT0
2213 904 M
90 -45 V
3 -1 V
-255 46 R
3 0 R
159 0 V
-162 0 R
1.000 UL
LT0
2213 904 M
93 -46 R
-93 46 R
3 0 R
0 -1 R
249 -40 R
0 -1 R
1.000 UL
LT2
1636 695 M
90 -24 V
-252 -2 R
3 0 R
159 26 V
1474 669 M
1.000 UL
LT2
1636 695 M
93 -24 R
-93 24 R
1.000 UL
LT2
2051 720 M
90 -23 V
-251 -1 R
3 0 R
156 24 V
1890 696 M
1.000 UL
LT2
2051 720 M
93 -23 R
-93 23 R
1.000 UL
LT0
940 1624 M
93 -260 R
3 -3 R
3 -1 V
3 -2 V
153 -89 V
940 1624 M
3 -2 R
0 -1 V
0 -1 R
3 -2 R
0 -1 V
0 -1 R
3 -2 R
0 -1 R
240 -334 R
0 -1 R
3 -3 R
0 -3 V
0 -1 R
3 0 R
0 -2 V
1.000 UL
LT0
1101 1470 M
87 -191 V
3 -4 V
0 -3 R
3 -1 V
940 1624 M
0 -1 R
3 -1 V
0 -2 R
3 -2 V
0 -1 R
3 -3 V
153 -144 V
940 1624 M
1.000 UL
LT2
2467 745 M
54 -13 R
36 -10 V
-251 -1 R
3 0 R
48 8 V
-51 -8 R
1.000 UL
LT2
2467 745 M
94 -23 R
-94 23 R
1.000 UL
LT0
1194 1269 M
84 -107 V
3 -3 V
3 -2 V
0 -2 R
3 -1 V
-254 210 R
1.000 UL
LT2
2884 769 M
78 -18 R
12 -4 V
-252 -1 R
1.000 UL
LT0
1958 955 M
87 -49 V
3 -1 V
3 -1 V
-254 58 R
3 0 R
156 -7 V
-159 7 R
1.000 UL
LT0
1958 955 M
93 -51 R
-93 51 R
3 0 R
0 -1 R
249 -49 R
0 -1 R
1.000 UL
LT0
2629 867 M
90 -40 V
3 -1 V
-255 36 R
3 0 R
159 5 V
-162 -5 R
1.000 UL
LT0
2629 867 M
93 -41 R
-93 41 R
3 0 R
0 -1 R
249 -32 R
0 -1 R
1.000 UL
LT2
965 667 M
90 -27 V
803 633 M
3 0 R
159 34 V
803 633 M
1.000 UL
LT2
965 667 M
93 -27 R
-93 27 R
1.000 UL
LT2
1381 694 M
90 -25 V
-252 -1 R
3 0 R
159 26 V
1219 668 M
1.000 UL
LT2
1381 694 M
93 -25 R
-93 25 R
1.000 UL
LT0
1704 1021 M
90 -58 V
3 -1 V
-255 77 R
6 -1 R
156 -17 V
-162 18 R
1.000 UL
LT0
1704 1021 M
93 -59 R
-93 59 R
3 0 R
0 -1 R
246 -64 R
0 -1 R
1.000 UL
LT2
1797 719 M
90 -23 V
-251 -1 R
3 0 R
159 24 V
1636 695 M
1.000 UL
LT2
1797 719 M
93 -23 R
-93 23 R
1.000 UL
LT0
1449 1115 M
90 -74 V
-251 112 R
3 -1 V
3 -1 V
156 -36 V
-162 38 R
1.000 UL
LT0
1194 1269 M
94 -116 R
-94 116 R
3 -1 R
0 -1 R
3 -1 R
0 -1 R
243 -145 R
0 -1 R
3 -1 R
0 -1 V
0 -1 R
3 0 R
0 -1 R
1.000 UL
LT2
2213 744 M
66 -16 R
24 -7 V
-252 -1 R
3 0 R
60 10 V
-63 -10 R
1.000 UL
LT2
2213 744 M
93 -23 R
-93 23 R
1.000 UL
LT0
1449 1115 M
93 -76 R
-93 76 R
249 -91 R
0 -1 R
3 0 R
3 -1 V
0 -1 R
1.000 UL
LT0
2374 906 M
90 -43 V
3 -1 V
-254 42 R
3 0 R
159 2 V
-162 -2 R
1.000 UL
LT0
2374 906 M
93 -44 R
-93 44 R
3 0 R
0 -1 R
249 -37 R
0 -1 R
1.000 UL
LT0
2119 953 M
90 -48 V
3 -1 V
-254 51 R
3 0 R
159 -2 V
-162 2 R
1.000 UL
LT0
2119 953 M
94 -49 R
-94 49 R
3 0 R
0 -1 R
249 -45 R
0 -1 R
1.000 UL
LT0
2791 874 M
90 -40 V
3 -1 V
-255 34 R
3 0 R
159 7 V
-162 -7 R
1.000 UL
LT2
710 663 M
93 -30 V
-93 30 R
3 0 V
3 0 R
1.000 UL
LT2
1126 693 M
90 -25 V
965 667 M
3 0 R
159 26 V
965 667 M
1.000 UL
LT2
1126 693 M
93 -25 R
-93 25 R
1.000 UL
LT2
1542 718 M
90 -23 V
-251 -1 R
3 0 R
159 24 V
1381 694 M
1.000 UL
LT2
1542 718 M
94 -23 R
-94 23 R
1.000 UL
LT2
1958 743 M
30 -7 R
3 -1 V
57 -15 V
-251 -1 R
3 0 R
126 20 V
1797 719 M
1.000 UL
LT2
1958 743 M
93 -23 R
-93 23 R
1.000 UL
LT0
1865 1012 M
90 -56 V
3 -1 V
-254 66 R
3 0 R
156 -9 V
-159 9 R
1.000 UL
LT0
1865 1012 M
93 -57 R
-93 57 R
3 0 R
0 -1 R
249 -57 R
0 -1 R
1.000 UL
LT0
1356 1215 M
87 -95 V
3 -2 V
0 -1 R
3 -1 V
-255 153 R
3 -2 R
3 -1 V
156 -51 V
-162 54 R
1.000 UL
LT0
1101 1470 M
93 -201 R
-93 201 R
3 -2 R
0 -1 R
249 -248 R
0 -1 V
0 -1 R
3 -1 R
0 -1 R
1.000 UL
LT0
2536 911 M
90 -43 V
3 -1 V
-255 39 R
3 0 R
159 5 V
-162 -5 R
1.000 UL
LT0
2536 911 M
93 -44 R
-93 44 R
3 0 R
0 -1 R
249 -35 R
0 -1 R
1.000 UL
LT0
1611 1092 M
87 -68 V
3 -1 V
-252 92 R
3 -1 V
159 -22 V
-162 23 R
1.000 UL
LT0
1611 1092 M
93 -71 R
-93 71 R
3 0 R
0 -1 R
3 0 R
0 -1 R
243 -77 R
0 -1 R
1.000 UL
LT0
1356 1215 M
93 -100 R
-93 100 R
3 -1 R
0 -1 R
3 0 R
0 -1 R
243 -116 R
0 -1 R
3 -1 R
0 -1 R
3 0 R
0 -1 R
1.000 UL
LT0
1263 1382 M
90 -163 V
0 -1 R
3 -2 V
-255 254 R
0 -1 R
3 -1 V
159 -86 V
-162 88 R
1.000 UL
LT2
871 694 M
90 -27 V
710 663 M
3 0 R
159 31 V
710 663 M
1.000 UL
LT2
871 694 M
94 -27 R
-94 27 R
1.000 UL
LT2
1288 718 M
90 -24 V
-252 -1 R
3 0 R
159 25 V
1126 693 M
1.000 UL
LT2
1288 718 M
93 -24 R
-93 24 R
1.000 UL
LT0
2281 954 M
90 -47 V
3 -1 V
-255 47 R
3 0 R
159 1 V
-162 -1 R
1.000 UL
LT0
2281 954 M
93 -48 R
-93 48 R
3 0 R
0 -1 R
249 -41 R
0 -1 R
1.000 UL
LT2
1704 742 M
90 -23 V
-252 -1 R
3 0 R
159 24 V
1542 718 M
1.000 UL
LT2
1704 742 M
93 -23 R
-93 23 R
1.000 UL
LT0
2026 1007 M
90 -53 V
3 -1 V
-254 59 R
3 0 R
159 -5 V
-162 5 R
1.000 UL
LT0
2026 1007 M
93 -54 R
-93 54 R
3 0 R
0 -1 R
249 -51 R
0 -1 R
1.000 UL
LT0
2697 917 M
90 -42 V
3 -1 V
-254 37 R
3 0 R
159 6 V
-162 -6 R
1.000 UL
LT2
617 692 M
93 -29 V
-93 29 R
3 0 V
3 0 R
1.000 UL
LT2
1033 718 M
90 -25 V
-252 1 R
3 0 R
159 24 V
871 694 M
1.000 UL
LT2
1033 718 M
93 -25 R
-93 25 R
1.000 UL
LT0
1772 1077 M
90 -64 V
3 -1 V
-254 80 R
3 0 R
3 -1 V
153 -14 V
-159 15 R
1.000 UL
LT0
1772 1077 M
93 -65 R
-93 65 R
3 0 R
0 -1 R
249 -68 R
0 -1 R
1.000 UL
LT2
1449 741 M
90 -23 V
-251 0 R
3 0 R
159 23 V
1288 718 M
1.000 UL
LT2
1449 741 M
93 -23 R
-93 23 R
1.000 UL
LT0
1518 1181 M
87 -85 V
3 -2 V
-252 121 R
3 -1 V
3 -1 V
156 -32 V
-162 34 R
1.000 UL
LT0
1263 1382 M
93 -167 R
-93 167 R
3 -1 R
0 -1 V
0 -1 R
249 -194 R
0 -2 V
0 -1 R
3 0 R
0 -1 R
1.000 UL
LT2
1865 764 M
1704 742 M
3 0 R
93 13 V
-96 -13 R
1.000 UL
LT0
1518 1181 M
93 -89 R
-93 89 R
249 -102 R
3 -1 V
0 -1 R
1.000 UL
LT0
2442 957 M
90 -45 V
3 -1 V
-254 43 R
3 0 R
159 3 V
-162 -3 R
1.000 UL
LT0
2442 957 M
94 -46 R
-94 46 R
3 0 R
0 -1 R
249 -38 R
0 -1 R
1.000 UL
LT2
778 719 M
90 -25 V
617 692 M
3 0 R
159 27 V
617 692 M
1.000 UL
LT2
778 719 M
93 -25 R
-93 25 R
1.000 UL
LT0
1424 1327 M
93 -142 V
0 -2 R
3 -1 V
-257 200 R
3 -2 R
156 -53 V
-159 55 R
1.000 UL
LT0
2188 1006 M
90 -51 V
3 -1 V
-255 53 R
3 0 R
159 -1 V
-162 1 R
1.000 UL
LT0
2188 1006 M
93 -52 R
-93 52 R
3 0 R
0 -1 R
249 -47 R
0 -1 R
1.000 UL
LT2
1194 740 M
90 -22 V
-251 0 R
3 0 R
159 22 V
1033 718 M
1.000 UL
LT2
1194 740 M
94 -22 R
-94 22 R
1.000 UL
LT2
1611 763 M
90 -21 V
-252 -1 R
3 0 R
159 22 V
1449 741 M
1.000 UL
LT2
1611 763 M
93 -21 R
-93 21 R
1.000 UL
LT0
1933 1069 M
90 -61 V
3 -1 V
-254 70 R
3 0 R
159 -8 V
-162 8 R
1.000 UL
LT0
1933 1069 M
93 -62 R
-93 62 R
3 0 R
0 -1 R
249 -61 R
0 -1 R
1.000 UL
LT0
2604 962 M
90 -44 V
3 -1 V
-255 40 R
3 0 R
159 5 V
-162 -5 R
1.000 UL
LT0
1679 1160 M
90 -81 V
-251 102 R
3 -1 V
156 -20 V
-159 21 R
1.000 UL
LT0
1679 1160 M
93 -83 R
-93 83 R
252 -89 R
3 -1 V
0 -1 R
1.000 UL
LT0
1424 1327 M
94 -146 R
-94 146 R
3 -1 R
0 -1 R
249 -162 R
0 -1 V
0 -1 R
3 0 R
0 -1 R
1.000 UL
LT2
523 722 M
93 -30 V
1 0 R
-94 30 R
6 0 V
3 0 R
1.000 UL
LT2
940 741 M
90 -23 V
-252 1 R
3 0 R
159 22 V
778 719 M
1.000 UL
LT2
940 741 M
93 -23 R
-93 23 R
1.000 UL
LT2
1356 762 M
90 -21 V
-252 -1 R
3 0 R
159 22 V
1194 740 M
1.000 UL
LT2
1356 762 M
93 -21 R
-93 21 R
1.000 UL
LT0
2349 1007 M
90 -49 V
3 -1 V
-254 49 R
3 0 R
159 1 V
-162 -1 R
1.000 UL
LT0
2349 1007 M
93 -50 R
-93 50 R
3 0 R
0 -1 R
249 -43 R
0 -1 R
1.000 UL
LT2
1772 784 M
1611 763 M
3 0 R
102 14 V
1611 763 M
1.000 UL
LT0
1586 1290 M
90 -127 V
0 -1 R
3 -1 V
-255 166 R
3 -1 V
159 -36 V
-162 37 R
1.000 UL
LT0
2094 1065 M
90 -58 V
3 -1 V
-254 63 R
3 0 R
159 -4 V
-162 4 R
1.000 UL
LT0
2094 1065 M
94 -59 R
-94 59 R
3 0 R
0 -1 R
249 -56 R
0 -1 R
1.000 UL
LT2
685 739 M
90 -20 V
-252 3 R
6 0 R
156 17 V
523 722 M
1.000 UL
LT2
685 739 M
93 -20 R
-93 20 R
6 0 R
3 0 R
1.000 UL
LT2
1101 761 M
90 -21 V
-251 1 R
3 0 R
159 20 V
940 741 M
1.000 UL
LT2
1101 761 M
93 -21 R
-93 21 R
1.000 UL
LT0
1840 1146 M
90 -75 V
-251 89 R
162 -14 V
-162 14 R
1.000 UL
LT0
1840 1146 M
93 -77 R
-93 77 R
3 0 R
0 -1 R
249 -79 R
0 -1 R
1.000 UL
LT2
1518 783 M
90 -20 V
-252 -1 R
3 0 R
159 21 V
1356 762 M
1.000 UL
LT2
1518 783 M
93 -20 R
-93 20 R
1.000 UL
LT0
1586 1290 M
93 -130 R
-93 130 R
3 -1 R
0 -1 R
249 -139 R
0 -1 V
0 -1 R
3 0 R
0 -1 R
1.000 UL
LT0
2511 1011 M
90 -48 V
3 -1 V
-255 45 R
3 0 R
159 4 V
-162 -4 R
1.000 UL
LT0
2256 1064 M
90 -56 V
3 -1 V
-255 58 R
3 0 R
159 -1 V
-162 1 R
1.000 UL
LT0
2256 1064 M
93 -57 R
-93 57 R
3 0 R
0 -1 R
249 -51 R
0 -1 R
1.000 UL
LT2
846 756 M
90 -15 V
685 739 M
6 0 R
156 17 V
685 739 M
1.000 UL
LT2
846 756 M
94 -15 R
-94 15 R
6 0 R
3 0 R
1.000 UL
LT2
1263 780 M
90 -18 V
-252 -1 R
3 0 R
159 19 V
1101 761 M
1.000 UL
LT2
1263 780 M
93 -18 R
-93 18 R
1.000 UL
LT2
1679 803 M
1518 783 M
3 0 R
120 16 V
1518 783 M
1.000 UL
LT0
1747 1264 M
90 -115 V
0 -1 R
3 -1 V
-254 143 R
3 -1 V
159 -25 V
-162 26 R
1.000 UL
LT0
2001 1137 M
90 -71 V
3 -1 V
-254 81 R
3 0 R
159 -9 V
-162 9 R
1.000 UL
LT0
2001 1137 M
93 -72 R
-93 72 R
3 0 R
0 -1 R
249 -71 R
0 -1 R
1.000 UL
LT0
1747 1264 M
93 -118 R
-93 118 R
252 -125 R
0 -1 R
3 0 R
0 -1 R
1.000 UL
LT2
1008 773 M
90 -12 V
846 756 M
6 0 R
156 17 V
846 756 M
1.000 UL
LT2
1008 773 M
93 -12 R
-93 12 R
1.000 UL
LT0
2417 1066 M
93 -54 V
3 -1 V
-257 53 R
3 0 R
156 2 V
-159 -2 R
1.000 UL
LT2
1424 800 M
93 -17 V
-254 -3 R
3 0 R
156 20 V
1263 780 M
1.000 UL
LT2
1424 800 M
94 -17 R
-94 17 R
1.000 UL
LT0
2163 1133 M
90 -68 V
3 -1 V
-255 73 R
3 0 R
159 -4 V
-162 4 R
1.000 UL
LT0
2163 1133 M
93 -69 R
-93 69 R
3 0 R
0 -1 R
246 -65 R
0 -1 R
1.000 UL
LT0
1908 1247 M
90 -108 V
3 -1 V
-254 126 R
162 -17 V
-162 17 R
1.000 UL
LT0
1908 1247 M
93 -110 R
-93 110 R
252 -112 R
0 -1 R
3 0 R
0 -1 R
1.000 UL
LT2
1169 791 M
93 -11 V
-254 -7 R
3 0 R
156 18 V
1008 773 M
1.000 UL
LT2
1169 791 M
94 -11 R
-94 11 R
6 0 R
3 0 R
6 1 R
3 0 R
231 8 R
3 0 R
1.000 UL
LT2
1586 818 M
18 -3 V
1424 800 M
3 0 R
159 18 V
1424 800 M
1.000 UL
LT2
1586 818 M
93 -15 R
-93 15 R
1.000 UL
LT0
2324 1131 M
90 -64 V
3 -1 V
-254 67 R
3 0 R
156 -2 V
-159 2 R
1.000 UL
LT0
2069 1235 M
93 -100 V
3 -1 V
-257 113 R
159 -12 V
-159 12 R
1.000 UL
LT0
2069 1235 M
94 -102 R
-94 102 R
252 -102 R
0 -1 R
3 0 R
0 -1 R
1.000 UL
LT2
1331 808 M
87 -8 V
-249 -9 R
6 0 R
9 1 V
147 16 V
1169 791 M
1.000 UL
LT2
1331 808 M
93 -8 R
-93 8 R
12 1 R
3 0 V
3 0 R
1.000 UL
LT2
1747 837 M
1586 818 M
3 0 R
9 1 V
-12 -1 R
1.000 UL
LT2
1493 825 M
87 -7 V
1331 808 M
6 0 R
6 1 V
3 0 R
147 16 V
1331 808 M
1.000 UL
LT2
1493 825 M
93 -7 R
-93 7 R
12 1 R
3 0 V
3 0 R
1.000 UL
LT0
2231 1227 M
90 -94 V
3 -1 V
-255 103 R
162 -8 V
-162 8 R
1.000 UL
LT2
1654 842 M
1493 825 M
6 0 R
6 1 V
3 0 R
66 7 V
-81 -8 R
1.000 UL
LT0
3124 1796 M
263 0 V
1.000 UL
LT2
3124 1696 M
263 0 V
1.000 UL
LTb
3077 595 M
1457 423 L
522 721 M
1457 423 L
522 721 M
3 0 V
2142 892 M
2799 682 M
278 -87 V
522 721 M
0 1 V
0 1 R
0 941 V
0 -943 R
3 0 V
2078 885 M
756 646 M
3 0 V
2312 811 M
990 572 M
6 0 V
2546 737 M
1223 498 M
6 0 V
2780 662 M
63 7 V
1457 423 M
62 6 V
3014 588 M
63 7 V
1396 442 M
61 -19 V
522 721 M
60 -20 V
1801 485 M
61 -19 V
927 764 M
2206 528 M
61 -19 V
1332 806 M
2611 571 M
61 -19 V
1737 849 M
3016 614 M
61 -19 V
2142 892 M
522 721 M
3 0 V
-3 188 R
63 0 V
-63 189 R
63 0 V
-63 189 R
63 0 V
-63 188 R
63 0 V
-63 189 R
63 0 V
stroke
grestore
end
showpage
}}%
\put(396,1664){\makebox(0,0)[r]{$5$}}%
\put(396,1475){\makebox(0,0)[r]{$4$}}%
\put(396,1287){\makebox(0,0)[r]{$3$}}%
\put(396,1098){\makebox(0,0)[r]{$2$}}%
\put(396,909){\makebox(0,0)[r]{$1$}}%
\put(396,721){\makebox(0,0)[r]{$0$}}%
\put(2501,435){\makebox(0,0){$w_B$}}%
\put(3124,564){\makebox(0,0)[l]{$0.4$}}%
\put(2719,521){\makebox(0,0)[l]{$0.3$}}%
\put(2314,478){\makebox(0,0)[l]{$0.2$}}%
\put(1909,435){\makebox(0,0)[l]{$0.1$}}%
\put(1504,392){\makebox(0,0)[l]{$0.0$}}%
\put(585,529){\makebox(0,0){$w_A$}}%
\put(1407,412){\makebox(0,0)[r]{$0.4$}}%
\put(1173,487){\makebox(0,0)[r]{$0.3$}}%
\put(940,561){\makebox(0,0)[r]{$0.2$}}%
\put(706,635){\makebox(0,0)[r]{$0.1$}}%
\put(472,710){\makebox(0,0)[r]{$0.0$}}%
\put(3074,1696){\makebox(0,0)[r]{$w_A w_B C(w_A, w_B)$}}%
\put(3074,1796){\makebox(0,0)[r]{$C(w_A, w_B)$}}%
\end{picture}%
\endgroup
 

%% file: fig6.tex
% GNUPLOT: LaTeX picture with Postscript
\begingroup%
  \makeatletter%
  \newcommand{\GNUPLOTspecial}{%
    \@sanitize\catcode`\%=14\relax\special}%
  \setlength{\unitlength}{0.1bp}%
{\GNUPLOTspecial{!
%!PS-Adobe-2.0
%%Title: fig6.tex
%%Creator: gnuplot 3.7 patchlevel 2
%%CreationDate: Tue Jun 18 19:52:24 2002
%%DocumentFonts: 
%%BoundingBox: 0 0 360 216
%%Orientation: Landscape
%%Pages: (atend)
%%EndComments
/gnudict 256 dict def
gnudict begin
/Color false def
/Solid false def
/gnulinewidth 5.000 def
/userlinewidth gnulinewidth def
/vshift -33 def
/dl {10 mul} def
/hpt_ 31.5 def
/vpt_ 31.5 def
/hpt hpt_ def
/vpt vpt_ def
/M {moveto} bind def
/L {lineto} bind def
/R {rmoveto} bind def
/V {rlineto} bind def
/vpt2 vpt 2 mul def
/hpt2 hpt 2 mul def
/Lshow { currentpoint stroke M
  0 vshift R show } def
/Rshow { currentpoint stroke M
  dup stringwidth pop neg vshift R show } def
/Cshow { currentpoint stroke M
  dup stringwidth pop -2 div vshift R show } def
/UP { dup vpt_ mul /vpt exch def hpt_ mul /hpt exch def
  /hpt2 hpt 2 mul def /vpt2 vpt 2 mul def } def
/DL { Color {setrgbcolor Solid {pop []} if 0 setdash }
 {pop pop pop Solid {pop []} if 0 setdash} ifelse } def
/BL { stroke userlinewidth 2 mul setlinewidth } def
/AL { stroke userlinewidth 2 div setlinewidth } def
/UL { dup gnulinewidth mul /userlinewidth exch def
      dup 1 lt {pop 1} if 10 mul /udl exch def } def
/PL { stroke userlinewidth setlinewidth } def
/LTb { BL [] 0 0 0 DL } def
/LTa { AL [1 udl mul 2 udl mul] 0 setdash 0 0 0 setrgbcolor } def
/LT0 { PL [] 1 0 0 DL } def
/LT1 { PL [4 dl 2 dl] 0 1 0 DL } def
/LT2 { PL [2 dl 3 dl] 0 0 1 DL } def
/LT3 { PL [1 dl 1.5 dl] 1 0 1 DL } def
/LT4 { PL [5 dl 2 dl 1 dl 2 dl] 0 1 1 DL } def
/LT5 { PL [4 dl 3 dl 1 dl 3 dl] 1 1 0 DL } def
/LT6 { PL [2 dl 2 dl 2 dl 4 dl] 0 0 0 DL } def
/LT7 { PL [2 dl 2 dl 2 dl 2 dl 2 dl 4 dl] 1 0.3 0 DL } def
/LT8 { PL [2 dl 2 dl 2 dl 2 dl 2 dl 2 dl 2 dl 4 dl] 0.5 0.5 0.5 DL } def
/Pnt { stroke [] 0 setdash
   gsave 1 setlinecap M 0 0 V stroke grestore } def
/Dia { stroke [] 0 setdash 2 copy vpt add M
  hpt neg vpt neg V hpt vpt neg V
  hpt vpt V hpt neg vpt V closepath stroke
  Pnt } def
/Pls { stroke [] 0 setdash vpt sub M 0 vpt2 V
  currentpoint stroke M
  hpt neg vpt neg R hpt2 0 V stroke
  } def
/Box { stroke [] 0 setdash 2 copy exch hpt sub exch vpt add M
  0 vpt2 neg V hpt2 0 V 0 vpt2 V
  hpt2 neg 0 V closepath stroke
  Pnt } def
/Crs { stroke [] 0 setdash exch hpt sub exch vpt add M
  hpt2 vpt2 neg V currentpoint stroke M
  hpt2 neg 0 R hpt2 vpt2 V stroke } def
/TriU { stroke [] 0 setdash 2 copy vpt 1.12 mul add M
  hpt neg vpt -1.62 mul V
  hpt 2 mul 0 V
  hpt neg vpt 1.62 mul V closepath stroke
  Pnt  } def
/Star { 2 copy Pls Crs } def
/BoxF { stroke [] 0 setdash exch hpt sub exch vpt add M
  0 vpt2 neg V  hpt2 0 V  0 vpt2 V
  hpt2 neg 0 V  closepath fill } def
/TriUF { stroke [] 0 setdash vpt 1.12 mul add M
  hpt neg vpt -1.62 mul V
  hpt 2 mul 0 V
  hpt neg vpt 1.62 mul V closepath fill } def
/TriD { stroke [] 0 setdash 2 copy vpt 1.12 mul sub M
  hpt neg vpt 1.62 mul V
  hpt 2 mul 0 V
  hpt neg vpt -1.62 mul V closepath stroke
  Pnt  } def
/TriDF { stroke [] 0 setdash vpt 1.12 mul sub M
  hpt neg vpt 1.62 mul V
  hpt 2 mul 0 V
  hpt neg vpt -1.62 mul V closepath fill} def
/DiaF { stroke [] 0 setdash vpt add M
  hpt neg vpt neg V hpt vpt neg V
  hpt vpt V hpt neg vpt V closepath fill } def
/Pent { stroke [] 0 setdash 2 copy gsave
  translate 0 hpt M 4 {72 rotate 0 hpt L} repeat
  closepath stroke grestore Pnt } def
/PentF { stroke [] 0 setdash gsave
  translate 0 hpt M 4 {72 rotate 0 hpt L} repeat
  closepath fill grestore } def
/Circle { stroke [] 0 setdash 2 copy
  hpt 0 360 arc stroke Pnt } def
/CircleF { stroke [] 0 setdash hpt 0 360 arc fill } def
/C0 { BL [] 0 setdash 2 copy moveto vpt 90 450  arc } bind def
/C1 { BL [] 0 setdash 2 copy        moveto
       2 copy  vpt 0 90 arc closepath fill
               vpt 0 360 arc closepath } bind def
/C2 { BL [] 0 setdash 2 copy moveto
       2 copy  vpt 90 180 arc closepath fill
               vpt 0 360 arc closepath } bind def
/C3 { BL [] 0 setdash 2 copy moveto
       2 copy  vpt 0 180 arc closepath fill
               vpt 0 360 arc closepath } bind def
/C4 { BL [] 0 setdash 2 copy moveto
       2 copy  vpt 180 270 arc closepath fill
               vpt 0 360 arc closepath } bind def
/C5 { BL [] 0 setdash 2 copy moveto
       2 copy  vpt 0 90 arc
       2 copy moveto
       2 copy  vpt 180 270 arc closepath fill
               vpt 0 360 arc } bind def
/C6 { BL [] 0 setdash 2 copy moveto
      2 copy  vpt 90 270 arc closepath fill
              vpt 0 360 arc closepath } bind def
/C7 { BL [] 0 setdash 2 copy moveto
      2 copy  vpt 0 270 arc closepath fill
              vpt 0 360 arc closepath } bind def
/C8 { BL [] 0 setdash 2 copy moveto
      2 copy vpt 270 360 arc closepath fill
              vpt 0 360 arc closepath } bind def
/C9 { BL [] 0 setdash 2 copy moveto
      2 copy  vpt 270 450 arc closepath fill
              vpt 0 360 arc closepath } bind def
/C10 { BL [] 0 setdash 2 copy 2 copy moveto vpt 270 360 arc closepath fill
       2 copy moveto
       2 copy vpt 90 180 arc closepath fill
               vpt 0 360 arc closepath } bind def
/C11 { BL [] 0 setdash 2 copy moveto
       2 copy  vpt 0 180 arc closepath fill
       2 copy moveto
       2 copy  vpt 270 360 arc closepath fill
               vpt 0 360 arc closepath } bind def
/C12 { BL [] 0 setdash 2 copy moveto
       2 copy  vpt 180 360 arc closepath fill
               vpt 0 360 arc closepath } bind def
/C13 { BL [] 0 setdash  2 copy moveto
       2 copy  vpt 0 90 arc closepath fill
       2 copy moveto
       2 copy  vpt 180 360 arc closepath fill
               vpt 0 360 arc closepath } bind def
/C14 { BL [] 0 setdash 2 copy moveto
       2 copy  vpt 90 360 arc closepath fill
               vpt 0 360 arc } bind def
/C15 { BL [] 0 setdash 2 copy vpt 0 360 arc closepath fill
               vpt 0 360 arc closepath } bind def
/Rec   { newpath 4 2 roll moveto 1 index 0 rlineto 0 exch rlineto
       neg 0 rlineto closepath } bind def
/Square { dup Rec } bind def
/Bsquare { vpt sub exch vpt sub exch vpt2 Square } bind def
/S0 { BL [] 0 setdash 2 copy moveto 0 vpt rlineto BL Bsquare } bind def
/S1 { BL [] 0 setdash 2 copy vpt Square fill Bsquare } bind def
/S2 { BL [] 0 setdash 2 copy exch vpt sub exch vpt Square fill Bsquare } bind def
/S3 { BL [] 0 setdash 2 copy exch vpt sub exch vpt2 vpt Rec fill Bsquare } bind def
/S4 { BL [] 0 setdash 2 copy exch vpt sub exch vpt sub vpt Square fill Bsquare } bind def
/S5 { BL [] 0 setdash 2 copy 2 copy vpt Square fill
       exch vpt sub exch vpt sub vpt Square fill Bsquare } bind def
/S6 { BL [] 0 setdash 2 copy exch vpt sub exch vpt sub vpt vpt2 Rec fill Bsquare } bind def
/S7 { BL [] 0 setdash 2 copy exch vpt sub exch vpt sub vpt vpt2 Rec fill
       2 copy vpt Square fill
       Bsquare } bind def
/S8 { BL [] 0 setdash 2 copy vpt sub vpt Square fill Bsquare } bind def
/S9 { BL [] 0 setdash 2 copy vpt sub vpt vpt2 Rec fill Bsquare } bind def
/S10 { BL [] 0 setdash 2 copy vpt sub vpt Square fill 2 copy exch vpt sub exch vpt Square fill
       Bsquare } bind def
/S11 { BL [] 0 setdash 2 copy vpt sub vpt Square fill 2 copy exch vpt sub exch vpt2 vpt Rec fill
       Bsquare } bind def
/S12 { BL [] 0 setdash 2 copy exch vpt sub exch vpt sub vpt2 vpt Rec fill Bsquare } bind def
/S13 { BL [] 0 setdash 2 copy exch vpt sub exch vpt sub vpt2 vpt Rec fill
       2 copy vpt Square fill Bsquare } bind def
/S14 { BL [] 0 setdash 2 copy exch vpt sub exch vpt sub vpt2 vpt Rec fill
       2 copy exch vpt sub exch vpt Square fill Bsquare } bind def
/S15 { BL [] 0 setdash 2 copy Bsquare fill Bsquare } bind def
/D0 { gsave translate 45 rotate 0 0 S0 stroke grestore } bind def
/D1 { gsave translate 45 rotate 0 0 S1 stroke grestore } bind def
/D2 { gsave translate 45 rotate 0 0 S2 stroke grestore } bind def
/D3 { gsave translate 45 rotate 0 0 S3 stroke grestore } bind def
/D4 { gsave translate 45 rotate 0 0 S4 stroke grestore } bind def
/D5 { gsave translate 45 rotate 0 0 S5 stroke grestore } bind def
/D6 { gsave translate 45 rotate 0 0 S6 stroke grestore } bind def
/D7 { gsave translate 45 rotate 0 0 S7 stroke grestore } bind def
/D8 { gsave translate 45 rotate 0 0 S8 stroke grestore } bind def
/D9 { gsave translate 45 rotate 0 0 S9 stroke grestore } bind def
/D10 { gsave translate 45 rotate 0 0 S10 stroke grestore } bind def
/D11 { gsave translate 45 rotate 0 0 S11 stroke grestore } bind def
/D12 { gsave translate 45 rotate 0 0 S12 stroke grestore } bind def
/D13 { gsave translate 45 rotate 0 0 S13 stroke grestore } bind def
/D14 { gsave translate 45 rotate 0 0 S14 stroke grestore } bind def
/D15 { gsave translate 45 rotate 0 0 S15 stroke grestore } bind def
/DiaE { stroke [] 0 setdash vpt add M
  hpt neg vpt neg V hpt vpt neg V
  hpt vpt V hpt neg vpt V closepath stroke } def
/BoxE { stroke [] 0 setdash exch hpt sub exch vpt add M
  0 vpt2 neg V hpt2 0 V 0 vpt2 V
  hpt2 neg 0 V closepath stroke } def
/TriUE { stroke [] 0 setdash vpt 1.12 mul add M
  hpt neg vpt -1.62 mul V
  hpt 2 mul 0 V
  hpt neg vpt 1.62 mul V closepath stroke } def
/TriDE { stroke [] 0 setdash vpt 1.12 mul sub M
  hpt neg vpt 1.62 mul V
  hpt 2 mul 0 V
  hpt neg vpt -1.62 mul V closepath stroke } def
/PentE { stroke [] 0 setdash gsave
  translate 0 hpt M 4 {72 rotate 0 hpt L} repeat
  closepath stroke grestore } def
/CircE { stroke [] 0 setdash 
  hpt 0 360 arc stroke } def
/Opaque { gsave closepath 1 setgray fill grestore 0 setgray closepath } def
/DiaW { stroke [] 0 setdash vpt add M
  hpt neg vpt neg V hpt vpt neg V
  hpt vpt V hpt neg vpt V Opaque stroke } def
/BoxW { stroke [] 0 setdash exch hpt sub exch vpt add M
  0 vpt2 neg V hpt2 0 V 0 vpt2 V
  hpt2 neg 0 V Opaque stroke } def
/TriUW { stroke [] 0 setdash vpt 1.12 mul add M
  hpt neg vpt -1.62 mul V
  hpt 2 mul 0 V
  hpt neg vpt 1.62 mul V Opaque stroke } def
/TriDW { stroke [] 0 setdash vpt 1.12 mul sub M
  hpt neg vpt 1.62 mul V
  hpt 2 mul 0 V
  hpt neg vpt -1.62 mul V Opaque stroke } def
/PentW { stroke [] 0 setdash gsave
  translate 0 hpt M 4 {72 rotate 0 hpt L} repeat
  Opaque stroke grestore } def
/CircW { stroke [] 0 setdash 
  hpt 0 360 arc Opaque stroke } def
/BoxFill { gsave Rec 1 setgray fill grestore } def
/Symbol-Oblique /Symbol findfont [1 0 .167 1 0 0] makefont
dup length dict begin {1 index /FID eq {pop pop} {def} ifelse} forall
currentdict end definefont
end
}}%
\begin{picture}(3600,2160)(0,0)%
{\GNUPLOTspecial{"
gnudict begin
gsave
0 0 translate
0.100 0.100 scale
0 setgray
newpath
1.000 UL
LTb
400 300 M
63 0 V
2987 0 R
-63 0 V
400 551 M
63 0 V
2987 0 R
-63 0 V
400 803 M
63 0 V
2987 0 R
-63 0 V
400 1054 M
63 0 V
2987 0 R
-63 0 V
400 1306 M
63 0 V
2987 0 R
-63 0 V
400 1557 M
63 0 V
2987 0 R
-63 0 V
400 1809 M
63 0 V
2987 0 R
-63 0 V
400 2060 M
63 0 V
2987 0 R
-63 0 V
400 300 M
0 63 V
0 1697 R
0 -63 V
706 300 M
0 31 V
0 1729 R
0 -31 V
885 300 M
0 31 V
0 1729 R
0 -31 V
1012 300 M
0 31 V
0 1729 R
0 -31 V
1111 300 M
0 31 V
0 1729 R
0 -31 V
1191 300 M
0 31 V
0 1729 R
0 -31 V
1259 300 M
0 31 V
0 1729 R
0 -31 V
1318 300 M
0 31 V
0 1729 R
0 -31 V
1370 300 M
0 31 V
0 1729 R
0 -31 V
1417 300 M
0 63 V
0 1697 R
0 -63 V
1723 300 M
0 31 V
0 1729 R
0 -31 V
1902 300 M
0 31 V
0 1729 R
0 -31 V
2029 300 M
0 31 V
0 1729 R
0 -31 V
2127 300 M
0 31 V
0 1729 R
0 -31 V
2208 300 M
0 31 V
0 1729 R
0 -31 V
2276 300 M
0 31 V
0 1729 R
0 -31 V
2335 300 M
0 31 V
0 1729 R
0 -31 V
2387 300 M
0 31 V
0 1729 R
0 -31 V
2433 300 M
0 63 V
0 1697 R
0 -63 V
2739 300 M
0 31 V
0 1729 R
0 -31 V
2918 300 M
0 31 V
0 1729 R
0 -31 V
3045 300 M
0 31 V
0 1729 R
0 -31 V
3144 300 M
0 31 V
0 1729 R
0 -31 V
3224 300 M
0 31 V
0 1729 R
0 -31 V
3293 300 M
0 31 V
0 1729 R
0 -31 V
3351 300 M
0 31 V
0 1729 R
0 -31 V
3403 300 M
0 31 V
0 1729 R
0 -31 V
3450 300 M
0 63 V
0 1697 R
0 -63 V
1.000 UL
LTb
400 300 M
3050 0 V
0 1760 V
-3050 0 V
400 300 L
1.000 UL
LT0
3087 913 M
263 0 V
400 1526 M
31 -2 V
31 -3 V
30 -2 V
31 -3 V
31 -3 V
31 -3 V
31 -4 V
30 -3 V
31 -4 V
31 -4 V
31 -5 V
31 -4 V
31 -6 V
30 -5 V
31 -6 V
31 -6 V
31 -7 V
31 -7 V
30 -7 V
31 -9 V
31 -8 V
31 -10 V
31 -9 V
30 -11 V
31 -11 V
31 -12 V
31 -13 V
31 -13 V
30 -15 V
31 -15 V
31 -16 V
31 -18 V
31 -18 V
30 -19 V
31 -20 V
31 -22 V
31 -22 V
31 -24 V
31 -25 V
30 -25 V
31 -27 V
31 -28 V
31 -29 V
31 -30 V
30 -30 V
31 -32 V
31 -31 V
31 -32 V
31 -32 V
30 -32 V
31 -32 V
31 -31 V
31 -31 V
31 -29 V
30 -28 V
31 -27 V
31 -26 V
31 -24 V
31 -22 V
30 -20 V
31 -19 V
31 -18 V
31 -15 V
31 -15 V
31 -13 V
30 -12 V
31 -11 V
31 -10 V
31 -9 V
31 -8 V
30 -8 V
31 -7 V
31 -6 V
31 -6 V
31 -6 V
30 -5 V
31 -5 V
31 -4 V
31 -4 V
31 -4 V
30 -3 V
31 -3 V
31 -3 V
31 -3 V
31 -3 V
30 -2 V
31 -2 V
31 -2 V
31 -2 V
31 -2 V
31 -2 V
30 -1 V
31 -2 V
31 -1 V
31 -1 V
31 -1 V
30 -1 V
31 -1 V
31 -1 V
1.000 UL
LT1
3087 813 M
263 0 V
400 404 M
31 6 V
31 6 V
30 7 V
31 7 V
31 7 V
31 7 V
31 7 V
30 8 V
31 8 V
31 8 V
31 8 V
31 9 V
31 8 V
30 8 V
31 9 V
31 8 V
31 9 V
31 8 V
30 8 V
31 8 V
31 7 V
31 7 V
31 7 V
30 6 V
31 6 V
31 5 V
31 5 V
31 4 V
30 3 V
31 3 V
31 2 V
31 1 V
31 0 V
30 0 V
31 -1 V
31 -2 V
31 -3 V
31 -3 V
31 -4 V
30 -5 V
31 -5 V
31 -6 V
31 -6 V
31 -7 V
30 -7 V
31 -7 V
31 -8 V
31 -8 V
31 -8 V
30 -9 V
31 -8 V
31 -9 V
31 -8 V
31 -8 V
30 -9 V
31 -8 V
31 -8 V
31 -8 V
31 -8 V
30 -7 V
31 -7 V
31 -7 V
31 -7 V
31 -7 V
31 -6 V
30 -6 V
31 -6 V
31 -5 V
31 -6 V
31 -5 V
30 -5 V
31 -4 V
31 -5 V
31 -4 V
31 -4 V
30 -3 V
31 -4 V
31 -3 V
31 -3 V
31 -3 V
30 -3 V
31 -2 V
31 -3 V
31 -2 V
31 -2 V
30 -2 V
31 -2 V
31 -2 V
31 -2 V
31 -1 V
31 -2 V
30 -1 V
31 -1 V
31 -2 V
31 -1 V
31 -1 V
30 -1 V
31 -1 V
31 -1 V
1.000 UL
LT2
3087 713 M
263 0 V
400 430 M
31 9 V
31 10 V
30 10 V
31 11 V
31 11 V
31 12 V
31 13 V
30 14 V
31 14 V
31 14 V
31 16 V
31 15 V
31 17 V
30 16 V
31 17 V
31 17 V
31 18 V
31 17 V
30 18 V
31 17 V
31 17 V
31 16 V
31 16 V
30 14 V
31 14 V
31 13 V
31 11 V
31 10 V
30 8 V
31 6 V
31 5 V
31 3 V
31 1 V
30 -1 V
31 -3 V
31 -5 V
31 -6 V
31 -8 V
31 -10 V
30 -11 V
31 -13 V
31 -14 V
31 -14 V
31 -16 V
30 -16 V
31 -17 V
31 -17 V
31 -18 V
31 -17 V
30 -18 V
31 -17 V
31 -17 V
31 -16 V
31 -17 V
30 -15 V
31 -16 V
31 -14 V
31 -14 V
31 -14 V
30 -13 V
31 -12 V
31 -11 V
31 -11 V
31 -10 V
31 -10 V
30 -9 V
31 -9 V
31 -8 V
31 -8 V
31 -7 V
30 -6 V
31 -7 V
31 -5 V
31 -6 V
31 -5 V
30 -5 V
31 -4 V
31 -4 V
31 -4 V
31 -3 V
30 -4 V
31 -3 V
31 -3 V
31 -2 V
31 -3 V
30 -2 V
31 -2 V
31 -2 V
31 -2 V
31 -2 V
31 -2 V
30 -1 V
31 -2 V
31 -1 V
31 -1 V
31 -1 V
30 -1 V
31 -1 V
31 -1 V
1.000 UL
LT3
3087 613 M
263 0 V
400 414 M
31 7 V
31 8 V
30 8 V
31 9 V
31 9 V
31 10 V
31 10 V
30 10 V
31 11 V
31 12 V
31 12 V
31 12 V
31 12 V
30 14 V
31 13 V
31 14 V
31 13 V
31 14 V
30 15 V
31 14 V
31 14 V
31 13 V
31 14 V
30 13 V
31 12 V
31 12 V
31 11 V
31 10 V
30 9 V
31 8 V
31 6 V
31 5 V
31 3 V
30 2 V
31 1 V
31 -1 V
31 -3 V
31 -4 V
31 -6 V
30 -7 V
31 -9 V
31 -10 V
31 -11 V
31 -11 V
30 -13 V
31 -13 V
31 -14 V
31 -15 V
31 -15 V
30 -14 V
31 -15 V
31 -15 V
31 -15 V
31 -15 V
30 -14 V
31 -14 V
31 -13 V
31 -13 V
31 -13 V
30 -12 V
31 -11 V
31 -11 V
31 -11 V
31 -10 V
31 -9 V
30 -9 V
31 -8 V
31 -8 V
31 -8 V
31 -6 V
30 -7 V
31 -6 V
31 -6 V
31 -5 V
31 -5 V
30 -5 V
31 -4 V
31 -4 V
31 -4 V
31 -4 V
30 -3 V
31 -3 V
31 -3 V
31 -2 V
31 -3 V
30 -2 V
31 -2 V
31 -2 V
31 -2 V
31 -2 V
31 -2 V
30 -1 V
31 -2 V
31 -1 V
31 -1 V
31 -1 V
30 -1 V
31 -1 V
31 -1 V
1.000 UL
LT4
3087 513 M
263 0 V
400 460 M
31 13 V
31 14 V
30 16 V
31 16 V
31 18 V
31 18 V
31 20 V
30 22 V
31 22 V
31 24 V
31 26 V
31 26 V
31 28 V
30 28 V
31 30 V
31 31 V
31 31 V
31 31 V
30 32 V
31 31 V
31 31 V
31 29 V
31 29 V
30 27 V
31 25 V
31 22 V
31 20 V
31 16 V
30 13 V
31 10 V
31 6 V
31 1 V
31 -2 V
30 -6 V
31 -10 V
31 -13 V
31 -17 V
31 -20 V
31 -23 V
30 -25 V
31 -27 V
31 -29 V
31 -30 V
31 -31 V
30 -32 V
31 -31 V
31 -32 V
31 -31 V
31 -30 V
30 -30 V
31 -28 V
31 -28 V
31 -26 V
31 -25 V
30 -23 V
31 -22 V
31 -21 V
31 -20 V
31 -18 V
30 -17 V
31 -16 V
31 -15 V
31 -13 V
31 -13 V
31 -12 V
30 -11 V
31 -10 V
31 -9 V
31 -9 V
31 -8 V
30 -7 V
31 -7 V
31 -7 V
31 -6 V
31 -5 V
30 -5 V
31 -5 V
31 -4 V
31 -4 V
31 -4 V
30 -3 V
31 -3 V
31 -3 V
31 -3 V
31 -3 V
30 -2 V
31 -2 V
31 -2 V
31 -2 V
31 -2 V
31 -2 V
30 -1 V
31 -2 V
31 -1 V
31 -1 V
31 -1 V
30 -1 V
31 -1 V
31 -1 V
1.000 UL
LT5
3087 413 M
263 0 V
400 423 M
31 8 V
31 9 V
30 10 V
31 11 V
31 11 V
31 12 V
31 13 V
30 13 V
31 15 V
31 15 V
31 16 V
31 18 V
31 18 V
30 20 V
31 21 V
31 22 V
31 24 V
31 24 V
30 27 V
31 28 V
31 29 V
31 31 V
31 32 V
30 34 V
31 36 V
31 38 V
31 39 V
31 41 V
30 42 V
31 44 V
31 45 V
31 47 V
31 48 V
30 49 V
31 50 V
31 51 V
31 50 V
31 51 V
31 50 V
30 50 V
31 48 V
31 46 V
31 44 V
31 40 V
30 38 V
31 33 V
31 29 V
31 24 V
31 19 V
30 14 V
31 7 V
31 3 V
31 -3 V
31 -8 V
30 -12 V
31 -16 V
31 -18 V
31 -21 V
31 -22 V
30 -22 V
31 -23 V
31 -21 V
31 -21 V
31 -19 V
31 -17 V
30 -16 V
31 -15 V
31 -12 V
31 -12 V
31 -10 V
30 -10 V
31 -8 V
31 -8 V
31 -7 V
31 -6 V
30 -6 V
31 -5 V
31 -5 V
31 -4 V
31 -4 V
30 -4 V
31 -4 V
31 -3 V
31 -3 V
31 -3 V
30 -2 V
31 -2 V
31 -3 V
31 -2 V
31 -1 V
31 -2 V
30 -2 V
31 -1 V
31 -2 V
31 -1 V
31 -1 V
30 -1 V
31 -1 V
31 -1 V
stroke
grestore
end
showpage
}}%
\put(3037,413){\makebox(0,0)[r]{$C(1,1)$}}%
\put(3037,513){\makebox(0,0)[r]{$n = 4$}}%
\put(3037,613){\makebox(0,0)[r]{$n = 3$}}%
\put(3037,713){\makebox(0,0)[r]{$n = 2$}}%
\put(3037,813){\makebox(0,0)[r]{$n = 1$}}%
\put(3037,913){\makebox(0,0)[r]{$C(1,1) / \rho$}}%
\put(1925,50){\makebox(0,0){$\rho$}}%
\put(100,1180){%
\special{ps: gsave currentpoint currentpoint translate
270 rotate neg exch neg exch translate}%
\makebox(0,0)[b]{\shortstack{$C(1, 1)/\rho$}}%
\special{ps: currentpoint grestore moveto}%
}%
\put(3450,200){\makebox(0,0){$100$}}%
\put(2433,200){\makebox(0,0){$10$}}%
\put(1417,200){\makebox(0,0){$1$}}%
\put(400,200){\makebox(0,0){$0.1$}}%
\put(350,2060){\makebox(0,0)[r]{$1.4$}}%
\put(350,1809){\makebox(0,0)[r]{$1.2$}}%
\put(350,1557){\makebox(0,0)[r]{$1.0$}}%
\put(350,1306){\makebox(0,0)[r]{$0.8$}}%
\put(350,1054){\makebox(0,0)[r]{$0.6$}}%
\put(350,803){\makebox(0,0)[r]{$0.4$}}%
\put(350,551){\makebox(0,0)[r]{$0.2$}}%
\put(350,300){\makebox(0,0)[r]{$0.0$}}%
\end{picture}%
\endgroup
 

%% file: fig9.tex
% GNUPLOT: LaTeX picture with Postscript
\begingroup%
  \makeatletter%
  \newcommand{\GNUPLOTspecial}{%
    \@sanitize\catcode`\%=14\relax\special}%
  \setlength{\unitlength}{0.1bp}%
{\GNUPLOTspecial{!
%!PS-Adobe-2.0
%%Title: fig9.tex
%%Creator: gnuplot 3.7 patchlevel 2
%%CreationDate: Tue Jun 18 19:54:59 2002
%%DocumentFonts: 
%%BoundingBox: 0 0 360 216
%%Orientation: Landscape
%%Pages: (atend)
%%EndComments
/gnudict 256 dict def
gnudict begin
/Color false def
/Solid false def
/gnulinewidth 5.000 def
/userlinewidth gnulinewidth def
/vshift -33 def
/dl {10 mul} def
/hpt_ 31.5 def
/vpt_ 31.5 def
/hpt hpt_ def
/vpt vpt_ def
/M {moveto} bind def
/L {lineto} bind def
/R {rmoveto} bind def
/V {rlineto} bind def
/vpt2 vpt 2 mul def
/hpt2 hpt 2 mul def
/Lshow { currentpoint stroke M
  0 vshift R show } def
/Rshow { currentpoint stroke M
  dup stringwidth pop neg vshift R show } def
/Cshow { currentpoint stroke M
  dup stringwidth pop -2 div vshift R show } def
/UP { dup vpt_ mul /vpt exch def hpt_ mul /hpt exch def
  /hpt2 hpt 2 mul def /vpt2 vpt 2 mul def } def
/DL { Color {setrgbcolor Solid {pop []} if 0 setdash }
 {pop pop pop Solid {pop []} if 0 setdash} ifelse } def
/BL { stroke userlinewidth 2 mul setlinewidth } def
/AL { stroke userlinewidth 2 div setlinewidth } def
/UL { dup gnulinewidth mul /userlinewidth exch def
      dup 1 lt {pop 1} if 10 mul /udl exch def } def
/PL { stroke userlinewidth setlinewidth } def
/LTb { BL [] 0 0 0 DL } def
/LTa { AL [1 udl mul 2 udl mul] 0 setdash 0 0 0 setrgbcolor } def
/LT0 { PL [] 1 0 0 DL } def
/LT1 { PL [4 dl 2 dl] 0 1 0 DL } def
/LT2 { PL [2 dl 3 dl] 0 0 1 DL } def
/LT3 { PL [1 dl 1.5 dl] 1 0 1 DL } def
/LT4 { PL [5 dl 2 dl 1 dl 2 dl] 0 1 1 DL } def
/LT5 { PL [4 dl 3 dl 1 dl 3 dl] 1 1 0 DL } def
/LT6 { PL [2 dl 2 dl 2 dl 4 dl] 0 0 0 DL } def
/LT7 { PL [2 dl 2 dl 2 dl 2 dl 2 dl 4 dl] 1 0.3 0 DL } def
/LT8 { PL [2 dl 2 dl 2 dl 2 dl 2 dl 2 dl 2 dl 4 dl] 0.5 0.5 0.5 DL } def
/Pnt { stroke [] 0 setdash
   gsave 1 setlinecap M 0 0 V stroke grestore } def
/Dia { stroke [] 0 setdash 2 copy vpt add M
  hpt neg vpt neg V hpt vpt neg V
  hpt vpt V hpt neg vpt V closepath stroke
  Pnt } def
/Pls { stroke [] 0 setdash vpt sub M 0 vpt2 V
  currentpoint stroke M
  hpt neg vpt neg R hpt2 0 V stroke
  } def
/Box { stroke [] 0 setdash 2 copy exch hpt sub exch vpt add M
  0 vpt2 neg V hpt2 0 V 0 vpt2 V
  hpt2 neg 0 V closepath stroke
  Pnt } def
/Crs { stroke [] 0 setdash exch hpt sub exch vpt add M
  hpt2 vpt2 neg V currentpoint stroke M
  hpt2 neg 0 R hpt2 vpt2 V stroke } def
/TriU { stroke [] 0 setdash 2 copy vpt 1.12 mul add M
  hpt neg vpt -1.62 mul V
  hpt 2 mul 0 V
  hpt neg vpt 1.62 mul V closepath stroke
  Pnt  } def
/Star { 2 copy Pls Crs } def
/BoxF { stroke [] 0 setdash exch hpt sub exch vpt add M
  0 vpt2 neg V  hpt2 0 V  0 vpt2 V
  hpt2 neg 0 V  closepath fill } def
/TriUF { stroke [] 0 setdash vpt 1.12 mul add M
  hpt neg vpt -1.62 mul V
  hpt 2 mul 0 V
  hpt neg vpt 1.62 mul V closepath fill } def
/TriD { stroke [] 0 setdash 2 copy vpt 1.12 mul sub M
  hpt neg vpt 1.62 mul V
  hpt 2 mul 0 V
  hpt neg vpt -1.62 mul V closepath stroke
  Pnt  } def
/TriDF { stroke [] 0 setdash vpt 1.12 mul sub M
  hpt neg vpt 1.62 mul V
  hpt 2 mul 0 V
  hpt neg vpt -1.62 mul V closepath fill} def
/DiaF { stroke [] 0 setdash vpt add M
  hpt neg vpt neg V hpt vpt neg V
  hpt vpt V hpt neg vpt V closepath fill } def
/Pent { stroke [] 0 setdash 2 copy gsave
  translate 0 hpt M 4 {72 rotate 0 hpt L} repeat
  closepath stroke grestore Pnt } def
/PentF { stroke [] 0 setdash gsave
  translate 0 hpt M 4 {72 rotate 0 hpt L} repeat
  closepath fill grestore } def
/Circle { stroke [] 0 setdash 2 copy
  hpt 0 360 arc stroke Pnt } def
/CircleF { stroke [] 0 setdash hpt 0 360 arc fill } def
/C0 { BL [] 0 setdash 2 copy moveto vpt 90 450  arc } bind def
/C1 { BL [] 0 setdash 2 copy        moveto
       2 copy  vpt 0 90 arc closepath fill
               vpt 0 360 arc closepath } bind def
/C2 { BL [] 0 setdash 2 copy moveto
       2 copy  vpt 90 180 arc closepath fill
               vpt 0 360 arc closepath } bind def
/C3 { BL [] 0 setdash 2 copy moveto
       2 copy  vpt 0 180 arc closepath fill
               vpt 0 360 arc closepath } bind def
/C4 { BL [] 0 setdash 2 copy moveto
       2 copy  vpt 180 270 arc closepath fill
               vpt 0 360 arc closepath } bind def
/C5 { BL [] 0 setdash 2 copy moveto
       2 copy  vpt 0 90 arc
       2 copy moveto
       2 copy  vpt 180 270 arc closepath fill
               vpt 0 360 arc } bind def
/C6 { BL [] 0 setdash 2 copy moveto
      2 copy  vpt 90 270 arc closepath fill
              vpt 0 360 arc closepath } bind def
/C7 { BL [] 0 setdash 2 copy moveto
      2 copy  vpt 0 270 arc closepath fill
              vpt 0 360 arc closepath } bind def
/C8 { BL [] 0 setdash 2 copy moveto
      2 copy vpt 270 360 arc closepath fill
              vpt 0 360 arc closepath } bind def
/C9 { BL [] 0 setdash 2 copy moveto
      2 copy  vpt 270 450 arc closepath fill
              vpt 0 360 arc closepath } bind def
/C10 { BL [] 0 setdash 2 copy 2 copy moveto vpt 270 360 arc closepath fill
       2 copy moveto
       2 copy vpt 90 180 arc closepath fill
               vpt 0 360 arc closepath } bind def
/C11 { BL [] 0 setdash 2 copy moveto
       2 copy  vpt 0 180 arc closepath fill
       2 copy moveto
       2 copy  vpt 270 360 arc closepath fill
               vpt 0 360 arc closepath } bind def
/C12 { BL [] 0 setdash 2 copy moveto
       2 copy  vpt 180 360 arc closepath fill
               vpt 0 360 arc closepath } bind def
/C13 { BL [] 0 setdash  2 copy moveto
       2 copy  vpt 0 90 arc closepath fill
       2 copy moveto
       2 copy  vpt 180 360 arc closepath fill
               vpt 0 360 arc closepath } bind def
/C14 { BL [] 0 setdash 2 copy moveto
       2 copy  vpt 90 360 arc closepath fill
               vpt 0 360 arc } bind def
/C15 { BL [] 0 setdash 2 copy vpt 0 360 arc closepath fill
               vpt 0 360 arc closepath } bind def
/Rec   { newpath 4 2 roll moveto 1 index 0 rlineto 0 exch rlineto
       neg 0 rlineto closepath } bind def
/Square { dup Rec } bind def
/Bsquare { vpt sub exch vpt sub exch vpt2 Square } bind def
/S0 { BL [] 0 setdash 2 copy moveto 0 vpt rlineto BL Bsquare } bind def
/S1 { BL [] 0 setdash 2 copy vpt Square fill Bsquare } bind def
/S2 { BL [] 0 setdash 2 copy exch vpt sub exch vpt Square fill Bsquare } bind def
/S3 { BL [] 0 setdash 2 copy exch vpt sub exch vpt2 vpt Rec fill Bsquare } bind def
/S4 { BL [] 0 setdash 2 copy exch vpt sub exch vpt sub vpt Square fill Bsquare } bind def
/S5 { BL [] 0 setdash 2 copy 2 copy vpt Square fill
       exch vpt sub exch vpt sub vpt Square fill Bsquare } bind def
/S6 { BL [] 0 setdash 2 copy exch vpt sub exch vpt sub vpt vpt2 Rec fill Bsquare } bind def
/S7 { BL [] 0 setdash 2 copy exch vpt sub exch vpt sub vpt vpt2 Rec fill
       2 copy vpt Square fill
       Bsquare } bind def
/S8 { BL [] 0 setdash 2 copy vpt sub vpt Square fill Bsquare } bind def
/S9 { BL [] 0 setdash 2 copy vpt sub vpt vpt2 Rec fill Bsquare } bind def
/S10 { BL [] 0 setdash 2 copy vpt sub vpt Square fill 2 copy exch vpt sub exch vpt Square fill
       Bsquare } bind def
/S11 { BL [] 0 setdash 2 copy vpt sub vpt Square fill 2 copy exch vpt sub exch vpt2 vpt Rec fill
       Bsquare } bind def
/S12 { BL [] 0 setdash 2 copy exch vpt sub exch vpt sub vpt2 vpt Rec fill Bsquare } bind def
/S13 { BL [] 0 setdash 2 copy exch vpt sub exch vpt sub vpt2 vpt Rec fill
       2 copy vpt Square fill Bsquare } bind def
/S14 { BL [] 0 setdash 2 copy exch vpt sub exch vpt sub vpt2 vpt Rec fill
       2 copy exch vpt sub exch vpt Square fill Bsquare } bind def
/S15 { BL [] 0 setdash 2 copy Bsquare fill Bsquare } bind def
/D0 { gsave translate 45 rotate 0 0 S0 stroke grestore } bind def
/D1 { gsave translate 45 rotate 0 0 S1 stroke grestore } bind def
/D2 { gsave translate 45 rotate 0 0 S2 stroke grestore } bind def
/D3 { gsave translate 45 rotate 0 0 S3 stroke grestore } bind def
/D4 { gsave translate 45 rotate 0 0 S4 stroke grestore } bind def
/D5 { gsave translate 45 rotate 0 0 S5 stroke grestore } bind def
/D6 { gsave translate 45 rotate 0 0 S6 stroke grestore } bind def
/D7 { gsave translate 45 rotate 0 0 S7 stroke grestore } bind def
/D8 { gsave translate 45 rotate 0 0 S8 stroke grestore } bind def
/D9 { gsave translate 45 rotate 0 0 S9 stroke grestore } bind def
/D10 { gsave translate 45 rotate 0 0 S10 stroke grestore } bind def
/D11 { gsave translate 45 rotate 0 0 S11 stroke grestore } bind def
/D12 { gsave translate 45 rotate 0 0 S12 stroke grestore } bind def
/D13 { gsave translate 45 rotate 0 0 S13 stroke grestore } bind def
/D14 { gsave translate 45 rotate 0 0 S14 stroke grestore } bind def
/D15 { gsave translate 45 rotate 0 0 S15 stroke grestore } bind def
/DiaE { stroke [] 0 setdash vpt add M
  hpt neg vpt neg V hpt vpt neg V
  hpt vpt V hpt neg vpt V closepath stroke } def
/BoxE { stroke [] 0 setdash exch hpt sub exch vpt add M
  0 vpt2 neg V hpt2 0 V 0 vpt2 V
  hpt2 neg 0 V closepath stroke } def
/TriUE { stroke [] 0 setdash vpt 1.12 mul add M
  hpt neg vpt -1.62 mul V
  hpt 2 mul 0 V
  hpt neg vpt 1.62 mul V closepath stroke } def
/TriDE { stroke [] 0 setdash vpt 1.12 mul sub M
  hpt neg vpt 1.62 mul V
  hpt 2 mul 0 V
  hpt neg vpt -1.62 mul V closepath stroke } def
/PentE { stroke [] 0 setdash gsave
  translate 0 hpt M 4 {72 rotate 0 hpt L} repeat
  closepath stroke grestore } def
/CircE { stroke [] 0 setdash 
  hpt 0 360 arc stroke } def
/Opaque { gsave closepath 1 setgray fill grestore 0 setgray closepath } def
/DiaW { stroke [] 0 setdash vpt add M
  hpt neg vpt neg V hpt vpt neg V
  hpt vpt V hpt neg vpt V Opaque stroke } def
/BoxW { stroke [] 0 setdash exch hpt sub exch vpt add M
  0 vpt2 neg V hpt2 0 V 0 vpt2 V
  hpt2 neg 0 V Opaque stroke } def
/TriUW { stroke [] 0 setdash vpt 1.12 mul add M
  hpt neg vpt -1.62 mul V
  hpt 2 mul 0 V
  hpt neg vpt 1.62 mul V Opaque stroke } def
/TriDW { stroke [] 0 setdash vpt 1.12 mul sub M
  hpt neg vpt 1.62 mul V
  hpt 2 mul 0 V
  hpt neg vpt -1.62 mul V Opaque stroke } def
/PentW { stroke [] 0 setdash gsave
  translate 0 hpt M 4 {72 rotate 0 hpt L} repeat
  Opaque stroke grestore } def
/CircW { stroke [] 0 setdash 
  hpt 0 360 arc Opaque stroke } def
/BoxFill { gsave Rec 1 setgray fill grestore } def
/Symbol-Oblique /Symbol findfont [1 0 .167 1 0 0] makefont
dup length dict begin {1 index /FID eq {pop pop} {def} ifelse} forall
currentdict end definefont
end
}}%
\begin{picture}(3600,2160)(0,0)%
{\GNUPLOTspecial{"
gnudict begin
gsave
0 0 translate
0.100 0.100 scale
0 setgray
newpath
1.000 UL
LTb
450 300 M
63 0 V
2937 0 R
-63 0 V
450 593 M
63 0 V
2937 0 R
-63 0 V
450 887 M
63 0 V
2937 0 R
-63 0 V
450 1180 M
63 0 V
2937 0 R
-63 0 V
450 1473 M
63 0 V
2937 0 R
-63 0 V
450 1767 M
63 0 V
2937 0 R
-63 0 V
450 2060 M
63 0 V
2937 0 R
-63 0 V
450 300 M
0 63 V
0 1697 R
0 -63 V
753 300 M
0 63 V
0 1697 R
0 -63 V
1056 300 M
0 63 V
0 1697 R
0 -63 V
1359 300 M
0 63 V
0 1697 R
0 -63 V
1662 300 M
0 63 V
0 1697 R
0 -63 V
1965 300 M
0 63 V
0 1697 R
0 -63 V
2268 300 M
0 63 V
0 1697 R
0 -63 V
2571 300 M
0 63 V
0 1697 R
0 -63 V
2874 300 M
0 63 V
0 1697 R
0 -63 V
3177 300 M
0 63 V
0 1697 R
0 -63 V
1.000 UL
LTb
450 300 M
3000 0 V
0 1760 V
-3000 0 V
450 300 L
1.000 UL
LT0
3087 1947 M
263 0 V
450 1767 M
60 -29 V
60 -29 V
60 -29 V
60 -29 V
60 -30 V
60 -29 V
60 -29 V
60 -29 V
60 -29 V
60 -29 V
60 -29 V
60 -29 V
60 -29 V
60 -29 V
60 -29 V
60 -29 V
60 -29 V
60 -29 V
60 -29 V
60 -29 V
60 -29 V
60 -29 V
60 -29 V
60 -29 V
60 -29 V
60 -29 V
60 -29 V
60 -29 V
60 -29 V
60 -30 V
60 -29 V
60 -29 V
60 -29 V
60 -29 V
60 -29 V
60 -29 V
60 -29 V
60 -29 V
60 -29 V
60 -29 V
60 -29 V
60 -29 V
60 -29 V
60 -29 V
60 -29 V
60 -29 V
60 -29 V
60 -29 V
60 -29 V
60 -29 V
1.000 UL
LT1
3087 1847 M
263 0 V
-2900 6 R
60 -39 V
60 -38 V
60 -38 V
60 -38 V
60 -37 V
60 -37 V
60 -37 V
60 -36 V
60 -36 V
60 -36 V
60 -35 V
60 -35 V
60 -34 V
60 -35 V
60 -34 V
60 -33 V
60 -33 V
60 -33 V
60 -33 V
60 -32 V
60 -32 V
60 -31 V
60 -32 V
60 -31 V
60 -30 V
60 -31 V
60 -30 V
60 -29 V
60 -30 V
60 -29 V
60 -29 V
60 -29 V
60 -28 V
60 -28 V
60 -28 V
60 -27 V
60 -27 V
60 -27 V
60 -27 V
60 -26 V
60 -27 V
60 -26 V
60 -25 V
60 -26 V
60 -25 V
60 -25 V
60 -24 V
60 -25 V
60 -24 V
60 -24 V
1.000 UL
LT2
3087 1747 M
263 0 V
450 887 M
60 -12 V
60 -12 V
60 -11 V
60 -12 V
60 -11 V
60 -12 V
60 -12 V
60 -11 V
60 -12 V
60 -11 V
60 -12 V
60 -12 V
60 -11 V
60 -12 V
60 -12 V
60 -11 V
60 -12 V
60 -11 V
60 -12 V
60 -12 V
60 -11 V
60 -12 V
60 -12 V
60 -11 V
60 -12 V
60 -11 V
60 -12 V
60 -12 V
60 -11 V
60 -12 V
60 -11 V
60 -12 V
60 -12 V
60 -11 V
60 -12 V
60 -12 V
60 -11 V
60 -12 V
60 -11 V
60 -12 V
60 -12 V
60 -11 V
60 -12 V
60 -11 V
60 -12 V
60 -12 V
60 -11 V
60 -12 V
60 -12 V
60 -11 V
1.000 UL
LT3
3087 1647 M
263 0 V
450 984 M
60 -16 V
60 -16 V
60 -15 V
60 -16 V
60 -15 V
60 -16 V
60 -15 V
60 -15 V
60 -15 V
60 -15 V
60 -15 V
60 -15 V
60 -14 V
60 -15 V
60 -14 V
60 -15 V
60 -14 V
60 -14 V
60 -14 V
60 -14 V
60 -14 V
60 -14 V
60 -14 V
60 -13 V
60 -14 V
60 -13 V
60 -14 V
60 -13 V
60 -13 V
60 -13 V
60 -13 V
60 -13 V
60 -13 V
60 -13 V
60 -12 V
60 -13 V
60 -13 V
60 -12 V
60 -12 V
60 -13 V
60 -12 V
60 -12 V
60 -12 V
60 -12 V
60 -12 V
60 -12 V
60 -11 V
60 -12 V
60 -12 V
60 -11 V
stroke
grestore
end
showpage
}}%
\put(3037,1647){\makebox(0,0)[r]{$T_\sigma/\sigma^2$, $\rho = 5$}}%
\put(3037,1747){\makebox(0,0)[r]{$S_{11}/\rho$, $\rho = 5$}}%
\put(3037,1847){\makebox(0,0)[r]{$T_\sigma/\sigma^2$, $\rho = 2$}}%
\put(3037,1947){\makebox(0,0)[r]{$S_{11}/\rho$, $\rho = 2$}}%
\put(1950,50){\makebox(0,0){$\delta$}}%
\put(100,1180){%
\special{ps: gsave currentpoint currentpoint translate
270 rotate neg exch neg exch translate}%
\makebox(0,0)[b]{\shortstack{$T_\sigma/\sigma^2$}}%
\special{ps: currentpoint grestore moveto}%
}%
\put(3177,200){\makebox(0,0){$0.9$}}%
\put(2874,200){\makebox(0,0){$0.8$}}%
\put(2571,200){\makebox(0,0){$0.7$}}%
\put(2268,200){\makebox(0,0){$0.6$}}%
\put(1965,200){\makebox(0,0){$0.5$}}%
\put(1662,200){\makebox(0,0){$0.4$}}%
\put(1359,200){\makebox(0,0){$0.3$}}%
\put(1056,200){\makebox(0,0){$0.2$}}%
\put(753,200){\makebox(0,0){$0.1$}}%
\put(450,200){\makebox(0,0){$0.0$}}%
\put(400,2060){\makebox(0,0)[r]{$0.60$}}%
\put(400,1767){\makebox(0,0)[r]{$0.50$}}%
\put(400,1473){\makebox(0,0)[r]{$0.40$}}%
\put(400,1180){\makebox(0,0)[r]{$0.30$}}%
\put(400,887){\makebox(0,0)[r]{$0.20$}}%
\put(400,593){\makebox(0,0)[r]{$0.10$}}%
\put(400,300){\makebox(0,0)[r]{$0.00$}}%
\end{picture}%
\endgroup
 

%% file: fig2.tex
% GNUPLOT: LaTeX picture with Postscript
\begingroup%
  \makeatletter%
  \newcommand{\GNUPLOTspecial}{%
    \@sanitize\catcode`\%=14\relax\special}%
  \setlength{\unitlength}{0.1bp}%
{\GNUPLOTspecial{!
%!PS-Adobe-2.0
%%Title: fig2.tex
%%Creator: gnuplot 3.7 patchlevel 2
%%CreationDate: Sun Jun 16 23:24:36 2002
%%DocumentFonts: 
%%BoundingBox: 0 0 360 216
%%Orientation: Landscape
%%Pages: (atend)
%%EndComments
/gnudict 256 dict def
gnudict begin
/Color false def
/Solid false def
/gnulinewidth 5.000 def
/userlinewidth gnulinewidth def
/vshift -33 def
/dl {10 mul} def
/hpt_ 31.5 def
/vpt_ 31.5 def
/hpt hpt_ def
/vpt vpt_ def
/M {moveto} bind def
/L {lineto} bind def
/R {rmoveto} bind def
/V {rlineto} bind def
/vpt2 vpt 2 mul def
/hpt2 hpt 2 mul def
/Lshow { currentpoint stroke M
  0 vshift R show } def
/Rshow { currentpoint stroke M
  dup stringwidth pop neg vshift R show } def
/Cshow { currentpoint stroke M
  dup stringwidth pop -2 div vshift R show } def
/UP { dup vpt_ mul /vpt exch def hpt_ mul /hpt exch def
  /hpt2 hpt 2 mul def /vpt2 vpt 2 mul def } def
/DL { Color {setrgbcolor Solid {pop []} if 0 setdash }
 {pop pop pop Solid {pop []} if 0 setdash} ifelse } def
/BL { stroke userlinewidth 2 mul setlinewidth } def
/AL { stroke userlinewidth 2 div setlinewidth } def
/UL { dup gnulinewidth mul /userlinewidth exch def
      dup 1 lt {pop 1} if 10 mul /udl exch def } def
/PL { stroke userlinewidth setlinewidth } def
/LTb { BL [] 0 0 0 DL } def
/LTa { AL [1 udl mul 2 udl mul] 0 setdash 0 0 0 setrgbcolor } def
/LT0 { PL [] 1 0 0 DL } def
/LT1 { PL [4 dl 2 dl] 0 1 0 DL } def
/LT2 { PL [2 dl 3 dl] 0 0 1 DL } def
/LT3 { PL [1 dl 1.5 dl] 1 0 1 DL } def
/LT4 { PL [5 dl 2 dl 1 dl 2 dl] 0 1 1 DL } def
/LT5 { PL [4 dl 3 dl 1 dl 3 dl] 1 1 0 DL } def
/LT6 { PL [2 dl 2 dl 2 dl 4 dl] 0 0 0 DL } def
/LT7 { PL [2 dl 2 dl 2 dl 2 dl 2 dl 4 dl] 1 0.3 0 DL } def
/LT8 { PL [2 dl 2 dl 2 dl 2 dl 2 dl 2 dl 2 dl 4 dl] 0.5 0.5 0.5 DL } def
/Pnt { stroke [] 0 setdash
   gsave 1 setlinecap M 0 0 V stroke grestore } def
/Dia { stroke [] 0 setdash 2 copy vpt add M
  hpt neg vpt neg V hpt vpt neg V
  hpt vpt V hpt neg vpt V closepath stroke
  Pnt } def
/Pls { stroke [] 0 setdash vpt sub M 0 vpt2 V
  currentpoint stroke M
  hpt neg vpt neg R hpt2 0 V stroke
  } def
/Box { stroke [] 0 setdash 2 copy exch hpt sub exch vpt add M
  0 vpt2 neg V hpt2 0 V 0 vpt2 V
  hpt2 neg 0 V closepath stroke
  Pnt } def
/Crs { stroke [] 0 setdash exch hpt sub exch vpt add M
  hpt2 vpt2 neg V currentpoint stroke M
  hpt2 neg 0 R hpt2 vpt2 V stroke } def
/TriU { stroke [] 0 setdash 2 copy vpt 1.12 mul add M
  hpt neg vpt -1.62 mul V
  hpt 2 mul 0 V
  hpt neg vpt 1.62 mul V closepath stroke
  Pnt  } def
/Star { 2 copy Pls Crs } def
/BoxF { stroke [] 0 setdash exch hpt sub exch vpt add M
  0 vpt2 neg V  hpt2 0 V  0 vpt2 V
  hpt2 neg 0 V  closepath fill } def
/TriUF { stroke [] 0 setdash vpt 1.12 mul add M
  hpt neg vpt -1.62 mul V
  hpt 2 mul 0 V
  hpt neg vpt 1.62 mul V closepath fill } def
/TriD { stroke [] 0 setdash 2 copy vpt 1.12 mul sub M
  hpt neg vpt 1.62 mul V
  hpt 2 mul 0 V
  hpt neg vpt -1.62 mul V closepath stroke
  Pnt  } def
/TriDF { stroke [] 0 setdash vpt 1.12 mul sub M
  hpt neg vpt 1.62 mul V
  hpt 2 mul 0 V
  hpt neg vpt -1.62 mul V closepath fill} def
/DiaF { stroke [] 0 setdash vpt add M
  hpt neg vpt neg V hpt vpt neg V
  hpt vpt V hpt neg vpt V closepath fill } def
/Pent { stroke [] 0 setdash 2 copy gsave
  translate 0 hpt M 4 {72 rotate 0 hpt L} repeat
  closepath stroke grestore Pnt } def
/PentF { stroke [] 0 setdash gsave
  translate 0 hpt M 4 {72 rotate 0 hpt L} repeat
  closepath fill grestore } def
/Circle { stroke [] 0 setdash 2 copy
  hpt 0 360 arc stroke Pnt } def
/CircleF { stroke [] 0 setdash hpt 0 360 arc fill } def
/C0 { BL [] 0 setdash 2 copy moveto vpt 90 450  arc } bind def
/C1 { BL [] 0 setdash 2 copy        moveto
       2 copy  vpt 0 90 arc closepath fill
               vpt 0 360 arc closepath } bind def
/C2 { BL [] 0 setdash 2 copy moveto
       2 copy  vpt 90 180 arc closepath fill
               vpt 0 360 arc closepath } bind def
/C3 { BL [] 0 setdash 2 copy moveto
       2 copy  vpt 0 180 arc closepath fill
               vpt 0 360 arc closepath } bind def
/C4 { BL [] 0 setdash 2 copy moveto
       2 copy  vpt 180 270 arc closepath fill
               vpt 0 360 arc closepath } bind def
/C5 { BL [] 0 setdash 2 copy moveto
       2 copy  vpt 0 90 arc
       2 copy moveto
       2 copy  vpt 180 270 arc closepath fill
               vpt 0 360 arc } bind def
/C6 { BL [] 0 setdash 2 copy moveto
      2 copy  vpt 90 270 arc closepath fill
              vpt 0 360 arc closepath } bind def
/C7 { BL [] 0 setdash 2 copy moveto
      2 copy  vpt 0 270 arc closepath fill
              vpt 0 360 arc closepath } bind def
/C8 { BL [] 0 setdash 2 copy moveto
      2 copy vpt 270 360 arc closepath fill
              vpt 0 360 arc closepath } bind def
/C9 { BL [] 0 setdash 2 copy moveto
      2 copy  vpt 270 450 arc closepath fill
              vpt 0 360 arc closepath } bind def
/C10 { BL [] 0 setdash 2 copy 2 copy moveto vpt 270 360 arc closepath fill
       2 copy moveto
       2 copy vpt 90 180 arc closepath fill
               vpt 0 360 arc closepath } bind def
/C11 { BL [] 0 setdash 2 copy moveto
       2 copy  vpt 0 180 arc closepath fill
       2 copy moveto
       2 copy  vpt 270 360 arc closepath fill
               vpt 0 360 arc closepath } bind def
/C12 { BL [] 0 setdash 2 copy moveto
       2 copy  vpt 180 360 arc closepath fill
               vpt 0 360 arc closepath } bind def
/C13 { BL [] 0 setdash  2 copy moveto
       2 copy  vpt 0 90 arc closepath fill
       2 copy moveto
       2 copy  vpt 180 360 arc closepath fill
               vpt 0 360 arc closepath } bind def
/C14 { BL [] 0 setdash 2 copy moveto
       2 copy  vpt 90 360 arc closepath fill
               vpt 0 360 arc } bind def
/C15 { BL [] 0 setdash 2 copy vpt 0 360 arc closepath fill
               vpt 0 360 arc closepath } bind def
/Rec   { newpath 4 2 roll moveto 1 index 0 rlineto 0 exch rlineto
       neg 0 rlineto closepath } bind def
/Square { dup Rec } bind def
/Bsquare { vpt sub exch vpt sub exch vpt2 Square } bind def
/S0 { BL [] 0 setdash 2 copy moveto 0 vpt rlineto BL Bsquare } bind def
/S1 { BL [] 0 setdash 2 copy vpt Square fill Bsquare } bind def
/S2 { BL [] 0 setdash 2 copy exch vpt sub exch vpt Square fill Bsquare } bind def
/S3 { BL [] 0 setdash 2 copy exch vpt sub exch vpt2 vpt Rec fill Bsquare } bind def
/S4 { BL [] 0 setdash 2 copy exch vpt sub exch vpt sub vpt Square fill Bsquare } bind def
/S5 { BL [] 0 setdash 2 copy 2 copy vpt Square fill
       exch vpt sub exch vpt sub vpt Square fill Bsquare } bind def
/S6 { BL [] 0 setdash 2 copy exch vpt sub exch vpt sub vpt vpt2 Rec fill Bsquare } bind def
/S7 { BL [] 0 setdash 2 copy exch vpt sub exch vpt sub vpt vpt2 Rec fill
       2 copy vpt Square fill
       Bsquare } bind def
/S8 { BL [] 0 setdash 2 copy vpt sub vpt Square fill Bsquare } bind def
/S9 { BL [] 0 setdash 2 copy vpt sub vpt vpt2 Rec fill Bsquare } bind def
/S10 { BL [] 0 setdash 2 copy vpt sub vpt Square fill 2 copy exch vpt sub exch vpt Square fill
       Bsquare } bind def
/S11 { BL [] 0 setdash 2 copy vpt sub vpt Square fill 2 copy exch vpt sub exch vpt2 vpt Rec fill
       Bsquare } bind def
/S12 { BL [] 0 setdash 2 copy exch vpt sub exch vpt sub vpt2 vpt Rec fill Bsquare } bind def
/S13 { BL [] 0 setdash 2 copy exch vpt sub exch vpt sub vpt2 vpt Rec fill
       2 copy vpt Square fill Bsquare } bind def
/S14 { BL [] 0 setdash 2 copy exch vpt sub exch vpt sub vpt2 vpt Rec fill
       2 copy exch vpt sub exch vpt Square fill Bsquare } bind def
/S15 { BL [] 0 setdash 2 copy Bsquare fill Bsquare } bind def
/D0 { gsave translate 45 rotate 0 0 S0 stroke grestore } bind def
/D1 { gsave translate 45 rotate 0 0 S1 stroke grestore } bind def
/D2 { gsave translate 45 rotate 0 0 S2 stroke grestore } bind def
/D3 { gsave translate 45 rotate 0 0 S3 stroke grestore } bind def
/D4 { gsave translate 45 rotate 0 0 S4 stroke grestore } bind def
/D5 { gsave translate 45 rotate 0 0 S5 stroke grestore } bind def
/D6 { gsave translate 45 rotate 0 0 S6 stroke grestore } bind def
/D7 { gsave translate 45 rotate 0 0 S7 stroke grestore } bind def
/D8 { gsave translate 45 rotate 0 0 S8 stroke grestore } bind def
/D9 { gsave translate 45 rotate 0 0 S9 stroke grestore } bind def
/D10 { gsave translate 45 rotate 0 0 S10 stroke grestore } bind def
/D11 { gsave translate 45 rotate 0 0 S11 stroke grestore } bind def
/D12 { gsave translate 45 rotate 0 0 S12 stroke grestore } bind def
/D13 { gsave translate 45 rotate 0 0 S13 stroke grestore } bind def
/D14 { gsave translate 45 rotate 0 0 S14 stroke grestore } bind def
/D15 { gsave translate 45 rotate 0 0 S15 stroke grestore } bind def
/DiaE { stroke [] 0 setdash vpt add M
  hpt neg vpt neg V hpt vpt neg V
  hpt vpt V hpt neg vpt V closepath stroke } def
/BoxE { stroke [] 0 setdash exch hpt sub exch vpt add M
  0 vpt2 neg V hpt2 0 V 0 vpt2 V
  hpt2 neg 0 V closepath stroke } def
/TriUE { stroke [] 0 setdash vpt 1.12 mul add M
  hpt neg vpt -1.62 mul V
  hpt 2 mul 0 V
  hpt neg vpt 1.62 mul V closepath stroke } def
/TriDE { stroke [] 0 setdash vpt 1.12 mul sub M
  hpt neg vpt 1.62 mul V
  hpt 2 mul 0 V
  hpt neg vpt -1.62 mul V closepath stroke } def
/PentE { stroke [] 0 setdash gsave
  translate 0 hpt M 4 {72 rotate 0 hpt L} repeat
  closepath stroke grestore } def
/CircE { stroke [] 0 setdash 
  hpt 0 360 arc stroke } def
/Opaque { gsave closepath 1 setgray fill grestore 0 setgray closepath } def
/DiaW { stroke [] 0 setdash vpt add M
  hpt neg vpt neg V hpt vpt neg V
  hpt vpt V hpt neg vpt V Opaque stroke } def
/BoxW { stroke [] 0 setdash exch hpt sub exch vpt add M
  0 vpt2 neg V hpt2 0 V 0 vpt2 V
  hpt2 neg 0 V Opaque stroke } def
/TriUW { stroke [] 0 setdash vpt 1.12 mul add M
  hpt neg vpt -1.62 mul V
  hpt 2 mul 0 V
  hpt neg vpt 1.62 mul V Opaque stroke } def
/TriDW { stroke [] 0 setdash vpt 1.12 mul sub M
  hpt neg vpt 1.62 mul V
  hpt 2 mul 0 V
  hpt neg vpt -1.62 mul V Opaque stroke } def
/PentW { stroke [] 0 setdash gsave
  translate 0 hpt M 4 {72 rotate 0 hpt L} repeat
  Opaque stroke grestore } def
/CircW { stroke [] 0 setdash 
  hpt 0 360 arc Opaque stroke } def
/BoxFill { gsave Rec 1 setgray fill grestore } def
/Symbol-Oblique /Symbol findfont [1 0 .167 1 0 0] makefont
dup length dict begin {1 index /FID eq {pop pop} {def} ifelse} forall
currentdict end definefont
end
}}%
\begin{picture}(3600,2160)(0,0)%
{\GNUPLOTspecial{"
gnudict begin
gsave
0 0 translate
0.100 0.100 scale
0 setgray
newpath
1.000 UL
LTb
450 300 M
63 0 V
2937 0 R
-63 0 V
450 501 M
63 0 V
2937 0 R
-63 0 V
450 702 M
63 0 V
2937 0 R
-63 0 V
450 903 M
63 0 V
2937 0 R
-63 0 V
450 1105 M
63 0 V
2937 0 R
-63 0 V
450 1306 M
63 0 V
2937 0 R
-63 0 V
450 1507 M
63 0 V
2937 0 R
-63 0 V
450 1708 M
63 0 V
2937 0 R
-63 0 V
450 1909 M
63 0 V
2937 0 R
-63 0 V
450 300 M
0 63 V
0 1697 R
0 -63 V
1200 300 M
0 63 V
0 1697 R
0 -63 V
1950 300 M
0 63 V
0 1697 R
0 -63 V
2700 300 M
0 63 V
0 1697 R
0 -63 V
3450 300 M
0 63 V
0 1697 R
0 -63 V
1.000 UL
LTb
450 300 M
3000 0 V
0 1760 V
-3000 0 V
450 300 L
1.000 UL
LT0
3087 1947 M
263 0 V
450 1901 M
30 -4 V
31 -4 V
30 -4 V
30 -5 V
30 -6 V
31 -6 V
30 -7 V
30 -8 V
31 -9 V
30 -11 V
30 -12 V
30 -14 V
31 -15 V
30 -17 V
30 -17 V
31 -19 V
30 -19 V
30 -21 V
30 -21 V
31 -22 V
30 -23 V
30 -23 V
31 -24 V
30 -25 V
30 -25 V
30 -25 V
31 -26 V
30 -26 V
30 -27 V
31 -27 V
30 -27 V
30 -28 V
30 -27 V
31 -28 V
30 -27 V
30 -28 V
31 -28 V
30 -28 V
30 -27 V
31 -27 V
30 -27 V
30 -27 V
30 -27 V
31 -26 V
30 -26 V
30 -25 V
31 -26 V
30 -24 V
30 -25 V
30 -23 V
31 -24 V
30 -22 V
30 -23 V
31 -21 V
30 -22 V
30 -20 V
30 -20 V
31 -20 V
30 -19 V
30 -18 V
31 -18 V
30 -17 V
30 -17 V
30 -16 V
31 -15 V
30 -15 V
30 -14 V
31 -14 V
30 -13 V
30 -13 V
30 -12 V
31 -12 V
30 -11 V
30 -10 V
31 -10 V
30 -10 V
30 -9 V
30 -9 V
31 -8 V
30 -8 V
30 -8 V
31 -7 V
30 -7 V
30 -6 V
30 -6 V
31 -6 V
30 -5 V
30 -5 V
31 -5 V
30 -4 V
30 -4 V
30 -4 V
31 -3 V
30 -4 V
30 -3 V
31 -3 V
30 -3 V
30 -3 V
30 -3 V
1.000 UL
LT1
3087 1847 M
263 0 V
450 1123 M
30 0 V
31 -1 V
30 -1 V
30 0 V
30 -1 V
31 -1 V
30 -1 V
30 -1 V
31 -2 V
30 -1 V
30 -2 V
30 -2 V
31 -3 V
30 -2 V
30 -3 V
31 -3 V
30 -3 V
30 -4 V
30 -3 V
31 -4 V
30 -4 V
30 -4 V
31 -4 V
30 -4 V
30 -5 V
30 -4 V
31 -5 V
30 -5 V
30 -5 V
31 -6 V
30 -5 V
30 -6 V
30 -5 V
31 -6 V
30 -6 V
30 -6 V
31 -7 V
30 -6 V
30 -6 V
31 -7 V
30 -7 V
30 -6 V
30 -7 V
31 -7 V
30 -7 V
30 -8 V
31 -7 V
30 -7 V
30 -8 V
30 -7 V
31 -8 V
30 -7 V
30 -8 V
31 -7 V
30 -8 V
30 -8 V
30 -8 V
31 -7 V
30 -8 V
30 -8 V
31 -8 V
30 -8 V
30 -7 V
30 -8 V
31 -8 V
30 -8 V
30 -8 V
31 -7 V
30 -8 V
30 -8 V
30 -7 V
31 -8 V
30 -7 V
30 -8 V
31 -7 V
30 -8 V
30 -7 V
30 -7 V
31 -7 V
30 -7 V
30 -7 V
31 -7 V
30 -7 V
30 -7 V
30 -7 V
31 -6 V
30 -7 V
30 -6 V
31 -6 V
30 -6 V
30 -6 V
30 -6 V
31 -6 V
30 -6 V
30 -5 V
31 -6 V
30 -6 V
30 -5 V
30 -6 V
1.000 UL
LT2
3087 1747 M
263 0 V
450 1658 M
30 -2 V
31 -1 V
30 -2 V
30 -2 V
30 -3 V
31 -2 V
30 -3 V
30 -4 V
31 -4 V
30 -5 V
30 -5 V
30 -7 V
31 -6 V
30 -8 V
30 -8 V
31 -8 V
30 -9 V
30 -10 V
30 -10 V
31 -10 V
30 -11 V
30 -11 V
31 -12 V
30 -12 V
30 -13 V
30 -13 V
31 -13 V
30 -13 V
30 -14 V
31 -15 V
30 -14 V
30 -15 V
30 -16 V
31 -15 V
30 -16 V
30 -16 V
31 -16 V
30 -17 V
30 -17 V
31 -17 V
30 -17 V
30 -17 V
30 -17 V
31 -18 V
30 -17 V
30 -18 V
31 -18 V
30 -17 V
30 -18 V
30 -18 V
31 -18 V
30 -17 V
30 -18 V
31 -18 V
30 -17 V
30 -18 V
30 -17 V
31 -18 V
30 -17 V
30 -17 V
31 -17 V
30 -16 V
30 -17 V
30 -16 V
31 -16 V
30 -16 V
30 -16 V
31 -15 V
30 -15 V
30 -15 V
30 -15 V
31 -14 V
30 -14 V
30 -14 V
31 -13 V
30 -14 V
30 -13 V
30 -12 V
31 -13 V
30 -12 V
30 -12 V
31 -11 V
30 -11 V
30 -11 V
30 -11 V
31 -10 V
30 -10 V
30 -9 V
31 -9 V
30 -9 V
30 -8 V
30 -9 V
31 -8 V
30 -7 V
30 -8 V
31 -7 V
30 -7 V
30 -7 V
30 -8 V
1.000 UL
LT3
3087 1647 M
263 0 V
450 1888 M
30 -3 V
31 -3 V
30 -3 V
30 -3 V
30 -4 V
31 -5 V
30 -5 V
30 -6 V
31 -7 V
30 -9 V
30 -9 V
30 -10 V
31 -12 V
30 -12 V
30 -14 V
31 -14 V
30 -15 V
30 -16 V
30 -16 V
31 -18 V
30 -18 V
30 -18 V
31 -19 V
30 -20 V
30 -20 V
30 -21 V
31 -21 V
30 -22 V
30 -22 V
31 -22 V
30 -23 V
30 -23 V
30 -24 V
31 -24 V
30 -24 V
30 -24 V
31 -25 V
30 -24 V
30 -25 V
31 -25 V
30 -25 V
30 -24 V
30 -25 V
31 -25 V
30 -25 V
30 -24 V
31 -25 V
30 -24 V
30 -24 V
30 -24 V
31 -24 V
30 -23 V
30 -23 V
31 -23 V
30 -22 V
30 -22 V
30 -22 V
31 -21 V
30 -21 V
30 -20 V
31 -20 V
30 -20 V
30 -19 V
30 -19 V
31 -18 V
30 -17 V
30 -18 V
31 -16 V
30 -16 V
30 -16 V
30 -15 V
31 -15 V
30 -14 V
30 -14 V
31 -13 V
30 -13 V
30 -12 V
30 -12 V
31 -11 V
30 -11 V
30 -11 V
31 -10 V
30 -9 V
30 -9 V
30 -9 V
31 -8 V
30 -8 V
30 -8 V
31 -7 V
30 -6 V
30 -6 V
30 -6 V
31 -6 V
30 -5 V
30 -5 V
31 -5 V
30 -5 V
30 -4 V
30 -5 V
1.000 UL
LT4
3087 1547 M
263 0 V
450 1912 M
30 -4 V
31 -3 V
30 -4 V
30 -5 V
30 -5 V
31 -5 V
30 -7 V
30 -7 V
31 -9 V
30 -10 V
30 -11 V
30 -13 V
31 -14 V
30 -15 V
30 -16 V
31 -18 V
30 -18 V
30 -19 V
30 -20 V
31 -21 V
30 -21 V
30 -22 V
31 -23 V
30 -23 V
30 -23 V
30 -25 V
31 -24 V
30 -25 V
30 -26 V
31 -25 V
30 -27 V
30 -26 V
30 -27 V
31 -26 V
30 -27 V
30 -28 V
31 -27 V
30 -27 V
30 -27 V
31 -27 V
30 -27 V
30 -26 V
30 -27 V
31 -26 V
30 -26 V
30 -26 V
31 -25 V
30 -25 V
30 -25 V
30 -24 V
31 -24 V
30 -23 V
30 -23 V
31 -23 V
30 -22 V
30 -21 V
30 -21 V
31 -21 V
30 -20 V
30 -19 V
31 -19 V
30 -18 V
30 -18 V
30 -17 V
31 -16 V
30 -16 V
30 -16 V
31 -15 V
30 -14 V
30 -14 V
30 -13 V
31 -13 V
30 -12 V
30 -11 V
31 -12 V
30 -10 V
30 -11 V
30 -9 V
31 -10 V
30 -8 V
30 -9 V
31 -8 V
30 -8 V
30 -7 V
30 -7 V
31 -6 V
30 -6 V
30 -6 V
31 -5 V
30 -5 V
30 -5 V
30 -4 V
31 -4 V
30 -4 V
30 -4 V
31 -3 V
30 -4 V
30 -3 V
30 -3 V
stroke
grestore
end
showpage
}}%
\put(3037,1547){\makebox(0,0)[r]{$\rho=5.0$}}%
\put(3037,1647){\makebox(0,0)[r]{$\rho=2.0$}}%
\put(3037,1747){\makebox(0,0)[r]{$\rho=1.0$}}%
\put(3037,1847){\makebox(0,0)[r]{$\rho=0.5$}}%
\put(3037,1947){\makebox(0,0)[r]{$w_A w_B$}}%
\put(1950,50){\makebox(0,0){$x$}}%
\put(100,1180){%
\special{ps: gsave currentpoint currentpoint translate
270 rotate neg exch neg exch translate}%
\makebox(0,0)[b]{\shortstack{$w_A w_B C(w_A, w_B)$}}%
\special{ps: currentpoint grestore moveto}%
}%
\put(3450,200){\makebox(0,0){$2.0$}}%
\put(2700,200){\makebox(0,0){$1.5$}}%
\put(1950,200){\makebox(0,0){$1.0$}}%
\put(1200,200){\makebox(0,0){$0.5$}}%
\put(450,200){\makebox(0,0){$0.0$}}%
\put(400,1909){\makebox(0,0)[r]{$0.16$}}%
\put(400,1708){\makebox(0,0)[r]{$0.14$}}%
\put(400,1507){\makebox(0,0)[r]{$0.12$}}%
\put(400,1306){\makebox(0,0)[r]{$0.10$}}%
\put(400,1105){\makebox(0,0)[r]{$0.08$}}%
\put(400,903){\makebox(0,0)[r]{$0.06$}}%
\put(400,702){\makebox(0,0)[r]{$0.04$}}%
\put(400,501){\makebox(0,0)[r]{$0.02$}}%
\put(400,300){\makebox(0,0)[r]{$0.00$}}%
\end{picture}%
\endgroup
 

%% file: fig3.tex
% GNUPLOT: LaTeX picture with Postscript
\begingroup%
  \makeatletter%
  \newcommand{\GNUPLOTspecial}{%
    \@sanitize\catcode`\%=14\relax\special}%
  \setlength{\unitlength}{0.1bp}%
{\GNUPLOTspecial{!
%!PS-Adobe-2.0
%%Title: fig3.tex
%%Creator: gnuplot 3.7 patchlevel 2
%%CreationDate: Sun Jun 16 23:24:37 2002
%%DocumentFonts: 
%%BoundingBox: 0 0 360 216
%%Orientation: Landscape
%%Pages: (atend)
%%EndComments
/gnudict 256 dict def
gnudict begin
/Color false def
/Solid false def
/gnulinewidth 5.000 def
/userlinewidth gnulinewidth def
/vshift -33 def
/dl {10 mul} def
/hpt_ 31.5 def
/vpt_ 31.5 def
/hpt hpt_ def
/vpt vpt_ def
/M {moveto} bind def
/L {lineto} bind def
/R {rmoveto} bind def
/V {rlineto} bind def
/vpt2 vpt 2 mul def
/hpt2 hpt 2 mul def
/Lshow { currentpoint stroke M
  0 vshift R show } def
/Rshow { currentpoint stroke M
  dup stringwidth pop neg vshift R show } def
/Cshow { currentpoint stroke M
  dup stringwidth pop -2 div vshift R show } def
/UP { dup vpt_ mul /vpt exch def hpt_ mul /hpt exch def
  /hpt2 hpt 2 mul def /vpt2 vpt 2 mul def } def
/DL { Color {setrgbcolor Solid {pop []} if 0 setdash }
 {pop pop pop Solid {pop []} if 0 setdash} ifelse } def
/BL { stroke userlinewidth 2 mul setlinewidth } def
/AL { stroke userlinewidth 2 div setlinewidth } def
/UL { dup gnulinewidth mul /userlinewidth exch def
      dup 1 lt {pop 1} if 10 mul /udl exch def } def
/PL { stroke userlinewidth setlinewidth } def
/LTb { BL [] 0 0 0 DL } def
/LTa { AL [1 udl mul 2 udl mul] 0 setdash 0 0 0 setrgbcolor } def
/LT0 { PL [] 1 0 0 DL } def
/LT1 { PL [4 dl 2 dl] 0 1 0 DL } def
/LT2 { PL [2 dl 3 dl] 0 0 1 DL } def
/LT3 { PL [1 dl 1.5 dl] 1 0 1 DL } def
/LT4 { PL [5 dl 2 dl 1 dl 2 dl] 0 1 1 DL } def
/LT5 { PL [4 dl 3 dl 1 dl 3 dl] 1 1 0 DL } def
/LT6 { PL [2 dl 2 dl 2 dl 4 dl] 0 0 0 DL } def
/LT7 { PL [2 dl 2 dl 2 dl 2 dl 2 dl 4 dl] 1 0.3 0 DL } def
/LT8 { PL [2 dl 2 dl 2 dl 2 dl 2 dl 2 dl 2 dl 4 dl] 0.5 0.5 0.5 DL } def
/Pnt { stroke [] 0 setdash
   gsave 1 setlinecap M 0 0 V stroke grestore } def
/Dia { stroke [] 0 setdash 2 copy vpt add M
  hpt neg vpt neg V hpt vpt neg V
  hpt vpt V hpt neg vpt V closepath stroke
  Pnt } def
/Pls { stroke [] 0 setdash vpt sub M 0 vpt2 V
  currentpoint stroke M
  hpt neg vpt neg R hpt2 0 V stroke
  } def
/Box { stroke [] 0 setdash 2 copy exch hpt sub exch vpt add M
  0 vpt2 neg V hpt2 0 V 0 vpt2 V
  hpt2 neg 0 V closepath stroke
  Pnt } def
/Crs { stroke [] 0 setdash exch hpt sub exch vpt add M
  hpt2 vpt2 neg V currentpoint stroke M
  hpt2 neg 0 R hpt2 vpt2 V stroke } def
/TriU { stroke [] 0 setdash 2 copy vpt 1.12 mul add M
  hpt neg vpt -1.62 mul V
  hpt 2 mul 0 V
  hpt neg vpt 1.62 mul V closepath stroke
  Pnt  } def
/Star { 2 copy Pls Crs } def
/BoxF { stroke [] 0 setdash exch hpt sub exch vpt add M
  0 vpt2 neg V  hpt2 0 V  0 vpt2 V
  hpt2 neg 0 V  closepath fill } def
/TriUF { stroke [] 0 setdash vpt 1.12 mul add M
  hpt neg vpt -1.62 mul V
  hpt 2 mul 0 V
  hpt neg vpt 1.62 mul V closepath fill } def
/TriD { stroke [] 0 setdash 2 copy vpt 1.12 mul sub M
  hpt neg vpt 1.62 mul V
  hpt 2 mul 0 V
  hpt neg vpt -1.62 mul V closepath stroke
  Pnt  } def
/TriDF { stroke [] 0 setdash vpt 1.12 mul sub M
  hpt neg vpt 1.62 mul V
  hpt 2 mul 0 V
  hpt neg vpt -1.62 mul V closepath fill} def
/DiaF { stroke [] 0 setdash vpt add M
  hpt neg vpt neg V hpt vpt neg V
  hpt vpt V hpt neg vpt V closepath fill } def
/Pent { stroke [] 0 setdash 2 copy gsave
  translate 0 hpt M 4 {72 rotate 0 hpt L} repeat
  closepath stroke grestore Pnt } def
/PentF { stroke [] 0 setdash gsave
  translate 0 hpt M 4 {72 rotate 0 hpt L} repeat
  closepath fill grestore } def
/Circle { stroke [] 0 setdash 2 copy
  hpt 0 360 arc stroke Pnt } def
/CircleF { stroke [] 0 setdash hpt 0 360 arc fill } def
/C0 { BL [] 0 setdash 2 copy moveto vpt 90 450  arc } bind def
/C1 { BL [] 0 setdash 2 copy        moveto
       2 copy  vpt 0 90 arc closepath fill
               vpt 0 360 arc closepath } bind def
/C2 { BL [] 0 setdash 2 copy moveto
       2 copy  vpt 90 180 arc closepath fill
               vpt 0 360 arc closepath } bind def
/C3 { BL [] 0 setdash 2 copy moveto
       2 copy  vpt 0 180 arc closepath fill
               vpt 0 360 arc closepath } bind def
/C4 { BL [] 0 setdash 2 copy moveto
       2 copy  vpt 180 270 arc closepath fill
               vpt 0 360 arc closepath } bind def
/C5 { BL [] 0 setdash 2 copy moveto
       2 copy  vpt 0 90 arc
       2 copy moveto
       2 copy  vpt 180 270 arc closepath fill
               vpt 0 360 arc } bind def
/C6 { BL [] 0 setdash 2 copy moveto
      2 copy  vpt 90 270 arc closepath fill
              vpt 0 360 arc closepath } bind def
/C7 { BL [] 0 setdash 2 copy moveto
      2 copy  vpt 0 270 arc closepath fill
              vpt 0 360 arc closepath } bind def
/C8 { BL [] 0 setdash 2 copy moveto
      2 copy vpt 270 360 arc closepath fill
              vpt 0 360 arc closepath } bind def
/C9 { BL [] 0 setdash 2 copy moveto
      2 copy  vpt 270 450 arc closepath fill
              vpt 0 360 arc closepath } bind def
/C10 { BL [] 0 setdash 2 copy 2 copy moveto vpt 270 360 arc closepath fill
       2 copy moveto
       2 copy vpt 90 180 arc closepath fill
               vpt 0 360 arc closepath } bind def
/C11 { BL [] 0 setdash 2 copy moveto
       2 copy  vpt 0 180 arc closepath fill
       2 copy moveto
       2 copy  vpt 270 360 arc closepath fill
               vpt 0 360 arc closepath } bind def
/C12 { BL [] 0 setdash 2 copy moveto
       2 copy  vpt 180 360 arc closepath fill
               vpt 0 360 arc closepath } bind def
/C13 { BL [] 0 setdash  2 copy moveto
       2 copy  vpt 0 90 arc closepath fill
       2 copy moveto
       2 copy  vpt 180 360 arc closepath fill
               vpt 0 360 arc closepath } bind def
/C14 { BL [] 0 setdash 2 copy moveto
       2 copy  vpt 90 360 arc closepath fill
               vpt 0 360 arc } bind def
/C15 { BL [] 0 setdash 2 copy vpt 0 360 arc closepath fill
               vpt 0 360 arc closepath } bind def
/Rec   { newpath 4 2 roll moveto 1 index 0 rlineto 0 exch rlineto
       neg 0 rlineto closepath } bind def
/Square { dup Rec } bind def
/Bsquare { vpt sub exch vpt sub exch vpt2 Square } bind def
/S0 { BL [] 0 setdash 2 copy moveto 0 vpt rlineto BL Bsquare } bind def
/S1 { BL [] 0 setdash 2 copy vpt Square fill Bsquare } bind def
/S2 { BL [] 0 setdash 2 copy exch vpt sub exch vpt Square fill Bsquare } bind def
/S3 { BL [] 0 setdash 2 copy exch vpt sub exch vpt2 vpt Rec fill Bsquare } bind def
/S4 { BL [] 0 setdash 2 copy exch vpt sub exch vpt sub vpt Square fill Bsquare } bind def
/S5 { BL [] 0 setdash 2 copy 2 copy vpt Square fill
       exch vpt sub exch vpt sub vpt Square fill Bsquare } bind def
/S6 { BL [] 0 setdash 2 copy exch vpt sub exch vpt sub vpt vpt2 Rec fill Bsquare } bind def
/S7 { BL [] 0 setdash 2 copy exch vpt sub exch vpt sub vpt vpt2 Rec fill
       2 copy vpt Square fill
       Bsquare } bind def
/S8 { BL [] 0 setdash 2 copy vpt sub vpt Square fill Bsquare } bind def
/S9 { BL [] 0 setdash 2 copy vpt sub vpt vpt2 Rec fill Bsquare } bind def
/S10 { BL [] 0 setdash 2 copy vpt sub vpt Square fill 2 copy exch vpt sub exch vpt Square fill
       Bsquare } bind def
/S11 { BL [] 0 setdash 2 copy vpt sub vpt Square fill 2 copy exch vpt sub exch vpt2 vpt Rec fill
       Bsquare } bind def
/S12 { BL [] 0 setdash 2 copy exch vpt sub exch vpt sub vpt2 vpt Rec fill Bsquare } bind def
/S13 { BL [] 0 setdash 2 copy exch vpt sub exch vpt sub vpt2 vpt Rec fill
       2 copy vpt Square fill Bsquare } bind def
/S14 { BL [] 0 setdash 2 copy exch vpt sub exch vpt sub vpt2 vpt Rec fill
       2 copy exch vpt sub exch vpt Square fill Bsquare } bind def
/S15 { BL [] 0 setdash 2 copy Bsquare fill Bsquare } bind def
/D0 { gsave translate 45 rotate 0 0 S0 stroke grestore } bind def
/D1 { gsave translate 45 rotate 0 0 S1 stroke grestore } bind def
/D2 { gsave translate 45 rotate 0 0 S2 stroke grestore } bind def
/D3 { gsave translate 45 rotate 0 0 S3 stroke grestore } bind def
/D4 { gsave translate 45 rotate 0 0 S4 stroke grestore } bind def
/D5 { gsave translate 45 rotate 0 0 S5 stroke grestore } bind def
/D6 { gsave translate 45 rotate 0 0 S6 stroke grestore } bind def
/D7 { gsave translate 45 rotate 0 0 S7 stroke grestore } bind def
/D8 { gsave translate 45 rotate 0 0 S8 stroke grestore } bind def
/D9 { gsave translate 45 rotate 0 0 S9 stroke grestore } bind def
/D10 { gsave translate 45 rotate 0 0 S10 stroke grestore } bind def
/D11 { gsave translate 45 rotate 0 0 S11 stroke grestore } bind def
/D12 { gsave translate 45 rotate 0 0 S12 stroke grestore } bind def
/D13 { gsave translate 45 rotate 0 0 S13 stroke grestore } bind def
/D14 { gsave translate 45 rotate 0 0 S14 stroke grestore } bind def
/D15 { gsave translate 45 rotate 0 0 S15 stroke grestore } bind def
/DiaE { stroke [] 0 setdash vpt add M
  hpt neg vpt neg V hpt vpt neg V
  hpt vpt V hpt neg vpt V closepath stroke } def
/BoxE { stroke [] 0 setdash exch hpt sub exch vpt add M
  0 vpt2 neg V hpt2 0 V 0 vpt2 V
  hpt2 neg 0 V closepath stroke } def
/TriUE { stroke [] 0 setdash vpt 1.12 mul add M
  hpt neg vpt -1.62 mul V
  hpt 2 mul 0 V
  hpt neg vpt 1.62 mul V closepath stroke } def
/TriDE { stroke [] 0 setdash vpt 1.12 mul sub M
  hpt neg vpt 1.62 mul V
  hpt 2 mul 0 V
  hpt neg vpt -1.62 mul V closepath stroke } def
/PentE { stroke [] 0 setdash gsave
  translate 0 hpt M 4 {72 rotate 0 hpt L} repeat
  closepath stroke grestore } def
/CircE { stroke [] 0 setdash 
  hpt 0 360 arc stroke } def
/Opaque { gsave closepath 1 setgray fill grestore 0 setgray closepath } def
/DiaW { stroke [] 0 setdash vpt add M
  hpt neg vpt neg V hpt vpt neg V
  hpt vpt V hpt neg vpt V Opaque stroke } def
/BoxW { stroke [] 0 setdash exch hpt sub exch vpt add M
  0 vpt2 neg V hpt2 0 V 0 vpt2 V
  hpt2 neg 0 V Opaque stroke } def
/TriUW { stroke [] 0 setdash vpt 1.12 mul add M
  hpt neg vpt -1.62 mul V
  hpt 2 mul 0 V
  hpt neg vpt 1.62 mul V Opaque stroke } def
/TriDW { stroke [] 0 setdash vpt 1.12 mul sub M
  hpt neg vpt 1.62 mul V
  hpt 2 mul 0 V
  hpt neg vpt -1.62 mul V Opaque stroke } def
/PentW { stroke [] 0 setdash gsave
  translate 0 hpt M 4 {72 rotate 0 hpt L} repeat
  Opaque stroke grestore } def
/CircW { stroke [] 0 setdash 
  hpt 0 360 arc Opaque stroke } def
/BoxFill { gsave Rec 1 setgray fill grestore } def
/Symbol-Oblique /Symbol findfont [1 0 .167 1 0 0] makefont
dup length dict begin {1 index /FID eq {pop pop} {def} ifelse} forall
currentdict end definefont
end
}}%
\begin{picture}(3600,2160)(0,0)%
{\GNUPLOTspecial{"
gnudict begin
gsave
0 0 translate
0.100 0.100 scale
0 setgray
newpath
1.000 UL
LTb
450 300 M
63 0 V
2937 0 R
-63 0 V
450 551 M
63 0 V
2937 0 R
-63 0 V
450 803 M
63 0 V
2937 0 R
-63 0 V
450 1054 M
63 0 V
2937 0 R
-63 0 V
450 1306 M
63 0 V
2937 0 R
-63 0 V
450 1557 M
63 0 V
2937 0 R
-63 0 V
450 1809 M
63 0 V
2937 0 R
-63 0 V
450 2060 M
63 0 V
2937 0 R
-63 0 V
450 300 M
0 63 V
0 1697 R
0 -63 V
1200 300 M
0 63 V
0 1697 R
0 -63 V
1950 300 M
0 63 V
0 1697 R
0 -63 V
2700 300 M
0 63 V
0 1697 R
0 -63 V
3450 300 M
0 63 V
0 1697 R
0 -63 V
1.000 UL
LTb
450 300 M
3000 0 V
0 1760 V
-3000 0 V
450 300 L
1.000 UL
LT0
3087 1947 M
263 0 V
450 1514 M
30 23 V
31 23 V
30 22 V
30 23 V
30 22 V
31 21 V
30 21 V
30 20 V
31 19 V
30 18 V
30 17 V
30 16 V
31 14 V
30 14 V
30 12 V
31 12 V
30 10 V
30 9 V
30 7 V
31 7 V
30 5 V
30 4 V
31 3 V
30 2 V
30 0 V
30 -1 V
31 -2 V
30 -4 V
30 -4 V
31 -6 V
30 -7 V
30 -9 V
30 -9 V
31 -11 V
30 -12 V
30 -13 V
31 -14 V
30 -15 V
30 -16 V
31 -18 V
30 -18 V
30 -19 V
30 -19 V
31 -21 V
30 -21 V
30 -23 V
31 -22 V
30 -24 V
30 -24 V
30 -24 V
31 -25 V
30 -25 V
30 -26 V
31 -26 V
30 -27 V
30 -26 V
30 -27 V
31 -27 V
30 -27 V
30 -27 V
31 -27 V
30 -27 V
30 -26 V
30 -27 V
31 -27 V
30 -26 V
30 -26 V
31 -26 V
30 -25 V
30 -25 V
30 -25 V
31 -25 V
30 -23 V
30 -24 V
31 -23 V
30 -23 V
30 -22 V
30 -21 V
31 -21 V
30 -21 V
30 -20 V
31 -20 V
30 -19 V
30 -18 V
30 -18 V
31 -17 V
30 -16 V
30 -16 V
31 -15 V
30 -15 V
30 -14 V
30 -13 V
31 -13 V
30 -12 V
30 -12 V
31 -12 V
30 -12 V
30 -11 V
30 -11 V
1.000 UL
LT1
3087 1847 M
263 0 V
450 1146 M
30 6 V
31 5 V
30 5 V
30 5 V
30 6 V
31 4 V
30 5 V
30 5 V
31 4 V
30 4 V
30 4 V
30 3 V
31 4 V
30 2 V
30 3 V
31 2 V
30 3 V
30 1 V
30 2 V
31 1 V
30 1 V
30 1 V
31 1 V
30 0 V
30 0 V
30 0 V
31 0 V
30 -1 V
30 -1 V
31 -1 V
30 -2 V
30 -2 V
30 -2 V
31 -2 V
30 -2 V
30 -3 V
31 -3 V
30 -3 V
30 -4 V
31 -4 V
30 -4 V
30 -4 V
30 -4 V
31 -5 V
30 -5 V
30 -5 V
31 -5 V
30 -6 V
30 -6 V
30 -6 V
31 -6 V
30 -6 V
30 -7 V
31 -7 V
30 -6 V
30 -8 V
30 -7 V
31 -7 V
30 -8 V
30 -8 V
31 -8 V
30 -8 V
30 -8 V
30 -8 V
31 -8 V
30 -9 V
30 -8 V
31 -9 V
30 -9 V
30 -9 V
30 -9 V
31 -9 V
30 -9 V
30 -9 V
31 -9 V
30 -10 V
30 -9 V
30 -9 V
31 -9 V
30 -10 V
30 -9 V
31 -10 V
30 -9 V
30 -9 V
30 -10 V
31 -9 V
30 -9 V
30 -9 V
31 -9 V
30 -9 V
30 -9 V
30 -9 V
31 -9 V
30 -9 V
30 -9 V
31 -8 V
30 -9 V
30 -9 V
30 -8 V
1.000 UL
LT2
3087 1747 M
263 0 V
450 1530 M
30 13 V
31 14 V
30 13 V
30 13 V
30 13 V
31 13 V
30 12 V
30 12 V
31 11 V
30 10 V
30 10 V
30 9 V
31 8 V
30 7 V
30 7 V
31 6 V
30 6 V
30 5 V
30 4 V
31 3 V
30 3 V
30 2 V
31 2 V
30 1 V
30 0 V
30 -1 V
31 -1 V
30 -2 V
30 -2 V
31 -3 V
30 -4 V
30 -5 V
30 -5 V
31 -6 V
30 -7 V
30 -7 V
31 -8 V
30 -8 V
30 -9 V
31 -10 V
30 -10 V
30 -11 V
30 -11 V
31 -12 V
30 -13 V
30 -13 V
31 -13 V
30 -14 V
30 -15 V
30 -15 V
31 -15 V
30 -16 V
30 -16 V
31 -16 V
30 -17 V
30 -17 V
30 -18 V
31 -18 V
30 -18 V
30 -18 V
31 -19 V
30 -18 V
30 -19 V
30 -20 V
31 -19 V
30 -19 V
30 -20 V
31 -19 V
30 -20 V
30 -19 V
30 -20 V
31 -20 V
30 -19 V
30 -20 V
31 -19 V
30 -20 V
30 -19 V
30 -19 V
31 -19 V
30 -19 V
30 -19 V
31 -19 V
30 -18 V
30 -18 V
30 -18 V
31 -18 V
30 -17 V
30 -17 V
31 -17 V
30 -16 V
30 -16 V
30 -16 V
31 -15 V
30 -15 V
30 -15 V
31 -15 V
30 -14 V
30 -15 V
30 -14 V
1.000 UL
LT3
3087 1647 M
263 0 V
450 1593 M
30 20 V
31 20 V
30 20 V
30 20 V
30 19 V
31 18 V
30 18 V
30 17 V
31 17 V
30 15 V
30 15 V
30 13 V
31 13 V
30 11 V
30 10 V
31 10 V
30 8 V
30 8 V
30 6 V
31 6 V
30 4 V
30 3 V
31 3 V
30 1 V
30 0 V
30 0 V
31 -2 V
30 -3 V
30 -4 V
31 -5 V
30 -6 V
30 -7 V
30 -8 V
31 -9 V
30 -10 V
30 -11 V
31 -12 V
30 -13 V
30 -13 V
31 -15 V
30 -15 V
30 -17 V
30 -17 V
31 -17 V
30 -19 V
30 -19 V
31 -20 V
30 -20 V
30 -21 V
30 -22 V
31 -22 V
30 -23 V
30 -23 V
31 -23 V
30 -24 V
30 -24 V
30 -25 V
31 -25 V
30 -25 V
30 -25 V
31 -25 V
30 -25 V
30 -26 V
30 -25 V
31 -26 V
30 -25 V
30 -26 V
31 -25 V
30 -25 V
30 -25 V
30 -25 V
31 -24 V
30 -25 V
30 -24 V
31 -23 V
30 -24 V
30 -23 V
30 -23 V
31 -22 V
30 -22 V
30 -22 V
31 -21 V
30 -21 V
30 -20 V
30 -20 V
31 -19 V
30 -19 V
30 -18 V
31 -17 V
30 -17 V
30 -16 V
30 -16 V
31 -15 V
30 -15 V
30 -15 V
31 -14 V
30 -14 V
30 -14 V
30 -14 V
1.000 UL
LT4
3087 1547 M
263 0 V
-2900 2 R
30 22 V
31 22 V
30 22 V
30 22 V
30 21 V
31 21 V
30 20 V
30 19 V
31 19 V
30 17 V
30 16 V
30 16 V
31 14 V
30 13 V
30 12 V
31 10 V
30 10 V
30 9 V
30 7 V
31 6 V
30 5 V
30 4 V
31 3 V
30 1 V
30 1 V
30 -1 V
31 -2 V
30 -4 V
30 -4 V
31 -6 V
30 -6 V
30 -9 V
30 -9 V
31 -10 V
30 -11 V
30 -13 V
31 -13 V
30 -15 V
30 -15 V
31 -17 V
30 -17 V
30 -19 V
30 -19 V
31 -20 V
30 -20 V
30 -22 V
31 -22 V
30 -22 V
30 -24 V
30 -23 V
31 -25 V
30 -25 V
30 -25 V
31 -25 V
30 -26 V
30 -26 V
30 -27 V
31 -26 V
30 -27 V
30 -26 V
31 -27 V
30 -27 V
30 -27 V
30 -26 V
31 -27 V
30 -26 V
30 -26 V
31 -26 V
30 -26 V
30 -25 V
30 -25 V
31 -25 V
30 -24 V
30 -24 V
31 -24 V
30 -23 V
30 -23 V
30 -22 V
31 -22 V
30 -21 V
30 -21 V
31 -20 V
30 -20 V
30 -19 V
30 -19 V
31 -18 V
30 -17 V
30 -17 V
31 -16 V
30 -15 V
30 -15 V
30 -14 V
31 -14 V
30 -13 V
30 -13 V
31 -13 V
30 -12 V
30 -12 V
30 -13 V
stroke
grestore
end
showpage
}}%
\put(3037,1547){\makebox(0,0)[r]{$\rho=5.0$}}%
\put(3037,1647){\makebox(0,0)[r]{$\rho=2.0$}}%
\put(3037,1747){\makebox(0,0)[r]{$\rho=1.0$}}%
\put(3037,1847){\makebox(0,0)[r]{$\rho=0.5$}}%
\put(3037,1947){\makebox(0,0)[r]{$w_A w_B$}}%
\put(1950,50){\makebox(0,0){$x$}}%
\put(100,1180){%
\special{ps: gsave currentpoint currentpoint translate
270 rotate neg exch neg exch translate}%
\makebox(0,0)[b]{\shortstack{$w_A w_B C(w_A, w_B)$}}%
\special{ps: currentpoint grestore moveto}%
}%
\put(3450,200){\makebox(0,0){$2.0$}}%
\put(2700,200){\makebox(0,0){$1.5$}}%
\put(1950,200){\makebox(0,0){$1.0$}}%
\put(1200,200){\makebox(0,0){$0.5$}}%
\put(450,200){\makebox(0,0){$0.0$}}%
\put(400,2060){\makebox(0,0)[r]{$0.14$}}%
\put(400,1809){\makebox(0,0)[r]{$0.12$}}%
\put(400,1557){\makebox(0,0)[r]{$0.10$}}%
\put(400,1306){\makebox(0,0)[r]{$0.08$}}%
\put(400,1054){\makebox(0,0)[r]{$0.06$}}%
\put(400,803){\makebox(0,0)[r]{$0.04$}}%
\put(400,551){\makebox(0,0)[r]{$0.02$}}%
\put(400,300){\makebox(0,0)[r]{$0.00$}}%
\end{picture}%
\endgroup
 

%% file: fig8.tex
% GNUPLOT: LaTeX picture with Postscript
\begingroup%
  \makeatletter%
  \newcommand{\GNUPLOTspecial}{%
    \@sanitize\catcode`\%=14\relax\special}%
  \setlength{\unitlength}{0.1bp}%
{\GNUPLOTspecial{!
%!PS-Adobe-2.0
%%Title: fig8.tex
%%Creator: gnuplot 3.7 patchlevel 2
%%CreationDate: Sun Jun 16 23:24:40 2002
%%DocumentFonts: 
%%BoundingBox: 0 0 360 216
%%Orientation: Landscape
%%Pages: (atend)
%%EndComments
/gnudict 256 dict def
gnudict begin
/Color false def
/Solid false def
/gnulinewidth 5.000 def
/userlinewidth gnulinewidth def
/vshift -33 def
/dl {10 mul} def
/hpt_ 31.5 def
/vpt_ 31.5 def
/hpt hpt_ def
/vpt vpt_ def
/M {moveto} bind def
/L {lineto} bind def
/R {rmoveto} bind def
/V {rlineto} bind def
/vpt2 vpt 2 mul def
/hpt2 hpt 2 mul def
/Lshow { currentpoint stroke M
  0 vshift R show } def
/Rshow { currentpoint stroke M
  dup stringwidth pop neg vshift R show } def
/Cshow { currentpoint stroke M
  dup stringwidth pop -2 div vshift R show } def
/UP { dup vpt_ mul /vpt exch def hpt_ mul /hpt exch def
  /hpt2 hpt 2 mul def /vpt2 vpt 2 mul def } def
/DL { Color {setrgbcolor Solid {pop []} if 0 setdash }
 {pop pop pop Solid {pop []} if 0 setdash} ifelse } def
/BL { stroke userlinewidth 2 mul setlinewidth } def
/AL { stroke userlinewidth 2 div setlinewidth } def
/UL { dup gnulinewidth mul /userlinewidth exch def
      dup 1 lt {pop 1} if 10 mul /udl exch def } def
/PL { stroke userlinewidth setlinewidth } def
/LTb { BL [] 0 0 0 DL } def
/LTa { AL [1 udl mul 2 udl mul] 0 setdash 0 0 0 setrgbcolor } def
/LT0 { PL [] 1 0 0 DL } def
/LT1 { PL [4 dl 2 dl] 0 1 0 DL } def
/LT2 { PL [2 dl 3 dl] 0 0 1 DL } def
/LT3 { PL [1 dl 1.5 dl] 1 0 1 DL } def
/LT4 { PL [5 dl 2 dl 1 dl 2 dl] 0 1 1 DL } def
/LT5 { PL [4 dl 3 dl 1 dl 3 dl] 1 1 0 DL } def
/LT6 { PL [2 dl 2 dl 2 dl 4 dl] 0 0 0 DL } def
/LT7 { PL [2 dl 2 dl 2 dl 2 dl 2 dl 4 dl] 1 0.3 0 DL } def
/LT8 { PL [2 dl 2 dl 2 dl 2 dl 2 dl 2 dl 2 dl 4 dl] 0.5 0.5 0.5 DL } def
/Pnt { stroke [] 0 setdash
   gsave 1 setlinecap M 0 0 V stroke grestore } def
/Dia { stroke [] 0 setdash 2 copy vpt add M
  hpt neg vpt neg V hpt vpt neg V
  hpt vpt V hpt neg vpt V closepath stroke
  Pnt } def
/Pls { stroke [] 0 setdash vpt sub M 0 vpt2 V
  currentpoint stroke M
  hpt neg vpt neg R hpt2 0 V stroke
  } def
/Box { stroke [] 0 setdash 2 copy exch hpt sub exch vpt add M
  0 vpt2 neg V hpt2 0 V 0 vpt2 V
  hpt2 neg 0 V closepath stroke
  Pnt } def
/Crs { stroke [] 0 setdash exch hpt sub exch vpt add M
  hpt2 vpt2 neg V currentpoint stroke M
  hpt2 neg 0 R hpt2 vpt2 V stroke } def
/TriU { stroke [] 0 setdash 2 copy vpt 1.12 mul add M
  hpt neg vpt -1.62 mul V
  hpt 2 mul 0 V
  hpt neg vpt 1.62 mul V closepath stroke
  Pnt  } def
/Star { 2 copy Pls Crs } def
/BoxF { stroke [] 0 setdash exch hpt sub exch vpt add M
  0 vpt2 neg V  hpt2 0 V  0 vpt2 V
  hpt2 neg 0 V  closepath fill } def
/TriUF { stroke [] 0 setdash vpt 1.12 mul add M
  hpt neg vpt -1.62 mul V
  hpt 2 mul 0 V
  hpt neg vpt 1.62 mul V closepath fill } def
/TriD { stroke [] 0 setdash 2 copy vpt 1.12 mul sub M
  hpt neg vpt 1.62 mul V
  hpt 2 mul 0 V
  hpt neg vpt -1.62 mul V closepath stroke
  Pnt  } def
/TriDF { stroke [] 0 setdash vpt 1.12 mul sub M
  hpt neg vpt 1.62 mul V
  hpt 2 mul 0 V
  hpt neg vpt -1.62 mul V closepath fill} def
/DiaF { stroke [] 0 setdash vpt add M
  hpt neg vpt neg V hpt vpt neg V
  hpt vpt V hpt neg vpt V closepath fill } def
/Pent { stroke [] 0 setdash 2 copy gsave
  translate 0 hpt M 4 {72 rotate 0 hpt L} repeat
  closepath stroke grestore Pnt } def
/PentF { stroke [] 0 setdash gsave
  translate 0 hpt M 4 {72 rotate 0 hpt L} repeat
  closepath fill grestore } def
/Circle { stroke [] 0 setdash 2 copy
  hpt 0 360 arc stroke Pnt } def
/CircleF { stroke [] 0 setdash hpt 0 360 arc fill } def
/C0 { BL [] 0 setdash 2 copy moveto vpt 90 450  arc } bind def
/C1 { BL [] 0 setdash 2 copy        moveto
       2 copy  vpt 0 90 arc closepath fill
               vpt 0 360 arc closepath } bind def
/C2 { BL [] 0 setdash 2 copy moveto
       2 copy  vpt 90 180 arc closepath fill
               vpt 0 360 arc closepath } bind def
/C3 { BL [] 0 setdash 2 copy moveto
       2 copy  vpt 0 180 arc closepath fill
               vpt 0 360 arc closepath } bind def
/C4 { BL [] 0 setdash 2 copy moveto
       2 copy  vpt 180 270 arc closepath fill
               vpt 0 360 arc closepath } bind def
/C5 { BL [] 0 setdash 2 copy moveto
       2 copy  vpt 0 90 arc
       2 copy moveto
       2 copy  vpt 180 270 arc closepath fill
               vpt 0 360 arc } bind def
/C6 { BL [] 0 setdash 2 copy moveto
      2 copy  vpt 90 270 arc closepath fill
              vpt 0 360 arc closepath } bind def
/C7 { BL [] 0 setdash 2 copy moveto
      2 copy  vpt 0 270 arc closepath fill
              vpt 0 360 arc closepath } bind def
/C8 { BL [] 0 setdash 2 copy moveto
      2 copy vpt 270 360 arc closepath fill
              vpt 0 360 arc closepath } bind def
/C9 { BL [] 0 setdash 2 copy moveto
      2 copy  vpt 270 450 arc closepath fill
              vpt 0 360 arc closepath } bind def
/C10 { BL [] 0 setdash 2 copy 2 copy moveto vpt 270 360 arc closepath fill
       2 copy moveto
       2 copy vpt 90 180 arc closepath fill
               vpt 0 360 arc closepath } bind def
/C11 { BL [] 0 setdash 2 copy moveto
       2 copy  vpt 0 180 arc closepath fill
       2 copy moveto
       2 copy  vpt 270 360 arc closepath fill
               vpt 0 360 arc closepath } bind def
/C12 { BL [] 0 setdash 2 copy moveto
       2 copy  vpt 180 360 arc closepath fill
               vpt 0 360 arc closepath } bind def
/C13 { BL [] 0 setdash  2 copy moveto
       2 copy  vpt 0 90 arc closepath fill
       2 copy moveto
       2 copy  vpt 180 360 arc closepath fill
               vpt 0 360 arc closepath } bind def
/C14 { BL [] 0 setdash 2 copy moveto
       2 copy  vpt 90 360 arc closepath fill
               vpt 0 360 arc } bind def
/C15 { BL [] 0 setdash 2 copy vpt 0 360 arc closepath fill
               vpt 0 360 arc closepath } bind def
/Rec   { newpath 4 2 roll moveto 1 index 0 rlineto 0 exch rlineto
       neg 0 rlineto closepath } bind def
/Square { dup Rec } bind def
/Bsquare { vpt sub exch vpt sub exch vpt2 Square } bind def
/S0 { BL [] 0 setdash 2 copy moveto 0 vpt rlineto BL Bsquare } bind def
/S1 { BL [] 0 setdash 2 copy vpt Square fill Bsquare } bind def
/S2 { BL [] 0 setdash 2 copy exch vpt sub exch vpt Square fill Bsquare } bind def
/S3 { BL [] 0 setdash 2 copy exch vpt sub exch vpt2 vpt Rec fill Bsquare } bind def
/S4 { BL [] 0 setdash 2 copy exch vpt sub exch vpt sub vpt Square fill Bsquare } bind def
/S5 { BL [] 0 setdash 2 copy 2 copy vpt Square fill
       exch vpt sub exch vpt sub vpt Square fill Bsquare } bind def
/S6 { BL [] 0 setdash 2 copy exch vpt sub exch vpt sub vpt vpt2 Rec fill Bsquare } bind def
/S7 { BL [] 0 setdash 2 copy exch vpt sub exch vpt sub vpt vpt2 Rec fill
       2 copy vpt Square fill
       Bsquare } bind def
/S8 { BL [] 0 setdash 2 copy vpt sub vpt Square fill Bsquare } bind def
/S9 { BL [] 0 setdash 2 copy vpt sub vpt vpt2 Rec fill Bsquare } bind def
/S10 { BL [] 0 setdash 2 copy vpt sub vpt Square fill 2 copy exch vpt sub exch vpt Square fill
       Bsquare } bind def
/S11 { BL [] 0 setdash 2 copy vpt sub vpt Square fill 2 copy exch vpt sub exch vpt2 vpt Rec fill
       Bsquare } bind def
/S12 { BL [] 0 setdash 2 copy exch vpt sub exch vpt sub vpt2 vpt Rec fill Bsquare } bind def
/S13 { BL [] 0 setdash 2 copy exch vpt sub exch vpt sub vpt2 vpt Rec fill
       2 copy vpt Square fill Bsquare } bind def
/S14 { BL [] 0 setdash 2 copy exch vpt sub exch vpt sub vpt2 vpt Rec fill
       2 copy exch vpt sub exch vpt Square fill Bsquare } bind def
/S15 { BL [] 0 setdash 2 copy Bsquare fill Bsquare } bind def
/D0 { gsave translate 45 rotate 0 0 S0 stroke grestore } bind def
/D1 { gsave translate 45 rotate 0 0 S1 stroke grestore } bind def
/D2 { gsave translate 45 rotate 0 0 S2 stroke grestore } bind def
/D3 { gsave translate 45 rotate 0 0 S3 stroke grestore } bind def
/D4 { gsave translate 45 rotate 0 0 S4 stroke grestore } bind def
/D5 { gsave translate 45 rotate 0 0 S5 stroke grestore } bind def
/D6 { gsave translate 45 rotate 0 0 S6 stroke grestore } bind def
/D7 { gsave translate 45 rotate 0 0 S7 stroke grestore } bind def
/D8 { gsave translate 45 rotate 0 0 S8 stroke grestore } bind def
/D9 { gsave translate 45 rotate 0 0 S9 stroke grestore } bind def
/D10 { gsave translate 45 rotate 0 0 S10 stroke grestore } bind def
/D11 { gsave translate 45 rotate 0 0 S11 stroke grestore } bind def
/D12 { gsave translate 45 rotate 0 0 S12 stroke grestore } bind def
/D13 { gsave translate 45 rotate 0 0 S13 stroke grestore } bind def
/D14 { gsave translate 45 rotate 0 0 S14 stroke grestore } bind def
/D15 { gsave translate 45 rotate 0 0 S15 stroke grestore } bind def
/DiaE { stroke [] 0 setdash vpt add M
  hpt neg vpt neg V hpt vpt neg V
  hpt vpt V hpt neg vpt V closepath stroke } def
/BoxE { stroke [] 0 setdash exch hpt sub exch vpt add M
  0 vpt2 neg V hpt2 0 V 0 vpt2 V
  hpt2 neg 0 V closepath stroke } def
/TriUE { stroke [] 0 setdash vpt 1.12 mul add M
  hpt neg vpt -1.62 mul V
  hpt 2 mul 0 V
  hpt neg vpt 1.62 mul V closepath stroke } def
/TriDE { stroke [] 0 setdash vpt 1.12 mul sub M
  hpt neg vpt 1.62 mul V
  hpt 2 mul 0 V
  hpt neg vpt -1.62 mul V closepath stroke } def
/PentE { stroke [] 0 setdash gsave
  translate 0 hpt M 4 {72 rotate 0 hpt L} repeat
  closepath stroke grestore } def
/CircE { stroke [] 0 setdash 
  hpt 0 360 arc stroke } def
/Opaque { gsave closepath 1 setgray fill grestore 0 setgray closepath } def
/DiaW { stroke [] 0 setdash vpt add M
  hpt neg vpt neg V hpt vpt neg V
  hpt vpt V hpt neg vpt V Opaque stroke } def
/BoxW { stroke [] 0 setdash exch hpt sub exch vpt add M
  0 vpt2 neg V hpt2 0 V 0 vpt2 V
  hpt2 neg 0 V Opaque stroke } def
/TriUW { stroke [] 0 setdash vpt 1.12 mul add M
  hpt neg vpt -1.62 mul V
  hpt 2 mul 0 V
  hpt neg vpt 1.62 mul V Opaque stroke } def
/TriDW { stroke [] 0 setdash vpt 1.12 mul sub M
  hpt neg vpt 1.62 mul V
  hpt 2 mul 0 V
  hpt neg vpt -1.62 mul V Opaque stroke } def
/PentW { stroke [] 0 setdash gsave
  translate 0 hpt M 4 {72 rotate 0 hpt L} repeat
  Opaque stroke grestore } def
/CircW { stroke [] 0 setdash 
  hpt 0 360 arc Opaque stroke } def
/BoxFill { gsave Rec 1 setgray fill grestore } def
/Symbol-Oblique /Symbol findfont [1 0 .167 1 0 0] makefont
dup length dict begin {1 index /FID eq {pop pop} {def} ifelse} forall
currentdict end definefont
end
}}%
\begin{picture}(3600,2160)(0,0)%
{\GNUPLOTspecial{"
gnudict begin
gsave
0 0 translate
0.100 0.100 scale
0 setgray
newpath
1.000 UL
LTb
450 300 M
63 0 V
2937 0 R
-63 0 V
450 520 M
63 0 V
2937 0 R
-63 0 V
450 740 M
63 0 V
2937 0 R
-63 0 V
450 960 M
63 0 V
2937 0 R
-63 0 V
450 1180 M
63 0 V
2937 0 R
-63 0 V
450 1400 M
63 0 V
2937 0 R
-63 0 V
450 1620 M
63 0 V
2937 0 R
-63 0 V
450 1840 M
63 0 V
2937 0 R
-63 0 V
450 2060 M
63 0 V
2937 0 R
-63 0 V
450 300 M
0 63 V
0 1697 R
0 -63 V
1050 300 M
0 63 V
0 1697 R
0 -63 V
1650 300 M
0 63 V
0 1697 R
0 -63 V
2250 300 M
0 63 V
0 1697 R
0 -63 V
2850 300 M
0 63 V
0 1697 R
0 -63 V
3450 300 M
0 63 V
0 1697 R
0 -63 V
1.000 UL
LTb
450 300 M
3000 0 V
0 1760 V
-3000 0 V
450 300 L
1.000 UL
LT0
3087 1947 M
263 0 V
450 1852 M
30 -3 V
31 -4 V
30 -4 V
30 -6 V
31 -8 V
30 -10 V
30 -13 V
30 -14 V
31 -16 V
30 -18 V
30 -20 V
31 -21 V
30 -22 V
30 -24 V
31 -25 V
30 -26 V
30 -28 V
30 -28 V
31 -30 V
30 -30 V
30 -31 V
31 -32 V
30 -32 V
30 -32 V
31 -33 V
30 -34 V
30 -33 V
30 -34 V
31 -33 V
30 -34 V
30 -33 V
31 -33 V
30 -33 V
30 -33 V
31 -32 V
30 -31 V
30 -31 V
31 -30 V
30 -30 V
30 -29 V
30 -28 V
31 -27 V
30 -27 V
30 -26 V
31 -24 V
30 -25 V
30 -23 V
31 -22 V
30 -21 V
30 -21 V
30 -20 V
31 -18 V
30 -18 V
30 -17 V
31 -17 V
30 -15 V
30 -15 V
31 -13 V
30 -13 V
30 -13 V
30 -11 V
31 -11 V
30 -11 V
30 -9 V
31 -9 V
30 -9 V
30 -7 V
31 -8 V
30 -6 V
30 -7 V
31 -6 V
30 -5 V
30 -5 V
30 -5 V
31 -4 V
30 -4 V
30 -4 V
31 -3 V
30 -3 V
30 -3 V
31 -2 V
30 -3 V
30 -2 V
30 -2 V
31 -2 V
30 -1 V
30 -2 V
31 -1 V
30 -1 V
30 -1 V
31 -1 V
30 -1 V
30 -1 V
30 0 V
31 -1 V
30 -1 V
30 0 V
31 -1 V
30 0 V
1.000 UL
LT1
3087 1847 M
263 0 V
450 2021 M
30 -3 V
31 -4 V
30 -5 V
30 -7 V
31 -9 V
30 -11 V
30 -14 V
30 -17 V
31 -18 V
30 -21 V
30 -22 V
31 -23 V
30 -25 V
30 -27 V
31 -28 V
30 -29 V
30 -31 V
30 -31 V
31 -33 V
30 -34 V
30 -34 V
31 -35 V
30 -35 V
30 -36 V
31 -36 V
30 -36 V
30 -37 V
30 -36 V
31 -36 V
30 -37 V
30 -36 V
31 -36 V
30 -35 V
30 -35 V
31 -35 V
30 -33 V
30 -34 V
31 -32 V
30 -32 V
30 -32 V
30 -30 V
31 -30 V
30 -29 V
30 -28 V
31 -27 V
30 -26 V
30 -25 V
31 -25 V
30 -23 V
30 -23 V
30 -21 V
31 -21 V
30 -20 V
30 -19 V
31 -18 V
30 -17 V
30 -17 V
31 -16 V
30 -14 V
30 -14 V
30 -14 V
31 -12 V
30 -12 V
30 -11 V
31 -11 V
30 -9 V
30 -10 V
31 -8 V
30 -8 V
30 -8 V
31 -7 V
30 -6 V
30 -6 V
30 -6 V
31 -5 V
30 -5 V
30 -5 V
31 -4 V
30 -4 V
30 -3 V
31 -3 V
30 -3 V
30 -3 V
30 -2 V
31 -3 V
30 -2 V
30 -1 V
31 -2 V
30 -2 V
30 -1 V
31 -1 V
30 -2 V
30 -1 V
30 0 V
31 -1 V
30 -1 V
30 -1 V
31 0 V
30 -1 V
1.000 UL
LT2
3087 1747 M
263 0 V
450 921 M
30 -2 V
31 -1 V
30 -2 V
30 -2 V
31 -3 V
30 -4 V
30 -5 V
30 -6 V
31 -7 V
30 -7 V
30 -8 V
31 -8 V
30 -9 V
30 -9 V
31 -10 V
30 -11 V
30 -11 V
30 -11 V
31 -12 V
30 -12 V
30 -13 V
31 -12 V
30 -13 V
30 -13 V
31 -13 V
30 -14 V
30 -13 V
30 -14 V
31 -13 V
30 -13 V
30 -14 V
31 -13 V
30 -13 V
30 -13 V
31 -13 V
30 -12 V
30 -13 V
31 -12 V
30 -12 V
30 -11 V
30 -12 V
31 -10 V
30 -11 V
30 -10 V
31 -10 V
30 -10 V
30 -9 V
31 -9 V
30 -9 V
30 -8 V
30 -8 V
31 -7 V
30 -8 V
30 -6 V
31 -7 V
30 -6 V
30 -6 V
31 -5 V
30 -6 V
30 -5 V
30 -4 V
31 -5 V
30 -4 V
30 -3 V
31 -4 V
30 -3 V
30 -4 V
31 -2 V
30 -3 V
30 -3 V
31 -2 V
30 -2 V
30 -2 V
30 -2 V
31 -2 V
30 -1 V
30 -2 V
31 -1 V
30 -1 V
30 -2 V
31 -1 V
30 0 V
30 -1 V
30 -1 V
31 -1 V
30 0 V
30 -1 V
31 -1 V
30 0 V
30 0 V
31 -1 V
30 0 V
30 -1 V
30 0 V
31 0 V
30 0 V
30 0 V
31 -1 V
30 0 V
1.000 UL
LT3
3087 1647 M
263 0 V
450 946 M
30 -1 V
31 -1 V
30 -2 V
30 -3 V
31 -3 V
30 -4 V
30 -6 V
30 -6 V
31 -6 V
30 -8 V
30 -8 V
31 -9 V
30 -9 V
30 -10 V
31 -11 V
30 -11 V
30 -11 V
30 -12 V
31 -13 V
30 -12 V
30 -13 V
31 -13 V
30 -14 V
30 -13 V
31 -14 V
30 -14 V
30 -14 V
30 -14 V
31 -14 V
30 -14 V
30 -13 V
31 -14 V
30 -14 V
30 -13 V
31 -13 V
30 -13 V
30 -13 V
31 -12 V
30 -13 V
30 -11 V
30 -12 V
31 -11 V
30 -11 V
30 -11 V
31 -10 V
30 -10 V
30 -10 V
31 -9 V
30 -9 V
30 -8 V
30 -9 V
31 -7 V
30 -8 V
30 -7 V
31 -7 V
30 -6 V
30 -6 V
31 -6 V
30 -5 V
30 -5 V
30 -5 V
31 -5 V
30 -4 V
30 -4 V
31 -4 V
30 -4 V
30 -3 V
31 -3 V
30 -3 V
30 -3 V
31 -2 V
30 -2 V
30 -3 V
30 -2 V
31 -1 V
30 -2 V
30 -2 V
31 -1 V
30 -1 V
30 -2 V
31 -1 V
30 -1 V
30 -1 V
30 0 V
31 -1 V
30 -1 V
30 -1 V
31 0 V
30 -1 V
30 0 V
31 -1 V
30 0 V
30 0 V
30 -1 V
31 0 V
30 0 V
30 0 V
31 0 V
30 -1 V
stroke
grestore
end
showpage
}}%
\put(3037,1647){\makebox(0,0)[r]{$T_\sigma/\sigma^2$, $\rho = 5$}}%
\put(3037,1747){\makebox(0,0)[r]{$S_{11}/\rho$, $\rho = 5$}}%
\put(3037,1847){\makebox(0,0)[r]{$T_\sigma/\sigma^2$, $\rho = 2$}}%
\put(3037,1947){\makebox(0,0)[r]{$S_{11}/\rho$, $\rho = 2$}}%
\put(1950,50){\makebox(0,0){$\delta$}}%
\put(100,1180){%
\special{ps: gsave currentpoint currentpoint translate
270 rotate neg exch neg exch translate}%
\makebox(0,0)[b]{\shortstack{$T_\sigma/\sigma^2$}}%
\special{ps: currentpoint grestore moveto}%
}%
\put(3450,200){\makebox(0,0){$5.0$}}%
\put(2850,200){\makebox(0,0){$4.0$}}%
\put(2250,200){\makebox(0,0){$3.0$}}%
\put(1650,200){\makebox(0,0){$2.0$}}%
\put(1050,200){\makebox(0,0){$1.0$}}%
\put(450,200){\makebox(0,0){$0.0$}}%
\put(400,2060){\makebox(0,0)[r]{$0.16$}}%
\put(400,1840){\makebox(0,0)[r]{$0.14$}}%
\put(400,1620){\makebox(0,0)[r]{$0.12$}}%
\put(400,1400){\makebox(0,0)[r]{$0.10$}}%
\put(400,1180){\makebox(0,0)[r]{$0.08$}}%
\put(400,960){\makebox(0,0)[r]{$0.06$}}%
\put(400,740){\makebox(0,0)[r]{$0.04$}}%
\put(400,520){\makebox(0,0)[r]{$0.02$}}%
\put(400,300){\makebox(0,0)[r]{$0.00$}}%
\end{picture}%
\endgroup
 

%% file: fig11.tex
% GNUPLOT: LaTeX picture with Postscript
\begingroup%
  \makeatletter%
  \newcommand{\GNUPLOTspecial}{%
    \@sanitize\catcode`\%=14\relax\special}%
  \setlength{\unitlength}{0.1bp}%
{\GNUPLOTspecial{!
%!PS-Adobe-2.0
%%Title: fig11.tex
%%Creator: gnuplot 3.7 patchlevel 2
%%CreationDate: Sun Jun 16 23:24:45 2002
%%DocumentFonts: 
%%BoundingBox: 0 0 360 216
%%Orientation: Landscape
%%Pages: (atend)
%%EndComments
/gnudict 256 dict def
gnudict begin
/Color false def
/Solid false def
/gnulinewidth 5.000 def
/userlinewidth gnulinewidth def
/vshift -33 def
/dl {10 mul} def
/hpt_ 31.5 def
/vpt_ 31.5 def
/hpt hpt_ def
/vpt vpt_ def
/M {moveto} bind def
/L {lineto} bind def
/R {rmoveto} bind def
/V {rlineto} bind def
/vpt2 vpt 2 mul def
/hpt2 hpt 2 mul def
/Lshow { currentpoint stroke M
  0 vshift R show } def
/Rshow { currentpoint stroke M
  dup stringwidth pop neg vshift R show } def
/Cshow { currentpoint stroke M
  dup stringwidth pop -2 div vshift R show } def
/UP { dup vpt_ mul /vpt exch def hpt_ mul /hpt exch def
  /hpt2 hpt 2 mul def /vpt2 vpt 2 mul def } def
/DL { Color {setrgbcolor Solid {pop []} if 0 setdash }
 {pop pop pop Solid {pop []} if 0 setdash} ifelse } def
/BL { stroke userlinewidth 2 mul setlinewidth } def
/AL { stroke userlinewidth 2 div setlinewidth } def
/UL { dup gnulinewidth mul /userlinewidth exch def
      dup 1 lt {pop 1} if 10 mul /udl exch def } def
/PL { stroke userlinewidth setlinewidth } def
/LTb { BL [] 0 0 0 DL } def
/LTa { AL [1 udl mul 2 udl mul] 0 setdash 0 0 0 setrgbcolor } def
/LT0 { PL [] 1 0 0 DL } def
/LT1 { PL [4 dl 2 dl] 0 1 0 DL } def
/LT2 { PL [2 dl 3 dl] 0 0 1 DL } def
/LT3 { PL [1 dl 1.5 dl] 1 0 1 DL } def
/LT4 { PL [5 dl 2 dl 1 dl 2 dl] 0 1 1 DL } def
/LT5 { PL [4 dl 3 dl 1 dl 3 dl] 1 1 0 DL } def
/LT6 { PL [2 dl 2 dl 2 dl 4 dl] 0 0 0 DL } def
/LT7 { PL [2 dl 2 dl 2 dl 2 dl 2 dl 4 dl] 1 0.3 0 DL } def
/LT8 { PL [2 dl 2 dl 2 dl 2 dl 2 dl 2 dl 2 dl 4 dl] 0.5 0.5 0.5 DL } def
/Pnt { stroke [] 0 setdash
   gsave 1 setlinecap M 0 0 V stroke grestore } def
/Dia { stroke [] 0 setdash 2 copy vpt add M
  hpt neg vpt neg V hpt vpt neg V
  hpt vpt V hpt neg vpt V closepath stroke
  Pnt } def
/Pls { stroke [] 0 setdash vpt sub M 0 vpt2 V
  currentpoint stroke M
  hpt neg vpt neg R hpt2 0 V stroke
  } def
/Box { stroke [] 0 setdash 2 copy exch hpt sub exch vpt add M
  0 vpt2 neg V hpt2 0 V 0 vpt2 V
  hpt2 neg 0 V closepath stroke
  Pnt } def
/Crs { stroke [] 0 setdash exch hpt sub exch vpt add M
  hpt2 vpt2 neg V currentpoint stroke M
  hpt2 neg 0 R hpt2 vpt2 V stroke } def
/TriU { stroke [] 0 setdash 2 copy vpt 1.12 mul add M
  hpt neg vpt -1.62 mul V
  hpt 2 mul 0 V
  hpt neg vpt 1.62 mul V closepath stroke
  Pnt  } def
/Star { 2 copy Pls Crs } def
/BoxF { stroke [] 0 setdash exch hpt sub exch vpt add M
  0 vpt2 neg V  hpt2 0 V  0 vpt2 V
  hpt2 neg 0 V  closepath fill } def
/TriUF { stroke [] 0 setdash vpt 1.12 mul add M
  hpt neg vpt -1.62 mul V
  hpt 2 mul 0 V
  hpt neg vpt 1.62 mul V closepath fill } def
/TriD { stroke [] 0 setdash 2 copy vpt 1.12 mul sub M
  hpt neg vpt 1.62 mul V
  hpt 2 mul 0 V
  hpt neg vpt -1.62 mul V closepath stroke
  Pnt  } def
/TriDF { stroke [] 0 setdash vpt 1.12 mul sub M
  hpt neg vpt 1.62 mul V
  hpt 2 mul 0 V
  hpt neg vpt -1.62 mul V closepath fill} def
/DiaF { stroke [] 0 setdash vpt add M
  hpt neg vpt neg V hpt vpt neg V
  hpt vpt V hpt neg vpt V closepath fill } def
/Pent { stroke [] 0 setdash 2 copy gsave
  translate 0 hpt M 4 {72 rotate 0 hpt L} repeat
  closepath stroke grestore Pnt } def
/PentF { stroke [] 0 setdash gsave
  translate 0 hpt M 4 {72 rotate 0 hpt L} repeat
  closepath fill grestore } def
/Circle { stroke [] 0 setdash 2 copy
  hpt 0 360 arc stroke Pnt } def
/CircleF { stroke [] 0 setdash hpt 0 360 arc fill } def
/C0 { BL [] 0 setdash 2 copy moveto vpt 90 450  arc } bind def
/C1 { BL [] 0 setdash 2 copy        moveto
       2 copy  vpt 0 90 arc closepath fill
               vpt 0 360 arc closepath } bind def
/C2 { BL [] 0 setdash 2 copy moveto
       2 copy  vpt 90 180 arc closepath fill
               vpt 0 360 arc closepath } bind def
/C3 { BL [] 0 setdash 2 copy moveto
       2 copy  vpt 0 180 arc closepath fill
               vpt 0 360 arc closepath } bind def
/C4 { BL [] 0 setdash 2 copy moveto
       2 copy  vpt 180 270 arc closepath fill
               vpt 0 360 arc closepath } bind def
/C5 { BL [] 0 setdash 2 copy moveto
       2 copy  vpt 0 90 arc
       2 copy moveto
       2 copy  vpt 180 270 arc closepath fill
               vpt 0 360 arc } bind def
/C6 { BL [] 0 setdash 2 copy moveto
      2 copy  vpt 90 270 arc closepath fill
              vpt 0 360 arc closepath } bind def
/C7 { BL [] 0 setdash 2 copy moveto
      2 copy  vpt 0 270 arc closepath fill
              vpt 0 360 arc closepath } bind def
/C8 { BL [] 0 setdash 2 copy moveto
      2 copy vpt 270 360 arc closepath fill
              vpt 0 360 arc closepath } bind def
/C9 { BL [] 0 setdash 2 copy moveto
      2 copy  vpt 270 450 arc closepath fill
              vpt 0 360 arc closepath } bind def
/C10 { BL [] 0 setdash 2 copy 2 copy moveto vpt 270 360 arc closepath fill
       2 copy moveto
       2 copy vpt 90 180 arc closepath fill
               vpt 0 360 arc closepath } bind def
/C11 { BL [] 0 setdash 2 copy moveto
       2 copy  vpt 0 180 arc closepath fill
       2 copy moveto
       2 copy  vpt 270 360 arc closepath fill
               vpt 0 360 arc closepath } bind def
/C12 { BL [] 0 setdash 2 copy moveto
       2 copy  vpt 180 360 arc closepath fill
               vpt 0 360 arc closepath } bind def
/C13 { BL [] 0 setdash  2 copy moveto
       2 copy  vpt 0 90 arc closepath fill
       2 copy moveto
       2 copy  vpt 180 360 arc closepath fill
               vpt 0 360 arc closepath } bind def
/C14 { BL [] 0 setdash 2 copy moveto
       2 copy  vpt 90 360 arc closepath fill
               vpt 0 360 arc } bind def
/C15 { BL [] 0 setdash 2 copy vpt 0 360 arc closepath fill
               vpt 0 360 arc closepath } bind def
/Rec   { newpath 4 2 roll moveto 1 index 0 rlineto 0 exch rlineto
       neg 0 rlineto closepath } bind def
/Square { dup Rec } bind def
/Bsquare { vpt sub exch vpt sub exch vpt2 Square } bind def
/S0 { BL [] 0 setdash 2 copy moveto 0 vpt rlineto BL Bsquare } bind def
/S1 { BL [] 0 setdash 2 copy vpt Square fill Bsquare } bind def
/S2 { BL [] 0 setdash 2 copy exch vpt sub exch vpt Square fill Bsquare } bind def
/S3 { BL [] 0 setdash 2 copy exch vpt sub exch vpt2 vpt Rec fill Bsquare } bind def
/S4 { BL [] 0 setdash 2 copy exch vpt sub exch vpt sub vpt Square fill Bsquare } bind def
/S5 { BL [] 0 setdash 2 copy 2 copy vpt Square fill
       exch vpt sub exch vpt sub vpt Square fill Bsquare } bind def
/S6 { BL [] 0 setdash 2 copy exch vpt sub exch vpt sub vpt vpt2 Rec fill Bsquare } bind def
/S7 { BL [] 0 setdash 2 copy exch vpt sub exch vpt sub vpt vpt2 Rec fill
       2 copy vpt Square fill
       Bsquare } bind def
/S8 { BL [] 0 setdash 2 copy vpt sub vpt Square fill Bsquare } bind def
/S9 { BL [] 0 setdash 2 copy vpt sub vpt vpt2 Rec fill Bsquare } bind def
/S10 { BL [] 0 setdash 2 copy vpt sub vpt Square fill 2 copy exch vpt sub exch vpt Square fill
       Bsquare } bind def
/S11 { BL [] 0 setdash 2 copy vpt sub vpt Square fill 2 copy exch vpt sub exch vpt2 vpt Rec fill
       Bsquare } bind def
/S12 { BL [] 0 setdash 2 copy exch vpt sub exch vpt sub vpt2 vpt Rec fill Bsquare } bind def
/S13 { BL [] 0 setdash 2 copy exch vpt sub exch vpt sub vpt2 vpt Rec fill
       2 copy vpt Square fill Bsquare } bind def
/S14 { BL [] 0 setdash 2 copy exch vpt sub exch vpt sub vpt2 vpt Rec fill
       2 copy exch vpt sub exch vpt Square fill Bsquare } bind def
/S15 { BL [] 0 setdash 2 copy Bsquare fill Bsquare } bind def
/D0 { gsave translate 45 rotate 0 0 S0 stroke grestore } bind def
/D1 { gsave translate 45 rotate 0 0 S1 stroke grestore } bind def
/D2 { gsave translate 45 rotate 0 0 S2 stroke grestore } bind def
/D3 { gsave translate 45 rotate 0 0 S3 stroke grestore } bind def
/D4 { gsave translate 45 rotate 0 0 S4 stroke grestore } bind def
/D5 { gsave translate 45 rotate 0 0 S5 stroke grestore } bind def
/D6 { gsave translate 45 rotate 0 0 S6 stroke grestore } bind def
/D7 { gsave translate 45 rotate 0 0 S7 stroke grestore } bind def
/D8 { gsave translate 45 rotate 0 0 S8 stroke grestore } bind def
/D9 { gsave translate 45 rotate 0 0 S9 stroke grestore } bind def
/D10 { gsave translate 45 rotate 0 0 S10 stroke grestore } bind def
/D11 { gsave translate 45 rotate 0 0 S11 stroke grestore } bind def
/D12 { gsave translate 45 rotate 0 0 S12 stroke grestore } bind def
/D13 { gsave translate 45 rotate 0 0 S13 stroke grestore } bind def
/D14 { gsave translate 45 rotate 0 0 S14 stroke grestore } bind def
/D15 { gsave translate 45 rotate 0 0 S15 stroke grestore } bind def
/DiaE { stroke [] 0 setdash vpt add M
  hpt neg vpt neg V hpt vpt neg V
  hpt vpt V hpt neg vpt V closepath stroke } def
/BoxE { stroke [] 0 setdash exch hpt sub exch vpt add M
  0 vpt2 neg V hpt2 0 V 0 vpt2 V
  hpt2 neg 0 V closepath stroke } def
/TriUE { stroke [] 0 setdash vpt 1.12 mul add M
  hpt neg vpt -1.62 mul V
  hpt 2 mul 0 V
  hpt neg vpt 1.62 mul V closepath stroke } def
/TriDE { stroke [] 0 setdash vpt 1.12 mul sub M
  hpt neg vpt 1.62 mul V
  hpt 2 mul 0 V
  hpt neg vpt -1.62 mul V closepath stroke } def
/PentE { stroke [] 0 setdash gsave
  translate 0 hpt M 4 {72 rotate 0 hpt L} repeat
  closepath stroke grestore } def
/CircE { stroke [] 0 setdash 
  hpt 0 360 arc stroke } def
/Opaque { gsave closepath 1 setgray fill grestore 0 setgray closepath } def
/DiaW { stroke [] 0 setdash vpt add M
  hpt neg vpt neg V hpt vpt neg V
  hpt vpt V hpt neg vpt V Opaque stroke } def
/BoxW { stroke [] 0 setdash exch hpt sub exch vpt add M
  0 vpt2 neg V hpt2 0 V 0 vpt2 V
  hpt2 neg 0 V Opaque stroke } def
/TriUW { stroke [] 0 setdash vpt 1.12 mul add M
  hpt neg vpt -1.62 mul V
  hpt 2 mul 0 V
  hpt neg vpt 1.62 mul V Opaque stroke } def
/TriDW { stroke [] 0 setdash vpt 1.12 mul sub M
  hpt neg vpt 1.62 mul V
  hpt 2 mul 0 V
  hpt neg vpt -1.62 mul V Opaque stroke } def
/PentW { stroke [] 0 setdash gsave
  translate 0 hpt M 4 {72 rotate 0 hpt L} repeat
  Opaque stroke grestore } def
/CircW { stroke [] 0 setdash 
  hpt 0 360 arc Opaque stroke } def
/BoxFill { gsave Rec 1 setgray fill grestore } def
/Symbol-Oblique /Symbol findfont [1 0 .167 1 0 0] makefont
dup length dict begin {1 index /FID eq {pop pop} {def} ifelse} forall
currentdict end definefont
end
}}%
\begin{picture}(3600,2160)(0,0)%
{\GNUPLOTspecial{"
gnudict begin
gsave
0 0 translate
0.100 0.100 scale
0 setgray
newpath
1.000 UL
LTb
500 300 M
63 0 V
2887 0 R
-63 0 V
500 496 M
63 0 V
2887 0 R
-63 0 V
500 691 M
63 0 V
2887 0 R
-63 0 V
500 887 M
63 0 V
2887 0 R
-63 0 V
500 1082 M
63 0 V
2887 0 R
-63 0 V
500 1278 M
63 0 V
2887 0 R
-63 0 V
500 1473 M
63 0 V
2887 0 R
-63 0 V
500 1669 M
63 0 V
2887 0 R
-63 0 V
500 1864 M
63 0 V
2887 0 R
-63 0 V
500 2060 M
63 0 V
2887 0 R
-63 0 V
500 300 M
0 63 V
0 1697 R
0 -63 V
1090 300 M
0 63 V
0 1697 R
0 -63 V
1680 300 M
0 63 V
0 1697 R
0 -63 V
2270 300 M
0 63 V
0 1697 R
0 -63 V
2860 300 M
0 63 V
0 1697 R
0 -63 V
3450 300 M
0 63 V
0 1697 R
0 -63 V
1.000 UL
LTb
500 300 M
2950 0 V
0 1760 V
-2950 0 V
500 300 L
1.000 UL
LT0
3087 1947 M
263 0 V
500 873 M
30 -2 V
30 -2 V
29 -3 V
30 -4 V
30 -5 V
30 -7 V
30 -8 V
29 -10 V
30 -10 V
30 -11 V
30 -13 V
30 -12 V
29 -14 V
30 -14 V
30 -14 V
30 -15 V
30 -15 V
29 -15 V
30 -15 V
30 -15 V
30 -15 V
30 -15 V
29 -15 V
30 -14 V
30 -13 V
30 -13 V
30 -13 V
29 -12 V
30 -11 V
30 -11 V
30 -9 V
30 -10 V
29 -8 V
30 -8 V
30 -7 V
30 -6 V
30 -6 V
29 -5 V
30 -4 V
30 -3 V
30 -3 V
30 -3 V
29 -2 V
30 -1 V
30 -1 V
30 0 V
30 -1 V
29 1 V
30 0 V
30 1 V
30 1 V
29 2 V
30 1 V
30 2 V
30 2 V
30 2 V
29 1 V
30 2 V
30 2 V
30 2 V
30 2 V
29 2 V
30 2 V
30 1 V
30 2 V
30 2 V
29 1 V
30 1 V
30 2 V
30 1 V
30 1 V
29 1 V
30 1 V
30 0 V
30 1 V
30 0 V
29 1 V
30 0 V
30 1 V
30 0 V
30 0 V
29 0 V
30 0 V
30 0 V
30 0 V
30 0 V
29 0 V
30 0 V
30 -1 V
30 0 V
30 0 V
29 0 V
30 0 V
30 -1 V
30 0 V
30 0 V
29 0 V
30 0 V
30 -1 V
1.000 UL
LT1
3087 1847 M
263 0 V
500 1539 M
30 -6 V
30 -5 V
29 -8 V
30 -9 V
30 -13 V
30 -16 V
30 -20 V
29 -23 V
30 -25 V
30 -26 V
30 -28 V
30 -30 V
29 -30 V
30 -30 V
30 -31 V
30 -32 V
30 -31 V
29 -30 V
30 -30 V
30 -29 V
30 -28 V
30 -26 V
29 -25 V
30 -24 V
30 -22 V
30 -20 V
30 -19 V
29 -16 V
30 -16 V
30 -13 V
30 -12 V
30 -10 V
29 -9 V
30 -8 V
30 -7 V
30 -5 V
30 -5 V
29 -4 V
30 -4 V
30 -2 V
30 -3 V
30 -3 V
29 -2 V
30 -2 V
30 -3 V
30 -3 V
30 -3 V
29 -3 V
30 -4 V
30 -4 V
30 -5 V
29 -5 V
30 -6 V
30 -6 V
30 -7 V
30 -7 V
29 -8 V
30 -8 V
30 -8 V
30 -9 V
30 -9 V
29 -9 V
30 -9 V
30 -9 V
30 -10 V
30 -9 V
29 -9 V
30 -10 V
30 -9 V
30 -8 V
30 -9 V
29 -8 V
30 -8 V
30 -8 V
30 -8 V
30 -7 V
29 -6 V
30 -7 V
30 -5 V
30 -6 V
30 -5 V
29 -5 V
30 -5 V
30 -4 V
30 -3 V
30 -4 V
29 -3 V
30 -3 V
30 -3 V
30 -2 V
30 -2 V
29 -2 V
30 -2 V
30 -1 V
30 -2 V
30 -1 V
29 -1 V
30 -1 V
30 -1 V
1.000 UL
LT2
3087 1747 M
263 0 V
500 1993 M
30 -3 V
30 -4 V
29 -6 V
30 -6 V
30 -10 V
30 -11 V
30 -14 V
29 -17 V
30 -18 V
30 -19 V
30 -21 V
30 -21 V
29 -23 V
30 -23 V
30 -24 V
30 -25 V
30 -25 V
29 -25 V
30 -26 V
30 -26 V
30 -27 V
30 -26 V
29 -27 V
30 -27 V
30 -28 V
30 -27 V
30 -28 V
29 -28 V
30 -28 V
30 -28 V
30 -29 V
30 -29 V
29 -29 V
30 -30 V
30 -29 V
30 -29 V
30 -30 V
29 -29 V
30 -29 V
30 -29 V
30 -29 V
30 -28 V
29 -27 V
30 -27 V
30 -27 V
30 -25 V
30 -25 V
29 -24 V
30 -23 V
30 -22 V
30 -21 V
29 -20 V
30 -19 V
30 -18 V
30 -17 V
30 -16 V
29 -15 V
30 -14 V
30 -13 V
30 -13 V
30 -11 V
29 -11 V
30 -10 V
30 -10 V
30 -9 V
30 -8 V
29 -8 V
30 -7 V
30 -6 V
30 -7 V
30 -6 V
29 -5 V
30 -5 V
30 -5 V
30 -4 V
30 -4 V
29 -4 V
30 -4 V
30 -3 V
30 -3 V
30 -3 V
29 -3 V
30 -2 V
30 -3 V
30 -2 V
30 -2 V
29 -1 V
30 -2 V
30 -1 V
30 -2 V
30 -1 V
29 -1 V
30 -1 V
30 -1 V
30 0 V
30 -1 V
29 -1 V
30 0 V
30 -1 V
1.000 UL
LT3
3087 1647 M
263 0 V
500 1995 M
30 -3 V
30 -3 V
29 -4 V
30 -6 V
30 -8 V
30 -10 V
30 -12 V
29 -13 V
30 -16 V
30 -17 V
30 -19 V
30 -20 V
29 -22 V
30 -22 V
30 -24 V
30 -25 V
30 -27 V
29 -27 V
30 -28 V
30 -29 V
30 -29 V
30 -30 V
29 -31 V
30 -31 V
30 -31 V
30 -31 V
30 -32 V
29 -32 V
30 -31 V
30 -32 V
30 -31 V
30 -32 V
29 -31 V
30 -30 V
30 -30 V
30 -30 V
30 -29 V
29 -29 V
30 -28 V
30 -27 V
30 -27 V
30 -26 V
29 -25 V
30 -25 V
30 -24 V
30 -23 V
30 -22 V
29 -22 V
30 -21 V
30 -20 V
30 -19 V
29 -18 V
30 -18 V
30 -16 V
30 -16 V
30 -16 V
29 -14 V
30 -14 V
30 -13 V
30 -13 V
30 -11 V
29 -11 V
30 -11 V
30 -10 V
30 -9 V
30 -9 V
29 -8 V
30 -7 V
30 -8 V
30 -6 V
30 -6 V
29 -6 V
30 -6 V
30 -5 V
30 -4 V
30 -4 V
29 -4 V
30 -4 V
30 -3 V
30 -4 V
30 -2 V
29 -3 V
30 -2 V
30 -2 V
30 -2 V
30 -2 V
29 -2 V
30 -1 V
30 -2 V
30 -1 V
30 -1 V
29 -2 V
30 -1 V
30 0 V
30 -1 V
30 -1 V
29 0 V
30 0 V
30 -1 V
stroke
grestore
end
showpage
}}%
\put(3037,1647){\makebox(0,0)[r]{$k = 5.0$}}%
\put(3037,1747){\makebox(0,0)[r]{$k = 2.0$}}%
\put(3037,1847){\makebox(0,0)[r]{$k = 1.0$}}%
\put(3037,1947){\makebox(0,0)[r]{$k = 0.5$}}%
\put(1975,50){\makebox(0,0){$\delta$}}%
\put(100,1180){%
\special{ps: gsave currentpoint currentpoint translate
270 rotate neg exch neg exch translate}%
\makebox(0,0)[b]{\shortstack{$T_P$}}%
\special{ps: currentpoint grestore moveto}%
}%
\put(3450,200){\makebox(0,0){$5.0$}}%
\put(2860,200){\makebox(0,0){$4.0$}}%
\put(2270,200){\makebox(0,0){$3.0$}}%
\put(1680,200){\makebox(0,0){$2.0$}}%
\put(1090,200){\makebox(0,0){$1.0$}}%
\put(500,200){\makebox(0,0){$0.0$}}%
\put(450,2060){\makebox(0,0)[r]{$0.08$}}%
\put(450,1864){\makebox(0,0)[r]{$0.07$}}%
\put(450,1669){\makebox(0,0)[r]{$0.06$}}%
\put(450,1473){\makebox(0,0)[r]{$0.05$}}%
\put(450,1278){\makebox(0,0)[r]{$0.04$}}%
\put(450,1082){\makebox(0,0)[r]{$0.03$}}%
\put(450,887){\makebox(0,0)[r]{$0.02$}}%
\put(450,691){\makebox(0,0)[r]{$0.01$}}%
\put(450,496){\makebox(0,0)[r]{$0.00$}}%
\put(450,300){\makebox(0,0)[r]{$-0.01$}}%
\end{picture}%
\endgroup
 

%% file: fig4.tex
% GNUPLOT: LaTeX picture with Postscript
\begingroup%
  \makeatletter%
  \newcommand{\GNUPLOTspecial}{%
    \@sanitize\catcode`\%=14\relax\special}%
  \setlength{\unitlength}{0.1bp}%
{\GNUPLOTspecial{!
%!PS-Adobe-2.0
%%Title: fig4.tex
%%Creator: gnuplot 3.7 patchlevel 2
%%CreationDate: Sun Jun 16 23:24:38 2002
%%DocumentFonts: 
%%BoundingBox: 0 0 360 216
%%Orientation: Landscape
%%Pages: (atend)
%%EndComments
/gnudict 256 dict def
gnudict begin
/Color false def
/Solid false def
/gnulinewidth 5.000 def
/userlinewidth gnulinewidth def
/vshift -33 def
/dl {10 mul} def
/hpt_ 31.5 def
/vpt_ 31.5 def
/hpt hpt_ def
/vpt vpt_ def
/M {moveto} bind def
/L {lineto} bind def
/R {rmoveto} bind def
/V {rlineto} bind def
/vpt2 vpt 2 mul def
/hpt2 hpt 2 mul def
/Lshow { currentpoint stroke M
  0 vshift R show } def
/Rshow { currentpoint stroke M
  dup stringwidth pop neg vshift R show } def
/Cshow { currentpoint stroke M
  dup stringwidth pop -2 div vshift R show } def
/UP { dup vpt_ mul /vpt exch def hpt_ mul /hpt exch def
  /hpt2 hpt 2 mul def /vpt2 vpt 2 mul def } def
/DL { Color {setrgbcolor Solid {pop []} if 0 setdash }
 {pop pop pop Solid {pop []} if 0 setdash} ifelse } def
/BL { stroke userlinewidth 2 mul setlinewidth } def
/AL { stroke userlinewidth 2 div setlinewidth } def
/UL { dup gnulinewidth mul /userlinewidth exch def
      dup 1 lt {pop 1} if 10 mul /udl exch def } def
/PL { stroke userlinewidth setlinewidth } def
/LTb { BL [] 0 0 0 DL } def
/LTa { AL [1 udl mul 2 udl mul] 0 setdash 0 0 0 setrgbcolor } def
/LT0 { PL [] 1 0 0 DL } def
/LT1 { PL [4 dl 2 dl] 0 1 0 DL } def
/LT2 { PL [2 dl 3 dl] 0 0 1 DL } def
/LT3 { PL [1 dl 1.5 dl] 1 0 1 DL } def
/LT4 { PL [5 dl 2 dl 1 dl 2 dl] 0 1 1 DL } def
/LT5 { PL [4 dl 3 dl 1 dl 3 dl] 1 1 0 DL } def
/LT6 { PL [2 dl 2 dl 2 dl 4 dl] 0 0 0 DL } def
/LT7 { PL [2 dl 2 dl 2 dl 2 dl 2 dl 4 dl] 1 0.3 0 DL } def
/LT8 { PL [2 dl 2 dl 2 dl 2 dl 2 dl 2 dl 2 dl 4 dl] 0.5 0.5 0.5 DL } def
/Pnt { stroke [] 0 setdash
   gsave 1 setlinecap M 0 0 V stroke grestore } def
/Dia { stroke [] 0 setdash 2 copy vpt add M
  hpt neg vpt neg V hpt vpt neg V
  hpt vpt V hpt neg vpt V closepath stroke
  Pnt } def
/Pls { stroke [] 0 setdash vpt sub M 0 vpt2 V
  currentpoint stroke M
  hpt neg vpt neg R hpt2 0 V stroke
  } def
/Box { stroke [] 0 setdash 2 copy exch hpt sub exch vpt add M
  0 vpt2 neg V hpt2 0 V 0 vpt2 V
  hpt2 neg 0 V closepath stroke
  Pnt } def
/Crs { stroke [] 0 setdash exch hpt sub exch vpt add M
  hpt2 vpt2 neg V currentpoint stroke M
  hpt2 neg 0 R hpt2 vpt2 V stroke } def
/TriU { stroke [] 0 setdash 2 copy vpt 1.12 mul add M
  hpt neg vpt -1.62 mul V
  hpt 2 mul 0 V
  hpt neg vpt 1.62 mul V closepath stroke
  Pnt  } def
/Star { 2 copy Pls Crs } def
/BoxF { stroke [] 0 setdash exch hpt sub exch vpt add M
  0 vpt2 neg V  hpt2 0 V  0 vpt2 V
  hpt2 neg 0 V  closepath fill } def
/TriUF { stroke [] 0 setdash vpt 1.12 mul add M
  hpt neg vpt -1.62 mul V
  hpt 2 mul 0 V
  hpt neg vpt 1.62 mul V closepath fill } def
/TriD { stroke [] 0 setdash 2 copy vpt 1.12 mul sub M
  hpt neg vpt 1.62 mul V
  hpt 2 mul 0 V
  hpt neg vpt -1.62 mul V closepath stroke
  Pnt  } def
/TriDF { stroke [] 0 setdash vpt 1.12 mul sub M
  hpt neg vpt 1.62 mul V
  hpt 2 mul 0 V
  hpt neg vpt -1.62 mul V closepath fill} def
/DiaF { stroke [] 0 setdash vpt add M
  hpt neg vpt neg V hpt vpt neg V
  hpt vpt V hpt neg vpt V closepath fill } def
/Pent { stroke [] 0 setdash 2 copy gsave
  translate 0 hpt M 4 {72 rotate 0 hpt L} repeat
  closepath stroke grestore Pnt } def
/PentF { stroke [] 0 setdash gsave
  translate 0 hpt M 4 {72 rotate 0 hpt L} repeat
  closepath fill grestore } def
/Circle { stroke [] 0 setdash 2 copy
  hpt 0 360 arc stroke Pnt } def
/CircleF { stroke [] 0 setdash hpt 0 360 arc fill } def
/C0 { BL [] 0 setdash 2 copy moveto vpt 90 450  arc } bind def
/C1 { BL [] 0 setdash 2 copy        moveto
       2 copy  vpt 0 90 arc closepath fill
               vpt 0 360 arc closepath } bind def
/C2 { BL [] 0 setdash 2 copy moveto
       2 copy  vpt 90 180 arc closepath fill
               vpt 0 360 arc closepath } bind def
/C3 { BL [] 0 setdash 2 copy moveto
       2 copy  vpt 0 180 arc closepath fill
               vpt 0 360 arc closepath } bind def
/C4 { BL [] 0 setdash 2 copy moveto
       2 copy  vpt 180 270 arc closepath fill
               vpt 0 360 arc closepath } bind def
/C5 { BL [] 0 setdash 2 copy moveto
       2 copy  vpt 0 90 arc
       2 copy moveto
       2 copy  vpt 180 270 arc closepath fill
               vpt 0 360 arc } bind def
/C6 { BL [] 0 setdash 2 copy moveto
      2 copy  vpt 90 270 arc closepath fill
              vpt 0 360 arc closepath } bind def
/C7 { BL [] 0 setdash 2 copy moveto
      2 copy  vpt 0 270 arc closepath fill
              vpt 0 360 arc closepath } bind def
/C8 { BL [] 0 setdash 2 copy moveto
      2 copy vpt 270 360 arc closepath fill
              vpt 0 360 arc closepath } bind def
/C9 { BL [] 0 setdash 2 copy moveto
      2 copy  vpt 270 450 arc closepath fill
              vpt 0 360 arc closepath } bind def
/C10 { BL [] 0 setdash 2 copy 2 copy moveto vpt 270 360 arc closepath fill
       2 copy moveto
       2 copy vpt 90 180 arc closepath fill
               vpt 0 360 arc closepath } bind def
/C11 { BL [] 0 setdash 2 copy moveto
       2 copy  vpt 0 180 arc closepath fill
       2 copy moveto
       2 copy  vpt 270 360 arc closepath fill
               vpt 0 360 arc closepath } bind def
/C12 { BL [] 0 setdash 2 copy moveto
       2 copy  vpt 180 360 arc closepath fill
               vpt 0 360 arc closepath } bind def
/C13 { BL [] 0 setdash  2 copy moveto
       2 copy  vpt 0 90 arc closepath fill
       2 copy moveto
       2 copy  vpt 180 360 arc closepath fill
               vpt 0 360 arc closepath } bind def
/C14 { BL [] 0 setdash 2 copy moveto
       2 copy  vpt 90 360 arc closepath fill
               vpt 0 360 arc } bind def
/C15 { BL [] 0 setdash 2 copy vpt 0 360 arc closepath fill
               vpt 0 360 arc closepath } bind def
/Rec   { newpath 4 2 roll moveto 1 index 0 rlineto 0 exch rlineto
       neg 0 rlineto closepath } bind def
/Square { dup Rec } bind def
/Bsquare { vpt sub exch vpt sub exch vpt2 Square } bind def
/S0 { BL [] 0 setdash 2 copy moveto 0 vpt rlineto BL Bsquare } bind def
/S1 { BL [] 0 setdash 2 copy vpt Square fill Bsquare } bind def
/S2 { BL [] 0 setdash 2 copy exch vpt sub exch vpt Square fill Bsquare } bind def
/S3 { BL [] 0 setdash 2 copy exch vpt sub exch vpt2 vpt Rec fill Bsquare } bind def
/S4 { BL [] 0 setdash 2 copy exch vpt sub exch vpt sub vpt Square fill Bsquare } bind def
/S5 { BL [] 0 setdash 2 copy 2 copy vpt Square fill
       exch vpt sub exch vpt sub vpt Square fill Bsquare } bind def
/S6 { BL [] 0 setdash 2 copy exch vpt sub exch vpt sub vpt vpt2 Rec fill Bsquare } bind def
/S7 { BL [] 0 setdash 2 copy exch vpt sub exch vpt sub vpt vpt2 Rec fill
       2 copy vpt Square fill
       Bsquare } bind def
/S8 { BL [] 0 setdash 2 copy vpt sub vpt Square fill Bsquare } bind def
/S9 { BL [] 0 setdash 2 copy vpt sub vpt vpt2 Rec fill Bsquare } bind def
/S10 { BL [] 0 setdash 2 copy vpt sub vpt Square fill 2 copy exch vpt sub exch vpt Square fill
       Bsquare } bind def
/S11 { BL [] 0 setdash 2 copy vpt sub vpt Square fill 2 copy exch vpt sub exch vpt2 vpt Rec fill
       Bsquare } bind def
/S12 { BL [] 0 setdash 2 copy exch vpt sub exch vpt sub vpt2 vpt Rec fill Bsquare } bind def
/S13 { BL [] 0 setdash 2 copy exch vpt sub exch vpt sub vpt2 vpt Rec fill
       2 copy vpt Square fill Bsquare } bind def
/S14 { BL [] 0 setdash 2 copy exch vpt sub exch vpt sub vpt2 vpt Rec fill
       2 copy exch vpt sub exch vpt Square fill Bsquare } bind def
/S15 { BL [] 0 setdash 2 copy Bsquare fill Bsquare } bind def
/D0 { gsave translate 45 rotate 0 0 S0 stroke grestore } bind def
/D1 { gsave translate 45 rotate 0 0 S1 stroke grestore } bind def
/D2 { gsave translate 45 rotate 0 0 S2 stroke grestore } bind def
/D3 { gsave translate 45 rotate 0 0 S3 stroke grestore } bind def
/D4 { gsave translate 45 rotate 0 0 S4 stroke grestore } bind def
/D5 { gsave translate 45 rotate 0 0 S5 stroke grestore } bind def
/D6 { gsave translate 45 rotate 0 0 S6 stroke grestore } bind def
/D7 { gsave translate 45 rotate 0 0 S7 stroke grestore } bind def
/D8 { gsave translate 45 rotate 0 0 S8 stroke grestore } bind def
/D9 { gsave translate 45 rotate 0 0 S9 stroke grestore } bind def
/D10 { gsave translate 45 rotate 0 0 S10 stroke grestore } bind def
/D11 { gsave translate 45 rotate 0 0 S11 stroke grestore } bind def
/D12 { gsave translate 45 rotate 0 0 S12 stroke grestore } bind def
/D13 { gsave translate 45 rotate 0 0 S13 stroke grestore } bind def
/D14 { gsave translate 45 rotate 0 0 S14 stroke grestore } bind def
/D15 { gsave translate 45 rotate 0 0 S15 stroke grestore } bind def
/DiaE { stroke [] 0 setdash vpt add M
  hpt neg vpt neg V hpt vpt neg V
  hpt vpt V hpt neg vpt V closepath stroke } def
/BoxE { stroke [] 0 setdash exch hpt sub exch vpt add M
  0 vpt2 neg V hpt2 0 V 0 vpt2 V
  hpt2 neg 0 V closepath stroke } def
/TriUE { stroke [] 0 setdash vpt 1.12 mul add M
  hpt neg vpt -1.62 mul V
  hpt 2 mul 0 V
  hpt neg vpt 1.62 mul V closepath stroke } def
/TriDE { stroke [] 0 setdash vpt 1.12 mul sub M
  hpt neg vpt 1.62 mul V
  hpt 2 mul 0 V
  hpt neg vpt -1.62 mul V closepath stroke } def
/PentE { stroke [] 0 setdash gsave
  translate 0 hpt M 4 {72 rotate 0 hpt L} repeat
  closepath stroke grestore } def
/CircE { stroke [] 0 setdash 
  hpt 0 360 arc stroke } def
/Opaque { gsave closepath 1 setgray fill grestore 0 setgray closepath } def
/DiaW { stroke [] 0 setdash vpt add M
  hpt neg vpt neg V hpt vpt neg V
  hpt vpt V hpt neg vpt V Opaque stroke } def
/BoxW { stroke [] 0 setdash exch hpt sub exch vpt add M
  0 vpt2 neg V hpt2 0 V 0 vpt2 V
  hpt2 neg 0 V Opaque stroke } def
/TriUW { stroke [] 0 setdash vpt 1.12 mul add M
  hpt neg vpt -1.62 mul V
  hpt 2 mul 0 V
  hpt neg vpt 1.62 mul V Opaque stroke } def
/TriDW { stroke [] 0 setdash vpt 1.12 mul sub M
  hpt neg vpt 1.62 mul V
  hpt 2 mul 0 V
  hpt neg vpt -1.62 mul V Opaque stroke } def
/PentW { stroke [] 0 setdash gsave
  translate 0 hpt M 4 {72 rotate 0 hpt L} repeat
  Opaque stroke grestore } def
/CircW { stroke [] 0 setdash 
  hpt 0 360 arc Opaque stroke } def
/BoxFill { gsave Rec 1 setgray fill grestore } def
/Symbol-Oblique /Symbol findfont [1 0 .167 1 0 0] makefont
dup length dict begin {1 index /FID eq {pop pop} {def} ifelse} forall
currentdict end definefont
end
}}%
\begin{picture}(3600,2160)(0,0)%
{\GNUPLOTspecial{"
gnudict begin
gsave
0 0 translate
0.100 0.100 scale
0 setgray
newpath
1.000 UL
LTb
450 300 M
63 0 V
2937 0 R
-63 0 V
450 593 M
63 0 V
2937 0 R
-63 0 V
450 887 M
63 0 V
2937 0 R
-63 0 V
450 1180 M
63 0 V
2937 0 R
-63 0 V
450 1473 M
63 0 V
2937 0 R
-63 0 V
450 1767 M
63 0 V
2937 0 R
-63 0 V
450 2060 M
63 0 V
2937 0 R
-63 0 V
450 300 M
0 63 V
0 1697 R
0 -63 V
950 300 M
0 63 V
0 1697 R
0 -63 V
1450 300 M
0 63 V
0 1697 R
0 -63 V
1950 300 M
0 63 V
0 1697 R
0 -63 V
2450 300 M
0 63 V
0 1697 R
0 -63 V
2950 300 M
0 63 V
0 1697 R
0 -63 V
3450 300 M
0 63 V
0 1697 R
0 -63 V
1.000 UL
LTb
450 300 M
3000 0 V
0 1760 V
-3000 0 V
450 300 L
1.000 UL
LT0
3087 1947 M
263 0 V
-2900 3 R
125 -8 V
124 -25 V
125 -40 V
125 -56 V
125 -70 V
124 -84 V
125 -95 V
125 -106 V
125 -114 V
124 -121 V
125 -125 V
1947 979 L
2072 854 L
2196 732 L
2321 619 L
2446 517 L
125 -87 V
124 -68 V
125 -45 V
125 -17 V
5 0 V
500 0 V
1.000 UL
LT1
3087 1847 M
263 0 V
450 955 M
125 -1 V
124 -1 V
125 -2 V
125 -3 V
125 -4 V
124 -5 V
125 -6 V
125 -6 V
125 -8 V
124 -10 V
125 -10 V
125 -12 V
125 -14 V
124 -15 V
125 -17 V
125 -19 V
125 -22 V
124 -23 V
125 -25 V
125 -24 V
5 -428 V
500 0 V
1.000 UL
LT2
3087 1747 M
263 0 V
450 1433 M
125 -2 V
124 -4 V
125 -7 V
125 -10 V
125 -14 V
124 -16 V
125 -19 V
125 -23 V
125 -27 V
124 -30 V
125 -34 V
125 -38 V
125 -43 V
124 -47 V
125 -52 V
125 -57 V
125 -61 V
124 -64 V
125 -66 V
125 -58 V
5 -461 V
500 0 V
1.000 UL
LT3
3087 1647 M
263 0 V
450 1918 M
125 -4 V
124 -11 V
125 -19 V
125 -27 V
125 -34 V
124 -43 V
125 -50 V
125 -58 V
125 -66 V
124 -75 V
125 -82 V
125 -90 V
125 -97 V
124 -105 V
125 -111 V
2446 931 L
2571 814 L
2695 699 L
2820 596 L
125 -76 V
5 -220 V
500 0 V
1.000 UL
LT4
3087 1547 M
263 0 V
450 2051 M
125 -7 V
124 -21 V
125 -34 V
125 -49 V
125 -61 V
124 -75 V
125 -86 V
125 -98 V
125 -108 V
124 -117 V
125 -125 V
125 -131 V
125 -134 V
2196 869 L
2321 736 L
2446 610 L
2571 496 L
124 -96 V
125 -68 V
125 -29 V
5 -3 V
500 0 V
stroke
grestore
end
showpage
}}%
\put(3037,1547){\makebox(0,0)[r]{$\rho=5.0$}}%
\put(3037,1647){\makebox(0,0)[r]{$\rho=2.0$}}%
\put(3037,1747){\makebox(0,0)[r]{$\rho=1.0$}}%
\put(3037,1847){\makebox(0,0)[r]{$\rho=0.5$}}%
\put(3037,1947){\makebox(0,0)[r]{$w_A w_B$}}%
\put(1950,50){\makebox(0,0){$x$}}%
\put(100,1180){%
\special{ps: gsave currentpoint currentpoint translate
270 rotate neg exch neg exch translate}%
\makebox(0,0)[b]{\shortstack{$w_A w_B C(w_A, w_B)$}}%
\special{ps: currentpoint grestore moveto}%
}%
\put(3450,200){\makebox(0,0){ 1.2}}%
\put(2950,200){\makebox(0,0){ 1}}%
\put(2450,200){\makebox(0,0){ 0.8}}%
\put(1950,200){\makebox(0,0){ 0.6}}%
\put(1450,200){\makebox(0,0){ 0.4}}%
\put(950,200){\makebox(0,0){ 0.2}}%
\put(450,200){\makebox(0,0){ 0}}%
\put(400,2060){\makebox(0,0)[r]{ 0.6}}%
\put(400,1767){\makebox(0,0)[r]{ 0.5}}%
\put(400,1473){\makebox(0,0)[r]{ 0.4}}%
\put(400,1180){\makebox(0,0)[r]{ 0.3}}%
\put(400,887){\makebox(0,0)[r]{ 0.2}}%
\put(400,593){\makebox(0,0)[r]{ 0.1}}%
\put(400,300){\makebox(0,0)[r]{ 0}}%
\end{picture}%
\endgroup
 

%% file: fig10.tex
% GNUPLOT: LaTeX picture with Postscript
\begingroup%
  \makeatletter%
  \newcommand{\GNUPLOTspecial}{%
    \@sanitize\catcode`\%=14\relax\special}%
  \setlength{\unitlength}{0.1bp}%
{\GNUPLOTspecial{!
%!PS-Adobe-2.0
%%Title: fig10.tex
%%Creator: gnuplot 3.7 patchlevel 2
%%CreationDate: Sun Jun 16 23:24:44 2002
%%DocumentFonts: 
%%BoundingBox: 0 0 360 216
%%Orientation: Landscape
%%Pages: (atend)
%%EndComments
/gnudict 256 dict def
gnudict begin
/Color false def
/Solid false def
/gnulinewidth 5.000 def
/userlinewidth gnulinewidth def
/vshift -33 def
/dl {10 mul} def
/hpt_ 31.5 def
/vpt_ 31.5 def
/hpt hpt_ def
/vpt vpt_ def
/M {moveto} bind def
/L {lineto} bind def
/R {rmoveto} bind def
/V {rlineto} bind def
/vpt2 vpt 2 mul def
/hpt2 hpt 2 mul def
/Lshow { currentpoint stroke M
  0 vshift R show } def
/Rshow { currentpoint stroke M
  dup stringwidth pop neg vshift R show } def
/Cshow { currentpoint stroke M
  dup stringwidth pop -2 div vshift R show } def
/UP { dup vpt_ mul /vpt exch def hpt_ mul /hpt exch def
  /hpt2 hpt 2 mul def /vpt2 vpt 2 mul def } def
/DL { Color {setrgbcolor Solid {pop []} if 0 setdash }
 {pop pop pop Solid {pop []} if 0 setdash} ifelse } def
/BL { stroke userlinewidth 2 mul setlinewidth } def
/AL { stroke userlinewidth 2 div setlinewidth } def
/UL { dup gnulinewidth mul /userlinewidth exch def
      dup 1 lt {pop 1} if 10 mul /udl exch def } def
/PL { stroke userlinewidth setlinewidth } def
/LTb { BL [] 0 0 0 DL } def
/LTa { AL [1 udl mul 2 udl mul] 0 setdash 0 0 0 setrgbcolor } def
/LT0 { PL [] 1 0 0 DL } def
/LT1 { PL [4 dl 2 dl] 0 1 0 DL } def
/LT2 { PL [2 dl 3 dl] 0 0 1 DL } def
/LT3 { PL [1 dl 1.5 dl] 1 0 1 DL } def
/LT4 { PL [5 dl 2 dl 1 dl 2 dl] 0 1 1 DL } def
/LT5 { PL [4 dl 3 dl 1 dl 3 dl] 1 1 0 DL } def
/LT6 { PL [2 dl 2 dl 2 dl 4 dl] 0 0 0 DL } def
/LT7 { PL [2 dl 2 dl 2 dl 2 dl 2 dl 4 dl] 1 0.3 0 DL } def
/LT8 { PL [2 dl 2 dl 2 dl 2 dl 2 dl 2 dl 2 dl 4 dl] 0.5 0.5 0.5 DL } def
/Pnt { stroke [] 0 setdash
   gsave 1 setlinecap M 0 0 V stroke grestore } def
/Dia { stroke [] 0 setdash 2 copy vpt add M
  hpt neg vpt neg V hpt vpt neg V
  hpt vpt V hpt neg vpt V closepath stroke
  Pnt } def
/Pls { stroke [] 0 setdash vpt sub M 0 vpt2 V
  currentpoint stroke M
  hpt neg vpt neg R hpt2 0 V stroke
  } def
/Box { stroke [] 0 setdash 2 copy exch hpt sub exch vpt add M
  0 vpt2 neg V hpt2 0 V 0 vpt2 V
  hpt2 neg 0 V closepath stroke
  Pnt } def
/Crs { stroke [] 0 setdash exch hpt sub exch vpt add M
  hpt2 vpt2 neg V currentpoint stroke M
  hpt2 neg 0 R hpt2 vpt2 V stroke } def
/TriU { stroke [] 0 setdash 2 copy vpt 1.12 mul add M
  hpt neg vpt -1.62 mul V
  hpt 2 mul 0 V
  hpt neg vpt 1.62 mul V closepath stroke
  Pnt  } def
/Star { 2 copy Pls Crs } def
/BoxF { stroke [] 0 setdash exch hpt sub exch vpt add M
  0 vpt2 neg V  hpt2 0 V  0 vpt2 V
  hpt2 neg 0 V  closepath fill } def
/TriUF { stroke [] 0 setdash vpt 1.12 mul add M
  hpt neg vpt -1.62 mul V
  hpt 2 mul 0 V
  hpt neg vpt 1.62 mul V closepath fill } def
/TriD { stroke [] 0 setdash 2 copy vpt 1.12 mul sub M
  hpt neg vpt 1.62 mul V
  hpt 2 mul 0 V
  hpt neg vpt -1.62 mul V closepath stroke
  Pnt  } def
/TriDF { stroke [] 0 setdash vpt 1.12 mul sub M
  hpt neg vpt 1.62 mul V
  hpt 2 mul 0 V
  hpt neg vpt -1.62 mul V closepath fill} def
/DiaF { stroke [] 0 setdash vpt add M
  hpt neg vpt neg V hpt vpt neg V
  hpt vpt V hpt neg vpt V closepath fill } def
/Pent { stroke [] 0 setdash 2 copy gsave
  translate 0 hpt M 4 {72 rotate 0 hpt L} repeat
  closepath stroke grestore Pnt } def
/PentF { stroke [] 0 setdash gsave
  translate 0 hpt M 4 {72 rotate 0 hpt L} repeat
  closepath fill grestore } def
/Circle { stroke [] 0 setdash 2 copy
  hpt 0 360 arc stroke Pnt } def
/CircleF { stroke [] 0 setdash hpt 0 360 arc fill } def
/C0 { BL [] 0 setdash 2 copy moveto vpt 90 450  arc } bind def
/C1 { BL [] 0 setdash 2 copy        moveto
       2 copy  vpt 0 90 arc closepath fill
               vpt 0 360 arc closepath } bind def
/C2 { BL [] 0 setdash 2 copy moveto
       2 copy  vpt 90 180 arc closepath fill
               vpt 0 360 arc closepath } bind def
/C3 { BL [] 0 setdash 2 copy moveto
       2 copy  vpt 0 180 arc closepath fill
               vpt 0 360 arc closepath } bind def
/C4 { BL [] 0 setdash 2 copy moveto
       2 copy  vpt 180 270 arc closepath fill
               vpt 0 360 arc closepath } bind def
/C5 { BL [] 0 setdash 2 copy moveto
       2 copy  vpt 0 90 arc
       2 copy moveto
       2 copy  vpt 180 270 arc closepath fill
               vpt 0 360 arc } bind def
/C6 { BL [] 0 setdash 2 copy moveto
      2 copy  vpt 90 270 arc closepath fill
              vpt 0 360 arc closepath } bind def
/C7 { BL [] 0 setdash 2 copy moveto
      2 copy  vpt 0 270 arc closepath fill
              vpt 0 360 arc closepath } bind def
/C8 { BL [] 0 setdash 2 copy moveto
      2 copy vpt 270 360 arc closepath fill
              vpt 0 360 arc closepath } bind def
/C9 { BL [] 0 setdash 2 copy moveto
      2 copy  vpt 270 450 arc closepath fill
              vpt 0 360 arc closepath } bind def
/C10 { BL [] 0 setdash 2 copy 2 copy moveto vpt 270 360 arc closepath fill
       2 copy moveto
       2 copy vpt 90 180 arc closepath fill
               vpt 0 360 arc closepath } bind def
/C11 { BL [] 0 setdash 2 copy moveto
       2 copy  vpt 0 180 arc closepath fill
       2 copy moveto
       2 copy  vpt 270 360 arc closepath fill
               vpt 0 360 arc closepath } bind def
/C12 { BL [] 0 setdash 2 copy moveto
       2 copy  vpt 180 360 arc closepath fill
               vpt 0 360 arc closepath } bind def
/C13 { BL [] 0 setdash  2 copy moveto
       2 copy  vpt 0 90 arc closepath fill
       2 copy moveto
       2 copy  vpt 180 360 arc closepath fill
               vpt 0 360 arc closepath } bind def
/C14 { BL [] 0 setdash 2 copy moveto
       2 copy  vpt 90 360 arc closepath fill
               vpt 0 360 arc } bind def
/C15 { BL [] 0 setdash 2 copy vpt 0 360 arc closepath fill
               vpt 0 360 arc closepath } bind def
/Rec   { newpath 4 2 roll moveto 1 index 0 rlineto 0 exch rlineto
       neg 0 rlineto closepath } bind def
/Square { dup Rec } bind def
/Bsquare { vpt sub exch vpt sub exch vpt2 Square } bind def
/S0 { BL [] 0 setdash 2 copy moveto 0 vpt rlineto BL Bsquare } bind def
/S1 { BL [] 0 setdash 2 copy vpt Square fill Bsquare } bind def
/S2 { BL [] 0 setdash 2 copy exch vpt sub exch vpt Square fill Bsquare } bind def
/S3 { BL [] 0 setdash 2 copy exch vpt sub exch vpt2 vpt Rec fill Bsquare } bind def
/S4 { BL [] 0 setdash 2 copy exch vpt sub exch vpt sub vpt Square fill Bsquare } bind def
/S5 { BL [] 0 setdash 2 copy 2 copy vpt Square fill
       exch vpt sub exch vpt sub vpt Square fill Bsquare } bind def
/S6 { BL [] 0 setdash 2 copy exch vpt sub exch vpt sub vpt vpt2 Rec fill Bsquare } bind def
/S7 { BL [] 0 setdash 2 copy exch vpt sub exch vpt sub vpt vpt2 Rec fill
       2 copy vpt Square fill
       Bsquare } bind def
/S8 { BL [] 0 setdash 2 copy vpt sub vpt Square fill Bsquare } bind def
/S9 { BL [] 0 setdash 2 copy vpt sub vpt vpt2 Rec fill Bsquare } bind def
/S10 { BL [] 0 setdash 2 copy vpt sub vpt Square fill 2 copy exch vpt sub exch vpt Square fill
       Bsquare } bind def
/S11 { BL [] 0 setdash 2 copy vpt sub vpt Square fill 2 copy exch vpt sub exch vpt2 vpt Rec fill
       Bsquare } bind def
/S12 { BL [] 0 setdash 2 copy exch vpt sub exch vpt sub vpt2 vpt Rec fill Bsquare } bind def
/S13 { BL [] 0 setdash 2 copy exch vpt sub exch vpt sub vpt2 vpt Rec fill
       2 copy vpt Square fill Bsquare } bind def
/S14 { BL [] 0 setdash 2 copy exch vpt sub exch vpt sub vpt2 vpt Rec fill
       2 copy exch vpt sub exch vpt Square fill Bsquare } bind def
/S15 { BL [] 0 setdash 2 copy Bsquare fill Bsquare } bind def
/D0 { gsave translate 45 rotate 0 0 S0 stroke grestore } bind def
/D1 { gsave translate 45 rotate 0 0 S1 stroke grestore } bind def
/D2 { gsave translate 45 rotate 0 0 S2 stroke grestore } bind def
/D3 { gsave translate 45 rotate 0 0 S3 stroke grestore } bind def
/D4 { gsave translate 45 rotate 0 0 S4 stroke grestore } bind def
/D5 { gsave translate 45 rotate 0 0 S5 stroke grestore } bind def
/D6 { gsave translate 45 rotate 0 0 S6 stroke grestore } bind def
/D7 { gsave translate 45 rotate 0 0 S7 stroke grestore } bind def
/D8 { gsave translate 45 rotate 0 0 S8 stroke grestore } bind def
/D9 { gsave translate 45 rotate 0 0 S9 stroke grestore } bind def
/D10 { gsave translate 45 rotate 0 0 S10 stroke grestore } bind def
/D11 { gsave translate 45 rotate 0 0 S11 stroke grestore } bind def
/D12 { gsave translate 45 rotate 0 0 S12 stroke grestore } bind def
/D13 { gsave translate 45 rotate 0 0 S13 stroke grestore } bind def
/D14 { gsave translate 45 rotate 0 0 S14 stroke grestore } bind def
/D15 { gsave translate 45 rotate 0 0 S15 stroke grestore } bind def
/DiaE { stroke [] 0 setdash vpt add M
  hpt neg vpt neg V hpt vpt neg V
  hpt vpt V hpt neg vpt V closepath stroke } def
/BoxE { stroke [] 0 setdash exch hpt sub exch vpt add M
  0 vpt2 neg V hpt2 0 V 0 vpt2 V
  hpt2 neg 0 V closepath stroke } def
/TriUE { stroke [] 0 setdash vpt 1.12 mul add M
  hpt neg vpt -1.62 mul V
  hpt 2 mul 0 V
  hpt neg vpt 1.62 mul V closepath stroke } def
/TriDE { stroke [] 0 setdash vpt 1.12 mul sub M
  hpt neg vpt 1.62 mul V
  hpt 2 mul 0 V
  hpt neg vpt -1.62 mul V closepath stroke } def
/PentE { stroke [] 0 setdash gsave
  translate 0 hpt M 4 {72 rotate 0 hpt L} repeat
  closepath stroke grestore } def
/CircE { stroke [] 0 setdash 
  hpt 0 360 arc stroke } def
/Opaque { gsave closepath 1 setgray fill grestore 0 setgray closepath } def
/DiaW { stroke [] 0 setdash vpt add M
  hpt neg vpt neg V hpt vpt neg V
  hpt vpt V hpt neg vpt V Opaque stroke } def
/BoxW { stroke [] 0 setdash exch hpt sub exch vpt add M
  0 vpt2 neg V hpt2 0 V 0 vpt2 V
  hpt2 neg 0 V Opaque stroke } def
/TriUW { stroke [] 0 setdash vpt 1.12 mul add M
  hpt neg vpt -1.62 mul V
  hpt 2 mul 0 V
  hpt neg vpt 1.62 mul V Opaque stroke } def
/TriDW { stroke [] 0 setdash vpt 1.12 mul sub M
  hpt neg vpt 1.62 mul V
  hpt 2 mul 0 V
  hpt neg vpt -1.62 mul V Opaque stroke } def
/PentW { stroke [] 0 setdash gsave
  translate 0 hpt M 4 {72 rotate 0 hpt L} repeat
  Opaque stroke grestore } def
/CircW { stroke [] 0 setdash 
  hpt 0 360 arc Opaque stroke } def
/BoxFill { gsave Rec 1 setgray fill grestore } def
/Symbol-Oblique /Symbol findfont [1 0 .167 1 0 0] makefont
dup length dict begin {1 index /FID eq {pop pop} {def} ifelse} forall
currentdict end definefont
end
}}%
\begin{picture}(3600,2160)(0,0)%
{\GNUPLOTspecial{"
gnudict begin
gsave
0 0 translate
0.100 0.100 scale
0 setgray
newpath
1.000 UL
LTb
450 300 M
63 0 V
2937 0 R
-63 0 V
450 551 M
63 0 V
2937 0 R
-63 0 V
450 803 M
63 0 V
2937 0 R
-63 0 V
450 1054 M
63 0 V
2937 0 R
-63 0 V
450 1306 M
63 0 V
2937 0 R
-63 0 V
450 1557 M
63 0 V
2937 0 R
-63 0 V
450 1809 M
63 0 V
2937 0 R
-63 0 V
450 2060 M
63 0 V
2937 0 R
-63 0 V
450 300 M
0 63 V
0 1697 R
0 -63 V
752 300 M
0 63 V
0 1697 R
0 -63 V
1053 300 M
0 63 V
0 1697 R
0 -63 V
1355 300 M
0 63 V
0 1697 R
0 -63 V
1656 300 M
0 63 V
0 1697 R
0 -63 V
1958 300 M
0 63 V
0 1697 R
0 -63 V
2259 300 M
0 63 V
0 1697 R
0 -63 V
2561 300 M
0 63 V
0 1697 R
0 -63 V
2862 300 M
0 63 V
0 1697 R
0 -63 V
3164 300 M
0 63 V
0 1697 R
0 -63 V
1.000 UL
LTb
450 300 M
3000 0 V
0 1760 V
-3000 0 V
450 300 L
1.000 UL
LT0
3087 1947 M
263 0 V
450 1809 M
60 -3 V
60 -9 V
60 -14 V
60 -18 V
60 -24 V
60 -27 V
60 -31 V
60 -35 V
60 -38 V
60 -41 V
60 -44 V
60 -45 V
60 -47 V
60 -49 V
60 -50 V
60 -52 V
60 -51 V
60 -52 V
60 -53 V
60 -52 V
60 -51 V
60 -51 V
60 -51 V
60 -49 V
60 -48 V
60 -46 V
60 -45 V
60 -42 V
60 -41 V
60 -39 V
60 -37 V
60 -34 V
60 -32 V
60 -30 V
60 -27 V
60 -25 V
60 -22 V
60 -20 V
60 -17 V
60 -15 V
60 -13 V
60 -11 V
60 -9 V
60 -6 V
60 -5 V
60 -4 V
60 -2 V
60 -1 V
60 -1 V
60 0 V
1.000 UL
LT1
3087 1847 M
263 0 V
450 2006 M
60 -5 V
60 -13 V
60 -19 V
60 -27 V
60 -31 V
60 -37 V
60 -40 V
60 -44 V
60 -47 V
60 -49 V
60 -51 V
60 -53 V
60 -53 V
60 -55 V
60 -55 V
60 -56 V
60 -55 V
60 -56 V
60 -55 V
60 -54 V
60 -53 V
60 -53 V
60 -52 V
60 -50 V
60 -49 V
60 -47 V
60 -46 V
60 -45 V
60 -42 V
60 -41 V
60 -39 V
60 -37 V
60 -35 V
60 -33 V
60 -31 V
60 -29 V
60 -26 V
60 -24 V
60 -23 V
60 -19 V
60 -17 V
60 -15 V
60 -13 V
60 -10 V
60 -8 V
60 -6 V
60 -4 V
60 -3 V
60 -1 V
60 0 V
1.000 UL
LT2
3087 1747 M
263 0 V
450 903 M
60 -1 V
60 -3 V
60 -6 V
60 -7 V
60 -9 V
60 -11 V
60 -13 V
60 -14 V
60 -15 V
60 -16 V
60 -18 V
60 -18 V
60 -19 V
60 -20 V
60 -20 V
60 -20 V
60 -21 V
60 -21 V
60 -20 V
60 -21 V
60 -21 V
60 -20 V
60 -20 V
60 -20 V
60 -19 V
60 -19 V
60 -18 V
60 -17 V
60 -16 V
60 -16 V
60 -14 V
60 -14 V
60 -13 V
60 -12 V
60 -11 V
60 -10 V
60 -8 V
60 -8 V
60 -7 V
60 -6 V
60 -6 V
60 -4 V
60 -3 V
60 -3 V
60 -2 V
60 -1 V
60 -1 V
60 -1 V
60 0 V
60 0 V
1.000 UL
LT3
3087 1647 M
263 0 V
450 966 M
60 -2 V
60 -4 V
60 -7 V
60 -9 V
60 -11 V
60 -14 V
60 -15 V
60 -16 V
60 -18 V
60 -19 V
60 -19 V
60 -21 V
60 -21 V
60 -22 V
60 -23 V
60 -22 V
60 -23 V
60 -22 V
60 -23 V
60 -22 V
60 -22 V
60 -22 V
60 -21 V
60 -21 V
60 -20 V
60 -20 V
60 -19 V
60 -18 V
60 -17 V
60 -16 V
60 -16 V
60 -15 V
60 -13 V
60 -13 V
60 -12 V
60 -11 V
60 -9 V
60 -9 V
60 -8 V
60 -7 V
60 -6 V
60 -4 V
60 -5 V
60 -3 V
60 -2 V
60 -2 V
60 -1 V
60 -1 V
60 0 V
60 0 V
stroke
grestore
end
showpage
}}%
\put(3037,1647){\makebox(0,0)[r]{$T_\sigma/\sigma^2$, $\rho = 5$}}%
\put(3037,1747){\makebox(0,0)[r]{$S_{11}/\rho$, $\rho = 5$}}%
\put(3037,1847){\makebox(0,0)[r]{$T_\sigma/\sigma^2$, $\rho = 2$}}%
\put(3037,1947){\makebox(0,0)[r]{$S_{11}/\rho$, $\rho = 2$}}%
\put(1950,50){\makebox(0,0){$\delta$}}%
\put(100,1180){%
\special{ps: gsave currentpoint currentpoint translate
270 rotate neg exch neg exch translate}%
\makebox(0,0)[b]{\shortstack{$T_\sigma/\sigma^2$}}%
\special{ps: currentpoint grestore moveto}%
}%
\put(3164,200){\makebox(0,0){$1.8$}}%
\put(2862,200){\makebox(0,0){$1.6$}}%
\put(2561,200){\makebox(0,0){$1.4$}}%
\put(2259,200){\makebox(0,0){$1.2$}}%
\put(1958,200){\makebox(0,0){$1.0$}}%
\put(1656,200){\makebox(0,0){$0.8$}}%
\put(1355,200){\makebox(0,0){$0.6$}}%
\put(1053,200){\makebox(0,0){$0.4$}}%
\put(752,200){\makebox(0,0){$0.2$}}%
\put(450,200){\makebox(0,0){$0.0$}}%
\put(400,2060){\makebox(0,0)[r]{$0.35$}}%
\put(400,1809){\makebox(0,0)[r]{$0.30$}}%
\put(400,1557){\makebox(0,0)[r]{$0.25$}}%
\put(400,1306){\makebox(0,0)[r]{$0.20$}}%
\put(400,1054){\makebox(0,0)[r]{$0.15$}}%
\put(400,803){\makebox(0,0)[r]{$0.10$}}%
\put(400,551){\makebox(0,0)[r]{$0.05$}}%
\put(400,300){\makebox(0,0)[r]{$0.00$}}%
\end{picture}%
\endgroup
 